\documentclass[twocolumn,amsmath,amssymb]{revtex4}

\usepackage{graphicx}
\usepackage{graphics}
\usepackage{subfigure}
\usepackage{dcolumn}
\usepackage{bm}
\usepackage{txfonts}

\usepackage{float}
\usepackage{tabu}
\usepackage{xcolor}
\usepackage{color}
\usepackage{makecell}

\begin{document}

\title{Outlearning Extortioners by Fair-minded Unbending Strategies}
\author{Xingru Chen$^{1,2}$}
\email{xingrucz@gmail.com}
\author{Feng Fu$^{2,3}$}
\email{fufeng@gmail.com}

\affiliation{ 
$^1$School of Sciences, Beijing University of Posts and Telecommunications, Beijing 100876, China\\
$^2$Department of Mathematics, Dartmouth College, Hanover, NH 03755, USA\\
$^3$Department of Biomedical Data Science, Geisel School of Medicine at Dartmouth, Lebanon, NH 03756, USA}

\date{\today}

\begin{abstract}
Recent theory shows that extortioners taking advantage of the zero-determinant (ZD) strategy can unilaterally claim an unfair share of the payoffs in the Iterated Prisoner's Dilemma. It is thus suggested that against a fixed extortioner, any adapting co-player should be subdued with full cooperation as their best response. In contrast, recent experiments demonstrate that human players often choose not to accede to extortion out of concern for fairness, actually causing extortioners to suffer more loss than themselves. In light of this, here we reveal fair-minded strategies that are \emph{unbending} to extortion such that any payoff-maximizing extortioner ultimately will concede in their own interest by offering a fair split in head-to-head matches. We find and characterize multiple general classes of such unbending strategies, including generous zero-determinant strategies and Win-Stay, Lose-Shift as particular examples. When against fixed unbending players, extortioners are forced with consequentially increasing losses whenever intending to demand more unfair share. Our analysis also pivots to the importance of payoff structure in determining the superiority of zero-determinant strategies and in particular their extortion ability. We show that an extortionate ZD player can be even outperformed by, for example, Win-Stay Lose-Shift, if the total payoff of unilateral cooperation is smaller than that of mutual defection. Unbending strategies can be used to outlearn evolutionary extortioners and catalyze the evolution of Tit-for-Tat-like strategies out of ZD players. Our work has implications for promoting fairness and resisting extortion so as to uphold a just and cooperative society.
\end{abstract}

%\pacs{89.75.Hc, 87.23.Kg, 02.50.Le}

 \maketitle

\section*{Introduction}

The Prisoner's Dilemma has been considered a central paradigm for understanding a wide variety of cooperation problems~\cite{axelrod1981evolution}. In this game, two players decide whether to cooperate (C) or defect (D). If both players choose to cooperate, they receive the same reward for mutual cooperation, $R$, and if they both defect, they receive the same punishment for mutual defection, $P$. However, if one cooperates but the other defects, the defector receives the temptation to defect, $T$, whereas the cooperator receives the sucker's payoff, $S$. These payoff values satisfy $T > R > P > S$~\cite{rapoport1965prisoner}. To shed insights into direct reciprocity (`I will if you will'), the Iterated Prisoner's Dilemma (IPD) games allow us to explore a wide range of simple and complex strategies that can be either cooperative or exploitive~\cite{hilbe2018partners}. 

Among common IPD strategies, Tit-for-Tat (TFT) and variants like generous TFT (GTFT) are cooperative and fair-minded in nature~\cite{nowak1992tit}: TFT-like players never defect in the first place unless their co-players had defected once (or more). Simple learning strategies such as Win-Stay, Lose-Shift (WSLS) are adaptive~\cite{macy2002learning}, and thus more robust to noise and error than TFT~\cite{nowak1993strategy}. WSLS deterministically keeps the current strategy if the resulting payoff is above a fixed aspiration level, or switches otherwise. An `equalizer' can unilaterally set any co-player's payoff level to the same arbitrary level within the range of $[P, R]$ -- hence the name~\cite{boerlijst1997equal}. Even more capable of a bilateral payoff control is the so-called zero-determinant (ZD) strategy, discovered by Press and Dyson~\cite{press2012iterated}. A ZD player X can unilaterally set a linear relation between X's own payoff, $s_X$, and that of the co-player Y, $s_Y$, regardless of Y's specific strategy. In recent years, the discovery of ZD strategies has aroused a new wave of interests in studying IPD games in light of Press and Dyson's finding~\cite{hilbe2013evolution,szolnoki2014defection,akin2015you,ichinose2018zero,hao2015extortion,chen2014robustness,hilbe2014extortion,mcavoy2016autocratic,becks2019extortion,harper2017reinforcement}.

Of particular interest is the existence of a continuous spectrum of ZD strategies that vary in their level of generosity, ranging from extortionate ZD to generous ZD~\cite{stewart2013extortion}. Undoubtedly, witting of ZD strategies enables players to gain the upper hand in IPD games, even allowing an implicit form of extortion~\cite{press2012iterated}. Self-serving extortioners can leverage ZD strategies to their advantage to the fullest extent, aiming to dominate any co-player preemptively. However, it is shown that two extortioners, both equipped with the knowledge of extortionate ZD, will neutralize each other in their interactions and lead to their own demise, both receiving $P$~\cite{press2012iterated}. Thus, without recognition mechanisms like green beards~\cite{adami2013evolutionary}, miserable outcomes of ZD's mutual destruction prevent them from being favored by natural selection unless they adapt to be more generous~\cite{stewart2013extortion}. 

Prior work almost invariably considers ZD fixed while their co-player tries to adapt to ZD's unilateral control. In reality, extortion can be met with resistance; \emph{unbending} individuals are willing to push back any attempt to extort out of concern for fairness. Indeed, recent experiments demonstrate that fixed computer ZD players are able to outcompete their human counterparts but at a huge cost in a way that human players are significantly less cooperative~\cite{hilbe2014extortion}. Thus the success of ZD's extortion attempt can be undermined and become less effective in reaching the fullest possible extent. Moreover, ZD players need to prescribe their strategies in a sophisticated way that explicitly depends on the underlying payoff matrix in the first place. It remains unknown how potential variations in the payoff matrix, which can arise from the uncertainty of evolving game environment~\cite{weitz2016oscillating}, will impact the pairwise dominance of ZD strategies and in particular their extortion ability, since not all Prisoner's Dilemma games are qualitatively the same~\cite{nowak1990evolution}.

These considerations lead us to reveal the previously unforeseen Achilles' heel of ZD strategies, specifically in one-on-one encounters. Namely, there exist simple strategies (including TFT-like strategies and WSLS as particular examples) that are unbending to extortion and can cause unfair demand to backfire on extortioners. When against a fixed unbending player, the best response of any payoff-maximizing extortioner, characterized by their prescribed smallest possible level of generosity, is to offer a fair split, thereby guaranteeing equal payoffs for both parties.

Moreover, we show that in interactions of more adversarial nature~\cite{d2015statistical}, characterized by the payoff structure condition $T+S < 2P$, ZD's dominance is drastically impaired, and extortioners tempting to dominate the co-player are likely to become victims of their own success. The fixed unbending strategies, discovered in the present study, are able to not only force greedy ZD co-players to be fair in their own interest, but also more importantly, steer adapting co-players (including those ZD co-players) towards fairness and cooperation in adaptive learning settings. Our work provides useful insights into understanding the important role played by unbending strategies as an enforcer and stabilizer of fairness and cooperation in dyadic interactions, of relevance and interest for studying direct reciprocity.

\section*{Results}

We begin with studying the effectiveness of ZD strategies in payoff control and extortion and how it depends on their prescribed parameter choices and the underlying payoff matrix. Doing so will provide a new perspective on understanding specific conditions required for intended extortion to be successful or lack thereof. These critical considerations ultimately lead us to reveal unbending strategies that are able to outlearn ZD players and foster fairness and cooperation in pairwise interactions.

%%%%%%%%%%%%%%%%%NEW FIG 1%%%%%%%%%%%%%%%%%

\begin{figure*}[htbp]
\centering
  \includegraphics[width=0.8\textwidth]{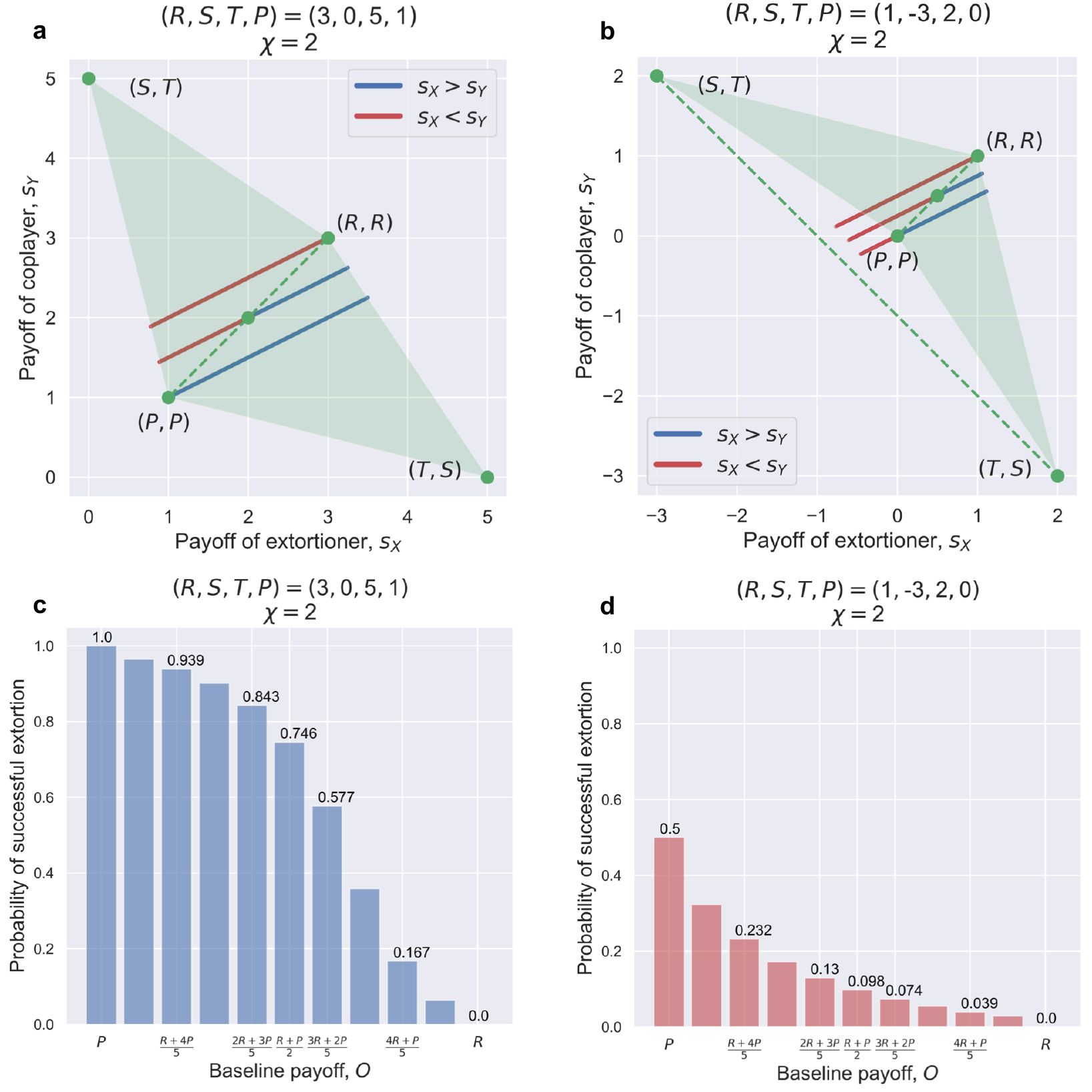}
        \caption{\textbf{Pairwise dominance and extortion ability of zero-determinant (ZD) strategies.}  The baseline payoff $O$ used by the zero-determinant (ZD) player X is regarded extortionate $P$ (least level of generosity), generous $R$ (maximum level of generosity), and in between $(P+R)/2$ (intermediate level of generosity). The optimality of extortionate ZD strategies (with $O = P$) nontrivially depends on both their co-player's strategy and the payoff structure. When playing against certain type of co-players (which we call unbending strategies), extortioners can maximize their prospective payoffs only if aiming for an equal split by letting the extortion factor $\chi \to 1$. Moreover, when $T+S < 2P$, extortioners can even be outperformed. In panels (\textbf{a}) and (\textbf{b}): we show the scatter plot of payoff pairs $(s_X, s_Y)$ of ZD players against random co-players uniformly drawn from all possible memory-one strategies $[0, 1]^4$ in (\textbf{a}) $T+S > 2P$ and in (\textbf{b}) $T+S < 2P$.  Shown in (\textbf{c}) and (\textbf{d}) is the probability that a ZD player X actually gets better payoff than their co-player Y ($s_X > s_Y$) who uses a random strategy uniformly drawn from memory-one strategies $[0,1]^4$, with respect to varying their baseline payoff $O\in [P, R]$. The parameter $O$ controls the level of generosity of a ZD player but also impacts their chance to outperform their co-players (`extortion ability'). Increasing $O$ above $P$ makes ZD less likely to be able to ensure the dominance over their co-players. Noticeably, the payoff structure plays an even more pronounced role than does the parameter $O$: (\textbf{a}) for $T+S > 2P$ the curvature is concave downward and ZD is able to maintain dominance for most of the time even using intermediate $O > P$ values, whereas concave upward for $T+S > 2P$ and ZD is more likely to lose dominance for any $P< O \le R$. In line with (\textbf{a}) and (\textbf{b}), extortion with $O = P$ always leads to superior payoff than co-players (a) when $T+S > 2P$, but not necessarily true for (\textbf{b}) $T+S < 2P$. Parameters: (\textbf{a})-(\textbf{d}) ZD player X's extortion factor $\chi = 2$, and $\phi$ is uniformly distributed and truncated at the admissible upper bound, (\textbf{a}) (\textbf{c}) $R = 3, S = 0, T = 5, P =1$, (\textbf{b}) (\textbf{d}) $R= 1, S = -3, T =2, P = 0$.  
         }
        \label{fig1}
\end{figure*}

%%%%%%%%%%%%%%%%%NEW FIG 1%%%%%%%%%%%%%%%%%

Following common practice~\cite{nowak1993strategy}, we denote memory-one IPD strategies by $\mathbf{p} = [p_1, p_2, p_3, p_4]$, where $p_i$, for $i = 1, \cdots, 4$, is the conditional probability to cooperate, respectively, after experiencing one of the four possible outcomes each round $\{CC, CD, DC, DD\}$, that is, written from the perspective of a focal player X (the first letter represents X's last move, and the second letter for the co-player Y's). Suppose that player X uses a ZD strategy $\mathbf{p}$ and the co-player Y uses an arbitrary strategy $\mathbf{q} = [q_1, q_2, q_3, q_4]$, and let $s_X$ denote the average payoff of player X and $s_Y$ that of player Y.  A general yet intuitive parameterization of memory-one ZD strategies are based on three control parameters $(O, \chi, \phi)$:

\begin{equation}
\left\{
\begin{array}{cl}
 p_1 = &   1 - (R - O)\phi(\chi - 1),\   \\
 p_2 = &   1 - \phi[(T - O)\chi + (O - S)],   \\
 p_3 = &  \phi[(O - S)\chi + (T - O)],    \\
 p_4 = &   (O - P)\phi(\chi - 1).   \\
\end{array}
\right.
\label{ZDs}
\end{equation}
including the extortion factor $\chi > 1$, the baseline payoff $O \in [P, R]$, and the normalization factor $\phi$ that ensures $\mathbf{p}$ to be a proper probability vector. A complete discussion of admissible ranges of these parameters can be found in the SI.

Regardless of Y's strategy $\mathbf{q}$, X unilaterally enforces a linear relative payoff relation of the form~\cite{press2012iterated,stewart2013extortion}:
\begin{equation}
s_X - O = \chi (s_Y - O),
\label{ZDlinear}
\end{equation}
which represents a straight line in the parametric plot of $(s_X, s_Y)$ with the slope $1/\chi$ (the reciprocal of the extortion factor $\chi$) (Figs. 1a and 1b). In this plane, the baseline payoff $O \in [P, R]$ determines the intercept to the line of equal payoffs, $s_X = s_Y$, and also dictates the level of generosity~\cite{stewart2013extortion}.

The payoff control as given in Eq.~\eqref{ZDlinear} enables an implicit form of extortion where ZD player X can prescribe their strategies in a way that they reciprocate cooperation less frequently than the co-player Y~\cite{press2012iterated,hilbe2013evolution}.  For ZD players, the way to attempt such dominance and extortion is to deliberately choose their parameters $O$ and $\chi$ in advance, which will in turn determine admissible values of $\phi$. The chosen values of $O$ and $\chi$ can be observed directly from pairwise payoff plots (Figs. 1a and 1b), and together with the underlying payoff matrix, they jointly determine the upper bound of admissible $\phi$ values. For example, a widely used parameterization of this ZD class, which is called extortionate ZD strategy~\cite{press2012iterated}, ensures that $s_X - P = \chi (s_Y - P)$ holds with $\chi>1$. The admissible range of $\phi$ for  extortionate ZD is given by

\begin{equation}
0 < \phi \leq \phi^{\text{upper}} = 
\begin{cases}\frac{1}{(T - P)\chi + (P - S)}, & T + S \geq 2P \\ \frac{1}{(P - S)\chi + (T - P)}. & T + S < 2P
\end{cases}
\label{phi_bound}
\end{equation}

Notably, the parameter $\phi$ has an upper bound that explicitly depends on the sign of $T + S - 2P$. We emphasize that this previously overlooked payoff structure condition, whether $T+S > 2P$ holds or not, surprisingly strikes out as an important condition for determining the optimality of ZD strategies and their extortion ability. As shown in Figs. 1c and 1d, as long as a ZD player uses the minimal $O = P$ and $\chi > 1$, they secure the most favorable position to dominate and get higher payoffs than their opponent as compared to other $O$ values, regardless of the underlying payoff matrix (Figs. 1c and 1d). Despite such contextual difference of ZD's extortion ability owing to the change in the underlying payoff structure, we still call this class of `extortionate ZD' as extortioner as in Ref.~\cite{hilbe2013evolution}, for the sake of consistency. First, the effect of varying their control parameter $O$ on their resulting level of generosity and extortion ability remains qualitatively consistent across IPD games of drastically different nature. Second, for ZD players with $O = P$, no matter what types of IPD games they are engaged in, the chosen value of $P$ characterizes the least level of generosity, and thus preemptively sets their extortion ability at maximum, even though these so-called extortioners will not always succeed in securing advantage as intended, particularly when $T+S < 2P$ (cf. Figs. 1c and 1d).

%%%%%%%%%%%%%%%%%NEW FIG 2%%%%%%%%%%%%%%%%%

\begin{figure*}[t]
\centering
  \includegraphics[width=\textwidth]{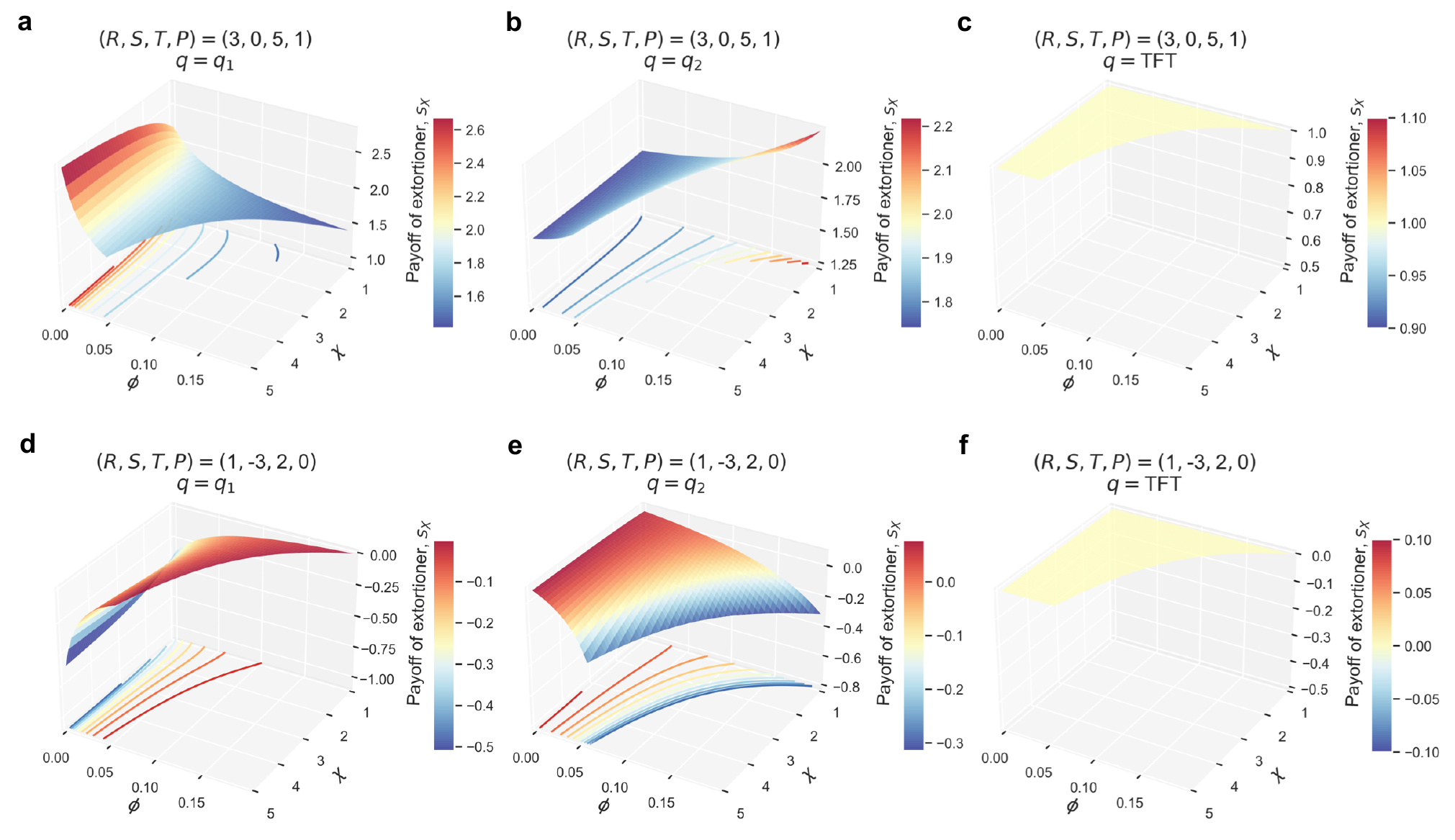}
        \caption{\textbf{Impacts of control parameters $(\phi, \chi)$ on the average payoff of a zero-determinant (ZD) player when playing against a fixed co-player.}  The ZD player X's payoff, $s_X$, is shown as a function of the normalization factor $\phi$ and the extortion factor $\chi$, along with contour lines projected on the $(\phi, \chi)$-plane: (\textbf{a})-(\textbf{c}) for $T + S > 2P$ and (\textbf{d})-(\textbf{f}) for $T + S < 2P$. The ZD player X's payoff is either monotonic or remains constant with respect to $\phi$ while X's payoff can exhibit non-monotonic behavior with respect to $\chi$. Despite being able to enforce a linear payoff relationship $s_X - P = \chi (s_Y-P)$, ZD player X that unilaterally uses a larger extortion factor $\chi$ does not necessarily lead to further payoff gains as demonstrated in (\textbf{a}). As long as `winning isn't everything' (actual payoff performance is concerned in wide-ranging scenarios), ZD can subtly tune their control parameters to optimize their own payoff performance against a fixed co-player. Parameters: (\textbf{a})-(\textbf{f}): X uses the most formidable ZD strategy with $O=P$ (also known as extortionate ZD), (\textbf{a})-(\textbf{c}) $R = 3, S = 0, T = 5, P =1$, the upper bound of $\phi = 1/[(T - P)\chi + (P - S)]$, (d)-(f) $R= 1, S = -3, T =2, P = 0$, the upper bound of $\phi = 1/[(P - S)\chi + (T - P)]$, 
co-player Y's strategy $q_1 = [0.05, 0.95, 0.05, 0.1] $, $q_2 = [0.4, 0.1, 0.9, 0.2]$, $q_3 = [1, 0, 1, 0]$. 
         }  
        \label{fig2}
\end{figure*}
%%%%%%%%%%%%%%%%%NEW FIG 2%%%%%%%%%%%%%%%%%

On the other hand, given the uncertainty of vastly possible strategies the co-player could use against ZD players, it is worthwhile to quantify the robustness of the dominance and performance of ZD strategies with particular respect to varying their baseline payoff $O$. It is likely that ZD players choose $O = P + \varepsilon$ deviating from $P$ for plausible reasons like the trembling hand ~\cite{bielefeld1988reexamination} or `blurred minds'~\cite{nowak1992tit}, and as a consequence they will respond with non-zero cooperation (i.e., $p_4 > 0$) after entering mutual defection state with their co-player. Even so, any ZD player using $O < R$ still has extortion ability to some extent unless they use the generous ZD with $O = R$ that ensures their average payoffs never above the co-player's~\cite{stewart2013extortion} (Fig. 1). It is thus reasonable to consider the extortion ability of ZD strategies as a continuous spectrum -- `the likelihood of getting better payoffs than any kind of opponent' -- instead of a binary character (either always or not at all). In doing so, we are able to quantify and compare the extortion ability of ZD strategies and how it depends on their control parameters $(O, \chi, \phi)$, and more remarkably, on the underlying payoff structure specified by the sign of $T+S - 2P$ (Figs. 1 and 2). 

We further note that $\phi$ is a hidden parameter, which has received little attention in prior studies. However, we find that albeit the normalization factor $\phi$ has no impact on the linear payoff relation, it can nontrivially affect the average payoff values that a ZD player will receive (Fig. 2). Mathematically, ZD's average payoff $s_X$ is given by the ratio of the determinants of two matrices, giving rise to a rational function~\cite{press2012iterated}. We can show that $s_X$ is a monotonic function of $\phi$ (Figs. S17 and S18 in the SI) but can have strict non-monotonicity with respect to $\chi$, exhibiting as a one-humped function of $\chi$ (see SI for derivation details). Figure 2 plots an extortionate ZD's average payoff $s_X$ (with the baseline payoff $O = P$) against a fixed co-player Y using a specific strategy as a function of the parameter space $(\phi, \chi)$. This result further demonstrates that ZD can unilaterally fine tune their control parameters, in particular the previously overlooked parameter $\phi$ to their own advantage (which would be boundary values of its admissible interval, either infinitely small or the upper bound).

Only if $T+S > 2P$ is an extortionate ZD unbeatable, ensuring no less payoffs than their opponent (the worst scenario is a tie, e.g., against TFT as shown in Figs. 2c and 2f). In this case, making the extortion factor $\chi$ excessively larger surely can help ZD impose a greater relative advantage over their opponent, but their actual average payoff can be seriously comprised (Fig. 2a). Even worse, when $T+S < 2P$, $s_X$ can drop below $P$ and due to $s_X - P = \chi (s_Y - P)$ we have $s_X < s_Y < P$ (Fig. 2e). In accordance with Fig. 1, the payoff structure can completely change the impact of varying $\phi$ and $\chi$ on ZD's performance (cf. Figs. 2a and 2d, cf. Figs. 2b and 2e).  This is one of the novel insights stemming from the present study, complementing the prior finding that ZD are disfavored in population dynamics settings~\cite{hilbe2013evolution,stewart2013extortion,adami2013evolutionary}. Altogether, these results are key to improving our understanding of previously unforeseen limitations of ZD strategies in head-to-head matches in IPD games.

When an individual is knowingly confronted with extortion and especially having known the limitations of ZD strategies (Figs. 1 and 2), should this player be subdued or otherwise unbending? Prior work demonstrates that if an individual accedes by fully cooperating with an extortioner who fixes their strategies, both their payoffs are maximized (Figs. 1a and 1b). Conversely, here we ask whether there exist \emph{unbending} players who choose to fix their strategies such that extortioners could maximize their payoffs only if they try to be fair by letting $\chi \to 1$. Otherwise, extortioners would have experienced a decline in their average payoffs if they ever demanded an unequal share by increasing $\chi$.

Motivated by these, we further explore unbending strategies that are able to force adapting ZD strategies, among these least generous ones with $O = P$ and hence equipped with the greatest level of extortion ability, to offer a fair split by letting $\chi \to 1$ in their own interest and guarantee equal pays for both sides. Considering that any ZD player can always modulate their hidden parameter $\phi$ to extreme values to favor their gains in the interactions (Fig. 2), we suppose unbending strategies, without loss of generality, will need to (i) neutralize the parameter $\phi$ in the first place such that both of their average payoffs are independent of $\phi$, $\partial s_X/\partial \phi = 0$, and (ii) guarantee that the derivative of $s_X$ with respect to $\chi$ is strictly negative, $\partial s_X/\partial \chi < 0$.

These required properties of unbending strategies lead us to search and identify general classes of strategy candidates that can counteract the adversary imposed by extortioners, provided that they can trigger the backfire of being extortionate. To put it simply, when confronted with a fixed unbending player, any extortionate ZD player is disciplined with payoff reductions in the way that a higher degree of extortion leads to a smaller average payoff. 

%%%%%%%%%%%%%%%%%NEW FIG 3 (ORIGINAL FIG 2)%%%%%%%%%%%%%%%%%

\begin{figure*}[t]
\centering
  \includegraphics[width=0.8\textwidth]{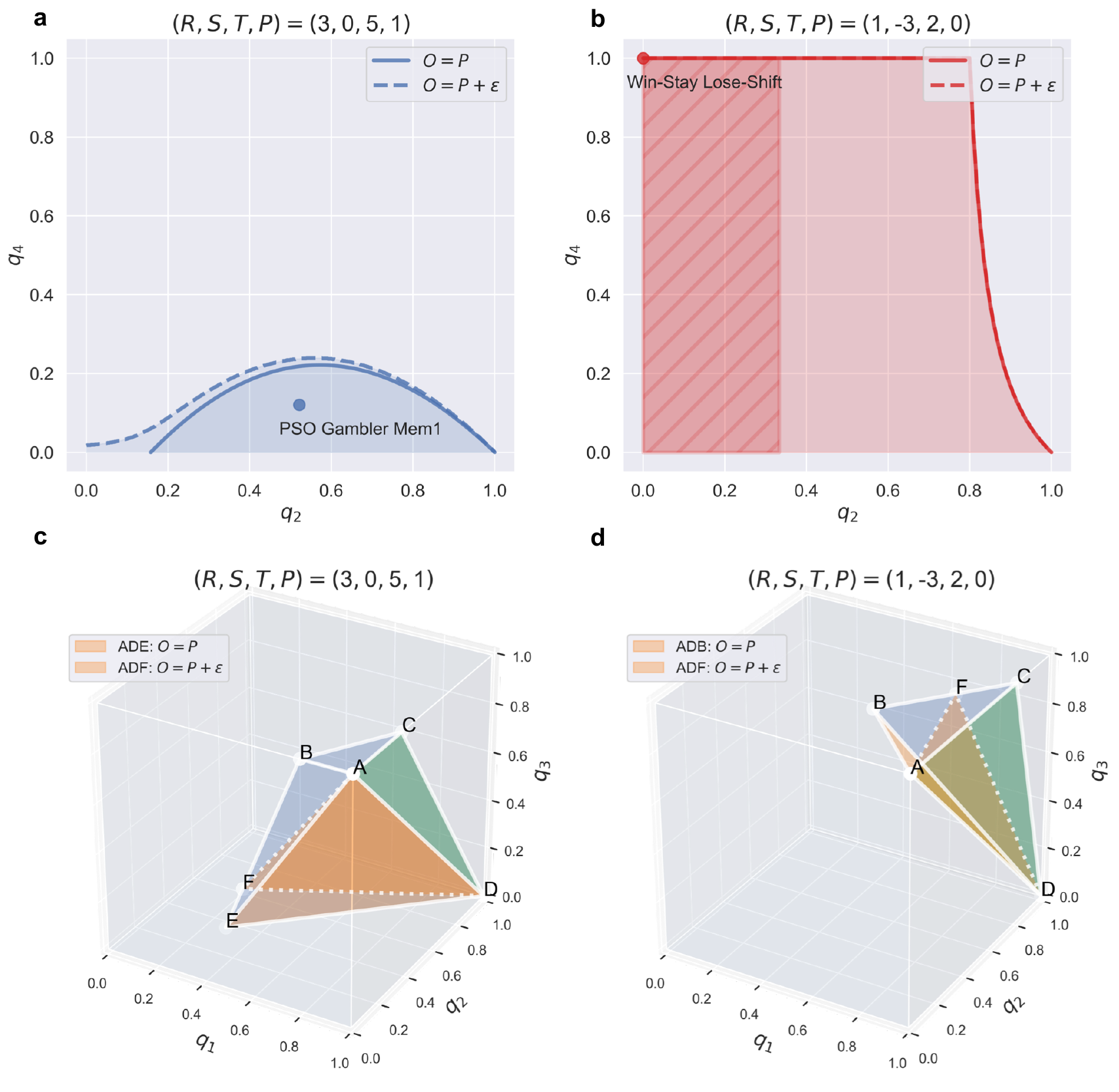}
        \caption{\textbf{Revealing strategies that are \emph{unbending} to extortioners in Iterated Prisoner's Dilemma games.} Shown is the strategy space of unbending players that are able to cause the monotonic decrease of an extortionate zero-determinant (ZD parameterized with $O = P$, namely, the least generous type regardless of the sign $T+S - 2P$) player's payoff with respect to the extortion factor $\chi$. Extortioners can demand an even more unfair share by unilaterally raising their extortion factor $\chi$. However, an unexpected drop in their prospective payoffs if intentionally being more extortionate is likely to compel self-interested extortioners who want to maximize their payoffs to be fair. In this sense, unbending strategies can be used to steer their co-players from extortion to fairness. The strategy space of unbending players depends on the sign of $T+S - 2P$, and we show two general classes of interest (see SI for the complete classification): one class has the form $[1, q_2, 0, q_4]$ with combinations of $q_2$ and $q_4$ shown in (\textbf{a}) and (\textbf{b}), and another class has the form $[q_1, q_2, q_3, q_4]$ where $q_4 = [T - R - P + S-(T + S - 2P)q_1 + (R - P)(q_2 + q_3)]/(2R - T - S)$. This latter class in fact contains all zero-determinant strategies that enforce a linear payoff relation $s_X - O = \chi (s_Y - O)$ with $O > P$. Particular examples of unbending strategies include (\textbf{a}) the memory-one PSO Gambler which is optimized by using particle swarm algorithms, (\textbf{b}) Win-Stay, Lose-Shift, and (\textbf{c}) and (\textbf{d}) all ZD strategies with $O>P$. The dashed lines show the altered boundary of unbending strategies against the ZD player X using $O = P + \varepsilon$ as opposed to $O=P$; the region of unbending strategies for class A ($T+S < 2P$) remains \emph{unchanged} as shown in panel (\textbf{b}). The shaded area in (\textbf{b}) shows the region where ZD player X, even though using minimal $O = P$, can be outperformed by unbending strategies (see Table S9 in the SI for details). This result is in line with Fig.~1, which shows that the payoff structure $T+S < 2P$ drastically hinders a ZD player's ability to extort and dominate their co-players, let alone those unbending ones.
Parameters: (\textbf{a}) (\textbf{b}) $R = 3, S = 0, T = 5, P =1$, $\varepsilon = 0.05$; (\textbf{c}) (\textbf{d}) $R= 1, S = -3, T =2, P = 0$, $\varepsilon = 0.5$. 
         }
        \label{fig3}
\end{figure*}

%%%%%%%%%%%%%%%%%NEW FIG 3 (ORIGINAL FIG 2)%%%%%%%%%%%%%%%%%

Thus, a potential candidate $\mathbf{q}$ of unbending strategies outlearning any extortionate ZD co-player $\mathbf{p}$ needs to mitigate the impact of $\chi$ and $\phi$, which are unilaterally controlled by the extortioner. To this end, we find that four classes of unbending strategies $\mathbf{q} = [q_1, q_2, q_3, q_4]$ that can make their average payoffs independent of $\phi$ (as detailed in the SI).

\begin{table}[H]
\begin{center}
\tabulinesep=1.5mm
\begin{tabular}{c c c c c}
\hline
& & & & \\
Class A: & $q_1 = 1$ and $q_3 = 0$ & & Class B: & $q_2 = q_3 = 0$ \\ 
& & & & \\ \hline
& & & & \\
Class C: & $q_1 = q_2 = q_3$ & & Class D: & $q_4 = h_D(q_1, q_2, q_3)$ \\
& & & & \\
\hline
\end{tabular}
\end{center}
\end{table}

%\begin{center}
%\begin{tabular}{ |ll|ll| } 
%\hline
%\text{A:} \,& $q_1 = 1, \, q_3 = 0$, & \text{B:}\, & $q_2 = q_3 = 0$,\\ \hline
%\text{C:} \,& $q_1 = q_2 = q_3$, & \text{D:}\, & $q_4 = h_D(q_1, q_2, q_3)$.\\
%\hline
%\end{tabular}
%\end{center}

Here in class D: $h_D = [T - R - P + S -(T + S - 2P)q_1 + (R - P)(q_2 + q_3)]/(2R - T - S)$, which is exactly the same linear relation satisfied by any ZD strategy. Class B only exists when $T+S < 2P$ and the maximum payoffs for both sides can receive is $R$ (which is an equal-pay outcome). Class C contains `willing' $[1, 1, 1, 0]$ on the boundary~\cite{denBerg_PRSB15}, against which an extortioner can only maximize their own payoffs by being fair ($\chi \to 1$), thereby ensuring equal payoffs ($s_X =s_Y \to R$) with unbending `willing' $(1-\delta,1-\delta,1-\delta,\varepsilon)$ for $\delta \to 0$, and $\varepsilon \to 0$ (Table S12 in the SI). The complete analysis and discussion of these two classes B and C can be found in the SI. 

Furthermore, the entire strategy space comprised of all admissible unbending strategies can be characterized by requiring the derivative $\partial s_X/\partial \chi < 0$ (Fig. 3). Again, the sign of $T+S -2P$ determines the geometry of the strategy space satisfying unbending properties (see Figs. 3a and 3b for class A, Figs. 3c and 3d for class D). Of particular interest, the memory-one PSO Gambler $\mathbf{q} = [1, 0.522, 0, 0.121]$, an optimized strategy using particle swarm algorithms in IPD games with the conventional payoff values~\cite{harper2017reinforcement}, belongs to class A of unbending strategies (Fig. 3a), and WSLS is an unbending strategy only if $T+S < 2P$ (Fig. 3b). 

Interestingly and coincidentally, we find that all ZD strategies with $O > P$ and $\chi > 1$ are unbending to extortionate ZD (Figs. 3c and 3d). It is worth noting that these planes specifying the boundary of class D have particular meanings. As shown in Fig. 3c, the shaded triangle ADE represents the set of extortionate ZD strategies with $O = P$ and $\chi > 1$, and the shaded area by the four-sided polygon BCDE represents the set of equalizer strategies, and all unbending strategies in class D are in between these two planes and bounded by the unit cube. Besides, the triangle ACD represents the set of generous ZD strategies with $O = R$, and the triangle ABD represents the set of ZD strategies with $O = (T + S)/2$. For $T + S < 2P$ (Fig. 3d), the strategy space degenerates into the region between the triangle ABD (extortionate ZD) and triangle BCD (equalizer). Hence, we conclude that class D contains all ZD strategies with $O > P$. 

We also have extended our search of fixed unbending strategies with respect to an even broader class of ZD strategies just with positive $\chi > 1$ (namely, using the baseline payoff $O = P+\varepsilon \ge P$ and still having extortion ability to some degree as shown in Figs. 1c and 1d), such that ZD's payoff is independent of the normalization factor $\phi$ and monotonically decreases with their extortion factor $\chi$. As shown in Fig. 3 (highlighted with dashed lines) and the SI, our classification of unbending strategies (especially nontrivial classes A and D) remains largely robust with respect to this important extension. Unexpectedly, we also find a set of nonlinear memory-one (non-ZD) strategies, when having the knowledge of the ZD co-player's baseline payoff $O$, will always be able to ensure equal payoffs $O$ for both (see Fig. S21 in the SI for details).

%%%%%%%%%%%%%%%%%NEW FIG 4 (ORIGINAL FIG 1)%%%%%%%%%%%%%%%%%

\begin{figure*}[t]
\centering
  \includegraphics[width=0.8\textwidth]{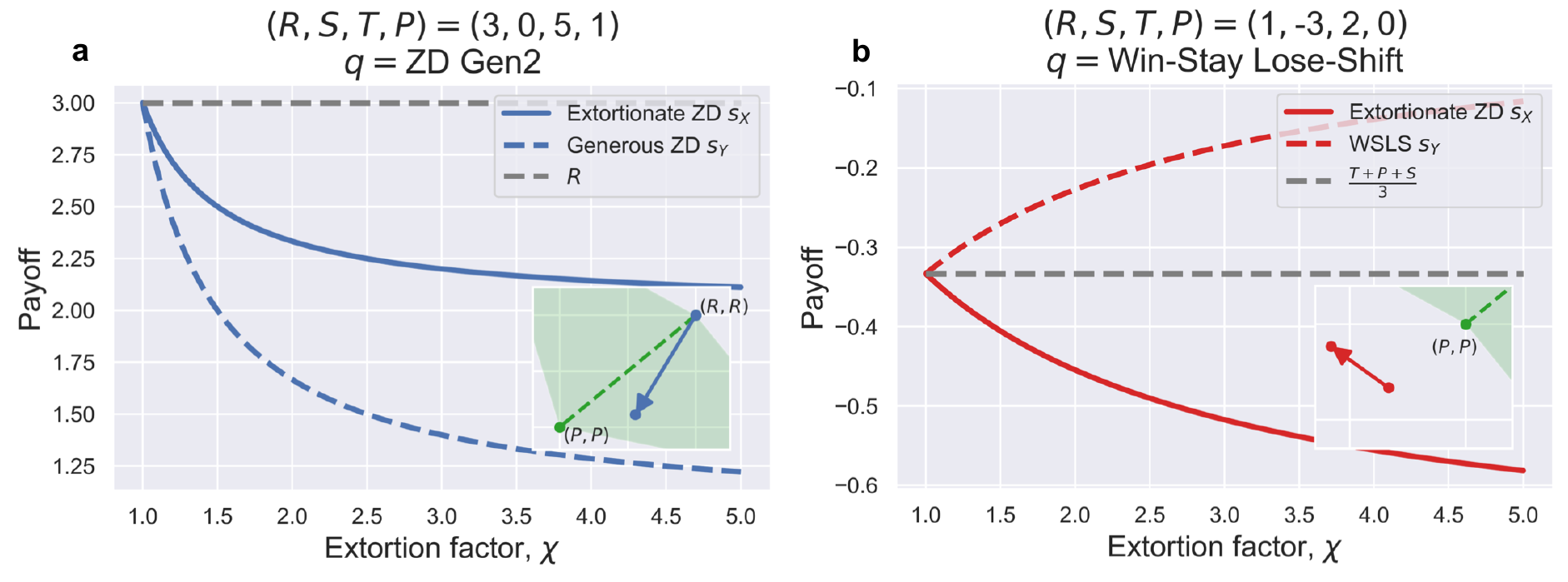}
        \caption{\textbf{Intuition for how unfair demand can backfire on extortioners.} As shown in panel (\textbf{a}), despite being able to enforce payoff control against generous ZD (with $\chi' = 2$), the prospective payoff of extortionate ZD monotonically decreases with their extortion factor $\chi$. The unfair extortion backfires on ZD which intended to demand a higher proportion but ended up with less payoff than what they would have obtained if being fairer otherwise. Even more, extortioners are outperformed by Win-Stay, Lose-Shift (WSLS), as shown in panel (\textbf{b}) when $T+S < 2P$. The payoffs of the extortioner and WSLS, $s_X$ and $s_Y$, are both less than $P$, but intended extortion inflicts the unprecedented opposite outcomes: extortioner suffers more whereas WSLS gains more. Targeted at extortioners, unbending strategies can be used to foster fairness in IPD games. The inset plots in (\textbf{a}) and (\textbf{b}) show the zoomed-in view of scatter plot of payoffs, in a fashion similar to Figs. 1a and 1b, with their arrows to indicate their directions of change starting from fair, equal split as ZD increases $\chi$ above one. Parameters: (\textbf{a}) $R = 3, S = 0, T = 5, P =1$, (\textbf{b}) $R= 1, S = -3, T =2, P = 0$.
         }
        \label{fig4}
\end{figure*}

%%%%%%%%%%%%%%%%%NEW FIG 4 (ORIGINAL FIG 1)%%%%%%%%%%%%%%%%%

To provide further intuition about why extortion against unbending players fails to yield better average payoffs, we consider the case where an extortioner $X$ with $(P, \chi,\phi)$ plays against a fixed generous ZD player $Y$ with $(R, \chi',\phi')$ which in fact belongs to class D of unbending strategies. Both of their resulting payoffs are \emph{independent} of their $\phi$ values, and the extortioner $X$ has an average payoff given by 

\begin{equation}
s_X(\chi) = \frac{P(\chi -1) + R\chi(\chi'-1)}{\chi \chi' -1}.
\end{equation}
We see that $s_X(\chi)$ is monotonically decreasing with $\chi$, as the derivative $s_X'(\chi) = \frac{(P-R)(\chi'-1)}{(\chi \chi' - 1)^2} < 0$ for $R>P$, $\chi>1$, and $\chi'>1$ (Fig. 4a). 

Geometrically visualizing this specific example, the generous ZD player $Y$ enforces a linear payoff relation as $s_Y -R = \chi'(s_X-R)$, whereas the extortioner $X$ enforces $s_X - P = \chi (s_Y - P)$, and the resulting payoff pair $(s_X, s_Y)$ lies in the intersection of these two straight lines. If the extortioner $X$ increases the extortion factor $\chi$, the intersection point will move down along the line of $s_Y - R = \chi'(s_X-R)$ (if the generous ZD player $Y$ remains unchanged). Therefore, the more unfair demand towards a fixed generous ZD player, the less payoff extortion yields. This previously unforeseen `backfire' is self-inflicted by the attempt to extort. For a self-interested individual who cares about how much they get, not just about monopolizing control of relative payoff, it does not pay to extort a generous ZD co-player, and unfair demand backfires on extortioners who would have received the maximum $R$ if trying to be fair by setting $\chi \to 1$ (Fig. 4a).

We now turn to explain the intuition behind the payoff structure of IPD games that can impact the dominance (optimality) of ZD strategies. It is well known that the condition $T+S < 2R$ is needed for mutual cooperation to fare better than alternating C and D pairs in the IPD. Yet another condition $T + S > 2P$ comes into sight if one ponders the condition under which the average payoff of any IPD strategy cannot be worse than $P$, the payoff for ending up with the deadlock of mutual defection. IPD is typically studied using the conventional values $R = 3$, $S = 0$, $T = 5$, $P = 1$, satisfying $2P < T+S < 2R$, and thus it ensures the average payoff of any IPD strategy cannot be less than $P$. Extortionate ZD players attain payoff control and extortion as desired $s_X - P = \chi (s_Y - P)$ in this scenario using the conventional payoff values (Figs. 1a and 1c), but the tide will turn against extortioners if the payoff structure satisfies $T+S < 2P$. In this latter case, the average payoff of extortionate ZD strategies can be lower than $P$ when facing off certain IPD strategies (Figs. 1b and 1d). 

As ZD strategies are explicitly dependent on the underlying payoff matrix whose elements are $(R, S, T, P)$, we discover that the particular payoff structure, which is governed by the sign of $T + S - 2P$, can fundamentally change the dominance of extortionate ZD strategies (Fig. 1). For example, when an extortionate ZD player is pitted against WSLS with $\mathbf{q} = [1, 0, 0, 1]$, the stationary distribution $\mathbf{v}$ of pairwise outcomes $\{CC, CD, DC, DD\}$ is, up to a positive normalization factor, given by:

\begin{equation}
v_{CC} =0, v_{CD} = \frac{T-P + \chi(P-S)}{\chi(T-P) + P-S}, v_{DC} = 1, v_{DD} = 1.
\end{equation}

Therefore, in order to gain an advantage, extortion ZD must ensure $v_{CD} < v_{DC}$. However, this condition cannot always be satisfied when $T+S <2P$ (the shaded region in Fig. 3b, see Tables S9 and S13 in the SI for details). On the contrary, the extortionate ZD player in fact reciprocates unilateral cooperation more frequently than WSLS if $v_{CD} > v_{DC}$ holds, which is equivalent to require $T+S < 2P$. Under this payoff structure condition, WSLS outperforms any extortionate ZD player (Fig. 4b); the more greedy extortion, the more ZD loses. Noteworthy, there is absolutely no mutual cooperation between WSLS and extortionate ZD players. Extortionate ZD does not fully cooperate after a mutual cooperation move, and thus ZD and WSLS will eventually end up with mutual defection from which ZD will never respond with cooperation while WSLS will always respond with cooperation; they will never be back to mutual cooperation. As a consequence, in the long run, no mutual cooperation between them can be established at all.

%%%%%%%%%%%%%%%%%NEW FIG 5%%%%%%%%%%%%%%%%%

\begin{figure*}[t]
\centering
  \includegraphics[width=0.8\textwidth]{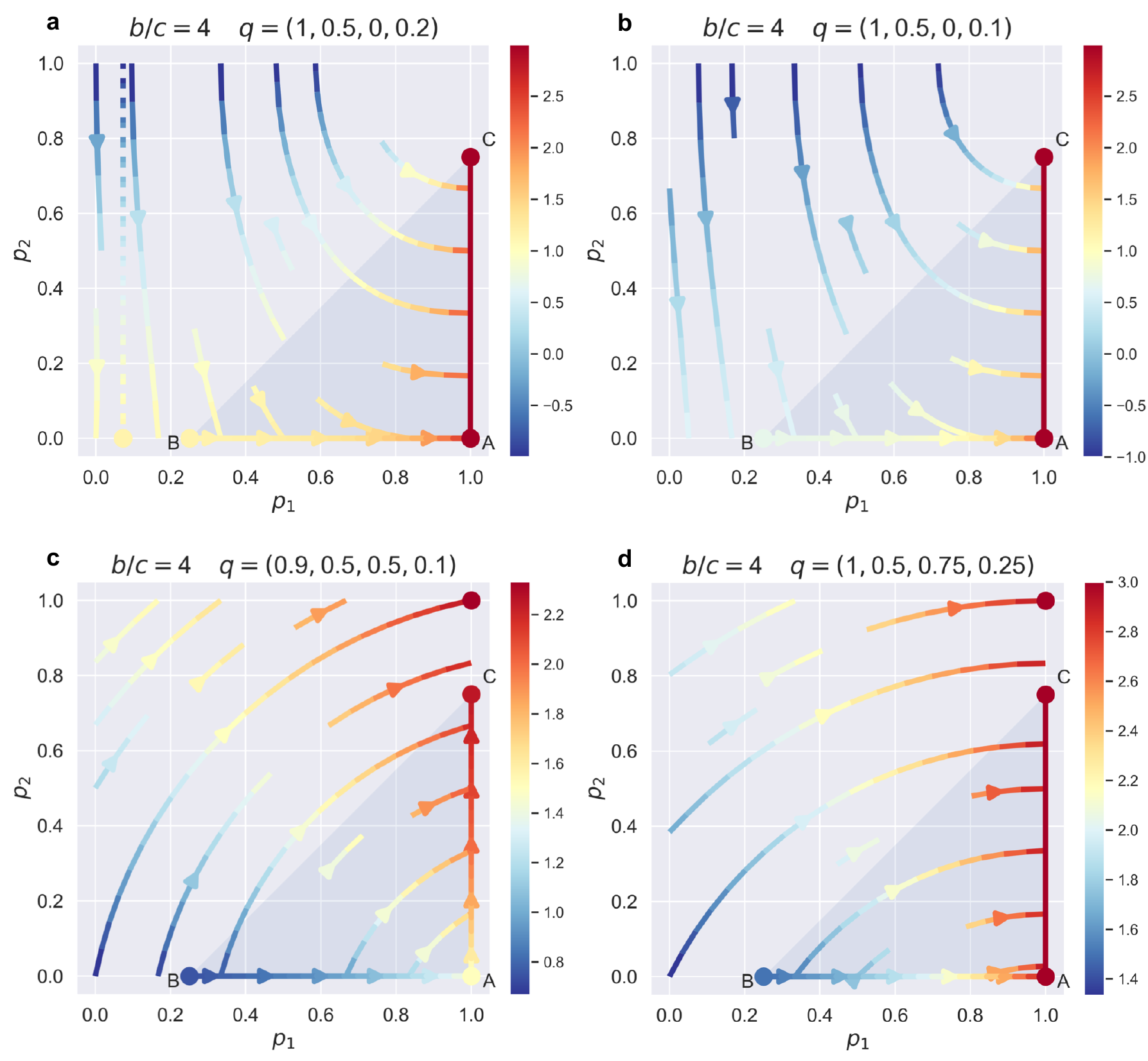}
        \caption{\textbf{Steering learning dynamics towards fairness and cooperation with unbending strategies.} Shown are the stream plots (vector fields) from the adaptive learning dynamics of a self-interested focal player X who uses a general reactive strategy $[p_1, p_2, p_1, p_2]$ against a fixed unbending co-player Y: (\textbf{a}) (\textbf{b}) from class A and (\textbf{c}) (\textbf{d}) from class D. Class A of unbending strategies are able to steer their co-player X ultimately to behave like generous TFT.  Panel (\textbf{a}) demonstrates that, depending on the specific unbending strategy player Y uses, the direction of change of player X's $p_1$ can exhibit bistability (separated by the dashed line in (\textbf{a})), which further depends on the initial state of X's strategy. Panel (\textbf{b}) shows that there exists a subset of class A that is able to direct the change of $p_1$ always towards full cooperation. In (\textbf{a}) and (\textbf{b}), on the edge $p_1 = 1$, the direction of change of $p_2$ is neutral, and the line segment of AC indicates the set of compliers; on the line segment of the edge $p_2  = 0$, indicated by BA, which represents a subset of extortionate ZD players (extortioners), the learning dynamics of the ZD player X always converges to TFT $(1, 0)$. Panels (\textbf{c}) and (\textbf{d}) show that class D of unbending strategies are able to steer adaptive learning dynamics of X globally to the cooperative edge $p_1 =1$ on which the direction of change of $p_2$ (\textbf{c}) either is increasing if Y uses a strategy from class D that is an intermediate ZD with $P < O < R$) (\textbf{d}) or remains neutral if Y uses a strategy from class D that happens to be the generous ZD with $O = R$. In both (\textbf{a}) and (\textbf{b}), at the edge $p_1 = 1$ reaches the global maximum for $X$'s payoff if $b/c > (\sqrt{5}+1)/2$, and in (\textbf{c}) only at $(1,1)$ does the global payoff maximum for X occur whereas so does the entire cooperative edge $p_1 = 1$ in (\textbf{d}). The shaded triangle ABC in each panel indicates the part of reactive strategies which belong to the subset of general ZD strategies with positive $\chi$. The color of the curves and dots corresponds to the payoff values of X in situ, as specified by the given colorbar. The Prisoner's Dilemma game is parameterized using a donation game $R = b-c, S = -c, T= b, P =0$ with $b = 4$ and $c = 1$.}
        \label{fig5}
\end{figure*}

%%%%%%%%%%%%%%%%%NEW FIG 5%%%%%%%%%%%%%%%%%

To further understand fixed unbending player's unprecedented \emph{steering} role in enforcing fairness and cooperation, we consider adaptive learning dynamics of a focal player $X$ using a much broader space of strategies, rather than being limited to extortionate ZD, in a donation game which is a simplified Prisoner's Dilemma~\cite{stewart2013extortion,hilbe2013evolution}. Under this donation game satisfying the `equal gains from switching' (i.e., $T+S = R+P$), the memory-one reactive strategies $\mathbf{p} = [p_1, p_2, p_1, p_2]$ is actually a subset of ZD strategies~\cite{hilbe2013evolution} (also see Fig. S19 in the SI). The shaded triangle in Fig. 5 indicates all such ZD strategies with positive $\chi>1$: BA represents extortionate ZD with $\chi > 1$; point A is TFT with $\chi \to 1$ and $O = P$; point B is equalizer with $O = P$ and $\chi \to \infty$; point C is generous TFT with $O =R$, and $\chi \to \infty$; BC represents the class of `equalizer' strategies. We find that if the benefit-to-cost ratio $b/c > (\sqrt{5}+1)/2$ (`golden ratio'), the cooperative edge $(1,p_2)$ is guaranteed to have the maximum average payoff value for player X when interacting with any fixed unbending player from class A (the region highlighted in Fig. 3a). Depending on the specific strategy of the unbending player $Y$ from class A (see Fig. S20 and Table S23 for more details in the SI), there could exist bistable learning outcomes of X's final strategies: X can converge to the all defection corner $(0, 0)$ or otherwise to the cooperative edge $(1,p_2)$ (Fig. 5a), but there is a subset of class A of unbending strategies that ensures the global convergence to the cooperative edge $(1,p_2)$ (Fig. 5b, and Fig. S20 in the SI).

The steered learning dynamics under the influence of the co-player Y from class D of unbending strategies is shown in Figs. 5c and 5d. Generally speaking, the learning dynamics of a focal player X against class D of unbending strategies (in other words, ZD strategies with a higher level of generosity than player X) adds useful insights by complementing previous results in Ref.~\cite{press2012iterated} that focuses on adapting co-player Y against an extortionate ZD player X. Here, we show that the final strategy of player X converges to the cooperative edge, reaching full cooperation if against an unbending strategy from class D with $O < R$ (still more generous than any extortionate ZD strategies on the edge BA with $O = P$) (Fig. 5c) or remains neutral on the cooperative edge once reaching there when against an unbending strategy from class D with $O = R$ (generous ZD) (Fig. 5d).

In the SI, we study the corresponding learning dynamics of a general ZD player within the parameter space $(O, \chi)$ (see Tables S25 and S26) and confirm qualitatively similar results as reported here. In particular, when against a fixed unbending player, an adapting extortioner with the intended extortion factor $\chi$ unexpectedly suffers greater payoff reductions than their counterpart who chooses not to accede unless offering a fair split. For this reason, any evolutionary extortioner who aspires to maximize their own payoff will be compelled from extortion to fairness by adjusting their $\chi$ values. Since there is no interference by the parameter $\phi$ as $\frac{\partial s_X(\chi,\phi)}{\partial \phi} = 0$, such reactive learning dynamics of extortioners is governed solely by the evolution of $\chi$ towards payoff optimization:

\begin{equation}
\frac{d\chi}{dt} = \omega \frac{\partial s_X(\chi,\phi)}{\partial \chi} <0,
\end{equation}
where the properly chosen timescale parameter $\omega$ guarantees that the state of Markov chains of game play reaches equilibrium faster than the learning dynamics (see SI). 

Thus a self-interested extortioner tends to adjust $\chi$ as small as possible and ultimately behaves like TFT by letting $\chi \to 1$,  thereby guaranteeing equal payoffs for both parties (see the change of direction on the edge of $O = P$ in Tables S25 and S26 in the SI). In evolving populations, natural selection favors generosity over extortion~\cite{stewart2013extortion}, and in head-to-head matches as demonstrated here, players with the knowledge of unbending strategies can outlearn extortioners and foster fairness and reciprocity in dyadic interactions.

Taken together, these results suggest that unbending strategies can not just outlearn self-interested extortionate ZD and force them to be fair and cooperative, but also steer the evolution of Tit-for-Tat-like strategies out of any focal player using a much broader strategy space (represented by the unit square $[0,1]^2$ in Fig. 5) including but not necessarily limited to extortionate ZD strategies (namely, the edge BA in Fig. 5).

\section*{Discussions}

It is thought that an evolutionary (adapting) player should be subdued to a fixed extortionate ZD player by fully cooperating as the best response~\cite{press2012iterated}. In contrast, recent experimental evidence suggests that human players often choose \emph{not} to accede to extortion out of concern for fairness~\cite{hilbe2014extortion,becks2019extortion}. Inspired by this empirical finding, here we show that there exist general classes of \emph{unbending} strategies such that the best response of any payoff-maximizing extortioner against a fixed unbending player is to be fair, thereby ensuring equal pays for both parties. From this perspective, the witting of unbending strategies has effectively turned the opponent's choices of whether or not adopting extortionate ZD strategy into an Ultimatum game~\cite{nowak2000fairness}: to demand unfair division via unilaterally setting a large $\chi$ value, or to guarantee fair share by letting $\chi \to 1$. In the former, extortion effort is sabotaged by unbending, and both sides will be hurt, whereas in the latter, both sides will get an equal split of the payoffs. Our results demonstrate that unbending strategies can be used to rein seemingly formidable extortionate ZD players, whose fair offer ultimately can be cultivated in their own interest.

In light of unbending strategies, there is no guarantee that extortioners will be able to subdue their opponent with certainty and get their own way as desired. Extortion cannot be successful unless their co-players give up resistance in the first place. The extortion ZD exerts on the co-player can backfire on them. For example, an extortionate ZD player will not be able to rein Tit-for-Tat-like players (a limiting case belonging to class D of unbending strategies) who are fair-minded but willing to punish defection by responding with defection. They will end up a tie both receiving $P$ (Fig. 2). Even if an extortionate ZD player $X$ does end up with dominance over the co-player Y, namely, $s_X > s_Y > P$, a higher ratio of relative payoff surplus, $\chi = (s_X - P)/(s_Y - P)$, does not necessarily translate to higher actual payoffs (Fig. 2a). Increasing $\chi$ appears to put ZD a more advantageous position, but such unfair demand would be pushed back by unbending players such as generous ZD and TFT-like players and hence does not always yield higher actual payoffs (Fig. 4a). As recently demonstrated in experiments involving human players against fixed machine extortioners~\cite{hilbe2014extortion}, human players respond to more extortionate ZD players with much lower cooperation levels, which are in essence passive punishment measures to counter ZD's intended extortion.

Our work highlights the importance of payoff structure in determining the optimality of ZD strategies in IPD games (Fig. 1). In particular, if the condition $T+S < 2P$ holds, which means the total payoff, $T+S$, from alternating C and D pairs of two players is worse than that of their mutual defection, $2P$, extortionate ZD players can be outperformed (Figs. 1b and 1d). This surprising finding is an important new insight that stems from the present study. Moreover, the sign of $T+S - 2P$ qualitatively determines the admissible strategy space of unbending players that can cause the backfire on extortioners (Fig. 3). Noticeably, only if $T+S < 2P$ is WSLS an unbending strategy, and in this case, WSLS dominates any extortionate ZD strategy (Fig. 4b).

The payoff condition $T+S < 2P$ implies a more adversarial nature in pairwise interactions than the conventional IPD games where $T + S > 2P$ typically holds~\cite{d2015statistical}. Intuitively, this means that the best response for a pair of individuals alternating between $(C, D)$ and $(D, C)$ is always to switch to mutual defection $(D, D)$ (cf. Figs. 1a and 1b). Unbending strategies (those highlighted in the shaded area in Fig. 3b) can outcompete seemingly invincible extortionate ZD players who would have the greatest potential to dominate by using the minimal $O = P$ and $\chi > 1$ (Fig. 1). As aforementioned, since ZD's strategy is parameterized explicitly by the underlying payoff matrix, variations in the payoff structure can have a previously unforeseen effect that will turn the tables on ZD (Fig. 4b): an extortionate ZD may become a victim of their own success in IPD games satisfying $T+S < 2P$ and more broadly, in social dilemmas of more adversarial nature as discussed in Ref.~\cite{d2015statistical}.

In the presence of errors and noises~\cite{fundenberg1990evolution,axelrod2012launching}, complex strategies informed by longer memory of past moves are likely at an advantage against simple memory-one strategies. Beyond pairwise interactions, higher-order ones in multi-person games~\cite{hilbe2014cooperation}, such as the public goods game, are also of relevance in studying reciprocity in group situations. Extensions incorporating these considerations are meaningful, but incur computational and theoretical challenges in search of robust optimal strategies. Nevertheless, the recent breakthrough in reinforcement (deep) learning of zero-sum games~\cite{balduzzi2019open}, like the Go~\cite{silver2017mastering}, can lend some insight into the study of non-zero-sum games where learning agents, despite being self-serving, can mutually foster cooperation for the greater good under certain conditions~\cite{barfuss2020caring}. Thus, the classic framework of IPD still has the potential to be used as a primary testbed for synergistically combining artificial intelligence (AI) and game theory in future work~\cite{harper2017reinforcement,noordman2019evolving,mcavoy2021selfish}, all with an eye towards helping us to enhance global cooperation in many challenging issues confronting our common humanity~\cite{dafoe2021cooperative}.

In summary, we have found and characterized general classes of unbending strategies that are fair-minded and can outlearn extortioners in their head-to-head encounters. When an extortionate ZD player attempts to demand an unfair greater share from an unbending player who instead uses a fixed strategy, the unbending player is able to restrain the extortioner from profiting more. The intent to extort an unbending player has unprecedented consequences: extortioners would fare worse than if being fairer, and they can even be outperformed by, for example, WSLS, if the payoff matrix satisfies $T+S<2P$. Such previously unforeseen backfires caused by unbending players can steer reactive learning dynamics of extortionate ZD players from extortion to fairness. Our work offers novel insights into fostering fairness and suppressing extortion for a more equitable and just society. 

\section*{Model \& Methods}

\paragraph{Model and analytical approach.} We use the same analytical approach invented by Press and Dyson~\cite{press2012iterated} to calculate expected payoffs of any two given players that are head-to-head in the Iterated Prisoner's Dilemma games. In this work, we focus on revealing strategies that are unbending to extortionate zero-determinant (ZD) players using explicit closed-form solutions (see details in the SI). The ZD strategies are usually parameterized by three important parameters, the extortion factor $\chi$, the normalization factor $\phi$, plus an additional baseline payoff $O \in [P, R]$ which controls the level of generosity~\cite{press2012iterated,stewart2013extortion}. Tuning the parameter $\phi$ of extortionate ZD strategies with $O = P$ and $\chi > 1$ does not affect the linear payoff relation $s_X - P = \chi(s_Y - P)$, but will impact the dependence of their own average payoffs on the extortion factor $\chi$ in a nontrivial way (see Fig. 2). Therefore, we restrict our search for unbending strategies that can neutralize the impact of this parameter $\phi$, that is, we find specific classes of strategies that are able to render the independence of their payoffs on $\phi$. Further, we narrow down the search of unbending strategies that can cause the `backfire' of extortion, namely, the expected payoffs of extortionate ZD strategies against a fixed unbending player are monotonically decreasing with $\chi$. Ultimately, these considerations lead us to discover multiple general classes of unbending strategies, against which attempt to extort and dominate, if any, does not pay off at all for ZD players using $O = P + \varepsilon \ge P$ (including but not limited to extortionate ZD). In some cases, extortionate ZD strategies can even be outperformed by unbending co-players, if the payoff matrix satisfies $2P < T+S$ (Fig. 3b). We also investigate how fixed unbending players can steer the learning dynamics of their adapting co-players who use a much broader range of memory-one strategies beyond the class of extortionate ZD towards fairness and cooperation. We detail our comprehensive analysis in the SI.

\paragraph{Data Availability.}
All data pertaining to the present work has been included in the main text and the SI.

\paragraph{Code Availability.}
Our Python Jupyter notebooks that can be used to reproduce results reported in this work are available at GitHub: \url{https://github.com/fufeng/unbending}.

\section*{Acknowledgements.}
%\xc{We are indebted to the editor and the three referees for their constructive comments which immensely helped improve this work.} We would like to thank Christian Hilbe and Joshua Plotkin for helpful suggestions and discussions. F.F. thanks Dan Rockmore for leading a stimulating brainstorming session on social justice during the Call for Engagement with the Wright Center for the Study of Computation and Just Communities during which this work was first presented. 
X.C. gratefully acknowledges the generous faculty startup fund support by BUPT. F.F. is supported by the Bill \& Melinda Gates Foundation (award no. OPP1217336), the NIH COBRE Program (grant no. 1P20GM130454), a Neukom CompX Faculty Grant, the Dartmouth Faculty Startup Fund, and the Walter \& Constance Burke Research Initiation Award. 

\section*{Author Contributions} 
X.C., \& F.F. conceived the model, performed analyses, and wrote the manuscript.

\section*{Competing Interests} The authors declare that they have no competing financial interests.

%\nocite{*}
%\bibliography{ref}
%\bibliographystyle{naturemag}

\end{document}

\maketitle

\tableofcontents

\medskip

\section{Introduction}

Years have passed since Press and Dyson discovered the class of zero-determinant (ZD) strategies, some of which seem to be able to dominate any evolutionary opponent in the Iterated Prisoner's Dilemma (IPD) game. Assume that there are two players X and Y. Label the four outcomes of a single round $1$, $2$, $3$ and $4$ for $\bm{xy} \in (\text{CC}, \text{CD}, \text{DC}, \text{DD})$. We can write X's strategy as $\bm{p} = (p_1, p_2, p_3, p_4)$, denoting the probabilities to cooperate under these outcomes. Analogously, Y's strategy is $\bm{q} = (q_1, q_2, q_3, q_4)$. When X applies a ZD strategy, its payoff $s_X$ and the opponent's payoff $s_Y$ satisfy the equation 
\begin{equation}
\alpha s_X + \beta s_Y + \gamma = 0. 
\end{equation}
\par

Further, consider the above linear relation between $s_X$ and $s_Y$ of the form:
\begin{equation}
s_X - O = \chi (s_Y - O),
\end{equation}
where $O$ is the baseline payoff for both players, $\chi$ is the extortion factor, and $\phi$ is the parameter that guarantees $0 \leq p_i \leq 1$ for $i \in \{1, 2, 3, 4\}$. Routine calculation gives the corresponding probabilities for player X:
\begin{equation}
\begin{cases}
p_1 = 1 - (R - O)\phi(\chi - 1),\\
p_2 =  1 - \phi[(T - O)\chi + (O - S)],\\
p_3 = \phi[(O - S)\chi + (T - O)],\\
p_4 = (O - P)\phi(\chi - 1).
\end{cases}
\label{GZD_strategy}
\end{equation}
Here, the range of $\chi$ is
\begin{equation}
\chi \geq 1 \qquad \text{or} \qquad \chi \leq \chi^{\text{upper}} = \begin{cases}-\frac{T - O}{O  - S}, & T + S \geq 2O \\ -\frac{O - S}{T - O}. & T + S < 2O \end{cases}
\end{equation}
Given that all the components of $p$ are between $0$ and $1$, $O$ has to be in the interval $[P, R]$. Choosing $P$ will lead to the meanest ZD strategy (extortioner) while choosing $R$ will generate the most generous one (complier). In our study, we let $O = P$ most of the time. \par

If X is an extortioner, we have
\begin{equation}
s_X - P = \chi (s_Y - P),
\label{linear_relation}
\end{equation}
which yields the extortionate ZD strategy with probabilities
\begin{equation}
\begin{cases}
p_1 = 1 - (R - P)\phi(\chi - 1),\\
p_2 =  1 - \phi[(T - P)\chi + (P - S)],\\
p_3 = \phi[(P - S)\chi + (T - P)],\\
p_4 = 0.
\end{cases}
\label{ZD_strategy}
\end{equation}
\par

Regarding the two parameters in Equation~\ref{ZD_strategy}, the extortion factor $\chi \geq 1$ by convention. Meanwhile, the legitimate range of $\phi$ is (a typo has been pointed out in Press and Dyson)
\begin{equation}
0 < \phi \leq \phi^{\text{upper}} = 
\begin{cases}\frac{1}{(T - P)\chi + (P - S)}, & T + S \geq 2P \\ \frac{1}{(P - S)\chi + (T - P)}. & T + S < 2P
\end{cases}
\label{phi_bound}
\end{equation}
Notice that it is also possible for player X to obtain a payoff greater than that of player Y when the extortion factor $\chi$ satisfies 
\begin{equation}
\chi \leq \chi^{\text{upper}} =  \begin{cases}-\frac{T - P}{P  - S}, & T + S \geq 2P \\ -\frac{P - S}{T - P}. & T + S < 2P \end{cases}
\label{chi_bound}
\end{equation}
In this case, the parameter $\phi$ needs to follow the inequality $\phi^{\text{lower}} \leq \phi < 0$, where $\phi^{\text{lower}}$ can be derived under the same restriction that $p_1$, $p_2$, $p_3$, and $p_4$ should be reasonable probabilistically.  Moreover, since $\chi$ is negative, it is straightforward to see that $s_X < P < s_Y$ (Figure~\ref{abn_wsls}) or $s_Y < P < s_X$ (Figure~\ref{monotonicity}). \par

The linear relation between the two payoffs $s_X$ and $s_Y$ convinces latecomers that \begin {enumerate*} [label=(\roman*\upshape)] \item the extortioner's payoff $s_X$ is always no less than the opponent's payoff $s_Y$ and \item $s_X$ is an increasing function (at least a non-decreasing function) with respect to the extortion factor $\chi$ (on the two branches of $\chi$). \end {enumerate*} \par

\begin{figure}[H]
\centering
   \subfloat[]{\includegraphics[width=0.5\linewidth]{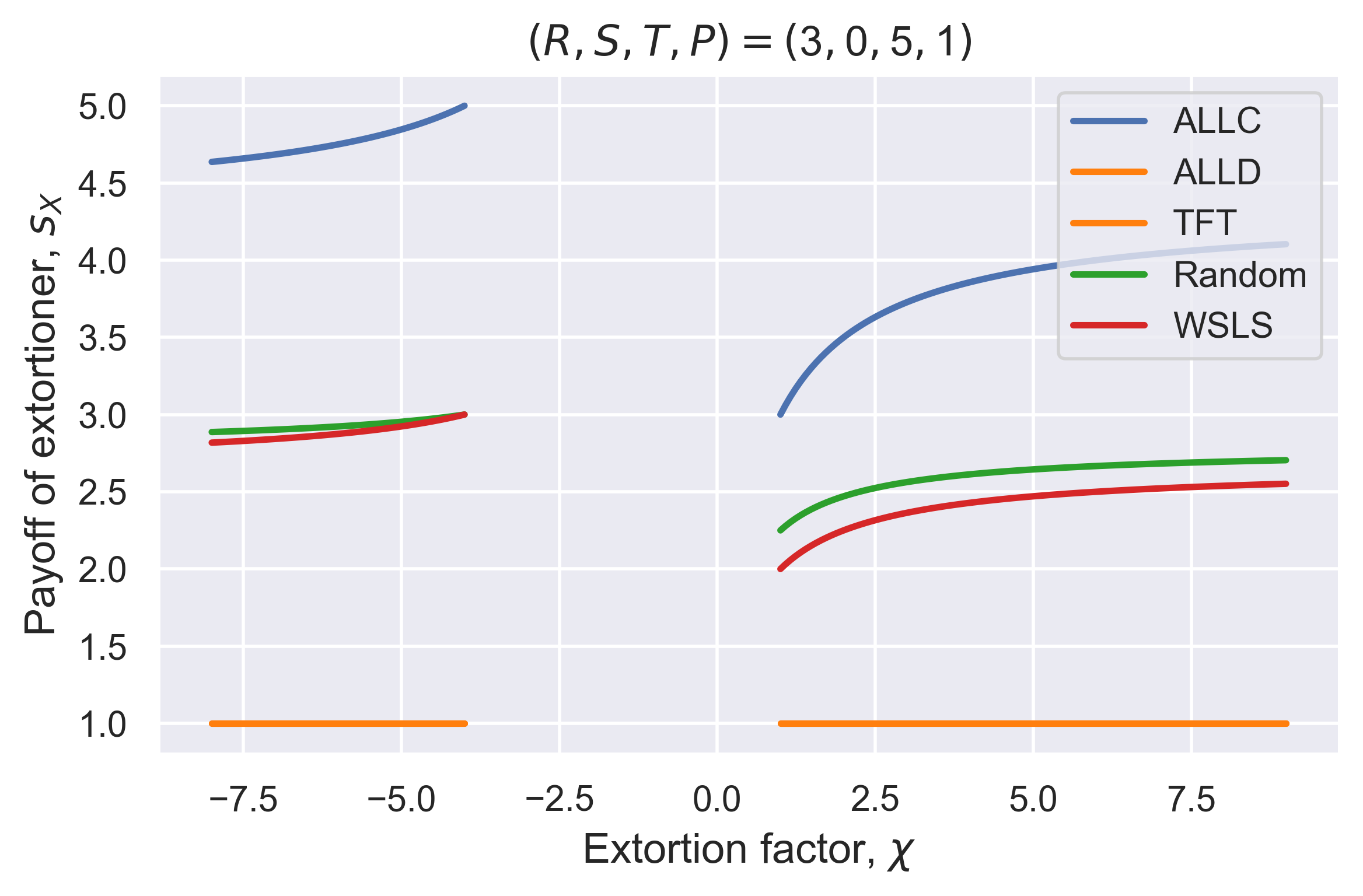}} 
    \subfloat[]{\includegraphics[width=0.5\linewidth]{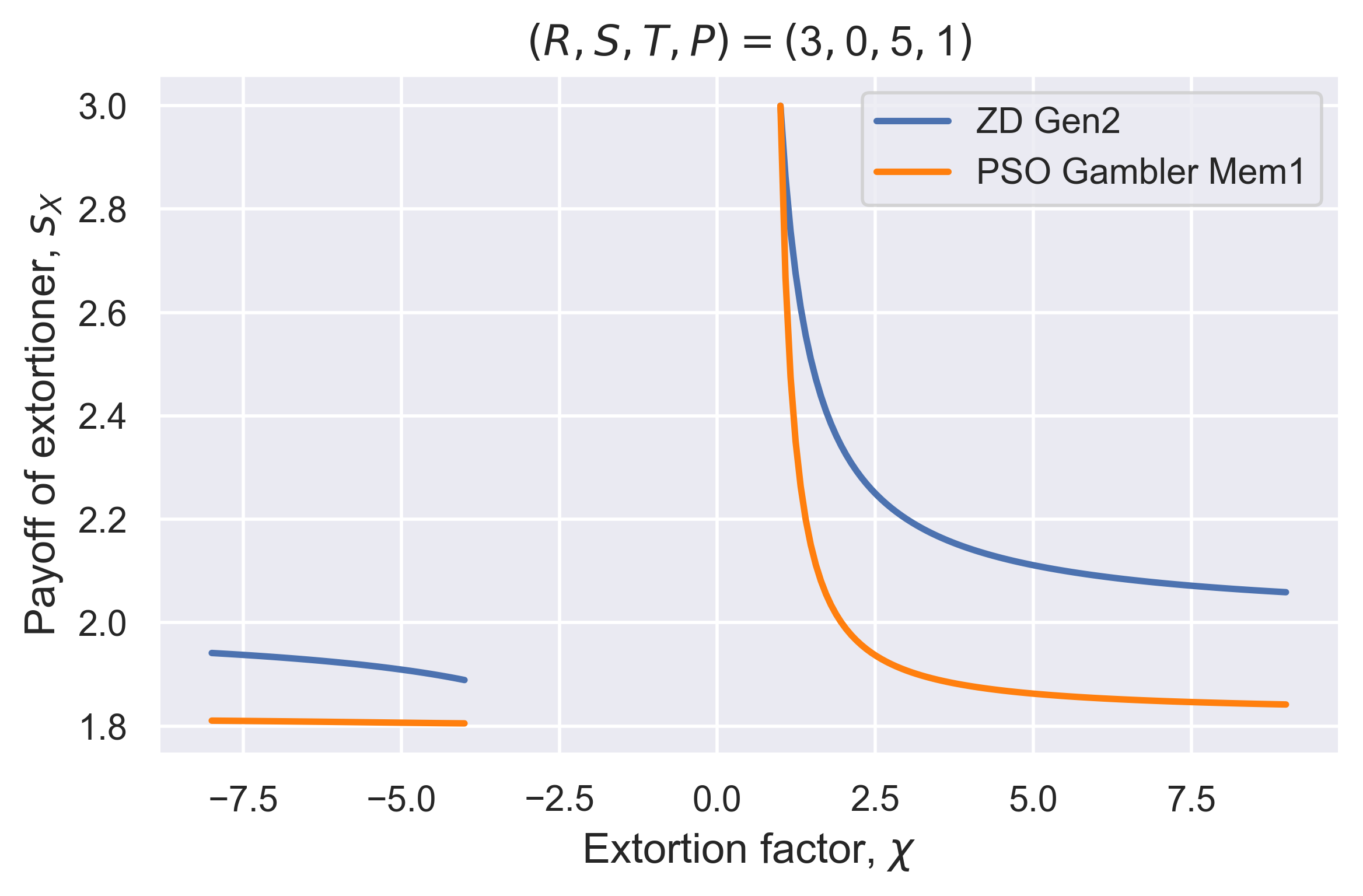}}
\caption{Extortioner's payoff against different strategies. We consider the conventional IPD game.}
\label{con_strategies}
\end{figure}

\begin{figure}[H]
\centering
  \includegraphics[width=0.8\linewidth]{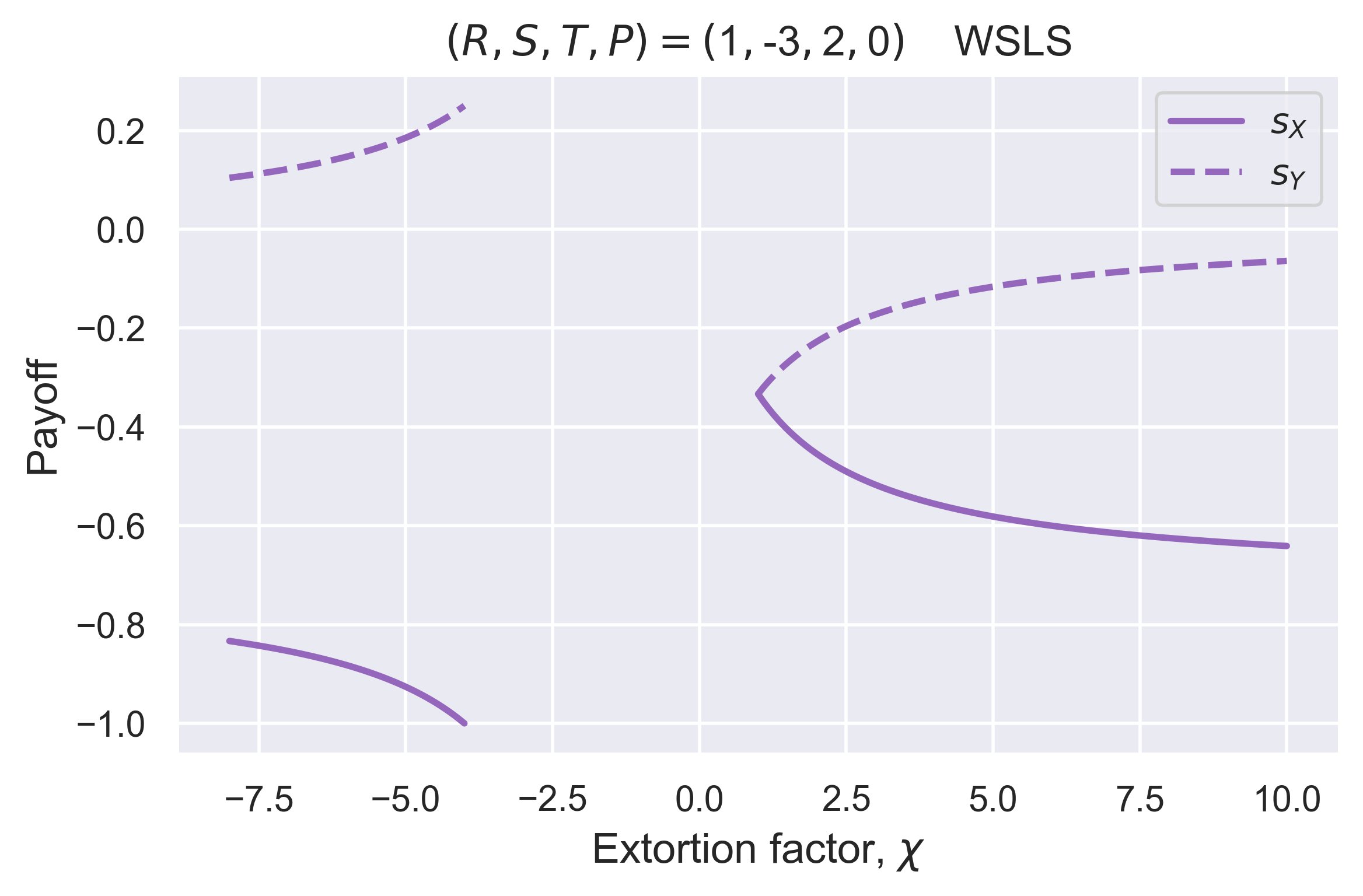}
  \caption{Payoffs of the two players X and Y. The extortionate ZD strategy is applied by X and the WSLS strategy by Y. We work on a IPD game with payments satisfying $T + S < 2P$.}
  \label{abn_wsls}
\end{figure}

The claim above holds at large. When we work on the conventional values of an IPD game, that is, $(R, S, T, P) = (3, 0, 5, 1)$, it is true for common strategies such as ALLC, whose $\bm{q} = (1, 1, 1, 1)$, Random, whose $\bm{q} = (0.5, 0.5, 0.5, 0.5)$, and Win-Stay Lose-Shift (WSLS), whose $\bm{q} = (1, 0, 0, 1)$ (see Figure \ref{con_strategies} (a)). Also, if we assume that $T + S > 2P$ and the opponent Y fixes its cooperation rate by taking an unconditional memory-one strategy with $\bm{q} = (q, q, q, q)$, routine calculation shows that the derivative of ZD's payoff will be
\begin{equation}
\msmall{
\frac{d s_X}{d \chi} = \frac{q(T - S)[(T - R)q + (P - S)(1 - q)][(R - P)q + (T + S -  2P)(1 - q)]}{f^2(\chi)},}
\end{equation}
where $f(\chi) = [(R - S)q + (T - P)(1 - q)]\chi + [(T - R)q + (P - S)(1 - q)]$ is a linear function of $\chi$. Apparently, the derivative is positive as long as $q$ is nonzero and hence $s_X$ is an increasing function of $\chi$ on both branches. 

However, the above results may not tell the complete story.  Figure~\ref{con_strategies}(b) indicates that even for the conventional IPD game, an extortioner's payoff does not necessarily increase with $\chi$. For another example, if $(R, S, T, P) = (1, -3, 2, 0)$, Figure~\ref{abn_wsls} shows that the well-known strategy WSLS, whose $\bm{q} = (1, 0, 0, 1)$, can not only cause $s_X$ to decrease with $\chi$ but also bludgeon the extortioner into accepting a payoff less than the punishment $P$. 

Therefore, given the four payoffs $R$, $S$, $T$ and $P$ satisfying $T > R > P > S$ and $T + S < 2R$, there exist strategies playing against which an extortioner's payoff may not always increase as $\chi$ increases on either branch. The monotonicity of $s_X$ is dependent on the opponent's strategy. In particular, we have found a set of memory-one strategies that cannot be extorted by any ZD strategy with extortion factor $\chi$ greater than $1$.  We refer to them as unbending strategies.
\begin{defn}
A memory-one strategy is unbending if for any extortionate ZD co-player with extortion factor $\chi > 1$, the co-player's payoff is independent of $\phi$ and monotonically decreases with respect to $\chi$. 
\end{defn}
\par

Contrary to the current consensus on ZD strategies, if $\bm{q}$ is an unbending strategy, it will force the payoff of the extortionate ZD co-player to decrease as $\chi$ increases from $1$ to $+\infty$ if $T + S \geq 2P$. It may even impose a payoff less than $P$ for the co-player should $T + S < 2P$. \par

\section{The role played by \texorpdfstring{$\phi$}{e} and the maximization of payoffs}

\begin{figure}[H]
    \centering
    \subfloat[$s_X(\bm{q}_1, \chi, \phi)$]{\includegraphics[width=0.5\linewidth]{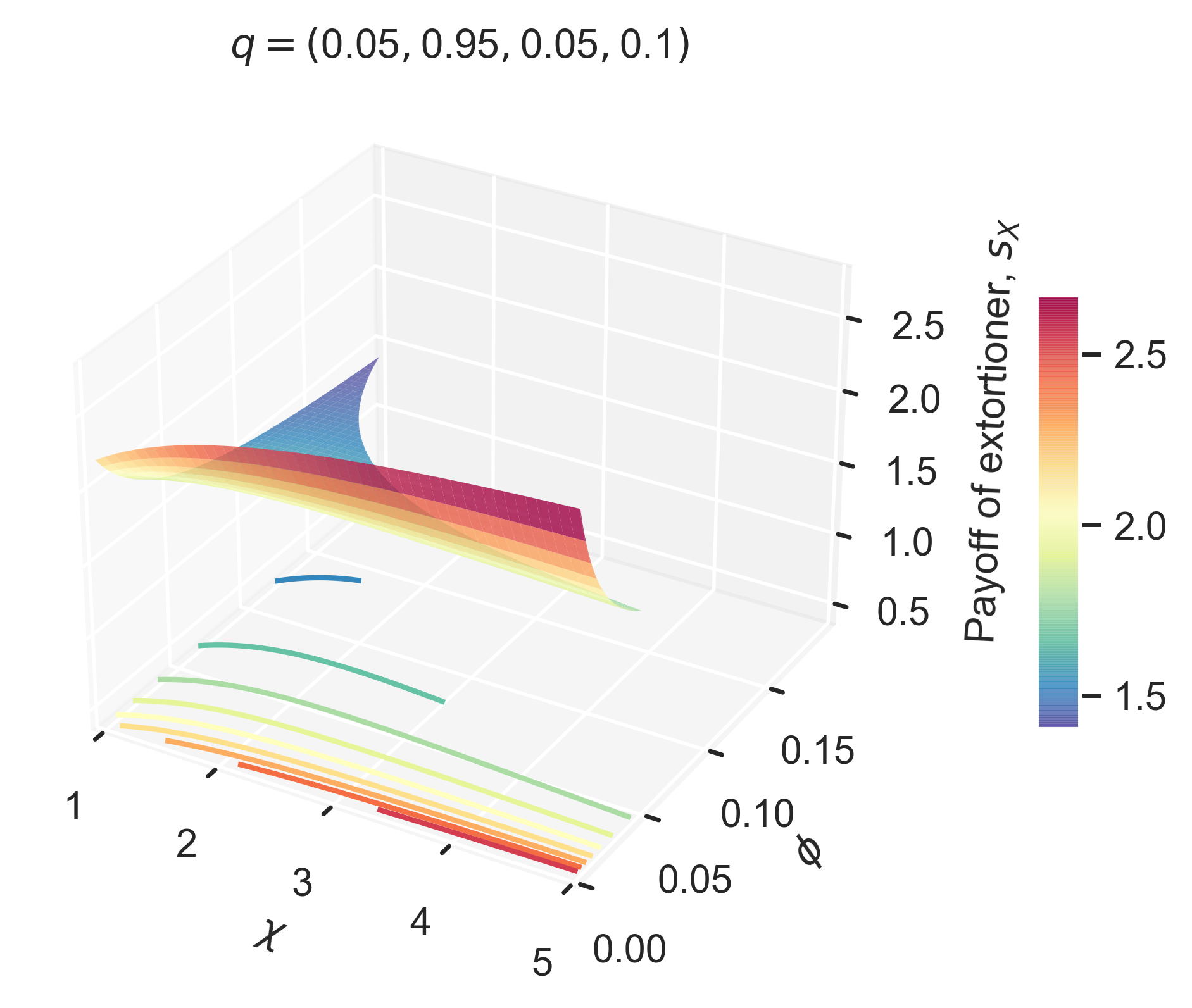}} 
    \subfloat[$s_Y(\bm{q}_1, \chi, \phi)$]{\includegraphics[width=0.5\linewidth]{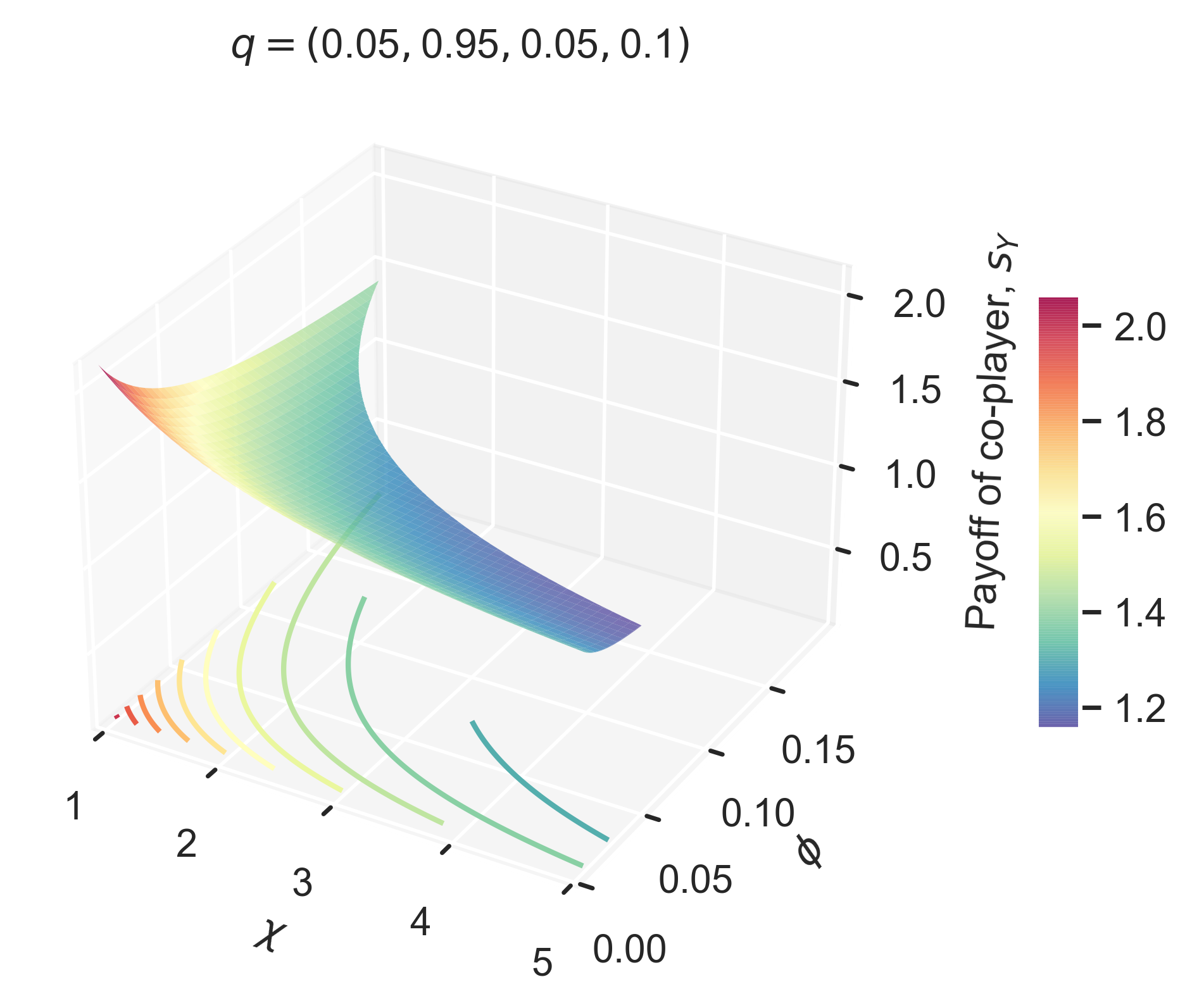}} 
    \\
     \subfloat[$s_X(\bm{q}_2, \chi, \phi)$]{\includegraphics[width=0.5\linewidth]{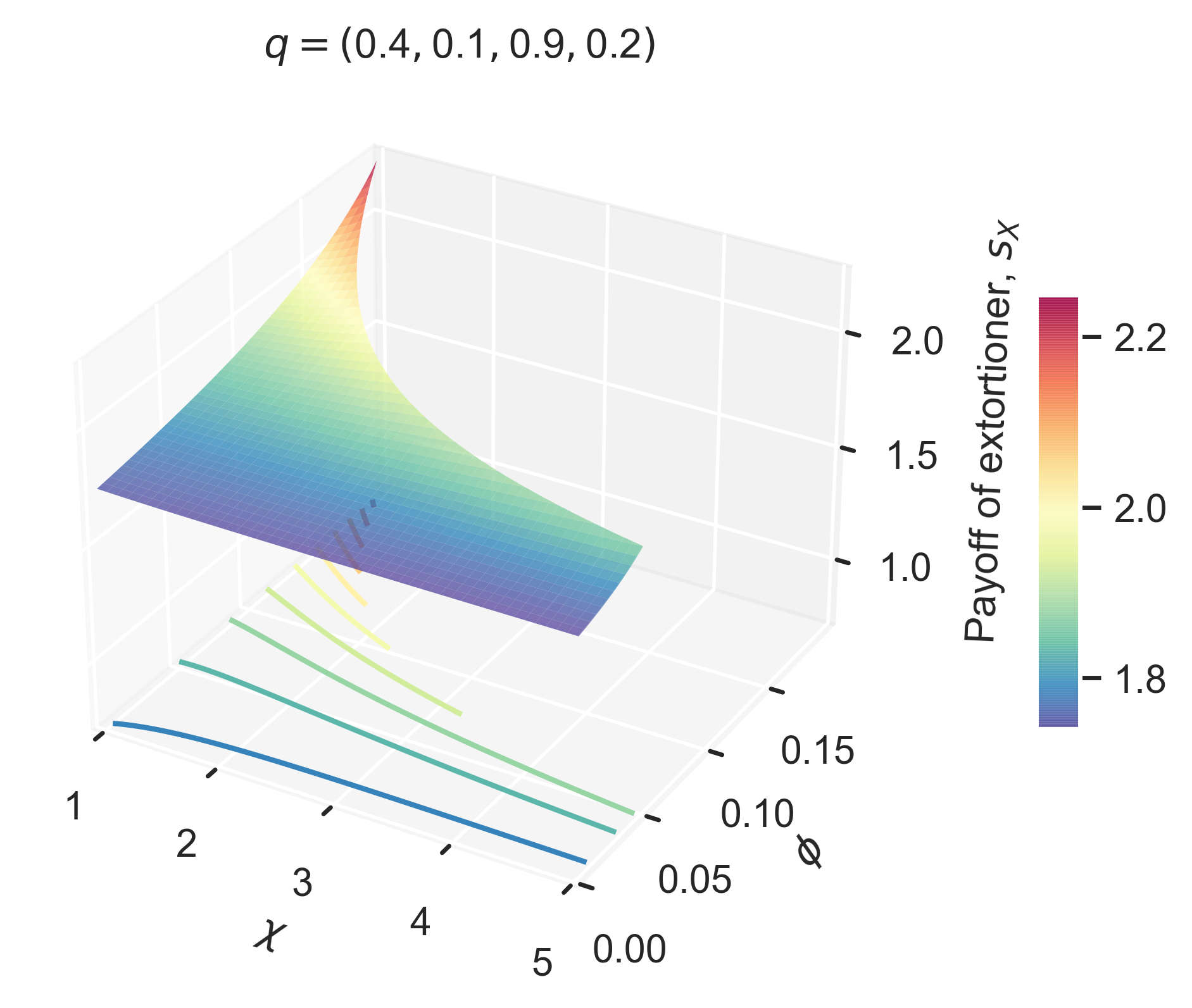}}
     \subfloat[$s_Y(\bm{q}_2, \chi, \phi)$]{\includegraphics[width=0.5\linewidth]{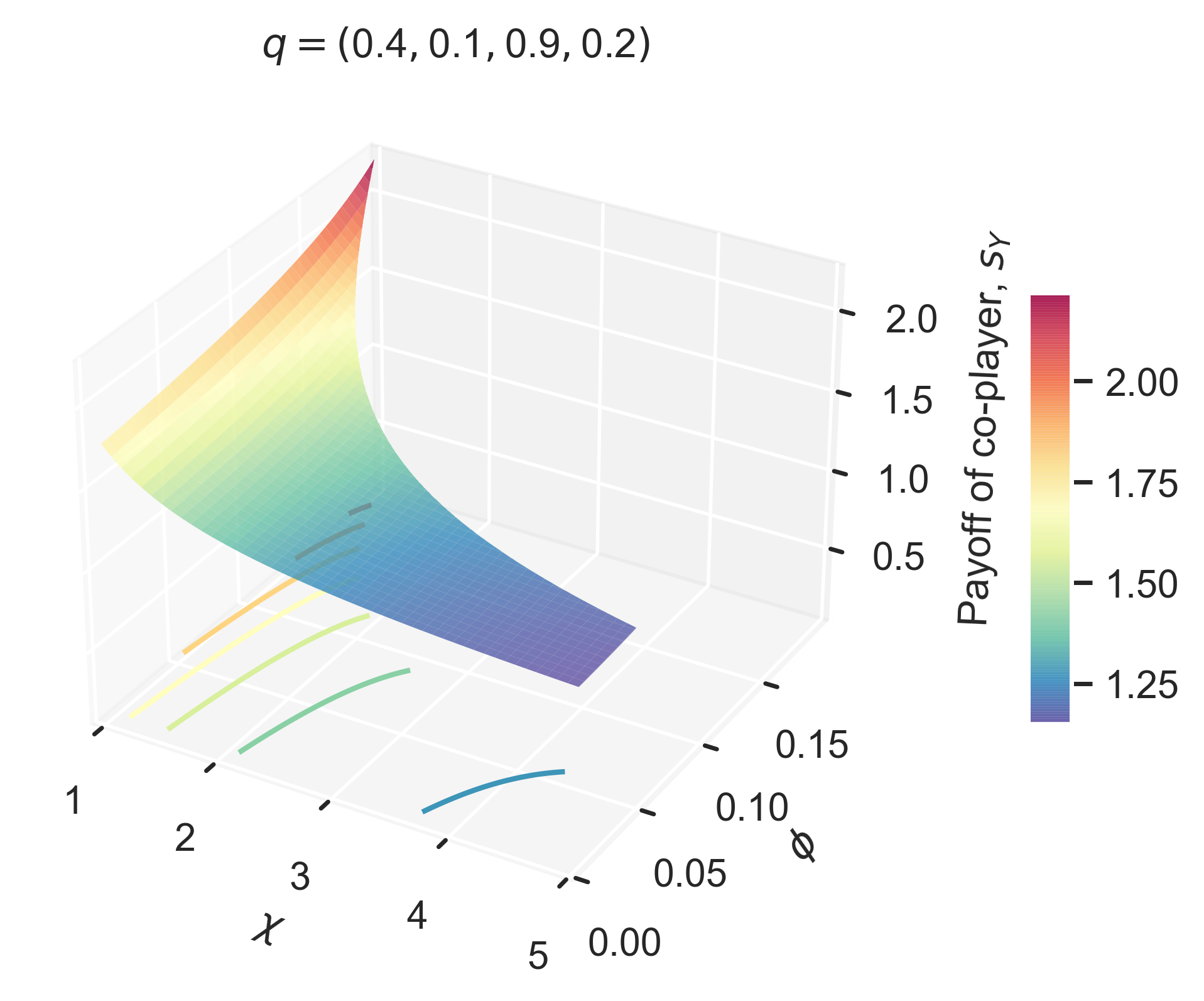}}
     \\
     \subfloat[$s_X(\text{TFT}, \chi, \phi)$]{\includegraphics[width=0.5\linewidth]{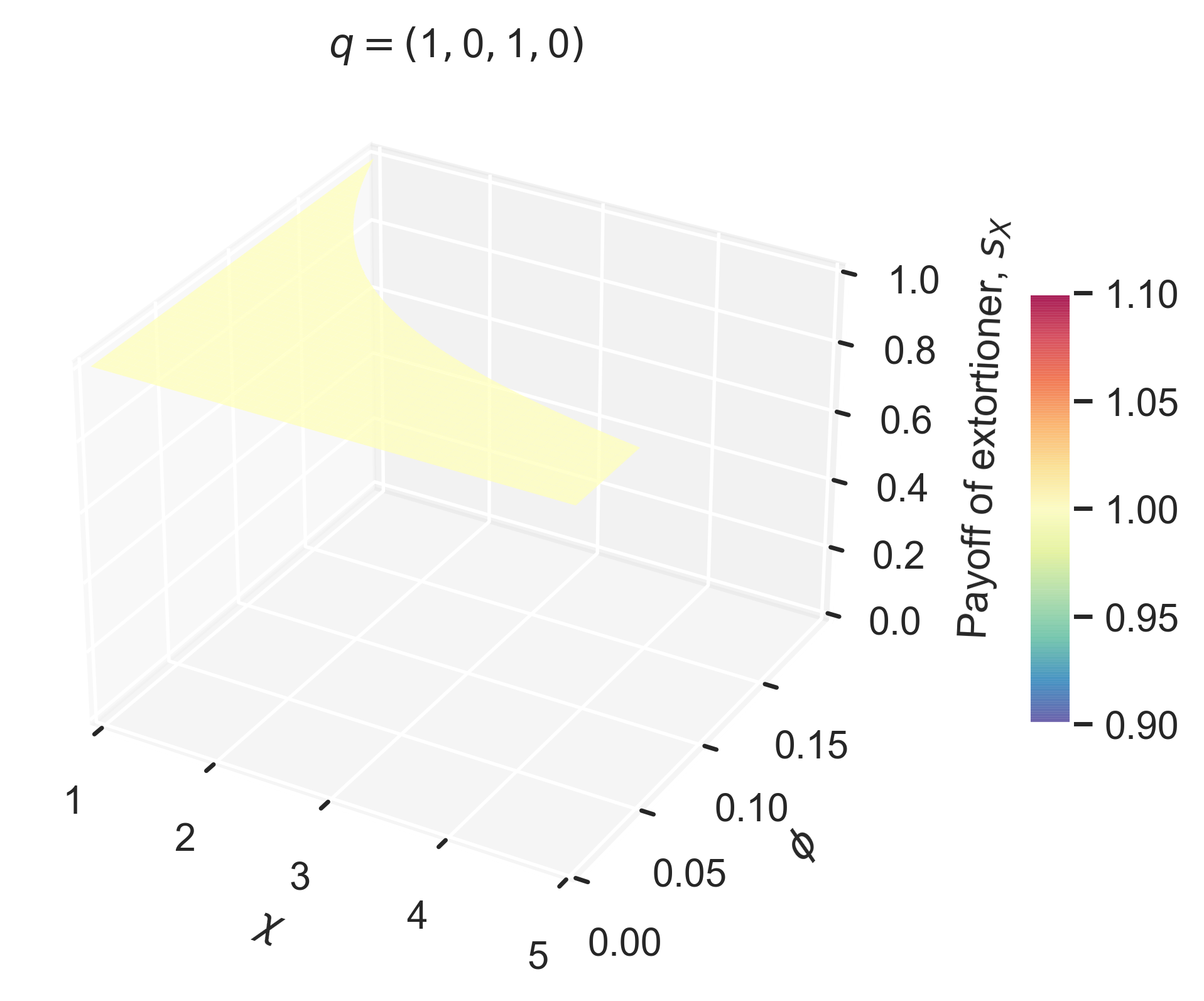}} 
     \subfloat[$s_Y(\text{TFT}, \chi, \phi)$]{\includegraphics[width=0.5\linewidth]{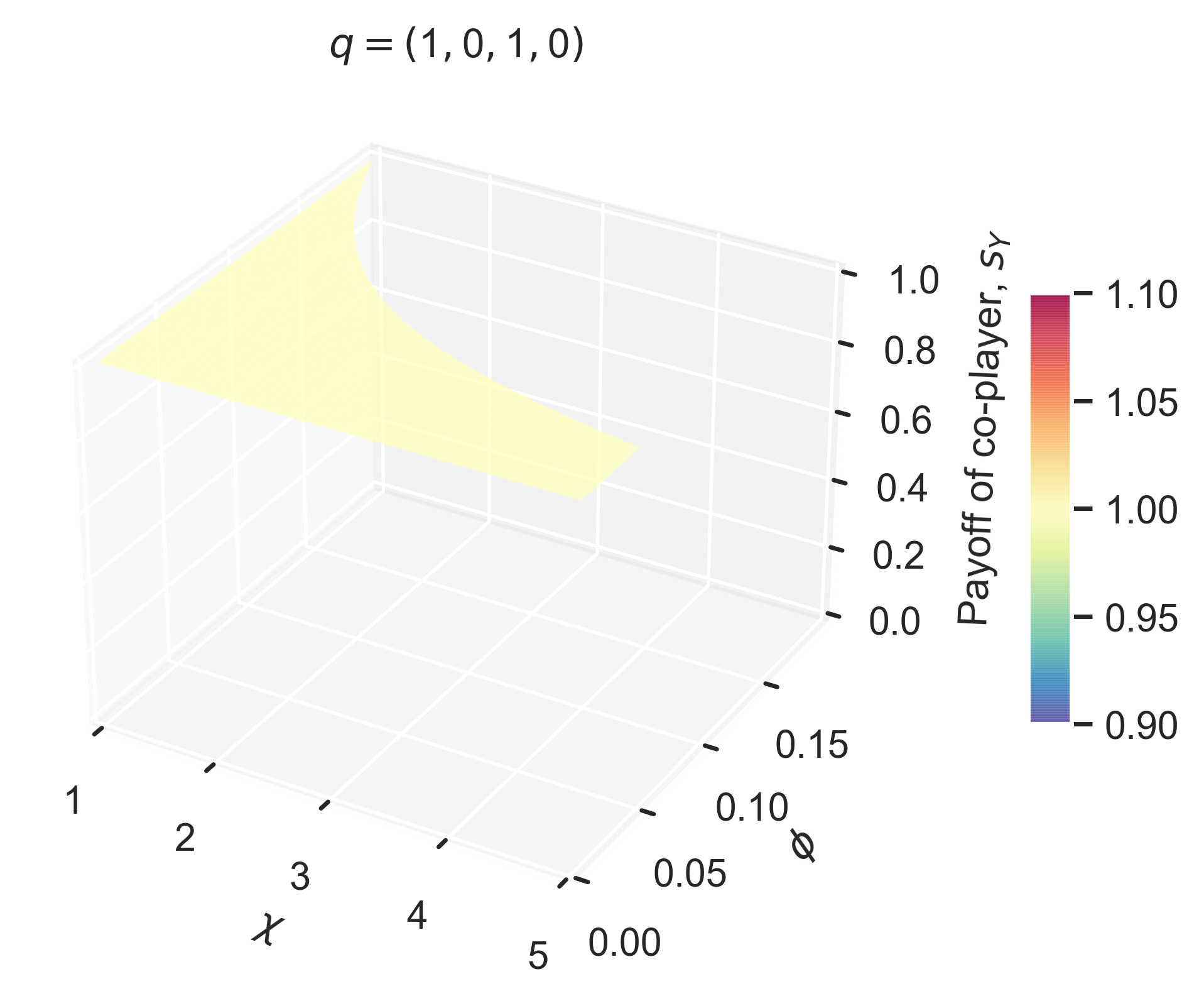}} 
     \caption{Surfaces of $s_X$ ($s_Y$). We work on the conventional IPD game where strategy $\bm{q}$ is applied by player Y. A point $(x, y, z)$ in the 3-dimensional space represents the corresponding tuple consisting of the extortion factor $\chi$, the parameter $\phi$, and extortioner's payoff $s_X$ (opponent's payoff $s_Y$). Its color indicates the value of $s_X$ ($s_Y$). The contour curves on the $xy$-plane are also given.}
     \label{phi_dependency}
\end{figure}

Recall that another parameter $\phi$ appears in the expression of $\bm{p}$ (see Equation~\ref{ZD_strategy}). If the opponent Y tries a common strategy such as ALLC, the payoff of X will be $(T - S)(R - P)\chi/[(R - S)\chi + T - R] + P$, which does not involve $\phi$. However, when X plays against a more general opponent, its payoff is oftentimes dependent on $\phi$. Take the conventional IPD game as an example (see Figure~\ref{phi_dependency}). It is apparent that $s_X$ can either increase or decrease with $\phi$. And even if we have fixed $\bm{q}$, it is possible to alter the monotonicity of $s_X$ by choosing different $\phi$. 

\begin{figure}[H]
    \centering
     \subfloat[$q_1 = q_2 = 1$]{\includegraphics[width=0.5\linewidth]{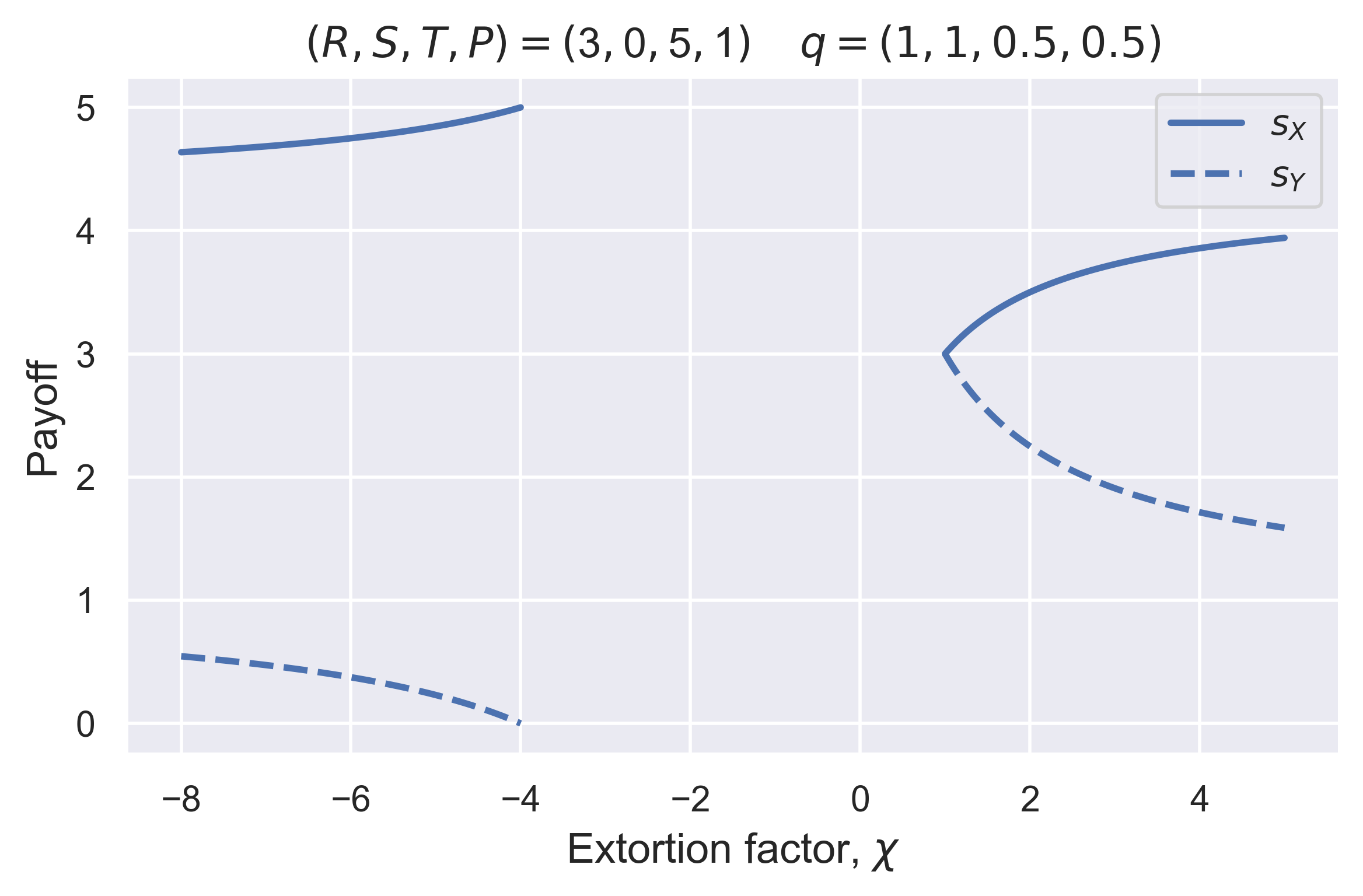}}
     \subfloat[$q_4 = 0$]{\includegraphics[width=0.5\linewidth]{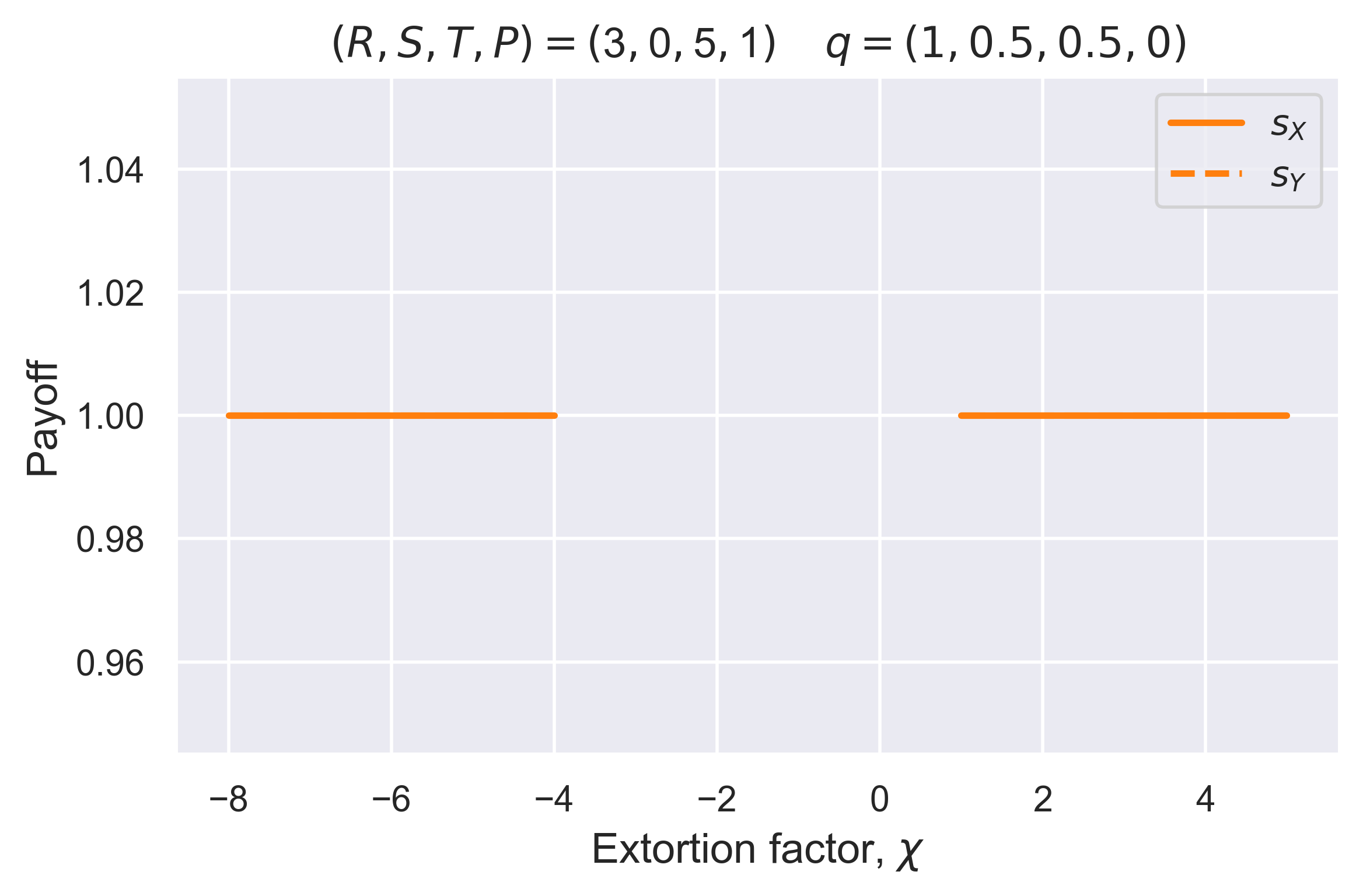}} 
     \\
     \subfloat[$q_1 = 1$ and $q_3 = 0$]{\includegraphics[width=0.5\linewidth]{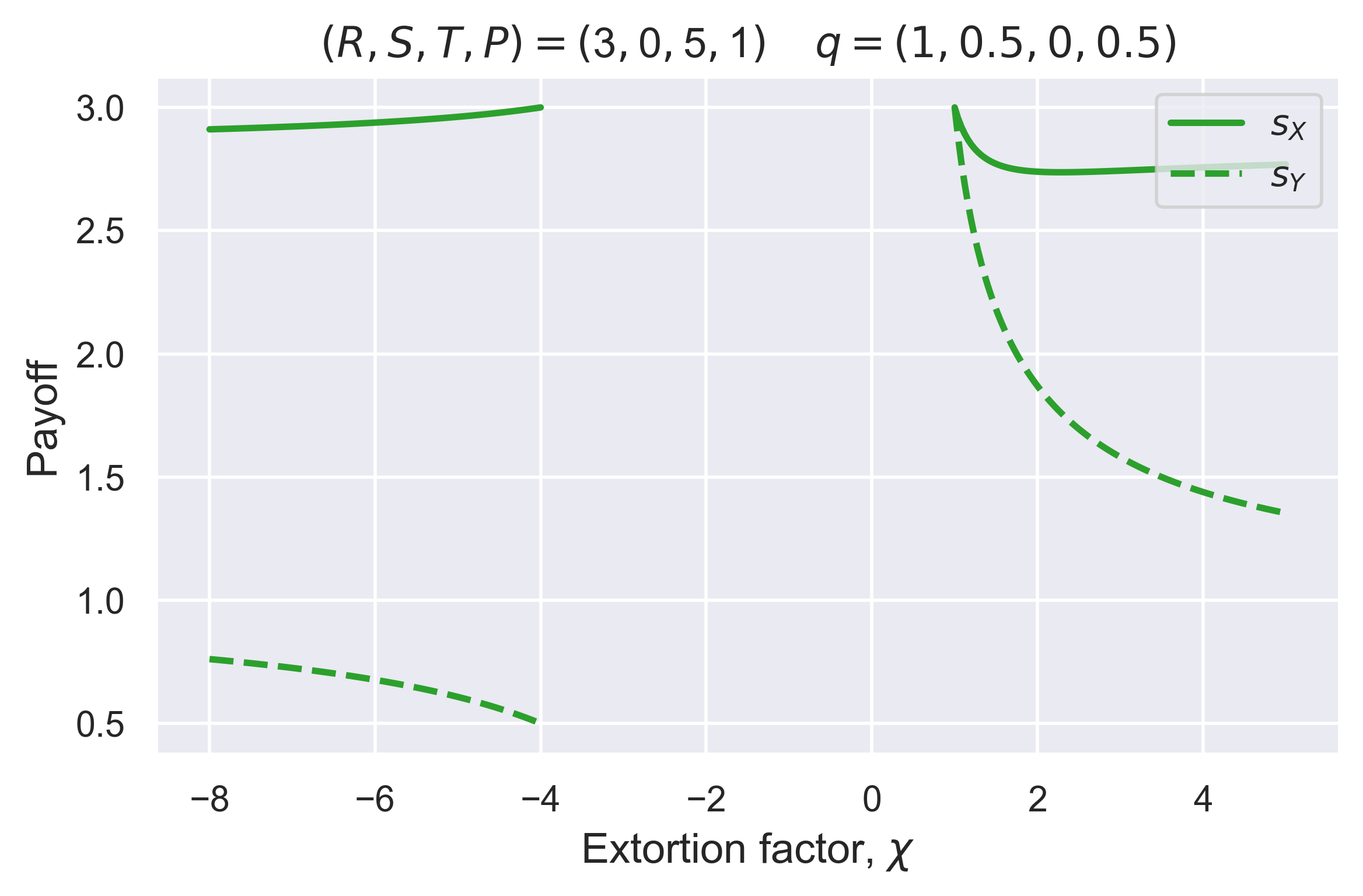}} 
     \subfloat[$q_2 = q_3 = 0$]{\includegraphics[width=0.5\linewidth]{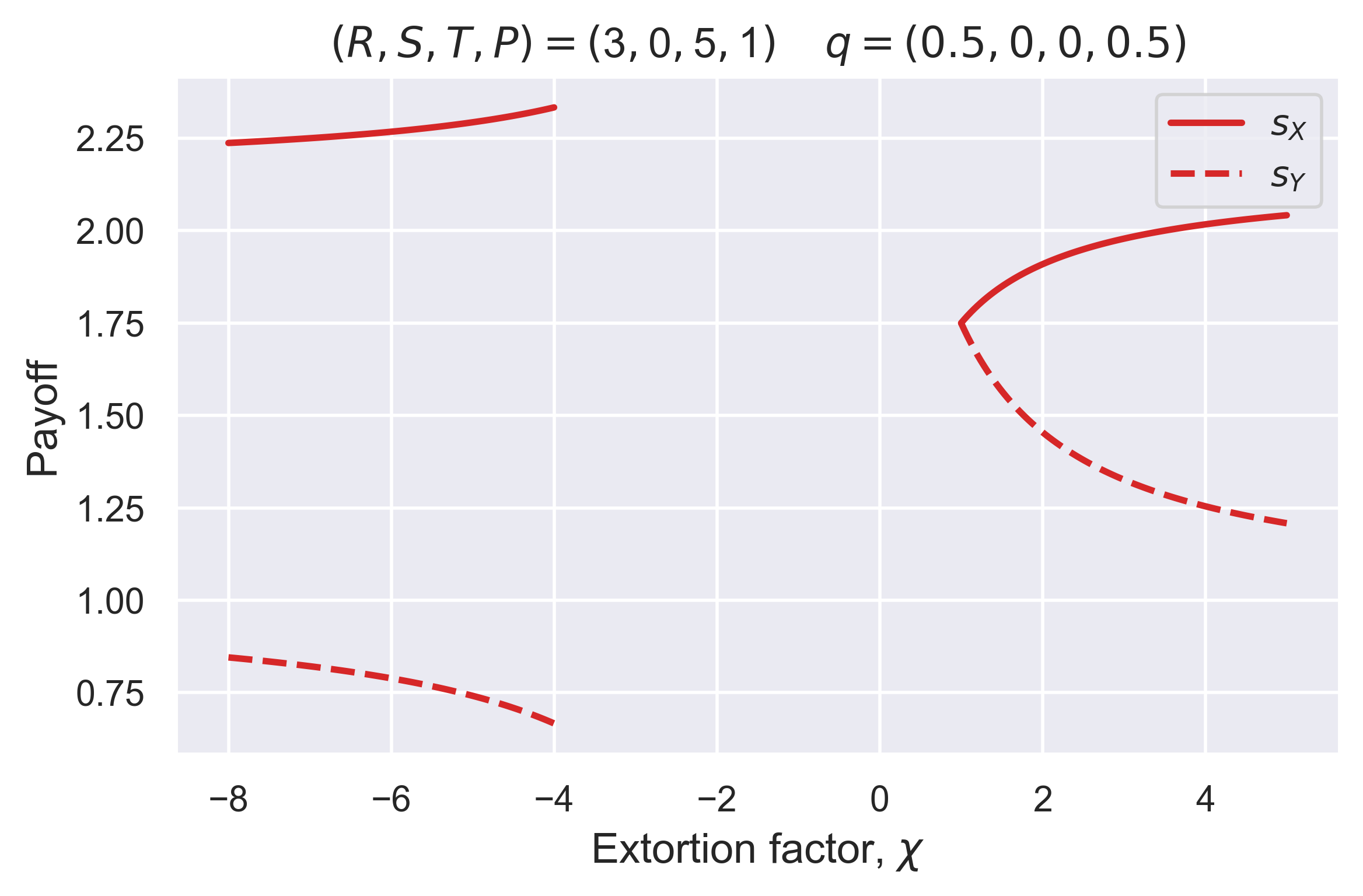}} 
     \\
     \subfloat[$q_1 = q_2 = q_3$]{\includegraphics[width=0.5\linewidth]{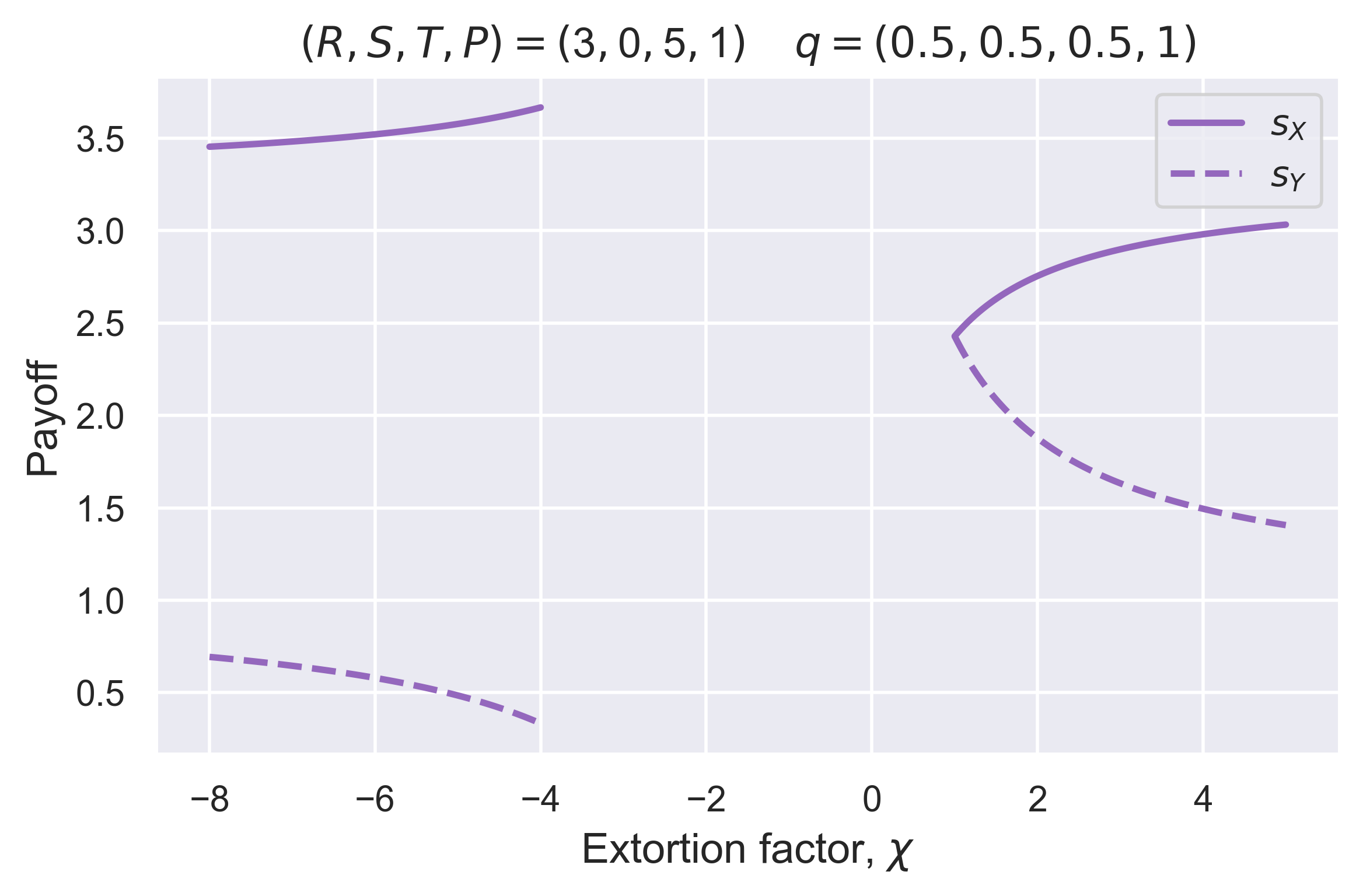}}
     \subfloat[$q_4 = h_D$]{\includegraphics[width=0.5\linewidth]{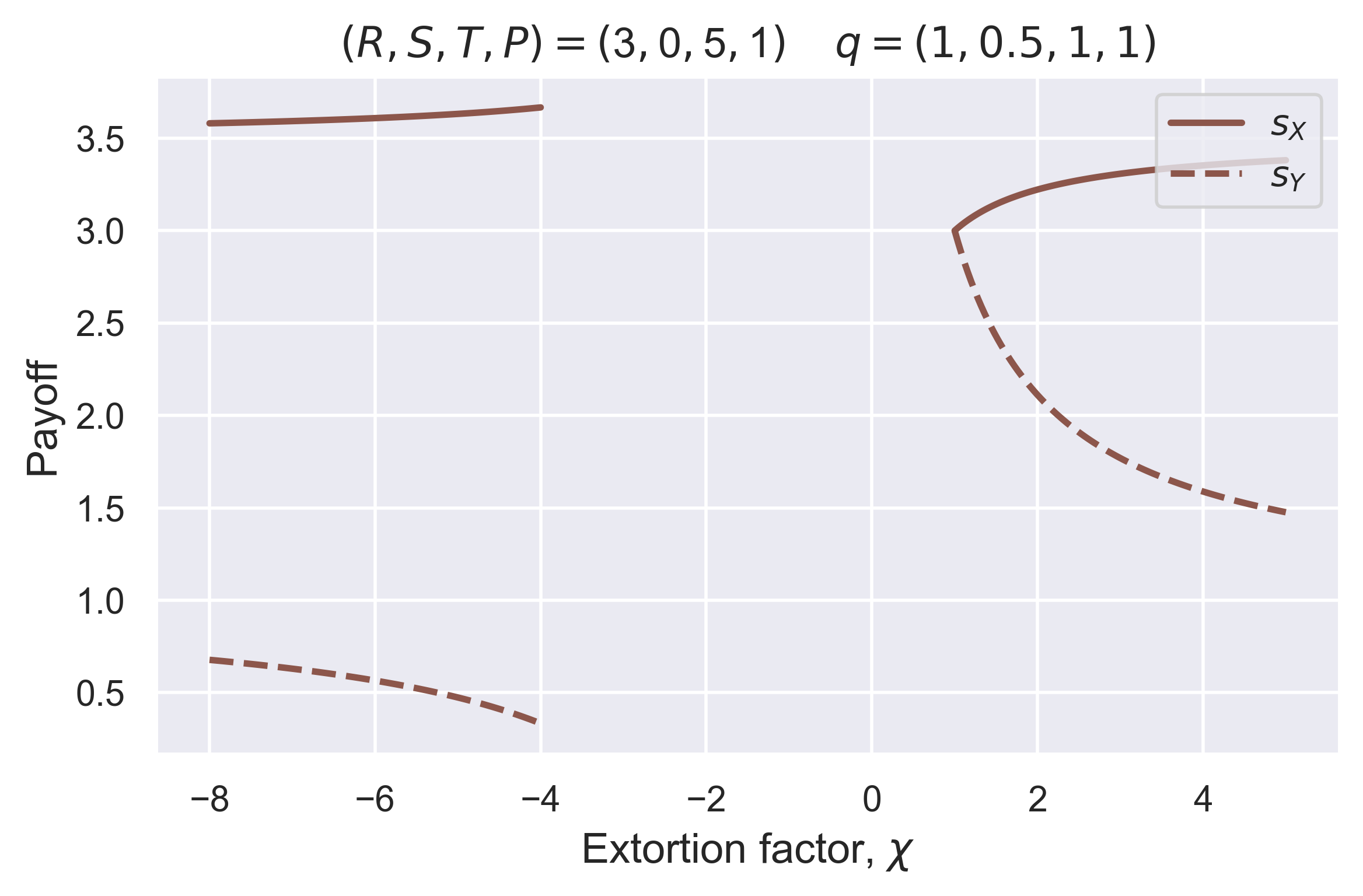}}
     \caption{Payoffs of the two players X and Y.  We work on the conventional IPD game where strategy $\bm{q}$ is applied by Y.  For every panel, $\bm{q}$ satisfies one of the six solutions in Table~\ref{phi_independency}.}
     \label{monotonicity}
\end{figure}

Therefore, unbending strategies should be a subset of strategies playing against which the extortioner's payoff is independent of $\phi$. To find the latter, we consider a necessary condition 
\begin{equation}
s_X(\bm{q}, \chi, \phi^{\text{upper}}) = s_X(\bm{q}, \chi, \frac{\phi^{\text{upper}}}{2}).
\end{equation}
The equation has six solutions, which further yield six expressions of $s_X$ (see Table~\ref{phi_independency}). These $s_X$'s are all free of $\phi$. Therefore, the condition is both necessary and sufficient. Examples are given in Figure~\ref{monotonicity}. 

\begin{table}[H]
\centering
\tabulinesep=1.5mm
\begin{tabu}{c c}
\hline
\rowcolor[HTML]{EFEFEF} 
Solution & $s_X$ \\
$q_1 = q_2 = 1$ & $\frac{(T - S)(R - P)\chi}{(R - S)\chi + (T - R)} + P$ \\
\rowcolor[HTML]{EFEFEF} 
$q_4 = 0$ & $P$ \\
\makecell[cc]{$q_1 = 1$ and \\ $q_3 = 0$} &  $\frac{(T - S)(R - P)\{[(P - S)q_2 + T + S - 2P]\chi + (T - P)q_2 - (T + S - 2P) \}q_4\chi}{f_A(\chi)} + P$\\
\rowcolor[HTML]{EFEFEF} 
$q_2 = q_3 = 0$ & $\frac{(T - S)(T + S - 2P)q_4\chi}{f_B(\chi)} + P$ \\
$q_1 = q_2 = q_3$ &  $\frac{(T - S)[-(T + S - R - P)q_1 + T + S - 2P]q_4\chi}{f_C(\chi)} + P$\\
\rowcolor[HTML]{EFEFEF} 
$q_4 = h_D$ & $\frac{(T - S)[-(T + S - 2P)q_1 + (R - P)(q_2 + q_3) + T + S - R - P]\chi}{f_D(\chi)} + P$ \\
\hline
\end{tabu}
\caption{The solutions of $\bm{q}$ for $s_X$ being independent of $\phi$ and the corresponding expressions of $s_X$. Here, $h_D$ is a multivariate linear function of $q_1$, $q_2$ and $q_3$.  Additionally, $f_A$ is a quadratic function of $\chi$, and $f_B$, $f_C$, and $f_D$ are linear functions of $\chi$.}
\label{phi_independency}
\end{table}

For a few side notes, we emphasize that if the opponent's strategy $\bm{q}$ fulfills any condition other than the third one in Table~\ref{phi_independency}, the extortioner's payoff will have the same monotonicity for $\chi < \chi^{\text{upper}}$ and $\chi > 1$. The proof is trivial. Whenever $s_X$ is a linear rational function of $\chi$, its curve is a hyperbola and is monotonically increasing or decreasing on both the left and the right branches. Besides, when $\chi = 1$, the IPD game is likely to enter an absorbing state. Given Equation~\ref{ZD_strategy}, we have $p_1 = 1$ when $\chi = 1$. If the opponent satisfies $q_1 = 1$, there will always be mutual cooperation and hence $s_X = s_Y = R$. On the other hand, we will always have $s_X = s_Y = P$ when $q_4 = 0$. \par

%As to the third requirement under which $s_X$ is a rational function of degree $2$, we will explore its monotonicity in detail in the section after the next.

Therefore, without loss of generality, we assume that $\chi > 1$ and $0 < q_4 \leq 1$ in the discussion that follows. \par

\section{The unbending strategies}

Given the above results, we are now ready to figure out the set of unbending strategies. \par

It is trivial to show that if $q_1 = q_2 = 1$ and $s_X = (T - S)(R - P)\chi/[(R - S)\chi + (T - R)] + P$, the derivative $ds_X/d\chi = (T - R)(T - S)(R - P)/[(R - S)\chi + (T - R)]^2$ will always be positive. Therefore, we only need to work on the last four solutions and explore the 4-dimensional space generated by $\bm{q} = (q_1, q_2, q_3, q_4)$ where $0 \leq q_i \leq 1$ for $i \in \{1, 2, 3\}$ and $0 < q_4 \leq 1$. We aim to obtain the set of $\bm{q}$'s against which player X has a monotonically decreasing payoff $s_X$ with respect to $\chi$. \par

Notice that $s_X - P$ can be rewritten as a quadratic or linear rational function of $\chi$ (recall Table~\ref{phi_independency}) under any of the four solutions:
\begin{equation}
s_X - P = \frac{\chi(a_1\chi + a_0)}{d_2\chi^2 + d_1\chi + d_0},
\end{equation}
which may sometimes degenerate to 
\begin{equation}
s_X - P= \frac{a_0\chi}{d_1\chi + d_0}.
\end{equation}
The derivative of $s_X$ is henceforth
\begin{equation}
s'_X = \frac{e_2\chi^2 + e_1\chi + e_0}{(d_2\chi^2 + d_1\chi + d_0)^2},
\end{equation}
or 
\begin{equation}
s'_X = \frac{e_0}{(d_1\chi + d_0)^2},
\end{equation}
where 
\begin{equation}
\begin{cases}
e_2 = a_1d_1 - a_0d_2,\\
e_1 = 2a_1d_0,\\
e_0 = a_0d_0.
\end{cases}
\end{equation}
\par

As a result, we only need to take into account the sign of $g(\chi) = e_2\chi^2 + e_1\chi + e_0$ to decide the monotonicity of $s_X$. If $s_X$ is a decreasing function of $\chi$, $g(\chi)$ has to satisfy one of the following arguments:

\begin{enumerate}[label = (\roman*)]
\item $e_2 = e_1 = 0$ and $e_0 < 0$.
\item $e_2 = 0$, $e_1 < 0$ and $g(1) = e_1 + e_0 \leq 0$.
\item $e_2 < 0$, $g(1) = e_2 + e_1 + e_0 \leq 0$, and $g(-\frac{e_1}{2e_2}) < 0$ if $-\frac{e_1}{2e_2} > 1$ (the highest point of the parabola opening downwards is below the x-axis).
\end{enumerate}
\par

The above criterion yields the (explicit) expressions of unbending strategies (see Table~\ref{four_classes}). 

\begin{table}[H]
\centering
\tabulinesep=1mm
\begin{tabu}{|c | c | c | c |}
\hline
\backslashbox{Class}{Case} & $T + S > 2P$ & $T + S = 2P$ & $T + S < 2P$ \\
\hline
\makecell[cc]{A \\ $q_1 = 1, q_3 = 0$} & $\begin{cases} q_a < q_2 < 1 \\ 0 < q_4 \leq h_A\end{cases}$ &
$\begin{cases} 0 < q_2 < 1 \\ 0 < q_4 \leq h_A (\ast)\end{cases}$ & $\begin{cases} 0 \leq q_2 < 1 \\ 0 < q_4 \leq h_A\end{cases}$ \\
\hline
\makecell[cc]{B \\ $q_2 = q_3 = 0$} & $\emptyset$ & $\emptyset$ & $\begin{cases} 0 \leq q_1 \leq 1 \\ 0 < q_4 \leq 1\end{cases}$ \\
\hline
\makecell[cc]{C \\ $q_1 = q_2 = q_3$} & $\begin{cases} q_C < q_1 < 1 \\ 0 < q_4 < h_C\end{cases}$ &
$\begin{cases} q_C < q_1 < 1 \\ 0 < q_4 < h_C\end{cases}$ & \makecell[cc]{$\begin{cases} q_C <  q_1 < 1 \\ 0 < q_4 < h_C \end{cases}$ \\ or \\ 
$\begin{cases}0 \leq q_1 < q_{c} \\ 0 < q_4 \leq 1\end{cases}$} \\
\hline
\makecell[cc]{D \\ $q_4 = h_D$} & \multicolumn{3}{c|}{$\begin{cases} 0 \leq q_1, q_2, q_3 \leq 1 \\ k(q_1, q_2, q_3) < 0 < K(q_1, q_2, q_3)\end{cases}$} \\
\hline
\end{tabu}
\caption{Four classes of unbending strategies.}
\label{four_classes}
\end{table}

The boundary functions in Table~\ref{four_classes} are given in Table~\ref{boundary_function} and the boundary values are in Table~\ref{boundary_value}. The $(\ast)$ after the piecewise function $h_A$ indicates that the equality can only hold if the value of the function equals $1$ and is not evaluated at the endpoint. \par

\begin{table}[H]
\centering
\tabulinesep=1mm
\begin{tabu}{|c | c | c|}
\hline
Class & \multicolumn{2}{c|}{Boundary Function}  \\ \hline 
\multirow{3}{*}{A} & $\msmall{T + S > 2P}$ & $h_A(q_2) = \begin{cases} 1, & e_{21} < -e_{20}\\ -\frac{e_{20}}{e_{21}}. & \text{otherwise}\end{cases}$
\\ \cline{2-3} 
& $\msmall{T + S = 2P}$ &  $h_A(q_2) = \begin{cases} 1, & q_2 < \frac{3(2R - T - S)}{4R - T - 3S}\\ \frac{(2R - T - S)(1 - q_2)}{2[(R - S)q_2 - (2R - T - S)]}. & \text{otherwise}\end{cases}$
\\ \cline{2-3} 
& $\msmall{T + S < 2P}$ &  $h_A(q_2) = \begin{cases}1, & q_2 <\frac{(R - P) + (2R - T - S)}{2R - P - S}\\ \frac{(R - P)(1 - q_2)}{(R - S)q_2 - (2R - T - S)}. & \text{otherwise}\end{cases}$
\\ \hline
C & \multicolumn{2}{c|}{$h_C(q_1) = \frac{[(R - S)q_1 - (P - S)](1 - q_1)}{(T - S) - (R - S)q_1}$}  \\ \hline
D & \multicolumn{2}{c|}{\makecell[cc]{$h_D = \frac{-(T + S - 2P)q_1 + (R - P)(q_2 + q_3) + T + S - R - P}{2R - T - S}$ \\ \\ $\begin{cases}k(q_1, q_2, q_3) = (T + S - 2P)(1 - q_1) - (R - P)(1 - q_2) + (R - P)q_3, \\ K(q_1, q_2, q_3) = (T - S)(1 - q_1) - (R - S)(1 - q_2) + (T - R)q_3. \end{cases}$}} \\ \hline
\end{tabu}
\caption{Boundary functions for unbending strategies. The expressions of $e_{21}$ and $e_{20}$ are given in the next section.}
\label{boundary_function}
\end{table}

\begin{table}[H]
\centering
\tabulinesep=1mm
\begin{tabu}{|c | c | c |}
\hline
Class & A & C  \\ \hline
Value & $q_a = \frac{(T + S - 2P)(P - S)}{(T - P)(T - S) - (P - S)^2}$ 
& $\begin{cases}q_c = \frac{2P - T - S}{R + P - T - S} \\ q_C = \frac{P - S}{R - S}\end{cases}$
%& $\begin{cases}q_d = \frac{T - R}{T - P}\\ q_D = \frac{2P - R - S}{P - S}\end{cases}$
 \\ \hline 
\end{tabu}
\caption{Boundary values for unbending strategies.}
\label{boundary_value}
\end{table}

We now discuss the four different classes in detail. \par

\section{Class A of unbending strategies}

The first class is under the restriction of $q_1 = 1$ and $q_3 = 0$ where the extortioner's payoff $s_X$ is a quadratic rational function of $\chi$ as shown in the previous section. We have
\begin{equation}
\mfootnotesize{
s_X = \frac{(T - S)(R - P)\{[(P - S)q_2 + T + S - 2P]\chi + (T - P)q_2 - (T + S - 2P) \}q_4\chi}{f_A(\chi)} + P.}
\end{equation}
The denominator $f_A(\chi) = d_{A2}\chi^2 + d_{A1}\chi + d_{A0}$ is a quadratic function of $\chi$ with
\begin{equation}
\mfootnotesize{
\begin{cases}
d_{A2} = [(T - R)(P - S)q_2 + (T - S)(R - P)]q_4 + (T - P)(R - P)(1 - q_2), \\
d_{A1} = [T(T - P) - S(P - S) - R(T + S - 2P)]q_2q_4 - (T + S - 2P)(R - P)(1 - q_2),\\
d_{A0} = [(T - P)(R - S)q_2 - (T - S)(R - P)]q_4 - (R - P)(P - S)(1 - q_2).\\
\end{cases}}
\end{equation}
\par

Notice that 
\begin{equation}
\mfootnotesize{
\begin{cases}
d_{A2} > 0, \\
f_A(1) = q_2q_4(T - S)^2 \geq 0, \\
2d_{A2} + d_{A1} = (T - S)\{[(T - S)q_2 + (R - P)(2 - q_2)]q_4 + (R - P)(1 - q_2)\} > 0.
\end{cases}}
\end{equation}
That is, $f_A(\chi)$ is a quadratic function which opens upwards and whose line of symmetry $\chi = -d_{A1}/2d_{A2}$ lies to the left of $\chi = 1$. Altogether we get $f_A(\chi) > 0$ for $\chi > 1$. \par

If $q_2 = 0$, we have
\begin{equation}
\mfootnotesize{
\begin{aligned}
s_X &= \frac{(T - S)(T + S - 2P)q_4\chi}{[(T - S)q_4 + T - P]\chi + (T - S)q_4 + P - S} + P,  \\
\frac{ds_X}{d\chi} &= \frac{(T - S)(T + S - 2P)[(T - S)q_4 + P - S]q_4}{\{[(T - S)q_4 + T - P]\chi + (T - S)q_4 + P - S\}^2}.
\end{aligned}}
\label{payoff_A_0}
\end{equation}
And if $T + S = 2P$, we obtain
\begin{equation}
\mfootnotesize{
s_X = \frac{(T - S)(2R - T - S)q_2q_4\chi}{d_1\chi + d_0} + \frac{T + S}{2},\qquad
\frac{ds_X}{d\chi} = \frac{(T - S)(2R - T - S)q_2q_4d_0}{(d_1\chi + d_0)^2},}
\label{payoff_A_eqn}
\end{equation}
where
\begin{equation}
\mfootnotesize{
\begin{cases}
d_{A1} = 2[(T - R)q_2 + (2R - T - S)]q_4 + (2R - T - S)(1 - q_2),\\
d_{A0} = 2[(R - S)q_2 - (2R - T - S)]q_4 - (2R - T - S)(1 - q_2).
\end{cases}}
\end{equation}
\par

The set of unbending strategies depends on the relation between $T + S$ and $2P$. We present the details of our analysis in three different cases:  $T + S > 2P$, $T + S = 2P$, and $T + S < 2P$. \par

\newcommand{\rom}[1]{\uppercase\expandafter{\romannumeral #1\relax}}

\subsection{Case \rom{1}: \texorpdfstring{$T + S > 2P$}{e}}

Before considering a general IPD game satisfying $T + S > 2P$, we first work on two special cases: the conventional IPD game with payments $(R, S, T, P) = (3, 0, 5, 1)$ and the donation game with benefit $b$ and cost $c$. \par

\subsubsection{The conventional IPD game}

\begin{table}[H]
\centering
\tabulinesep=1.5mm
\begin{tabu}{c c c c}
\hline
\rowcolor[HTML]{EFEFEF} 
Payoff Matrix & Class A & $s_X$ & $\frac{ds_X}{d\chi}$ \\
$\begin{bmatrix} 3 & 0 \\ 5 & 1\end{bmatrix}$
&
$\begin{cases}
\frac{3}{19} < q_2 < 1\\
0 < q_4 \leq \frac{2(1 - q_2)(19q_2 - 3)}{3q_2^2 - q_2 + 30}
\end{cases}$
&
$\frac{10[(q_2 + 3)\chi + 4q_2 - 3]q_4\chi}{f_A(\chi)} + 1$  
&
$\frac{10q_4g_A(\chi)}{f^2_A(\chi)}$ \\ 
\hline
\end{tabu}
\caption{Summary of the conventional IPD game if the extortioner's co-player applies an unbending strategy from Class A.}
\label{summary_A_con}
\end{table}

Here, $f_A(\chi)$ is defined as before and $g_A(\chi) = e_{2}\chi^2 + e_{1}\chi + e_{0}$. We have
\begin{equation}
\begin{cases}
d_{A2} = 2(q_2 + 5)q_4 + 8(1 - q_2),\\
d_{A1} =11q_2q_4 - 6(1 - q_2),\\
d_{A0} = 2(6q_2 - 5)q_4 - 2(1 - q_2), \\
\end{cases}
\end{equation}
and
\begin{equation}
\begin{cases}
e_2 = (3q_2^2 - q_2 + 30)q_4 - 2(1 - q_2)(19q_2 - 3),\\
e_1 = 4(q_2 + 3)[(6q_2 - 5)q_4 - (1 - q_2)],\\
e_0 = 2(4q_2 - 3)[(6q_2 - 5)q_4 - (1 - q_2)].
\end{cases}
\end{equation}
In particular, if $q_2 = 0$, we obtain
\begin{equation}
s_X = \frac{15q_4\chi}{(5q_4 + 4)\chi + 5q_4 + 1} + 1, \qquad \frac{ds_X}{d\chi} = \frac{15q_4(5q_4 + 1)}{[(5q_4 + 4)\chi + 5q_4 + 1]^2}.
\end{equation}
\par

\begin{figure}[H]
\centering
\includegraphics[width=0.5\linewidth]{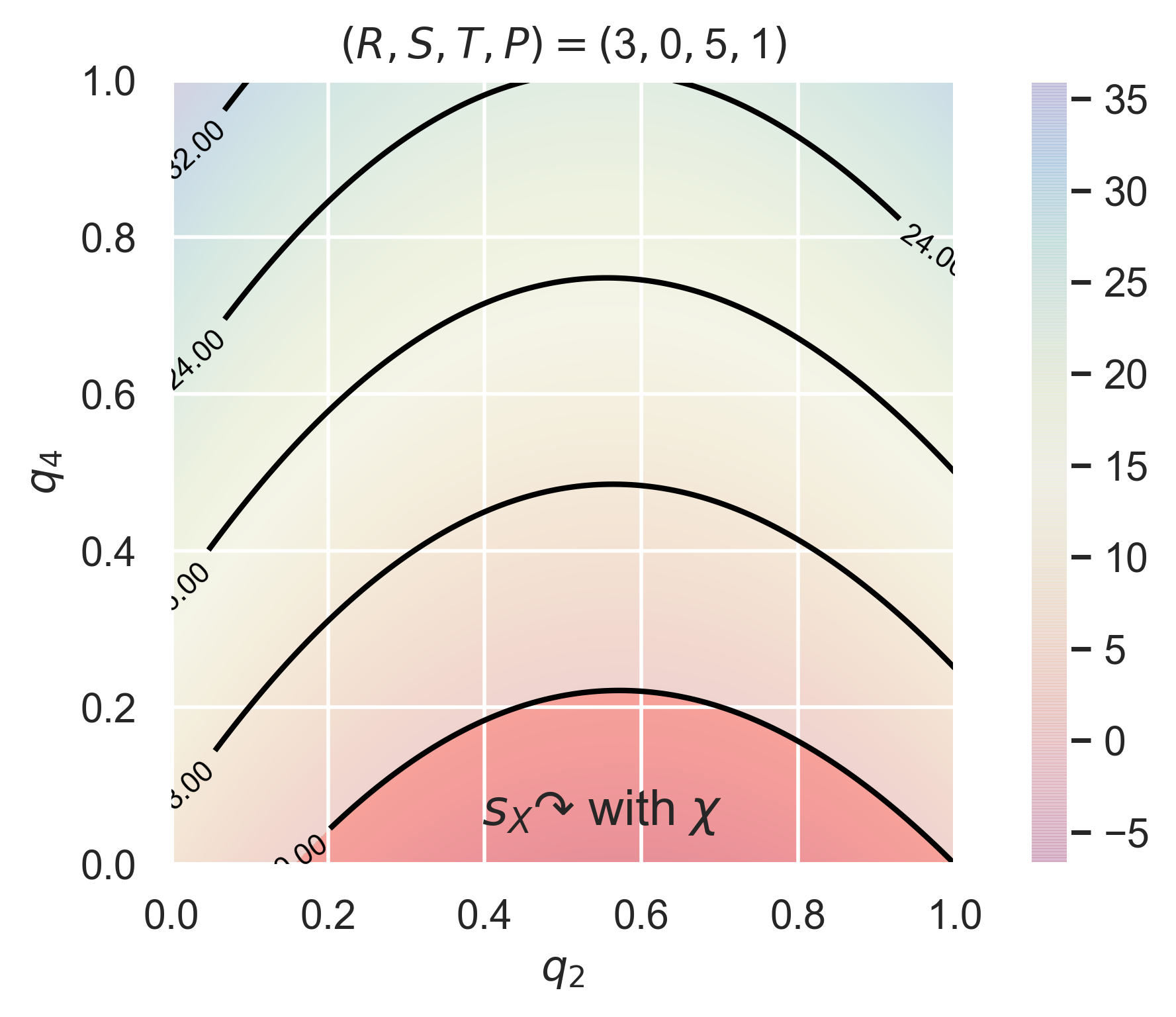}
\caption{Contour curves of $e_2$ for the conventional IPD game. The red region is where $e_2 \leq 0$.}
\label{target_A_con}
\end{figure}

The visualization of Class A, which corresponds to the second column in Table~\ref{summary_A_con}, is given in Figure~\ref{target_A_con}. 

No more than fundamental calculus is needed to show that $e_2 \leq 0$ is the necessary and sufficient condition for the extortioner's payoff to be monotonically decreasing with $\chi$ (that is, $ds_X/d\chi < 0$). Therefore, we will not elaborate on the proof. The explicit expression of $e_2 \leq 0$ under the restriction that $0 < q_2, q_4 \leq 1$ is given in the second column of Table~\ref{summary_A_con}. \par

\subsubsection{The donation game}

\begin{table}[H]
\centering
\tabulinesep=1.5mm
\begin{tabu}{c c}
\hline
\rowcolor[HTML]{EFEFEF} 
Payoff Matrix  & Class A \\
$\begin{bmatrix} b - c & -c \\ b & 0 \end{bmatrix}$
&
$\begin{cases}
q_a < q_2 < 1 \\
0 < q_4 \leq h_A(q_2)
\end{cases}$ \\
\rowcolor[HTML]{EFEFEF} 
$s_X$ & $\frac{ds_X}{d\chi}$ \\  
$\frac{(b + c)(b - c)[(cq_2 + b - c)\chi + bq_2 - (b - c)]q_4\chi}{f_A(\chi)}$ 
&
$\frac{(b + c)(b - c)q_4g_A(\chi)}{f^2_A(\chi)}$ \\ 
\hline
\end{tabu}
\caption{Summary of the donation game if the extortioner's co-player applies an unbending strategy from Class A. The inequalities $b, c > 0$ and $b > c$ always hold.}
\label{summary_A_dg}
\end{table}

The definitions of $f_A(\chi)$ and $g_A(\chi)$ are the same as before. Now we have
\begin{equation}
\begin{cases}
d_{A2} = [c^2q_2 + b^2 - c^2]q_4 + b(b - c)(1 - q_2), \\
d_{A1} = 2bcq_2q_4 - (b - c)^2(1 - q_2),\\
d_{A0} = [b^2q_2 - b^2 + c^2]q_4 - c(b - c)(1 - q_2),\\
\end{cases}
\end{equation}
and
\begin{equation}
e_i = e_{i1}q_4 + e_{i0}, \qquad i \in \{0, 1, 2\},
\end{equation}
where
\begin{equation}
\begin{cases}
e_{21} = w_2q_2^2 + w_1q_2 + w_0,\\
e_{20} = (b - c)(1 - q_2)(uq_2 + v),\\
e_{11} = 2(cq_2 + b - c)(b^2q_2 - b^2 + c^2),\\
e_{10} = -2c(b - c)(1 - q_2)(cq_2 + b - c),\\
e_{01} = (bq_2 - b + c)(b^2q_2 - b^2 + c^2),\\
e_{00} = -c(b - c)(1 - q_2)[bq_2 - b + c],\\
\end{cases}
\end{equation}
and
\begin{equation}
\begin{cases}
w_2 = bc^2,\\
w_1 = -(b - c)(b^2 - bc - c^2),\\
w_0 = (b + c)(b - c)^2,\\
u = -b^2 - bc + c^2,\\
v = c(b - c).\\
\end{cases}
\end{equation}
\\
The expression of the boundary value $q_a$ is
\begin{equation}
q_a = -\frac{v}{u} = \frac{c(b - c)}{b^2 + bc - c^2},
\end{equation} 
and that of the boundary function $h_A(q_2)$ is
\begin{equation}
 h_A(q_2) = -\frac{e_{20}}{e_{21}} = \frac{(b - c)(1 - q_2)[(b^2 + bc - c^2)q_2 - c(b - c)]}{bc^2q_2^2 - (b - c)(b^2 - bc - c^2)q_2 + (b + c)(b - c)^2}.
\end{equation}
Notice that $e_{21} \geq w_1q_2 + w_0 > w_0(1 - q_2) \geq 0$. In other words, $e_{21}$ is positive all the time. \par

In particular, if $q_2 = 0$, we have
\begin{equation}
s_X = \frac{(b + c)(b - c)q_4\chi}{[(b + c)q_4 + b]\chi + (b + c)q_4 + c}.
\end{equation}
\par

\begin{figure}[H]
\centering
\subfloat[$r = 2$]{\includegraphics[width=0.5\linewidth]{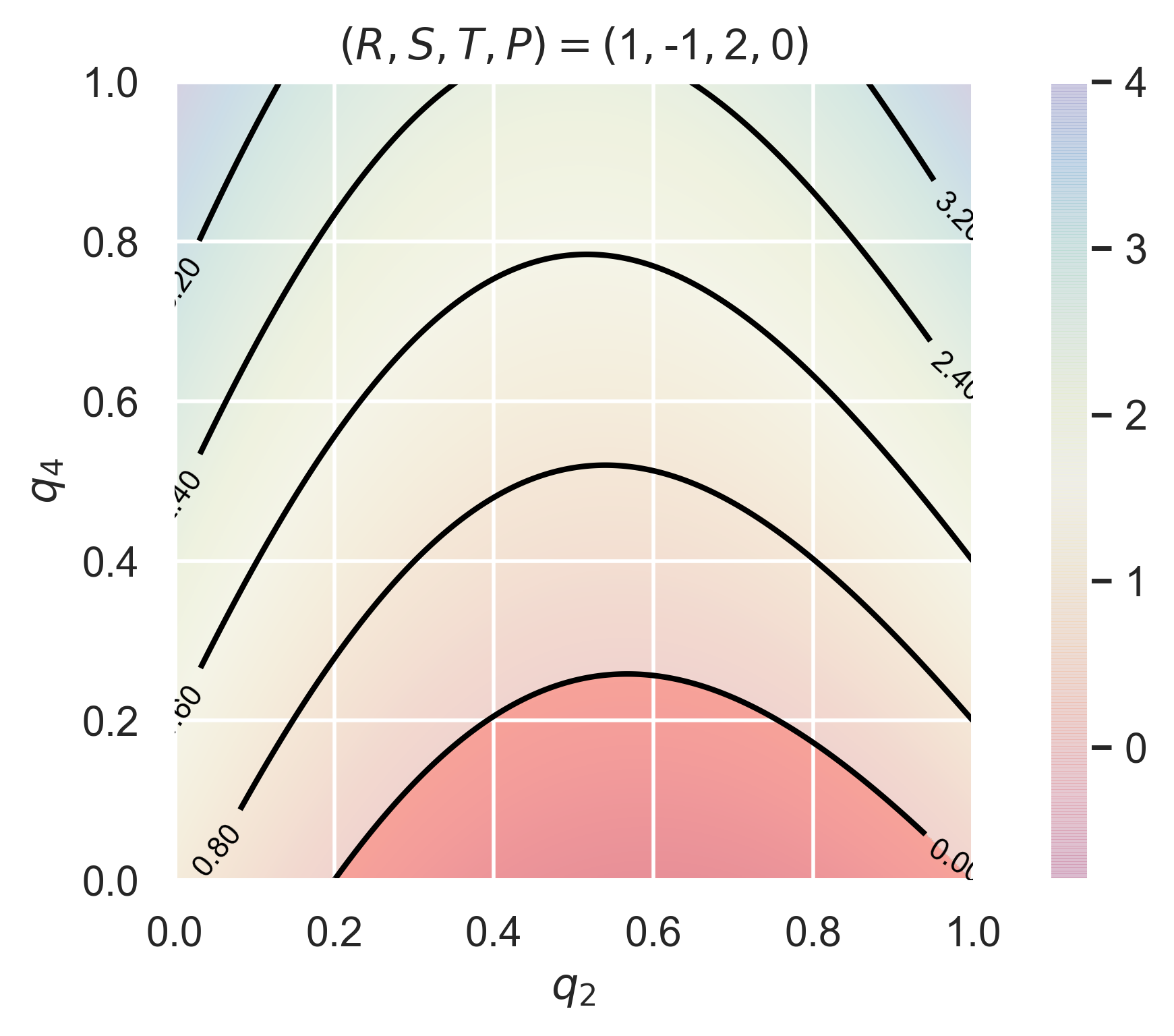}} 
\subfloat[$r = 4$]{\includegraphics[width=0.5\linewidth]{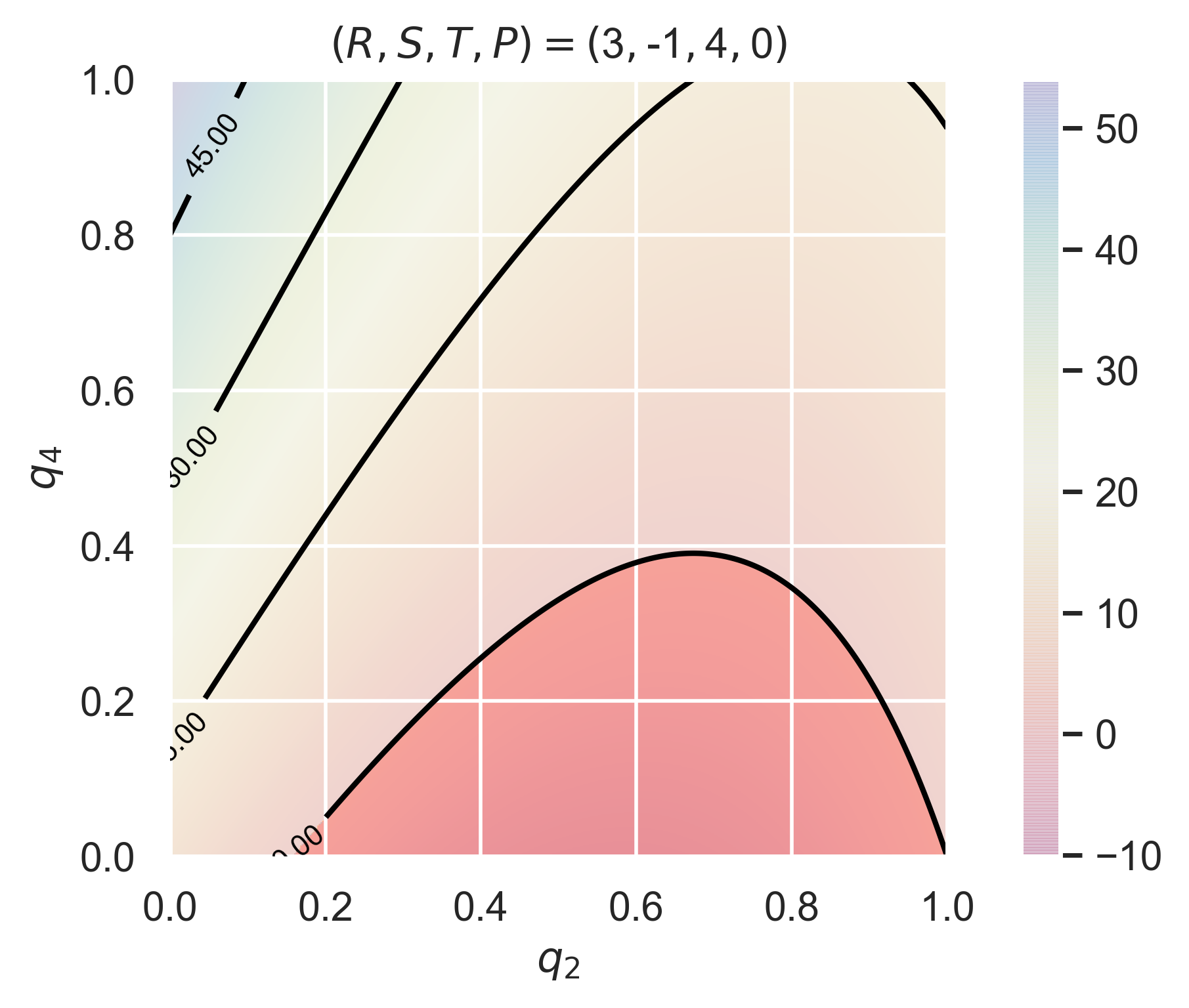}}
\\
\subfloat[$r = 8$]{\includegraphics[width=0.5\linewidth]{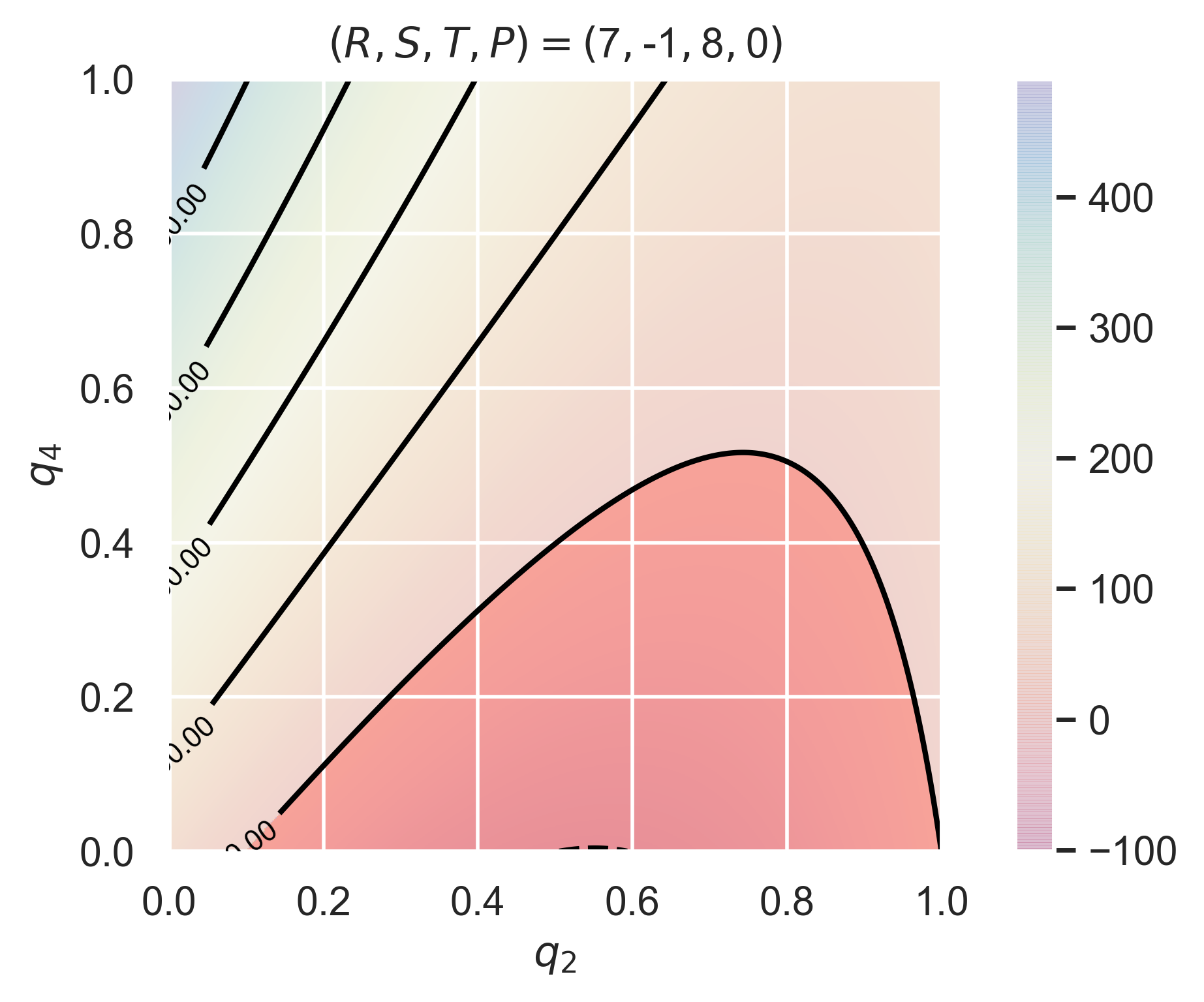}} 
\subfloat[$r = 16$]{\includegraphics[width=0.5\linewidth]{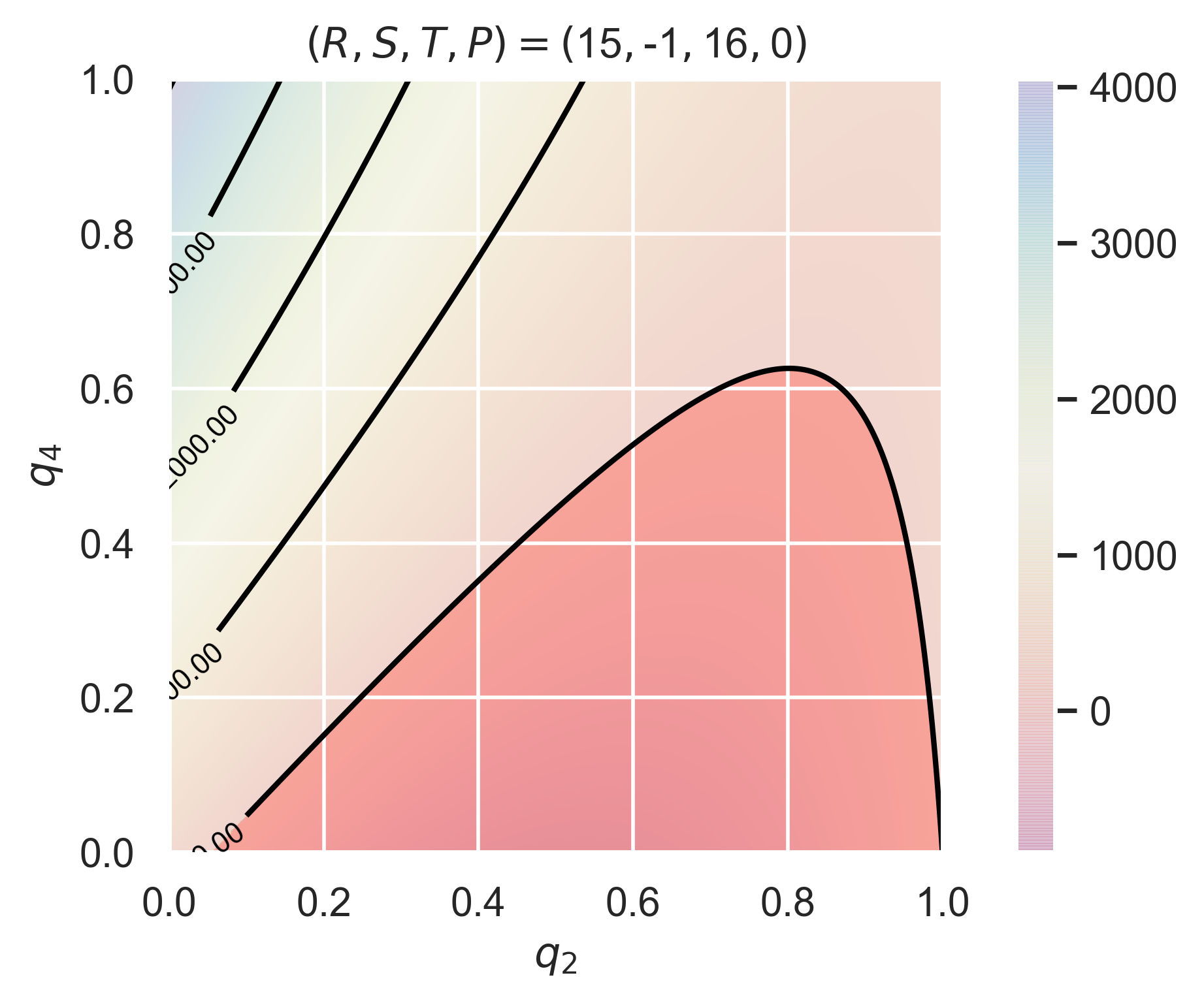}}
\caption{Contour curves of $e_2$ for the donation game with different benefit-cost ratios $r = b/c$. The red regions are where $e_2 \leq 0$. As the ratio increases, the area of the region increases accordingly.}
\label{target_A_dg}
\end{figure}

The visualization of Class A, which corresponds to the second column in Table~\ref{summary_A_dg}, is given in Figure~\ref{target_A_dg}. 

As before, the set of inequalities regarding $q_2$ and $q_4$ is the explicit expression of $e_2 \leq 0$. We now show that $e_2 \leq 0$ is the necessary and sufficient condition for the extortioner's payoff to be monotonically decreasing with $\chi$. Notice that the monotonicity is decided by the sign of $g_A(\chi)$, a quadratic function of $\chi$. Therefore, $e_2 \leq 0$ is a necessary condition for $s_X$ to decrease with respect to $\chi$. It suffices to show that $e_2 \leq 0$ is also a sufficient condition. \par

If $e_2 = 0$, we have $q_4 = h_A(q_2)$ and $g_A(\chi)$ being a linear function of $\chi$. Given that the slope $e_1$ satisfies
\begin{equation}
e_1 = -\frac{2b(b + c)^2(b - c)^2q_2(1 - q_2)^2(cq_2 + b - c)}{e_{21}} < 0,
\end{equation}
and the value of the function at $\chi = 1$ satisfies
\begin{equation}
g_A(1) = e_1 + e_0 = -\frac{b(b + c)^2(b - c)^2q_2(1 - q_2)^2[(b + 2c)q_2 + b - c]}{e_{21}} < 0,
\end{equation}
we find that $g_A(\chi) < 0$ for $\chi > 1$. \par

Otherwise, if $e_2 < 0$, we immediately have $0 < q_4 < h_A(q_2)$ and $-\frac{u}{v} < q_2 < 1$. Further calculation yields that 
\begin{equation}
g_A(1) = e_2 + e_1 + e_0 = (b + c)^2q_2[(bq_2 - b + c)q_4 - (b - c)(1 - q_2)],
\end{equation} 
of which the last factor on the right-hand side takes its maximum at $q_4 = 0$ or $q_4 = h_A(q_2)$. It is straightforward to show that 
\begin{equation}
\max\{- (b - c)(1 - q_2), - \frac{b(b - c)^2(1 - q_2)^2[(b + 2c)q_2 + b - c]}{e_{21}}\} < 0.
\end{equation}
Hence, $g_A(1) < 0$.\par

Moreover, we consider the sign of 
\begin{equation}
2e_2 + e_1 = 2b(b + c)q_2[cq_2q_4 - (b - c)(1 - q_2)].
\end{equation}
The last factor on the right-hand side takes its maximum at $q_4 = h_A(q_2)$ with value
\begin{equation}
-\frac{(b + c)(b - c)^2(1 - q_2)^2(cq_2 + b - c)}{e_{21}} < 0.
\end{equation}
Therefore, $2e_2 + e_1 < 0$ and thus $-e_1/2e_2 < 1$. \par

Given the above results, we conclude that $g_A(\chi) < 0$ for $\chi > 1$.\par

\subsubsection{The general IPD game}

\begin{table}[H]
\centering
\tabulinesep=1.5mm
\begin{tabu}{c c}
\hline
\rowcolor[HTML]{EFEFEF} 
Payoff Matrix  & Class A \\
$\begin{bmatrix} R & S \\ T & P \end{bmatrix}$
&
$\begin{cases}
q_a < q_2 < 1 \\
0 < q_4 \leq h_A(q_2)
\end{cases}$ \\
\rowcolor[HTML]{EFEFEF} 
$s_X$ & $s'_X$ \\  
$\frac{(T - S)(R - P)\{[(P - S)q_2 + T + S - 2P]\chi + (T - P)q_2 - (T + S - 2P)\}q_4\chi}{f_A(\chi)}$ 
&
$ \frac{(T - S)(R - P)q_4g_A(\chi)}{f_A^2(\chi)}$ \\ 
\hline
\end{tabu}
\caption{Summary of the general IPD game with $T + S > 2P$.}
\label{summary_A_general}
\end{table}

Likewise, $g_A(\chi) = e_2 \chi^2 + e_1 \chi + e_0$, where
\begin{equation}
e_i = e_{i1}q_4 + e_{i0}. \qquad i \in \{0, 1, 2\}
\end{equation}
More specifically, we have
\begin{equation}
\begin{cases}
e_{21} = w_2q_2^2 + w_1q_2 + w_0,\\
e_{20} =(R - P)(1 - q_2)(uq_2 + v),\\
e_{11} = 2[(P - S)q_2 + (T + S - 2P)][(T - P)(R - S)q_2 - (T - S)(R - P)],\\
e_{10} = -2(R - P)(P - S)(1 - q_2)[(P - S)q_2 + (T + S - 2P)],\\
e_{01} = [(T - P)q_2 - (T + S - 2P)][(T - P)(R - S)q_2 - (T - S)(R - P)],\\
e_{00} = -(R - P)(P - S)(1 - q_2)[(T - P)q_2 - (T + S - 2P)],\\
\end{cases}
\label{coefficient_A}
\end{equation}
and
\begin{equation}
\begin{cases}
w_2 = (R - S)(P - S)^2,\\
w_1 = (T - P)^2(T + P - 2R) - (P - S)^3,\\
w_0 = (T - S)(R - P)(T + S - 2P),\\
u =-[(T - P)(T - S) - (P - S)^2],\\
v = (P - S)(T + S - 2P). \\
\end{cases}
\label{parameter_uvw}
\end{equation}
\par

The expression of the boundary value $q_a$ is
\begin{equation}
q_a = -\frac{v}{u} = \frac{(T + S - 2P)(P - S)}{(T - P)(T - S) - (P - S)^2},
\end{equation}
and that of the boundary function $h_A(q_2)$ is
\begin{equation}
h_A(q_2) = 
\begin{cases}
1, & e_{21} \leq -e_{20}\\
h(q_4) = -\frac{e_{20}}{e_{21}}. & \text{otherwise}
\end{cases}
\end{equation}
\par

\begin{figure}[H]
\centering
\includegraphics[width=0.5\linewidth]{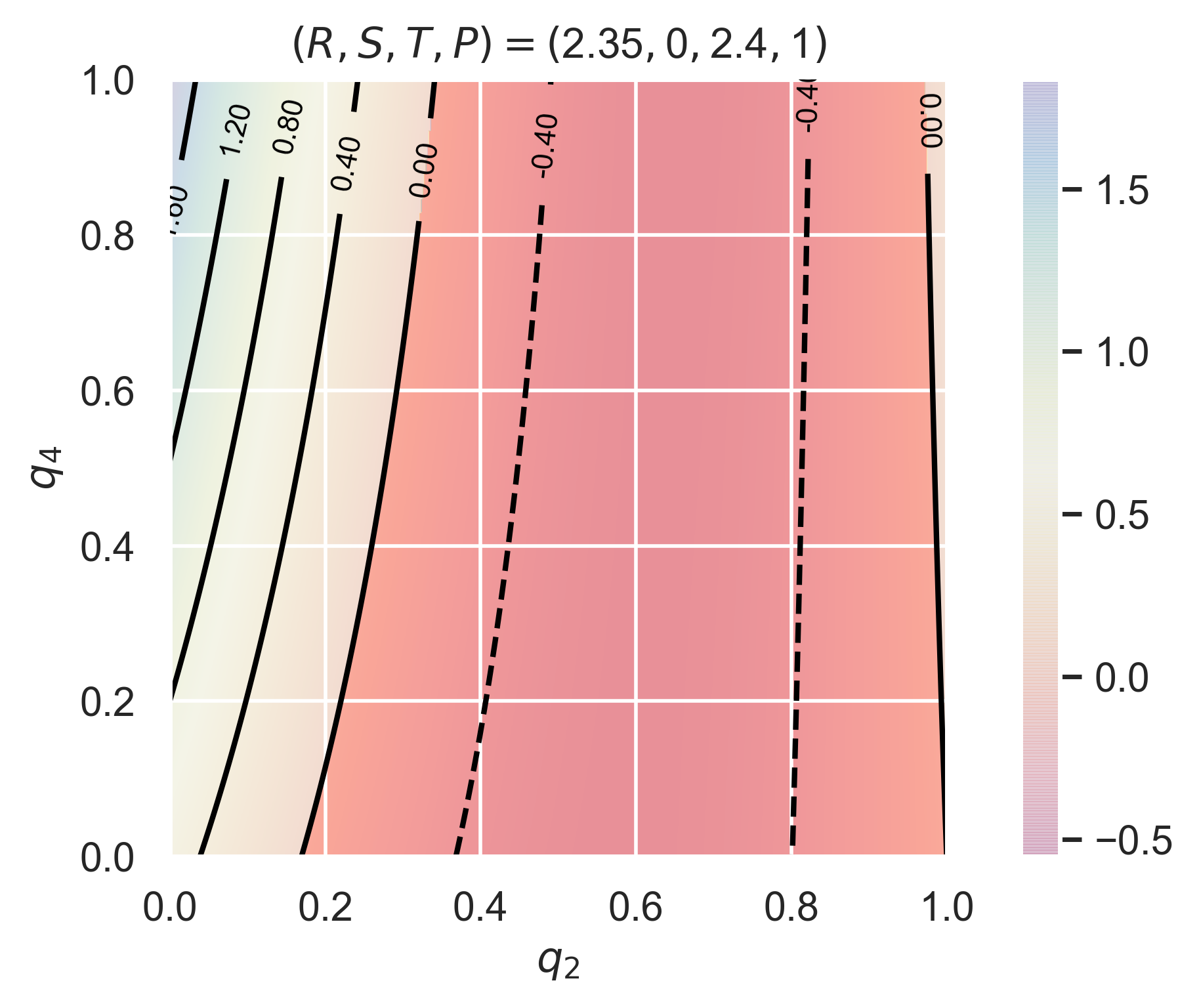}
\caption{Contour curves of $e_2$ for the general IPD game with $T + S > 2P$. Here we have $(R, S, T, P) = (2.35, 0, 2.4, 1)$. The red region is where $e_2 \leq 0$.}
\label{target_A_con_x}
\end{figure}

The visualization of Class A, which corresponds to the second column in Table~\ref{summary_A_general} (the explicit expression of $e_2 \leq 0$), is given in Figure~\ref{target_A_con_x}. 

Notice that $e_{20} > 0$ implies $e_{21} > 0$. That is, if $0 < q_2 < q_a$, $e_{21} > 0$. The contrapositive statement also holds ($e_{21} \leq 0$ implies $e_{20} \leq 0$ ). The proof is trivial. Consider $e_{21}$ as a bivariate function of $R$ and $q_2$ (linear with respect to $R$). Letting $R = P$ and $R = T$, we have
\begin{equation}
\begin{cases}
e_{21}(P, q_2) = q_2[(P - S)^3q_2 + (T - P)^3 - (P - S)^3] > 0, \\
e_{21}(T, q_2) = (T - S)(1 - q_2)[-(P - S)q_2 + (T - P)(T + S - 2P)].
\end{cases}
\end{equation}
The last factor of $e_{21}(T, q_2)$ takes its minimum $-(T + S - 2P)^2(T - S)^2/u$ at $q_2 = q_a$, which is positive. \par

We now show that $e_2 \leq 0$ is the necessary and sufficient condition for the extortioner's payoff to be monotonically decreasing with respect to $\chi$. As before, $g_A(\chi)$ is a quadratic function of $\chi$. Thus, $e_2 \leq 0$ is a necessary condition for $s_X$ to decrease. It suffices to show that $e_2 \leq 0$ is also a sufficient condition. \par

If $e_2 = 0$, we have $q_4 = h_A(q_2)$ and $g_A(\chi)$ being a linear function of $\chi$. Given that the slope $e_1$ satisfies
\begin{equation}
\mfootnotesize{
e_1 = -\frac{2(T - S)^2(R - P)(R - S)(T + S - 2P)q_2(1 - q_2)^2[(P - S)q_2 + T + S - 2P]}{e_{21}} < 0,}
\end{equation}
and the value of the function at $\chi = 1$ satisfies
\begin{equation}
\mfootnotesize{
\begin{aligned}
g_A(1) &= e_1 + e_0 \\
&= -\frac{(T - S)^2(R - P)(R - S)(T + S - 2P)q_2(1 - q_2)^2[(T + P - 2S)q_2 + T + S - 2P]}{e_{21}} \\
&< 0,
\end{aligned}}
\end{equation}
we determine that $g_A(\chi) < 0$ for $\chi > 1$. \par

Otherwise, if $e_2 < 0$, we immediately have $0 < q_4 < h_A(q_2)$ and $q_a < q_2 < 1$. \par

Further calculation shows that 
\begin{equation}
g_A(1) = e_2 + e_1 + e_0 = (T - S)^2q_2\Gamma(q_4),
\end{equation}
of which the last factor on the right-hand side 
\begin{equation}
\Gamma(q_4) = [(R - S)q_2 - (2R - T - S)]q_4 - (R - P)(1 - q_2)
\end{equation}
takes its maximum at $q_4 = 0$ or $q_4 = h_A(q_2)$.\par

Moreover, we have
\begin{equation}
2e_2 + e_1 < e_1 = 2[(P - S)q_2 + (T + S - 2P)]\gamma(q_4),
\end{equation}
of which the last factor on the right-hand side 
\begin{equation}
\gamma(q_4) = [(T - P)(R - S)q_2 - (T - S)(R - P)]q_4 - (R - P)(P - S)(1 - q_2)
\end{equation}
also takes its maximum at $q_4 = 0$ or $q_4 = h_A(q_2)$.\par

Now we consider the following three different cases.
\begin{enumerate}[label = (\roman*)]
\item $e_{21} > 0 > e_{20}$ and $e_{21} + e_{20} > 0$, under which $h_A(q_2) = -e_{20}/e_{21}$. \par

We obtain
\begin{equation}
\mfootnotesize{
\begin{aligned}
\Gamma(0) &= -(R - P)(1 - q_2) < 0, \\
\Gamma(-\frac{e_{20}}{e_{21}}) &= -\frac{(R - P)(R - S)(T + S - 2P)(1 - q_2)^2[(T + P - 2S)q_2 + T + S - 2P]}{e_{21}}\} < 0,
\end{aligned}}
\end{equation}
and
\begin{equation}
\mfootnotesize{
\begin{aligned}
\gamma(0) &= -(R - P)(P - S)(1 - q_2) < 0, \\
\gamma(-\frac{e_{20}}{e_{21}}) &=  -\frac{(T - S)^2(R - P)(R - S)(T + S - 2P)q_2(1 - q_2)^2}{e_{21}}\} < 0.
\end{aligned}}
\end{equation}

\item $e_{21} > 0 > e_{20}$ and $e_{21} + e_{20} \leq 0$, under which $h_A(q_2) = 1$. \par

Notice that now $-\frac{e_{20}}{e_{21}} \geq 1$. Therefore,
\begin{equation}
\mfootnotesize{
\Gamma(1) = (2R - P - S)q_2 - (R - P) - (2R - T - S) \leq \max\{\Gamma(0), \Gamma(-\frac{e_{20}}{e_{21}})\} < 0,}
\end{equation}
and
\begin{equation}
\mfootnotesize{
\gamma(1) = [(T - S)(R - S) - (P - S)^2]q_2 - (R - P)(T + P - 2S) \leq \max\{\gamma(0), \gamma(-\frac{e_{20}}{e_{21}})\} < 0.}
\end{equation}

\item $e_{21} \leq 0$ and $e_{20} < 0$, under which $h_A(q_2) = 1$. \par

Consider the corresponding range of $q_2$ in this case. Routine calculation shows that
\begin{equation}
\mfootnotesize{
e_{21}(\frac{2R - T - S}{R - S}) = \frac{(T - R)(T - S)(T + S - 2P)(2R - T - S + R - P)}{R - S} > 0,}
\end{equation}
\begin{equation}
\mfootnotesize{
e_{21}(\frac{(T - S)(R - P)}{(T - P)(R - S)}) = \frac{(T - R)(T - S)^3(R - P)(T + S - 2P)}{(T - P)^2(R - S)} > 0,}
\end{equation}
and
\begin{equation}
\mfootnotesize{
\begin{aligned}
-\frac{w_1}{2w_2} 
&= -\frac{(T - P)^2(T + P - 2R) - (P - S)^3}{2(R - S)(P - S)^2} < -\frac{(P - S)^2(T + P - 2R) - (P - S)^3}{2(R - S)(P - S)^2} \\
&= \frac{2R - T - S}{2(R - S)} < \frac{2R - T - S}{R - S} < \frac{(T - S)(R - P)}{(T - P)(R - S)}.
\end{aligned}}
\end{equation}
The last inequality holds as 
\begin{equation}
\mfootnotesize{
\frac{(T - S)(R - P)}{(T - P)(R - S)} - \frac{2R - T - S}{R - S} = \frac{(T - R)(T + S - 2P)}{(T - P)(R - S)}.}
\end{equation}
We determine that $q_2$ has to be less than these two values where $e_{21}$ is evaluated. Therefore, both $\Gamma(q_4)$ and $\gamma(q_4)$ are decreasing functions of $q_4$ and are negative for $0 < q_4 \leq 1$.
\end{enumerate}
\par

To conclude, $e_2 < 0$ always implies $g_A(1) < 0$ and $2e_2 + e_1 < 0$ (that is, $-e_1/2e_2 < 1$). As a result, we have $g_A(\chi) < 0$ for $\chi > 1$. \par

\subsection{Case \rom{2}: \texorpdfstring{$T + S = 2P$}{e}}

The derivative of $s_X$ has been given in Equation~\ref{payoff_A_eqn}, the sign of which is decided by $d_{A0}$.\par

\begin{figure}[H]
\centering
\includegraphics[width=0.5\linewidth]{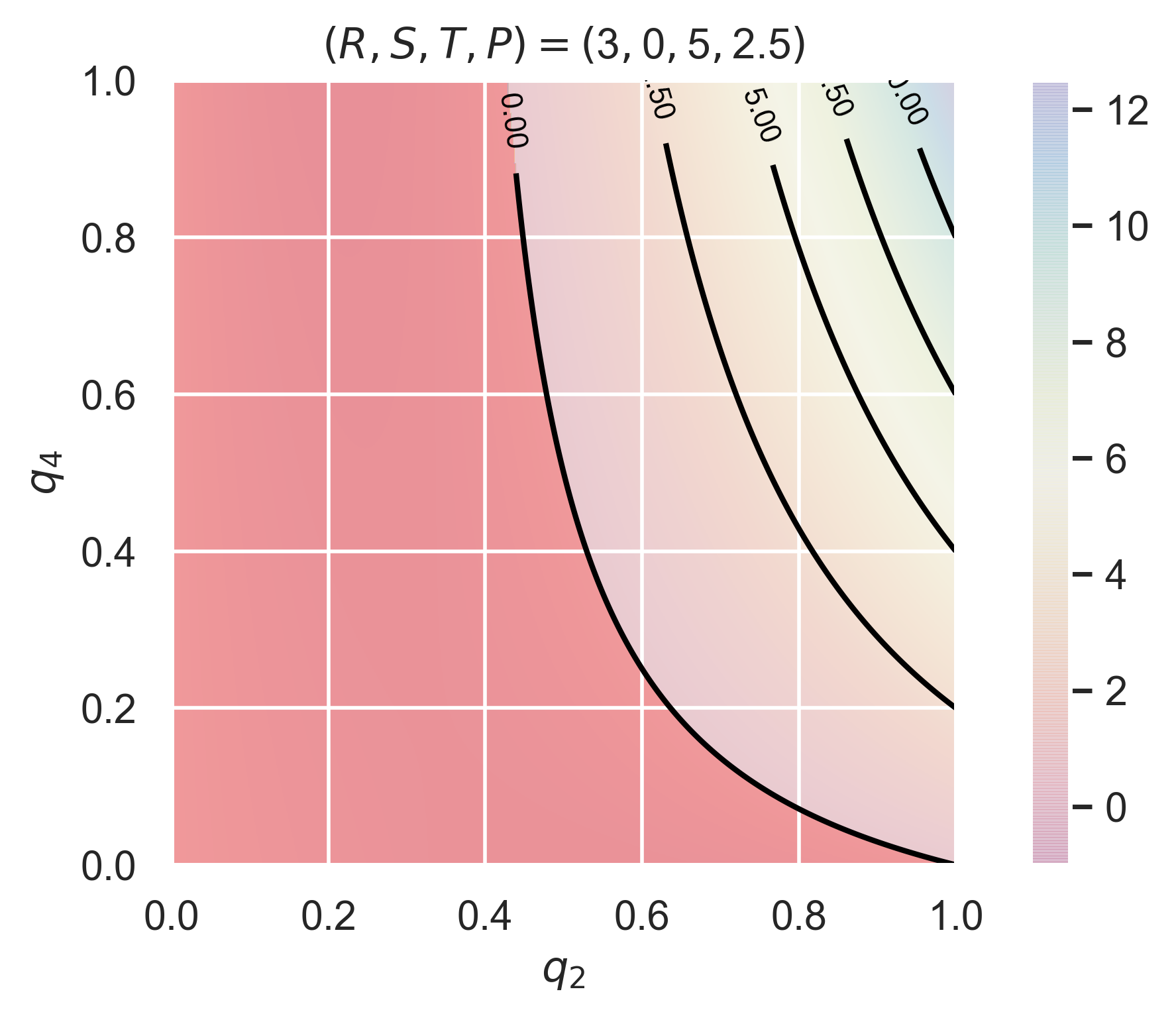}
\caption{Contour curves of $e_2 = g_A(1)/4 = (T - S)^2q_2d_0/8$ for the general IPD game with $T + S = 2P$. Here we have $(R, S, T, P) = (3, 0, 5, 2.5)$. The red region is where $e_2 \leq 0$, or equivalently, $g_A(1) \leq 0$.}
\label{target_A_eqn}
\end{figure}

Solving $d_{A0} < 0$, we obtain $0 < q_2 < 1$ and
\begin{equation} 
0 < q_4 
\begin{cases}
\leq 1, & 0 < q_2 < \frac{3(2R - T - S)}{4R - T - 3S}\\
< \frac{(1 - q_2)(2R - T - S)}{2[(R - S)q_2 - (2R - T - S)]}. & \frac{3(2R - T - S)}{4R - T - 3S} \leq q_2 < 1
\end{cases}
\end{equation}
It is the same as the inequality in Table~\ref{four_classes}. An example is given in Figure~\ref{target_A_eqn}. 

\subsection{Case \rom{3}: \texorpdfstring{$T + S < 2P$}{e}}

The set of $q_2$ and $q_4$ is:
\begin{equation}
\begin{cases}
0 \leq q_2 < 1,\\
0 < q_4 \leq h_A(q_2),\\
\end{cases}
\label{result_A_abn}
\end{equation}

where
\begin{equation}
h_A(q_2) = 
\begin{cases}
1, & 0 \leq q_2 < \frac{(R - P) + (2R - T - S)}{2R - P - S}\\
\frac{(R - P)(1 - q_2)}{(R - S)q_2 - (2R - T - S)}. & \frac{(R - P) + (2R - T - S)}{2R - P - S} \leq q_2 < 1
\end{cases}
\end{equation}
\par

\begin{figure}[H]
\centering
\includegraphics[width=0.5\linewidth]{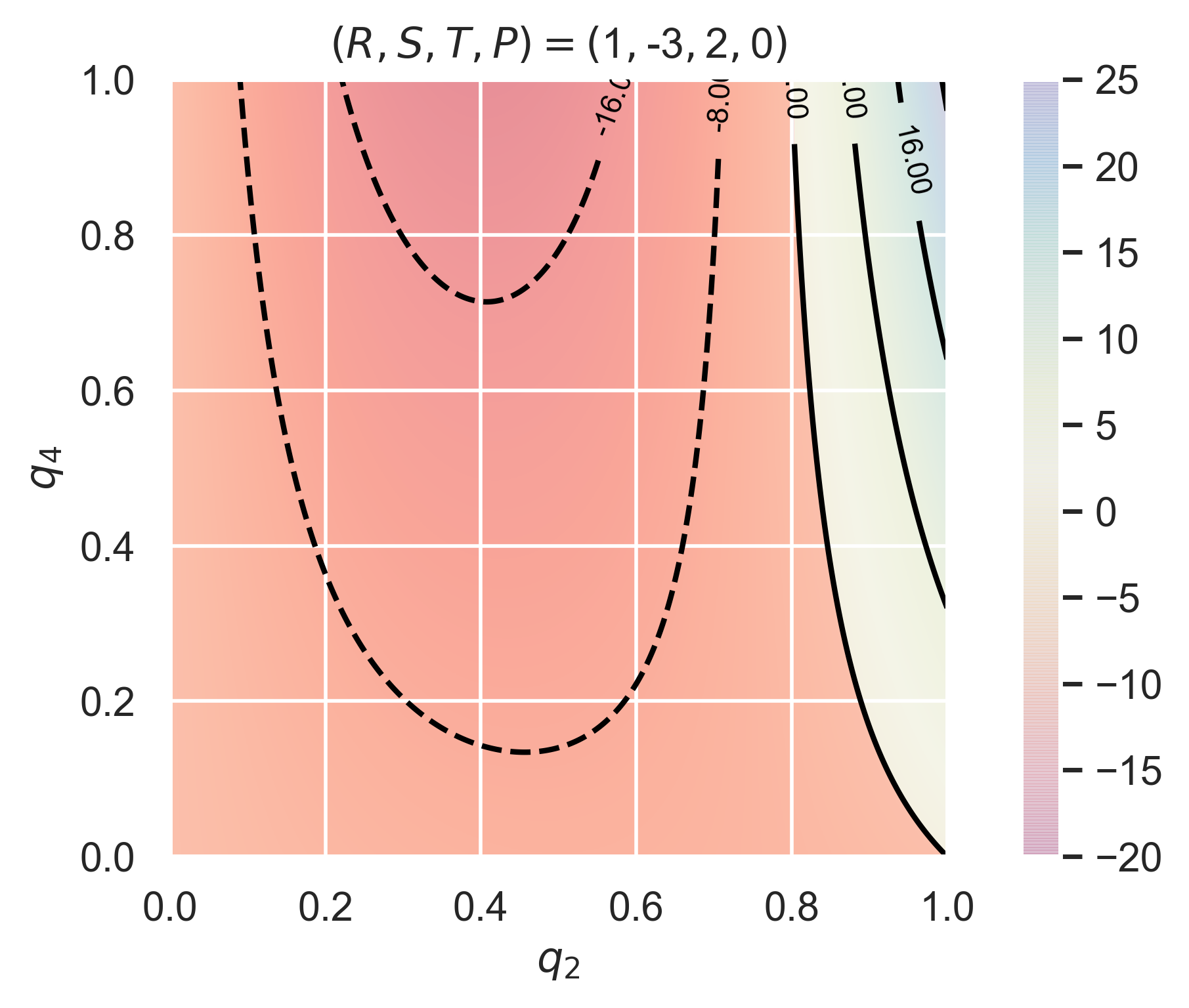}
\caption{Contour curves of $g_A(1)$ for the general IPD game with $T + S < 2P$. Here we have $(R, S, T, P) = (1, -3, 2, 0)$. The red region is where $g_A(1) \leq 0$.}
\label{target_A_abn}
\end{figure}

The visualization (the explicit expression of $g_A(1) \leq 0$) is given in Figure~\ref{target_A_abn}. 

We now show that $g_A(1) \leq 0$ is the necessary and sufficient condition for ZD's payoff to be monotonically decreasing with respect to $\chi$. \par

Recall Equation~\ref{payoff_A_0}. If $q_2 = 0$, we immediately have $\frac{ds_X}{d\chi} < 0$. \par

Otherwise, the sign of $\frac{ds_X}{d\chi}$ is decided by $g_A(\chi)$ while $g_A(1) \leq 0$ is a necessary condition for $s_X$ to decrease. It suffices to show that $g_A(1) \leq 0$, or equivalently, $\Gamma(q_4) \leq 0$ is also a sufficient condition.\par

Here, we let 
\begin{equation}
2e_2 + e_1 = 2(T - S)q_2\gamma(q_4),
\end{equation}
of which the last factor on the right-hand side is
\begin{equation}
\mfootnotesize{
\gamma(q_4) = [(R - S)(P - S)q_2 - (T - S)(R + P - T - S)]q_4 - (T - P)(R - P)(1 - q_2).}
\end{equation}
\par
Notice that 
\begin{equation}
\mfootnotesize{
(T - P)\Gamma - \gamma = (R - S)(2P - T - S)(1 - q_2)q_4 > 0.}
\end{equation}
We get $\gamma < 0$.\par

On the other hand, since
\begin{equation}
\mfootnotesize{
\frac{(T - P)^2q_2\Gamma - e_2}{(2P - T - S)(1 - q_2)} = (T - S)[(R - S)q_2 + R - P]q_4 + (R - P)(P - S)(1 - q_2) > 0,}
\end{equation}
we have $e_2 < 0$.\par

To conclude, $g_A(1) \leq 0$ always implies $e_2 < 0$ and $2e_2 + e_1 < 0$. As a result, we have $g_A(\chi) < 0$ for $\chi > 1$. \par

\section{Class B of unbending strategies}

The second class is located in the subspace where $q_2 = q_3 = 0$.  The extortioner's payoff $s_X$ is a rational function of degree $1$. We have

\begin{equation}
\begin{aligned}
s_X &= \frac{(T - S)(T + S - 2P)q_4\chi}{f_B(\chi)} + P, \\
\frac{ds_X}{d\chi} &= \frac{(T - S)(T + S - 2P)[(T - S)q_4 + P - S]q_4\chi}{f_B^2(\chi)},
\end{aligned}
\end{equation}
where $f_B(\chi) = [(T - S)q_4 + T - P]\chi + (T - S)q_4 + P - S$. Apparently, $f_B(\chi) > 0$ for $\chi > 1$.\par

Further, we can draw the following conclusions.
\begin{enumerate}[label=(\roman*)]
\item If $T + S > 2P$, $s_X$ is greater than $P$ and increases with $\chi$.
\item If $T + S = 2P$, $s_X = P$.
\item If $T + S < 2P$, $s_X$ is less than $P$ and decreases with $\chi$. The necessary and sufficient condition $q_1$ and $q_4$ need to satisfy is simply
\begin{equation} 
\begin{cases}
0 \leq q_1 \leq 1,\\
0 < q_4 \leq 1.
\end{cases}
\label{target_B}
\end{equation}
\end{enumerate}
\par

\section{Class C of unbending strategies}

The third class requires $q_1 = q_2 = q_3$. The extortioner's payoff $s_X$ is also a rational function of degree $1$. Now we have
\begin{equation}
\begin{cases}
s_X = \frac{(T - S)a_{C0}q_4\chi}{f_C(\chi)} + P,\\
\frac{ds_X}{d\chi} = \frac{(T - S)a_{C0}d_{C0}q_4}{f_C^2(\chi)},\\
\end{cases}
\label{payoff_C}
\end{equation}
where $f_C(\chi) = d_{C1}\chi + d_{C0}$ and
\begin{equation}
\begin{cases}
a_{C0} = (R + P - T - S)q_1 + T + S - 2P = (R - P)q_1 + (T + S - 2P)(1 - q_1),\\
d_{C1} = [(T - S) - (T - R)q_1]q_4  + [(T - P) - (T - R)q_1](1 - q_1),\\
d_{C0} = [(T - S) - (R - S)q_1]q_4 + [(P - S) - (R - S)q_1](1 - q_1).
\end{cases}
\end{equation}
\par

Given that
\begin{equation}
\begin{cases}
d_{C1} > 0, \\
d_{C1} + d_{C0} = (T - S)[(1 - q_1)^2 + q_4(2 - q_1)] > 0,
\end{cases}
\end{equation}
we obtain that $f_C(\chi) > 0$ for $\chi > 1$. \par

The monotonicity of $s_X$ is decided by the sign of $a_{C0}d_{C0}$, both being linear functions of $q_4$. It is straightforward to get

\begin{equation}
\mfootnotesize{
d_{C0} > 0 \qquad \Leftrightarrow \qquad
\begin{cases}
0  \leq  q_1  \leq q_C,\\
0 < q_4 \leq 1,
\end{cases}
\text{or}\qquad
\begin{cases}
q_C  <  q_1  \leq  1,\\
h_C(q_1) < q_4 \leq 1,
\end{cases}}
\end{equation}
and
\begin{equation}
\mfootnotesize{
d_{C0} < 0  \qquad \Leftrightarrow \qquad
\begin{cases}
q_C < q_1 < 1,\\
0 < q_4 < h_C(q_1).\\
\end{cases}}
\end{equation}
Here 
\begin{equation}
\mfootnotesize{
q_C = \frac{P - S}{R - S}, \qquad h_C(q_1) = \frac{[(R - S)q_1 - (P - S)](1 - q_1)}{(T - S) - (R - S)q_1}.}
\end{equation}
On the other hand, the sign of $a_{C0}$ is dependent on that of $T + S - 2P$. We now discuss it in detail. \par

\subsection{Case \rom{1}: \texorpdfstring{$T + S > 2P$}{e}}

\begin{figure}[H]
\centering
  \includegraphics[width=0.5\linewidth]{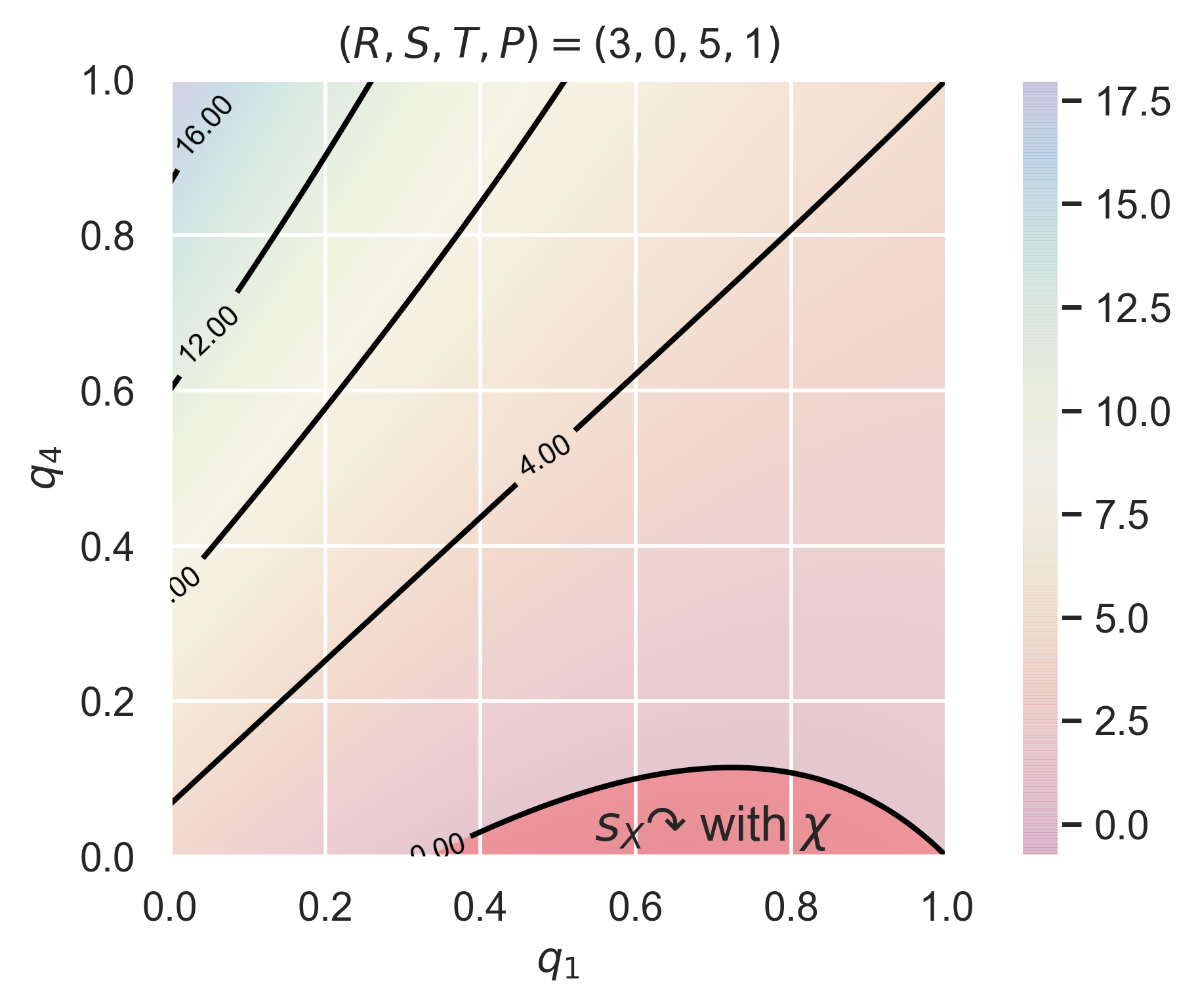}
  \caption{Contour curves of $a_{C0}d_{C0}$ for the general IPD game with $T + S > 2P$. Here we use $(R, S, T, P) = (3, 0, 5, 1)$. The red region is where $a_{C0}d_{C0} \leq 0$, or equivalently, $d_{C0} \leq 0$.}
  \label{target_C_con}
\end{figure}

Give that $a_{C0}$ is always positive, the sign of $a_{C0}d_{C0}$ is the same as that of $d_{C0}$. Hence the set of $q_1$ and $q_4$ is just
\begin{equation}
\begin{cases}
q_C < q_1 < 1,\\
0 < q_4 < h_C(q_1).\\
\end{cases}
\label{result_C_con}
\end{equation}
We can visualize the result in Figure~\ref{target_C_con}.

\subsection{Case \texorpdfstring{$T + S = 2P$}{e}}

\begin{figure}[H]
    \centering
    \subfloat[Original]{\includegraphics[width=0.5\linewidth]{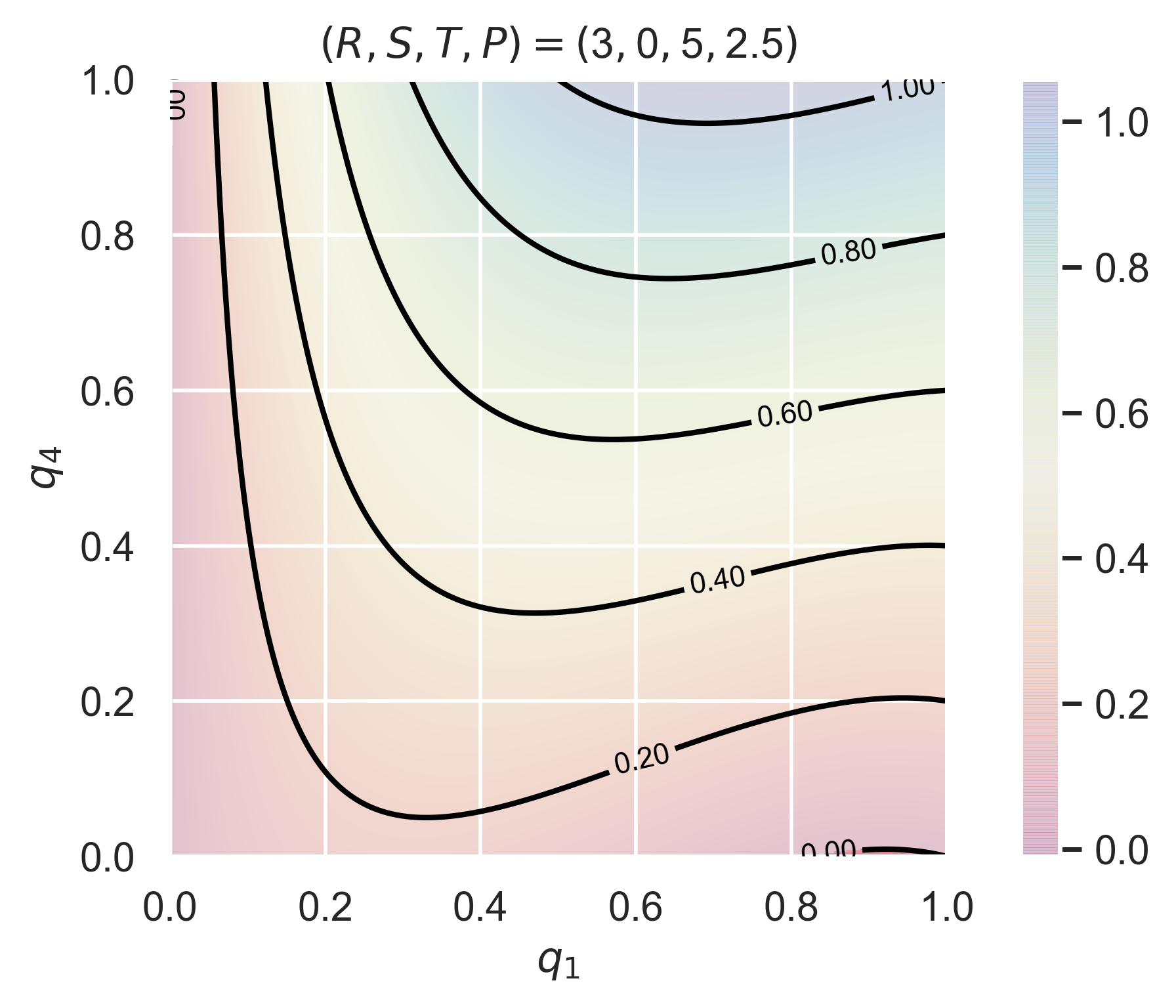}} 
     \subfloat[Zoomed]{\includegraphics[width=0.5\linewidth]{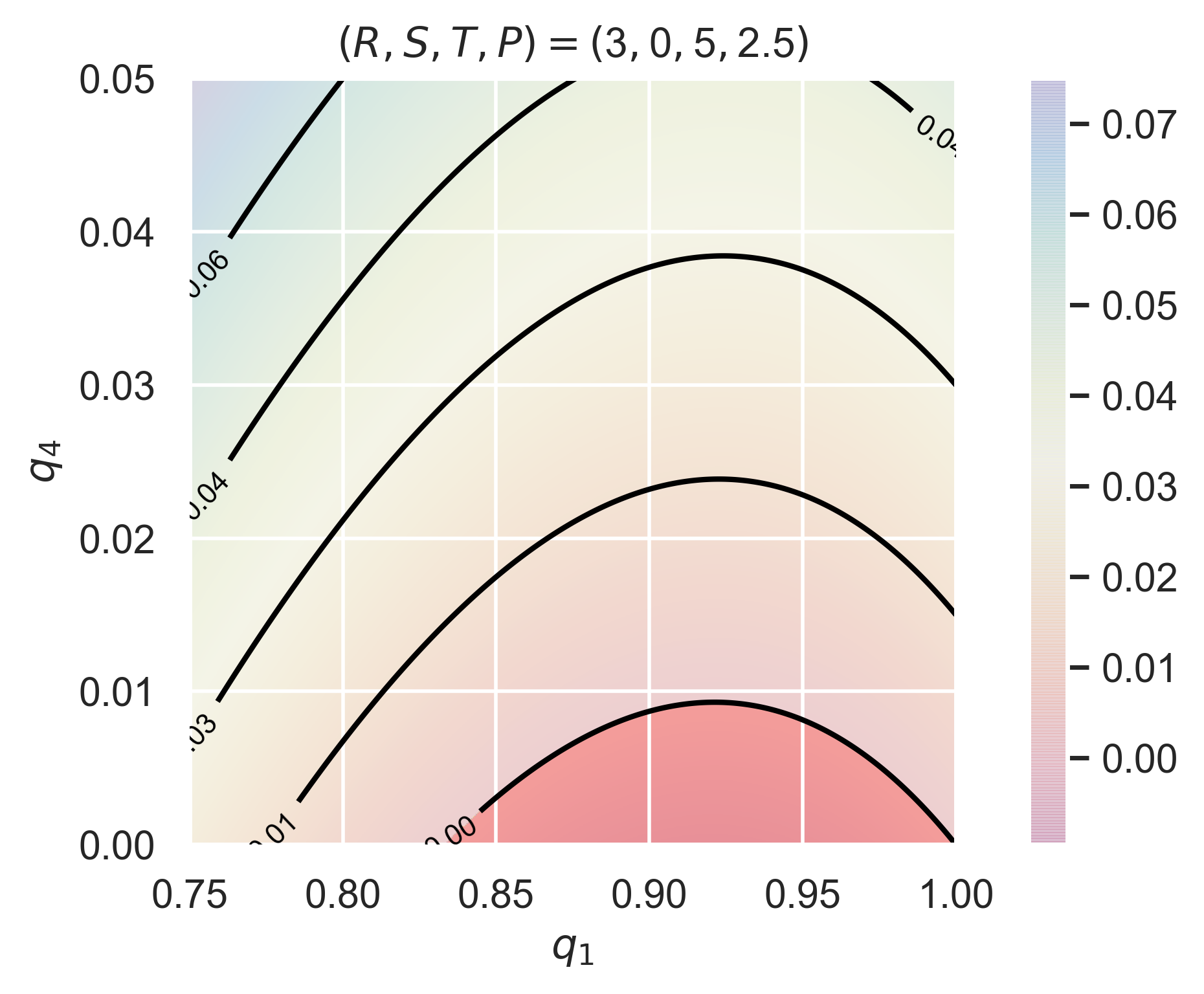}}
      \caption{Contour curves of $a_{C0}d_{C0}$ for the general IPD game with $T + S = 2P$. Here we use $(R, S, T, P) = (3, 0, 5, 2.5)$. The red region is where $a_{C0}d_{C0} \leq 0$.}
     \label{target_C_eqn}
\end{figure}

In this case, $a_{C0}$ is positive except when $q_1 = 0$. We obtain the same set of $q_1$ and $q_4$ as Equation~\ref{result_C_con}. To be more specific, 
\begin{equation}
q_C = \frac{T - S}{2(R - S)}, \qquad h_C(q_1) = \frac{[2(R - S)q_1 - (T - S)](1 - q_1)}{2[(T - S) - (R - S)q_1]}.
\end{equation}
The visualization is given in Figure~\ref{target_C_eqn} and we may zoom in to observe the region more closely.

\subsection{Case \rom{3}: \texorpdfstring{$T + S < 2P$}{e}}

 The sign of $a_{C0}$ now depends on the value of $q_2$. It is trivial to get 
\begin{equation}
\begin{cases}
a_{C0} < 0, & 0 \leq q_1 < q_c,\\
a_{C0} > 0, & q_c < q_1 \leq 1
\end{cases}
\end{equation}
where $q_c = (2P - T - S)/(R + P - T - S)$. Notice that
\begin{equation}
q_C - q_c = \frac{(T - P)(R - P)}{(R - S)(R + P - T - S)} > 0.
\end{equation}
\par

\begin{figure}[H]
\centering
  \includegraphics[width=0.5\linewidth]{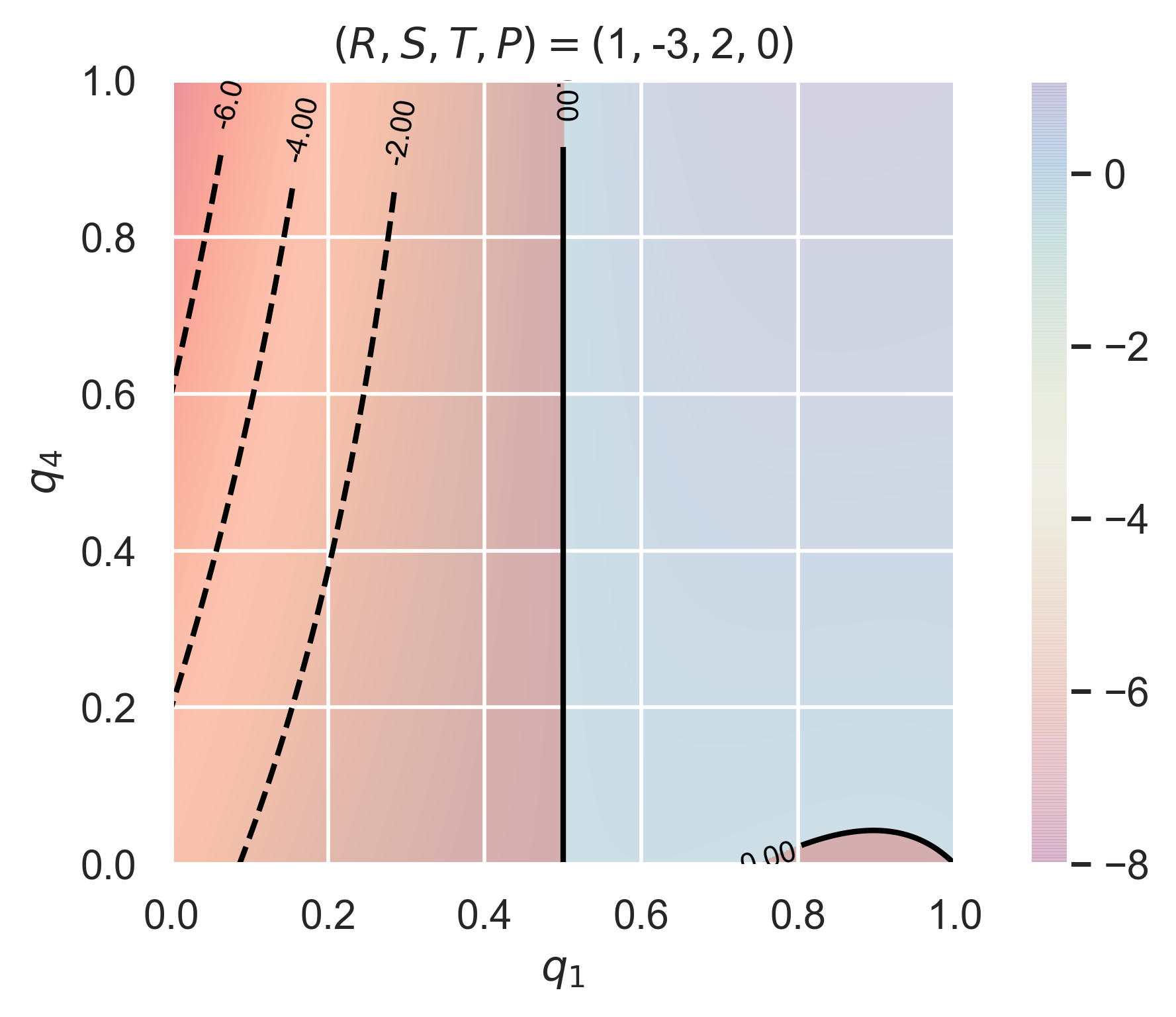}
  \caption{Contour curves of $a_{C0}d_{C0}$ for the general IPD game with $T + S < 2P$. Here we use $(R, S, T, P) = (1, -3, 2, 0)$. The two red regions are where $a_{C0}d_{C0} \leq 0$.}
  \label{target_C_abn}
\end{figure}

To guarantee that the product of $a_{C0}$ and $d_{C0}$ is negative, we need
\begin{equation}
\begin{cases}
0 \leq q_1 < q_c, \\
0 < q_4 \leq 1, 
\end{cases}
\qquad \text{or} \qquad 
\begin{cases}
q_C < q_1 < 1,\\
0 < q_4 < h_C(q_1).\\
\end{cases}
\end{equation}
An example is given in Figure~\ref{target_C_abn}.

\section{Class D of unbending strategies}

The last class requires $q_4 = h_D(q_1, q_2, q_3) = a_{D0}/(2R - T - S)$, under which we obtain
\begin{equation}
\begin{cases}
s_X = \frac{(T - S)a_{D0}\chi}{f_D(\chi)} + P,\\
\frac{ds_X}{d\chi} = \frac{(T - S)a_{D0}d_{D0}}{f_D^2(\chi)}.
\end{cases}
\end{equation}
Here, $f_D(\chi) = d_{D1}\chi + d_{D0}$ and 
\begin{equation}
\begin{cases}
a_{D0} = -(T + S - 2P)q_1 + (R - P)(q_2 + q_3) + T + S - R - P, \\
d_{D1} = -(T - S)q_1 + (T - R)q_2 + (R - S)q_3 + R - S, \\
d_{D0} = -(T - S)q_1 + (R - S)q_2 + (T - R)q_3 + T - R.
\end{cases}
\end{equation}
\\
Simple calculation shows that
\begin{equation}
\mfootnotesize{
\begin{cases}
(R - P)d_{D1} - (T - R)a_{D0} = (2R - T - S)[(T - P)(1 - q_1) + (R - P)q_3] \geq 0, \\
(R - P)(d_{D1} + d_{D0}) - (T - S)a_{D0} = (T - S)(2R - T - S)(1 - q_1) \geq 0.
\end{cases}}
\end{equation}
Further, $0 < q_4 \leq 1$ implies $0 < a_{D0} \leq 2R - T - S$. We conclude that $f_D(\chi) > 0$ for $\chi > 1$. \par

Likewise, to determine the monotonicity of $s_X$, we need to figure out the sign of $a_{D0}d_{D0}$. The factor $d_{D0}$ should be negative for $ds_X/d\chi$ to be negative. Moreover, routine calculation yields
\begin{equation}
\mfootnotesize{
(R - S)[(2R - T - S) - a_{D0}] + (R - P)d_{D0} = (2R - T - S)[(P - S)(1 - q_1) - (R - P)q_3 + (R - S)] > 0.}
\end{equation}
In other words, if $d_{D0} < 0 < a_{D0}$, we will always have $a_{D0} \leq 2R - T - S$. Therefore, the implicit condition for $s_X$ to be a decreasing function is 
\begin{equation}
d_{D0} < 0 < a_{D0}.
\label{inequality_D}
\end{equation}
\par

Notice that both $a_{D0} = 0$ and $d_{D0} = 0$ take the form of the equation of a plane in the 3-dimensional space and the above inequality corresponds to the space between the two planes whilst inside a unit cube. \par

We list a few critical points on the two planes as well as their line of intersection.
\begin{itemize}
\item $a_{D0} = 0$:  $(1, 1, 0)$, $(1, 0, 1)$.
\item $d_{D0} = 0$:  $(1, 1, 0)$, $(0, 0, -1)$.
\item $d_{D0} = a_{D0} = 0: \begin{bmatrix}x\\y\\z\end{bmatrix} = \begin{bmatrix}R - P\\T - P\\-(P - S)\end{bmatrix}t + \begin{bmatrix}1\\1\\0\end{bmatrix}$.
\end{itemize}
The projection of their intersection onto the plane $q_3 = 1$ is $(T - P)q_1 - (R - P)q_2 - (T - R) = 0$. \par

For ease of notation, we use the implicit expression in Equation~\ref{inequality_D} together with the default restrictions on the probabilities to represent the set of $q_1$, $q_2$, and $q_3$:
\begin{equation}
\begin{cases}
0 \leq q_1, q_2, q_3 \leq 1, \\
d_{D0} < 0 < a_{D0},
\end{cases}
\end{equation} 
where $a_{D0}$ and $d_{D0}$ can also be rewritten as
\begin{equation}
\mfootnotesize{
\begin{cases}
a_{D0} = (T + S - 2P)(1 - q_1) - (R - P)(1 - q_2) + (R - P)q_3, \\
d_{D0} = (T - S)(1 - q_1) - (R - S)(1 - q_2) + (T - R)q_3.
\end{cases}}
\end{equation}

\subsection{Case \rom{1}:  \texorpdfstring{$T + S > 2P$}{e}}

\begin{figure}[H]
\centering
  \includegraphics[width=0.8\linewidth]{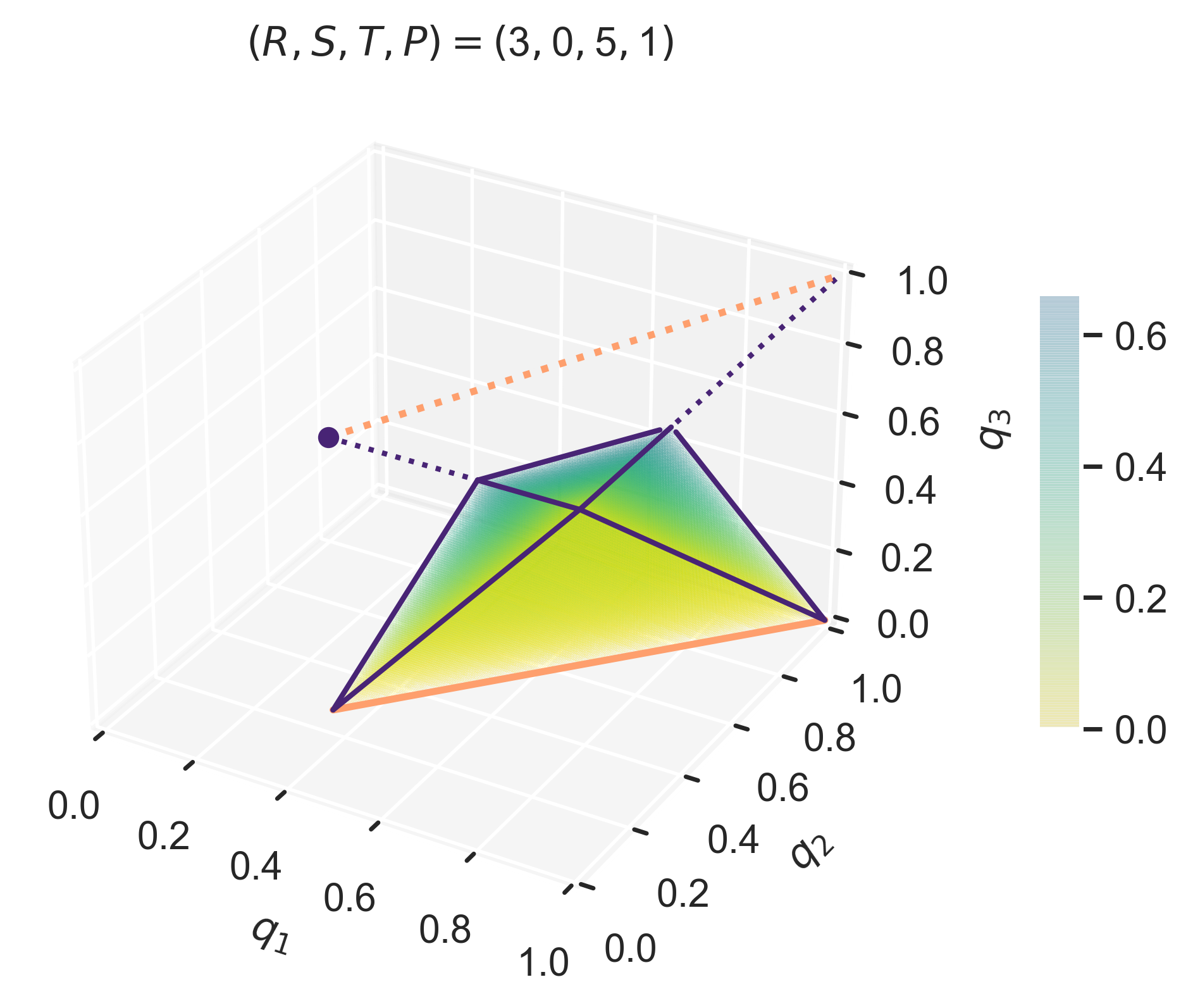}
  \caption{Set of points satisfying $d_{D0} < 0 < a_{D0}$. The color map indicates the value of $q_4$. The intersection of the two planes $a_{D0} = 0$ and $d_{D0} = 0$ is colored in pink. The projection of the set onto the plane $q_3 = 1$ is marked by dashed lines. In addition, the $x$-component of the navy point corresponds to the lower bound of $q_1$. Here, we use $(R, S, T, P) = (3, 0, 5, 1)$.}
  \label{target_D_con}
\end{figure}

An example of the possible set of $q_1$, $q_2$, and $q_3$ is given in Figure~\ref{target_D_con}. A point $(q_1, q_2, q_3)$ in the 3-dimensional space, lying between the two planes $a_{D0} = 0$ and $d_{D0} = 0$, corresponds to the strategy $\bm{q} = (q_1, q_2, q_3, h_D(q_1, q_2, q_3))$.

\subsection{Case \rom{2}: \texorpdfstring{$T + S = 2P$}{e}}

\begin{figure}[H]
\centering
  \includegraphics[width=0.8\linewidth]{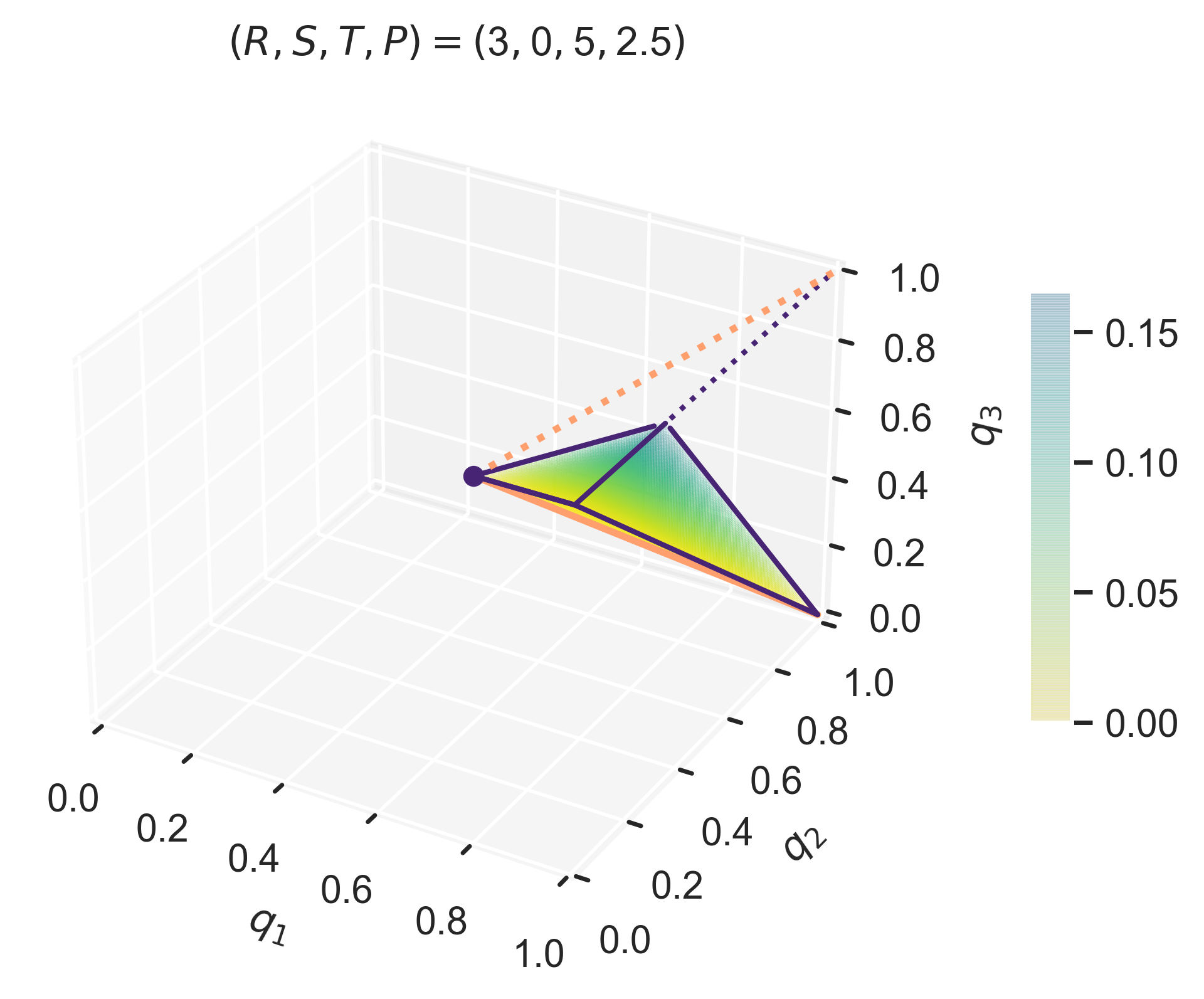}
  \caption{Set of points satisfying $d_{D0} < 0 < a_{D0}$. The color map indicates the value of $q_4$. The intersection of the two planes $a_{D0} = 0$ and $d_{D0} = 0$ is colored in pink. The projection of the set onto the plane $q_3 = 1$ is marked by dashed lines. In addition, the $x$-component of the navy point corresponds to the lower bound of $q_1$. Here, we use $(R, S, T, P) = (3, 0, 5, 2.5)$.}
  \label{target_D_eqn}
\end{figure}

The visualization of the set of $q_1$, $q_2$, and $q_3$ is given in Figure~\ref{target_D_eqn}. 

\subsection{Case \rom{3}: \texorpdfstring{$T + S < 2P$}{e}}

\begin{figure}[H]
\centering
  \includegraphics[width=0.8\linewidth]{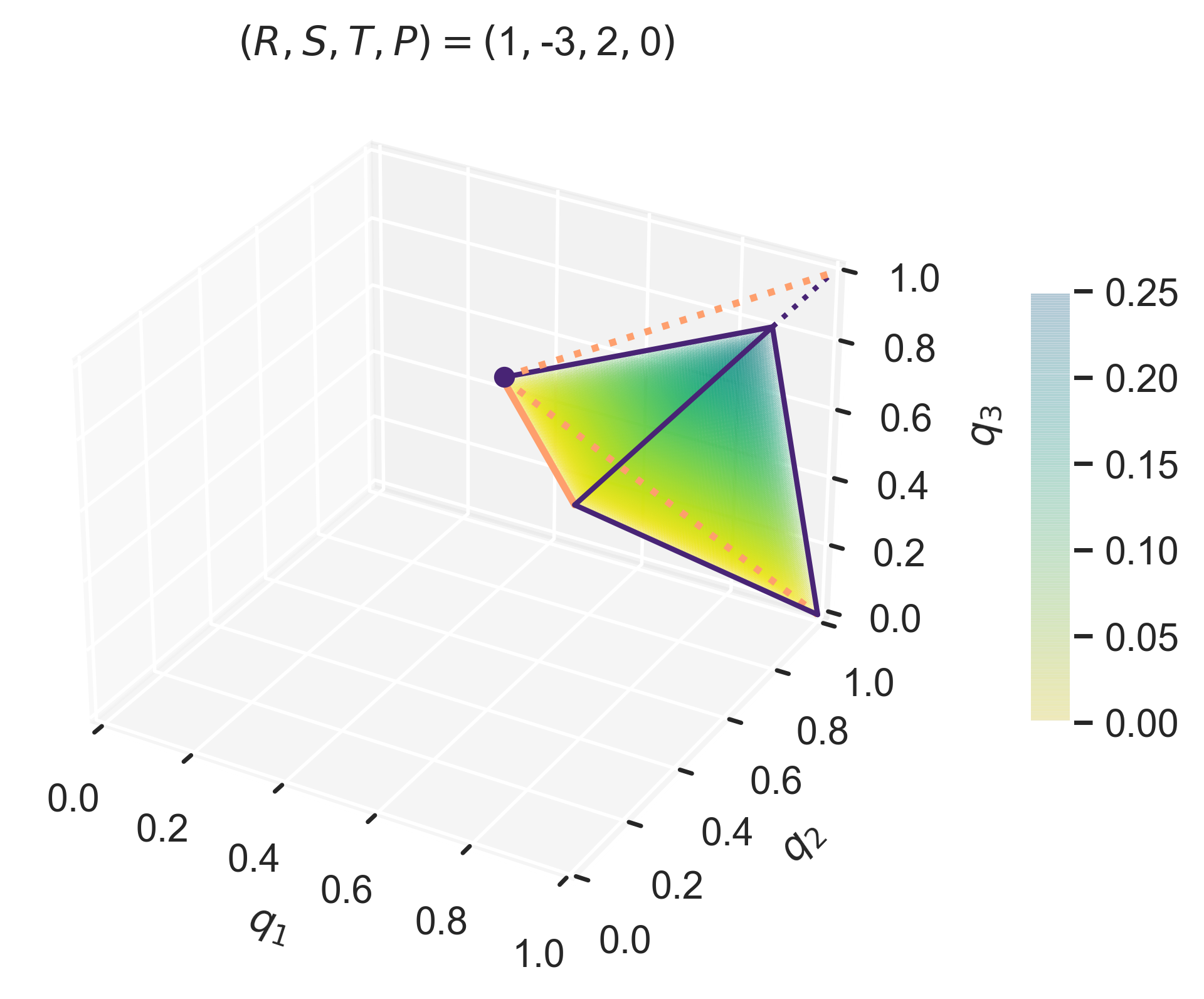}
  \caption{Set of points satisfying $d_{D0} < 0 < a_{D0}$. The color map indicates the value of $q_4$. The intersection of plane $a_{D0} = 0$ and plane $d_{D0} = 0$ (dashed line) and that of $a_{D0} = 0$ and $q_3 = 1$ (solid line) are both colored in pink. The projection of the set onto the plane $q_3 = 1$ is marked by dashed lines. In addition, the $x$-component of the navy point corresponds to the lower bound of $q_1$. Here, we use $(R, S, T, P) = (1, -3, 2, 0)$.}
  \label{target_D_abn}
\end{figure}

For the last case, an example is given in Figure~\ref{target_D_abn}.

\section{The impact of unbending strategies on extortionate ZD strategies: dominance, average payoffs, and best response}

The four classes of unbending strategies have been comprehensively studied so far. We further answer three significant questions in the scenario where X plays an extortionate ZD strategy and Y plays an unbending strategy:
\begin{itemize}
\item Is X's payoff $s_X$ always greater than the punishment $P$?
\item Is X's payoff $s_X$ always greater than Y's payoff $s_Y$?
\item Does Y's payoff $s_Y$ always decrease with respect to $\chi$?
\end{itemize}

\subsection{Comparing \texorpdfstring{$s_X$}{e} and \texorpdfstring{$P$}{e}}
\begin{table}[H]
\centering
\footnotesize
\tabulinesep=1.5mm
\begin{tabu}{c c c}
\hline
\rowcolor[HTML]{EFEFEF} 
Class & $s_X$ & $a_i$ \\
A & $\frac{(T - S)(R - P)q_4\chi(a_{A1}\chi + a_{A0})}{f_A(\chi)} + P$ & 
\makecell[cc]{$a_{A1} = (P - S)q_2 + T + S - 2P$ \\ $a_{A0} = (T - P)q_2 - (T + S - 2P)$} \\
\rowcolor[HTML]{EFEFEF} 
B & $\frac{(T - S)a_{B0}q_4\chi}{f_B(\chi)} + P$ & $T + S - 2P$ \\
C & $\frac{(T - S)a_{C0}q_4\chi}{f_C(\chi)} + P$ & $(R - P)q_1 +  (T + S - 2P)(1 - q_1)$ \\
\rowcolor[HTML]{EFEFEF} 
D & $\frac{(T - S)a_{D0}\chi}{f_D(\chi)} + P$ & $ (T + S - 2P)(1 - q_1) - (R - P)(1 - q_2) + (R - P)q_3$ \\
\hline
\end{tabu}
\caption{Payoff of an extortionate ZD strategy playing against an unbending strategy. The expressions of $f_A(\chi)$, $f_B(\chi)$, $f_C(\chi)$, and $f_D(\chi)$ have been given previously and we have verified that they are all positive.}
\label{summary_payoff}
\end{table}

Table~\ref{summary_payoff} lists the expressions of $s_X$ when playing against unbending strategies from different classes. The sign of $s_X - P$ is identical to that of $a_0$ or $a_1\chi + a_0$. It is worth mentioning that for Class A, if we let $q_2 = 0$, the expression can be further simplified as

\begin{equation}
\mfootnotesize{
\frac{(T - S)(T + S - 2P)q_4\chi}{[(T - S)q_4 + (T - P)]\chi + (T - S)q_4 + (P - S)} + P.}
\end{equation}
\par

After some routine calculation, we present the possible signs of $s_X - P$ in Table~\ref{summary_signs}. Besides, given the linear relation in Equation~\ref{linear_relation}, we know that $s_X - P$ and $s_Y - P$ always share the same sign. It is clear from the table that $s_X$ is greater than $P$ under most circumstances yet the opposite may happen as well. 
\begin{table}[H]
\centering
\footnotesize
\tabulinesep=1.5mm
\begin{tabu}{|c | c | c | c |}
\hline
\backslashbox{Class}{Case} & $T + S > 2P$ & $T + S = 2P$ & $T + S < 2P$ \\ \hline
A & $\bm{+}$ & $\bm{+}$ & $\begin{cases}\bm{-}, & q_2 = 0 \\ \bm{+} \to \bm{-}, & 0 < q_2 < q_{A}\\ \bm{+}. & q_{A} \leq q_2 < 1 \end{cases}$ \\ \hline
B & $ \emptyset$ & $\emptyset$ & $\bm{-}$ \\ \hline
C & $\bm{+}$ & $\bm{+}$ & $\begin{cases}\bm{-}, & 0 \leq q_1 < q_c\\  \bm{+}. & q_C < q_1 < 1\end{cases}$ \\ \hline
D & $\bm{+}$ & $\bm{+}$ & $\bm{+}$ \\
\hline
\end{tabu}
\caption{Sign of $s_X - P$ ($s_Y - P$ or $s_X - s_Y$) as $\chi$ increases from $1$ to $+\infty$. Here, $q_{A} = (2P - T - S)/(P - S)$.}
\label{summary_signs}
\end{table}

\subsection{Comparing \texorpdfstring{$s_X$}{e} and \texorpdfstring{$s_Y$}{e}}

Notice that the linear relation in Equation~\ref{linear_relation} also implies that $s_X - P > 0$ is equivalent to $s_X > s_Y$. Therefore, from Table~\ref{summary_signs}, we conclude that the extortioner's payoff is likely to be less than its opponent's payoff if $T + S < 2P$. More precisely, the IPD game needs to meet one of the following four conditions:
\begin{enumerate}[label=(\roman*)]
\item the opponent plays a strategy from Class A with $q_2 = 0$ (a subset of Class B), 
\item the opponent plays a strategy from Class A with $0 < q_2 < q_{A}$ and the extortion factor $\chi > -a_{A0}/a_{A1}$,
\item the opponent plays a strategy from Class B, or
\item the opponent plays a strategy from Class C with $0 \leq q_1 < q_c$.
\end{enumerate}

\subsection{Monotonicity of \texorpdfstring{$s_Y$}{e}}

For the completeness of our study, we also discuss the monotonicity of $s_Y$. The expressions of $s_Y$ are similar to those in Table~\ref{summary_payoff} and we only need to remove $\chi$ from the numerators. \par

Recall that if $s_X$ decreases with $\chi$ and $s_X - P > 0$, $s_Y$ will decrease with $\chi$ as well (once again, we use the linear relation). Now consider the four special circumstances from the last section. \par

For (i), we get
\begin{equation}
\mfootnotesize{
\frac{ds_Y}{d\chi} = -\frac{(T - S)(T + S - 2P)[(T - S)q_4 + (T - P)]q_4}{\{[(T - S)q_4 + (T - P)]\chi + (T - S)q_4 + (P - S)\}^2}.}
\end{equation}
It follows immediately that $s_Y$ increases with respect to $\chi$ for $\chi > 1$. \par

For (ii), we have
\begin{equation}
\mfootnotesize{
s'_Y = \frac{(T - S)(R - P)q_4(-a_{A1}d_{A2}\chi^2 - 2a_{A0}d_{A2}\chi + a_{A1}d_{A0} - a_{A0}d_{A1})}{f_A^2(\chi)}.}
\end{equation}
Given that $a_{A1} < 0$ and $d_{A2} > 0$, the last factor of the numerator is a quadratic function opening upwards. Additionally, we evaluate its value at $\chi = 1$ and get
\begin{equation}
\mfootnotesize{
-(T - S)^2q_2\{[(T - R)q_2 + (2R - T - S)]q_4 + (R - P)(1 - q_2)\} < 0.}
\end{equation}
Hence $s_Y$ first decreases then increases with $\chi$ for $\chi > 1$. \par

As to (iii) and (iv), notice that $s_Y - P$ is hyperbolic with a negative numerator and a positive denominator. It is straightforward to show that $s_Y$ is an increasing function of $\chi$. \par

We summarize the above results in Table~\ref{summary_monotonicity}.
\begin{table}[H]
\centering
\footnotesize
\tabulinesep=1.5mm
\begin{tabu}{|c | c | c | c |}
\hline
\backslashbox{Class}{Case} & $T + S > 2P$ & $T + S = 2P$ & $T + S < 2P$ \\ \hline
A & $\bm{\searrow}$ & $\bm{\searrow}$ & $\begin{cases}\bm{\nearrow}, & q_2 = 0 \\ \bm{\searrow} \to \bm{\nearrow}, & 0 < q_2 < q_{A}\\ \bm{\searrow}. & q_{A} \leq q_2 < 1 \end{cases}$\\ \hline
B & $ \emptyset$ & $\emptyset$ & $\bm{\nearrow}$\\ \hline
C & $\bm{\searrow}$ & $\bm{\searrow}$ & 
$\begin{cases}\bm{\nearrow}, & 0 \leq q_1 < q_c\\  \bm{\searrow}. & q_C < q_1 < 1\end{cases}$\\ \hline
D & $\bm{\searrow}$ & $\bm{\searrow}$ & $\bm{\searrow}$ \\ \hline
\end{tabu}
\caption{Monotonicity of $s_Y$.}
\label{summary_monotonicity}
\end{table}

\subsection{Maximum values of \texorpdfstring{$s_X$}{e} and \texorpdfstring{$s_Y$}{e}}

Now that we have figured out the relationship between $s_X$ and $P$ ($s_Y$ and $P$) and the monotonicity of $s_Y$, we can finally identify the maximum values of the two payoffs $s_X$ and $s_Y$ as well as the corresponding unbending strategies. \par

Given that $s_X$ is a decreasing function of $\chi$, player X must compromise with its opponent Y by reducing the extortion factor $\chi$ to $1$ to get the optimal payoff, which is given in Table~\ref{summary_maximum_o}. We also have $s_X(1) = s_Y(1)$. 
\begin{table}[H]
\centering
\tabulinesep=1.5mm
\begin{tabu}{c  c  c}
\hline
\rowcolor[HTML]{EFEFEF} 
Class & \multicolumn{2}{c}{$s_X(1)$ and $s_Y(1)$} \\
A & \multicolumn{2}{c}{$R$} \\
\rowcolor[HTML]{EFEFEF} 
B & $q_1 =1$: $R$ \qquad & \qquad $0 \leq q_1 < 1$: $\frac{(T + S - 2P)q_4}{2q_4 + 1} + P$ \\
C & \multicolumn{2}{c}{$\frac{[(R - P)q_1 +  (T + S - 2P)(1 - q_1)]q_4}{(1 - q_1)^2 + (2 - q_1)q_4} + P$} \\
\rowcolor[HTML]{EFEFEF} 
D & \multicolumn{2}{c}{$\frac{(T + S - 2P)(1 - q_1) - (R - P)(1 - q_2) + (R - P)q_3}{2(1 - q_1) - (1 - q_2) + q_3} + P$} \\
\hline
\end{tabu}
\caption{Payoff of the fair extortioner (or its opponent) with extortion factor $\chi \to 1$.}
\label{summary_maximum_o}
\end{table}

Furthermore, $s_X(1)$ can be maximized if player Y adopts a proper strategy. After some calculation, we present the final results in Table~\ref{summary_maximum}.
\begin{table}[H]
\centering
\footnotesize
\tabulinesep=1.5mm
\begin{tabu}{c c c}
\hline
\rowcolor[HTML]{EFEFEF} 
Class & Maximum of $s_X(1)$ & Opponent's strategy $q$  \\
A & $R$ & $(1, q_2, 0, q_4)$ \\
\rowcolor[HTML]{EFEFEF} 
%B & $\displaystyle\lim_{\varepsilon \to 0}\frac{(T + S - 2P)\varepsilon}{2\varepsilon + 1} + P = P$ & $[q_1, 0, 0, \varepsilon]$ \\
B & $R$ & $(1, 0, 0, q_4)$ \\
C & $\displaystyle\lim_{\delta \to 0}\frac{[(R - P)(1 - \delta) + (T + S - 2P)\delta]q_4}{\delta^2 + (1 + \delta)q_4} + P = R $ &
$(1 - \delta, 1 - \delta, 1 - \delta, q_4)$ \\
\rowcolor[HTML]{EFEFEF} 
D & $R$ & $(1, q_2, q_3, \frac{(R - P)(q_2 + q_3 - 1)}{2R - T - S})$ \\
\hline
\end{tabu}
\caption{Maximum value of $s_X(1)$ and the corresponding unbending strategy. Notice that for Class B and Class C, $s_X(1)$ can get arbitrarily close to the maximum value $P$ (if $0 \leq q_1 < 1$) and $R$. And for any strategy $\bm{q}$, the four components $q_i$'s still need to satisfy the requirements of being an unbending strategy in Table~\ref{four_classes}.}
\label{summary_maximum}
\end{table}

\subsection{Figures}

To sum up, we visualize the two payoffs $s_X$ (Table~\ref{summary_examples_X}) and $s_Y$ (Table~\ref{summary_examples_Y}) in an IPD game where X plays an extortionate ZD strategy and Y plays an unbending strategy. We also show the parametric curve of $(s_X, s_Y)$ (Table~\ref{summary_examples_quad}), which can be zoomed in for a closer observation (Table~\ref{summary_examples_quad_zoom}). All three cases ($T + S > 2P$, $T + S = 2P$, and $T + S < 2P$) are included. \par

We emphasize the following facts which can be observed directly from these figures.
\begin{itemize}
\item Both the two payoffs $s_X$ and $s_Y$ are independent of the parameter $\phi$.
\item The payoff $s_X$ is greater than the punishment $P$ most of the time. Nevertheless, it may be \begin {enumerate*} [label=(\roman*\upshape)] \item less than $P$ if the extortion factor $\chi$ exceeds the threshold value $-a_{A0}/a_{A1}$, or \item always less than $P$ for $\chi$ greater than one.\end{enumerate*}
\item  The ``extortioner" is dominant most of the time ($s_X > s_Y$) yet it may be beaten by an unbending strategy ($s_X < s_Y$) if $T + S < 2P$ and
\begin{enumerate}[label=(\roman*)]
\item the opponent plays a strategy from Class A with $q_2 = 0$ (a subset of Class B),
\item the opponent plays a strategy from Class A with $\begin{cases}\chi > -a_{A0}/a_{A1} \\ 0 < q_2 < q_{A}\end{cases}$, or 
\item the opponent plays a strategy from Class B, or
\item the opponent plays a strategy from Class C with $0 \leq q_1 < q_c$.
\end{enumerate}
\item The payoff $s_X$ is a decreasing function of $\chi$. The other payoff $s_Y$ decreases with $\chi$ as well under most circumstances but it may \begin {enumerate*} [label=(\roman*\upshape)] \item first decreases then increases with $\chi$, or \item always increases with $\chi$.\end{enumerate*}
\item The relation between $T + S$ and $2P$ determines the shape and convexity of the quadrilateral with four vertices $(P, P)$, $(T, S)$, $(R, R)$ and $(S, T)$ (when $T + S = 2P$ it degenerates to a triangle, where the vertex $(P, P)$ lies on the edge pointing from $(T, S)$ to $(S, T)$). if $T + S > 2P$, the quadrilateral is convex and the parametric curve of $(s_X, s_Y)$ stays within it. Whereas if $T + S < 2P$, the quadrilateral is concave and it is possible for a fraction of, or even the entire parametric curve lies outside it.
\end{itemize}

\begin{table}[H]
\centering
\footnotesize
\tabulinesep=1.5mm
\begin{tabu}{| c | c | c | c |}
\hline
Class & \makecell[cc]{$T + S > 2P$\\ $(R, S, T, P) = (3, 0, 5, 1)$} & \makecell[cc]{$T + S = 2P$\\ $(R, S, T, P) = (3, 0, 5, 2.5)$}  
&  \makecell[cc]{$T + S < 2P$\\ $(R, S, T, P) = (1, -3, 2, 0)$} \\
\hline
\multirow{2}{*}{A} & \multirow{2}{*}{\includegraphics[width=2.5cm]{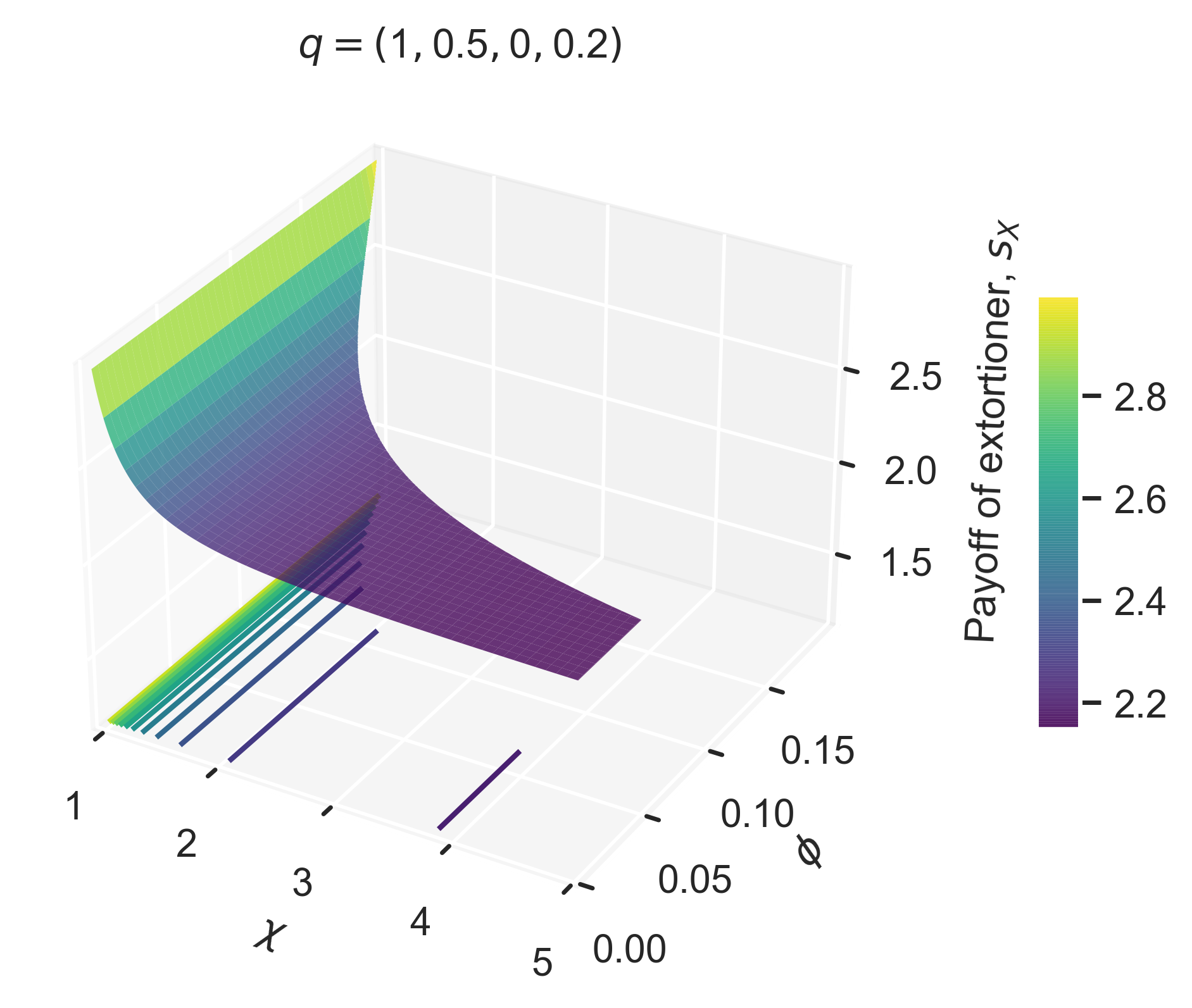}} & \multirow{2}{*}{\includegraphics[width=2.5cm]{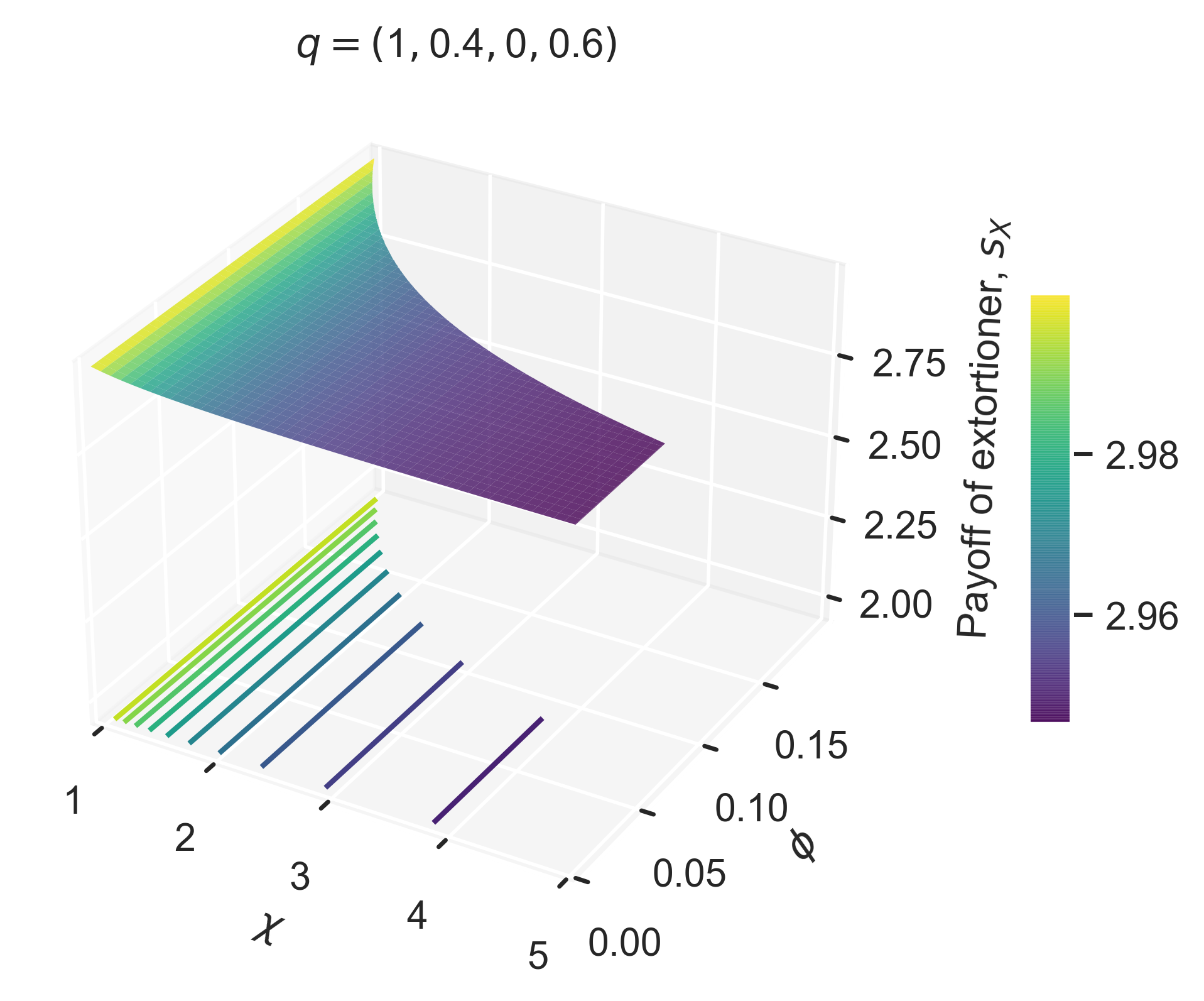}} &
$q_2 < q_A \qquad q_2 > q_A$ \\
& & & \includegraphics[width = 2cm]{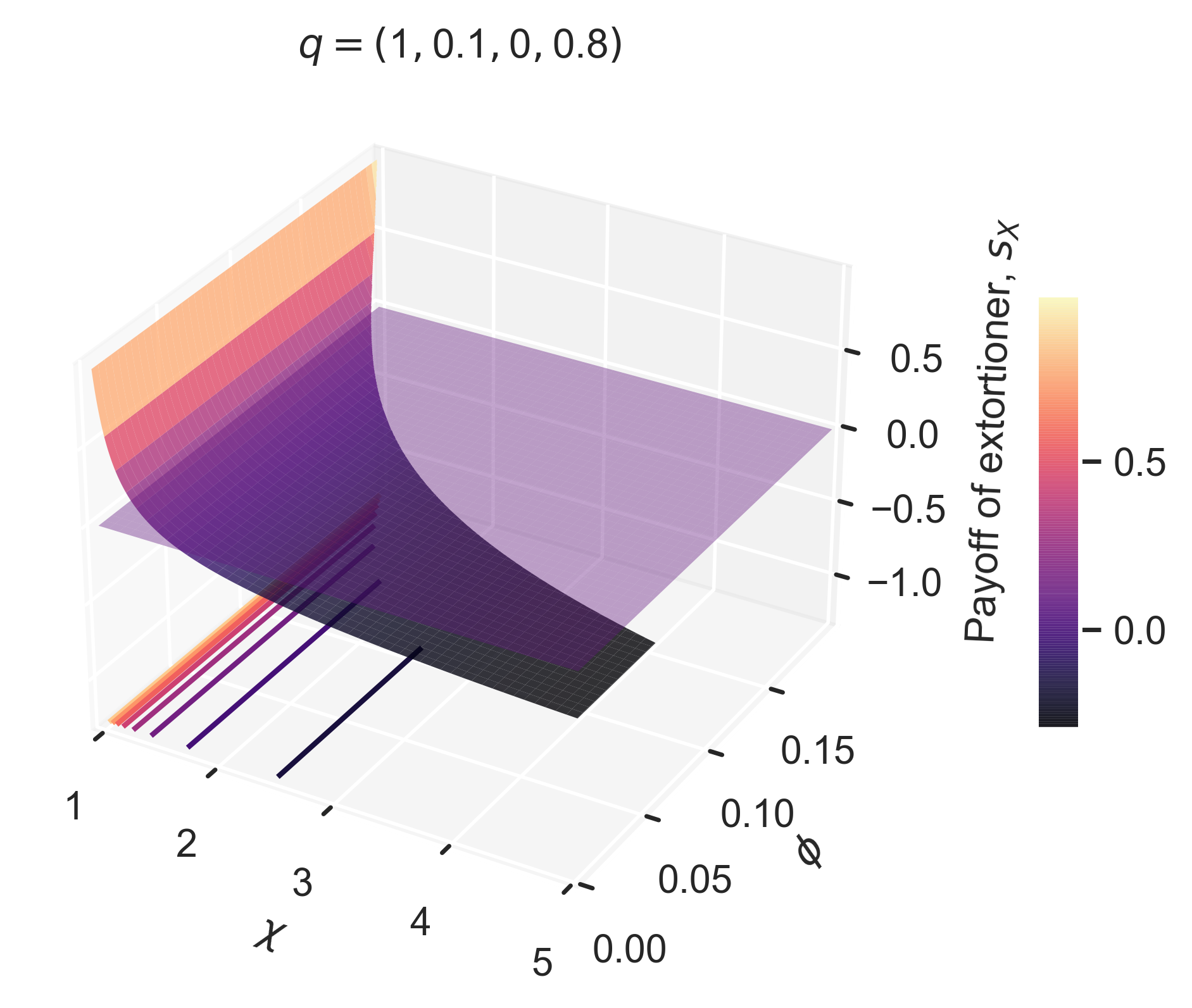}\includegraphics[width = 2cm]{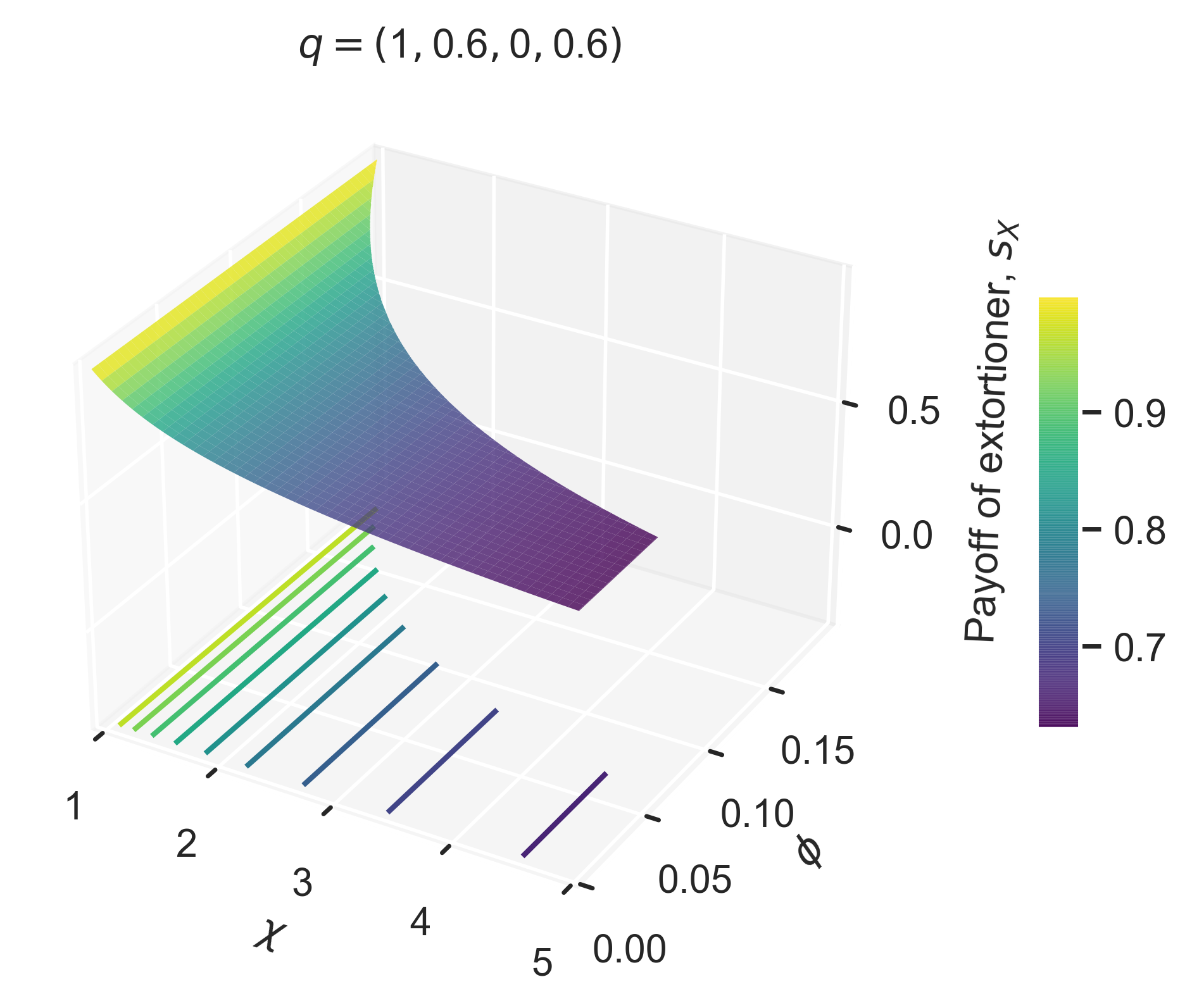} \\
\hline
B & & & \includegraphics[width=2.5cm]{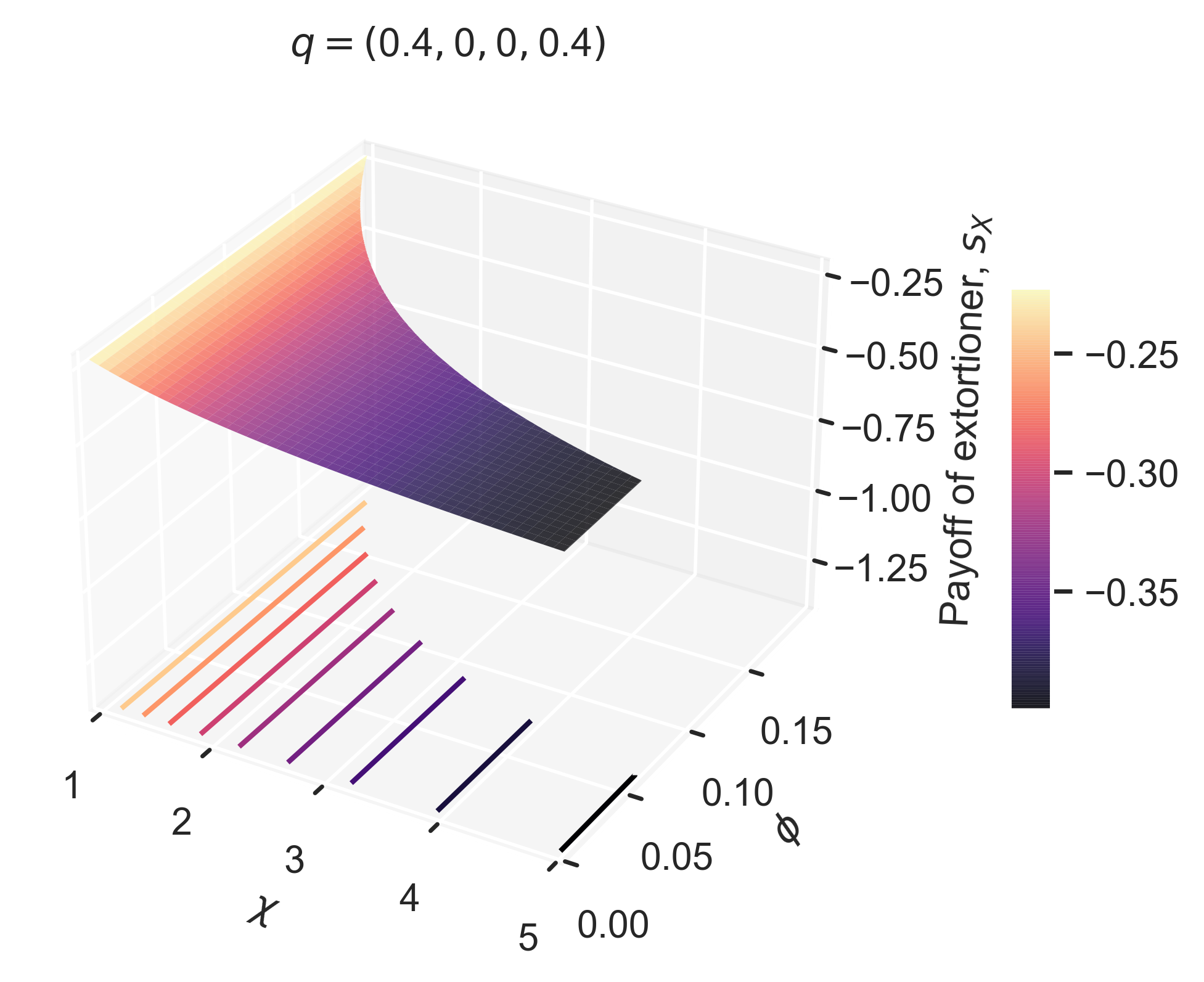} \\
\hline
\multirow{2}{*}{C} & \multirow{2}{*}{\includegraphics[width=2.5cm]{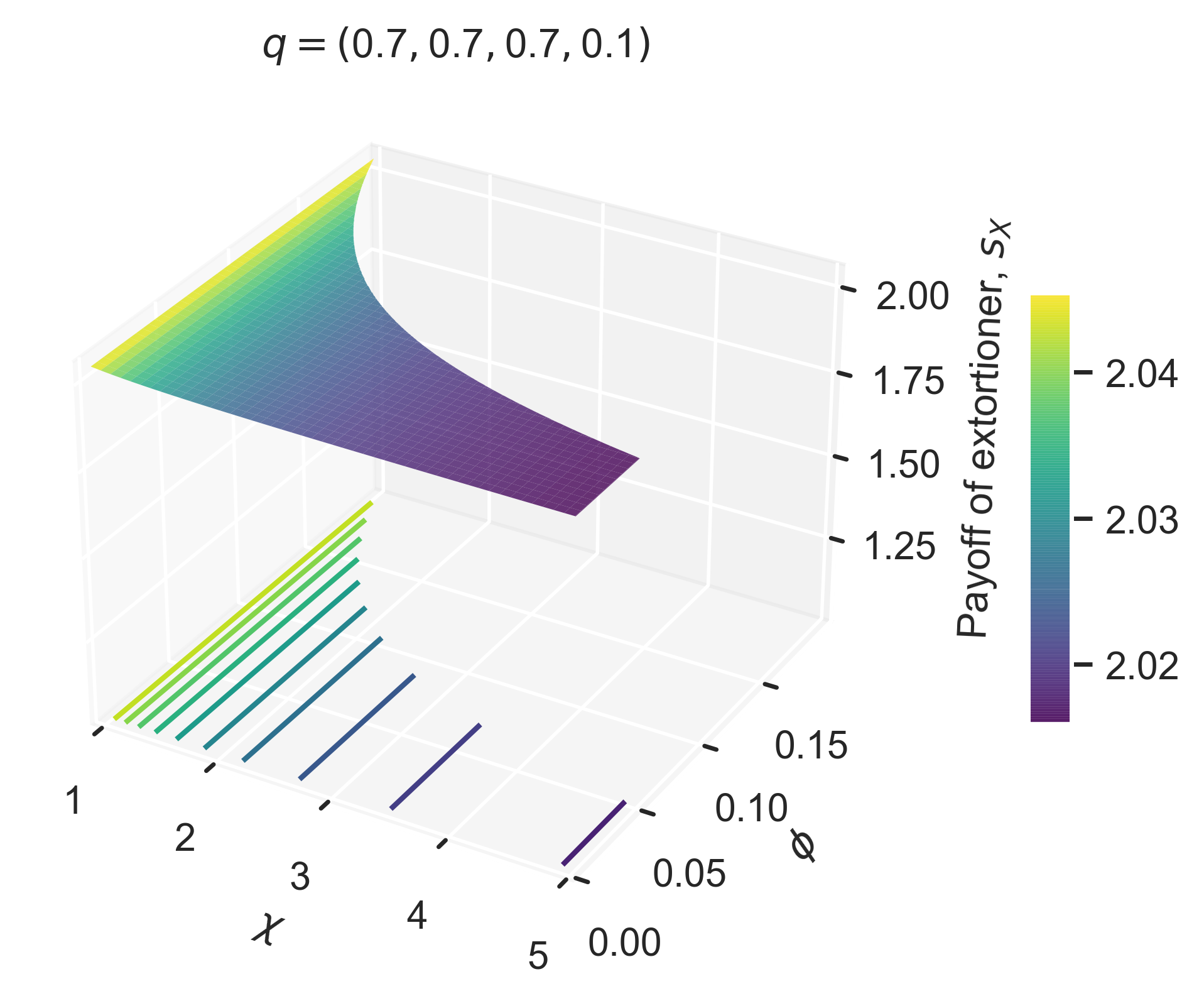}} 
& \multirow{2}{*}{\includegraphics[width=2.5cm]{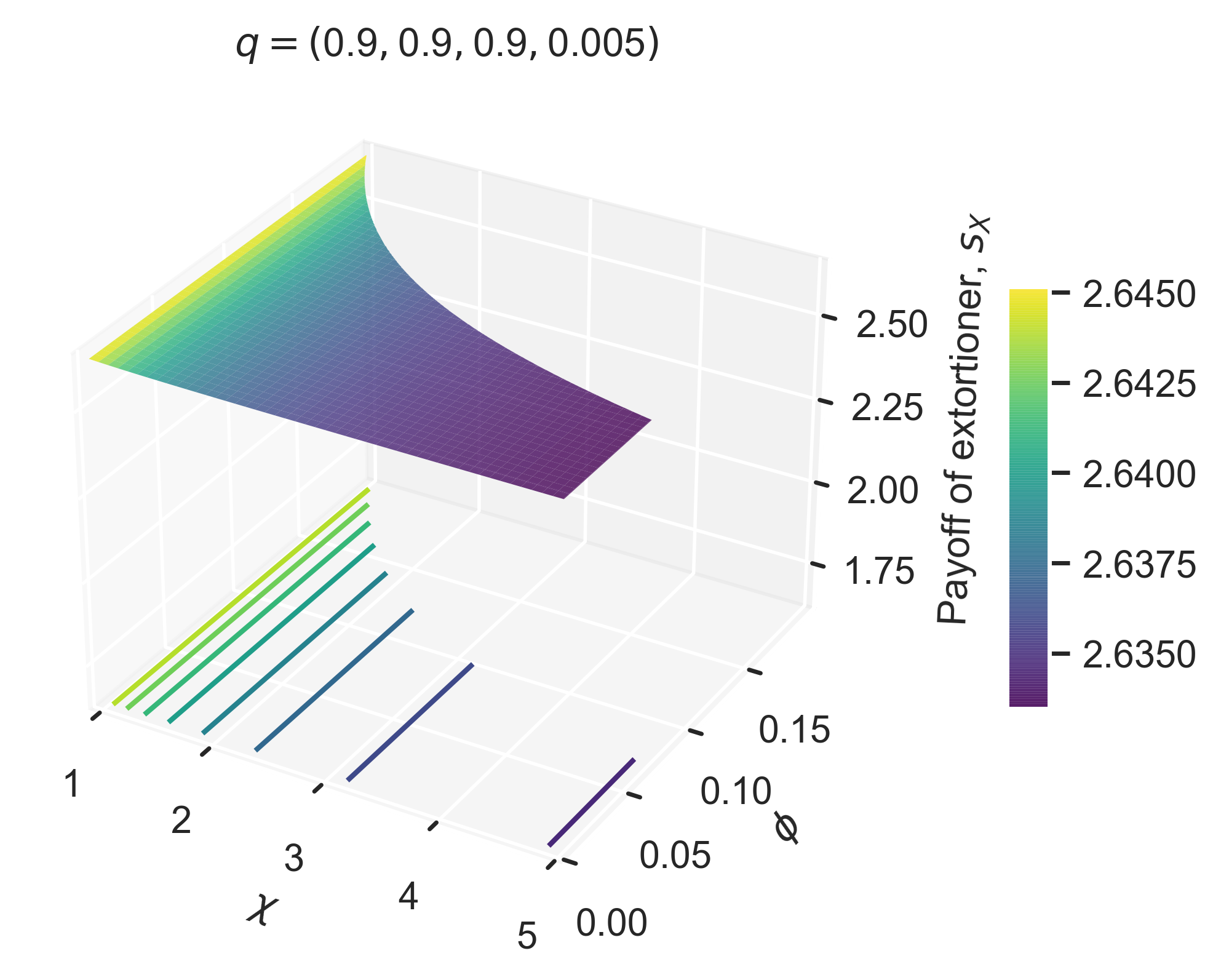}} 
& $q_1 < q_c \qquad q_1 > q_C$ \\
& & & \includegraphics[width=2cm]{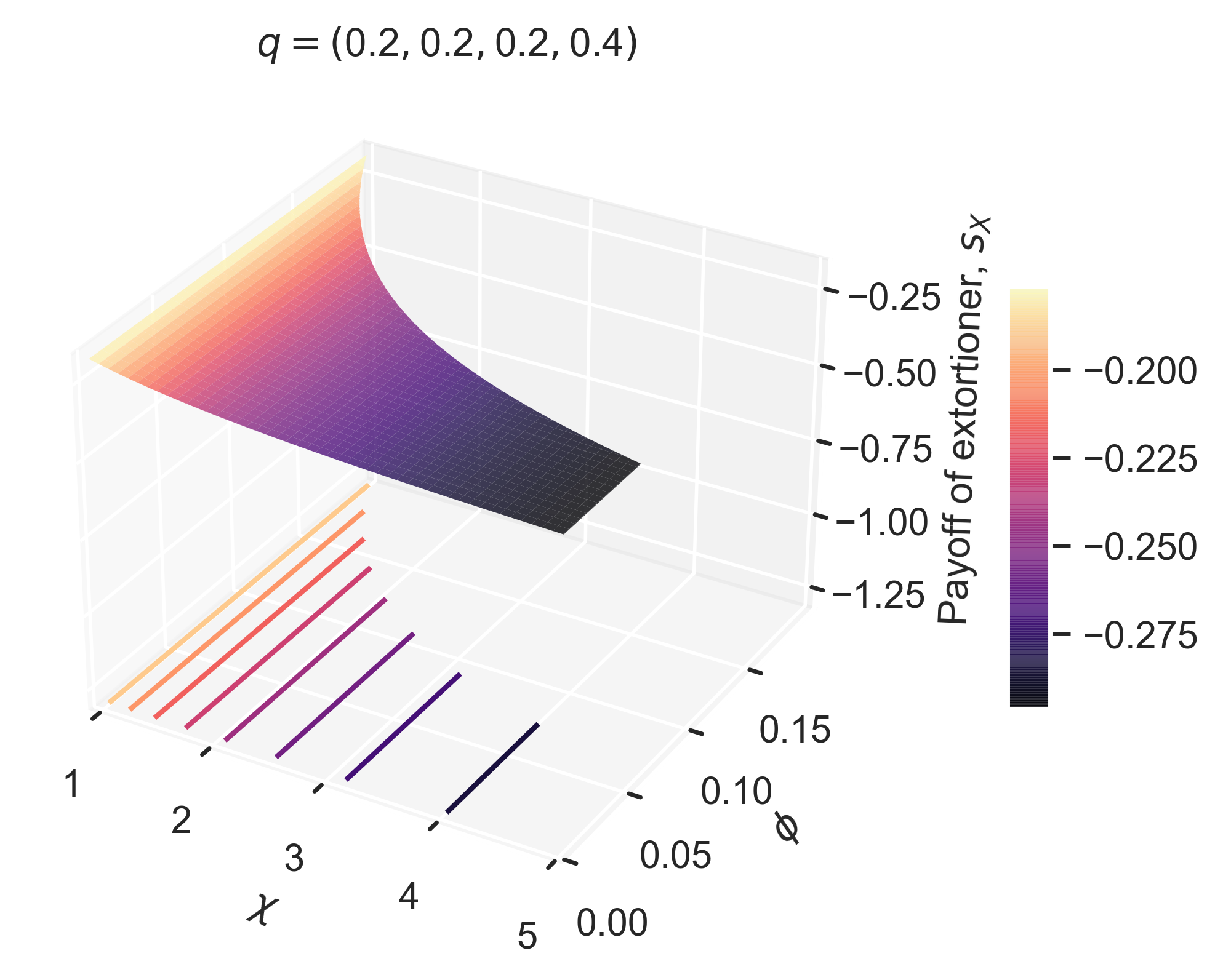}\includegraphics[width=2cm]{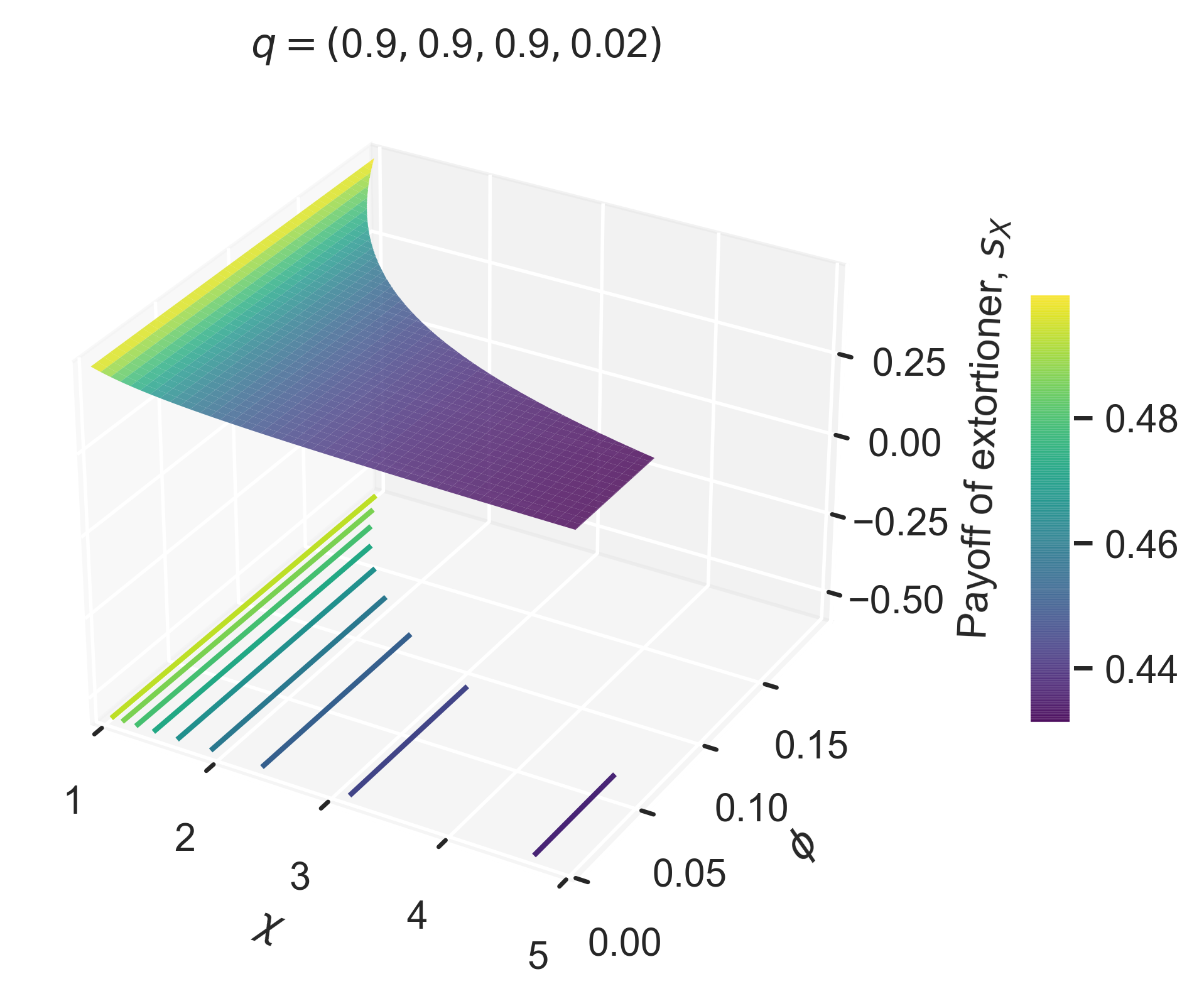} \\
\hline
D & \includegraphics[width=2.5cm]{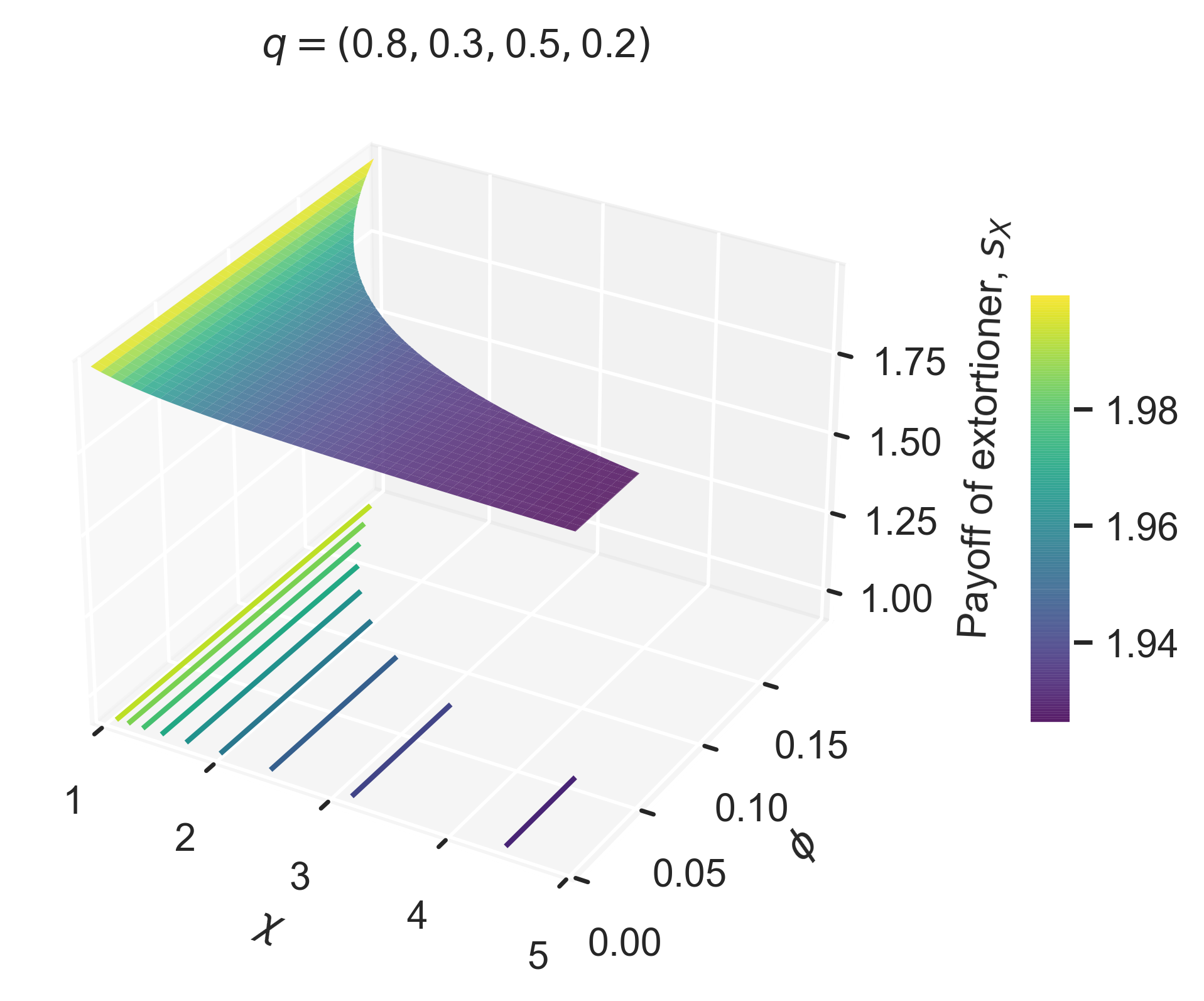} & \includegraphics[width=2.5cm]{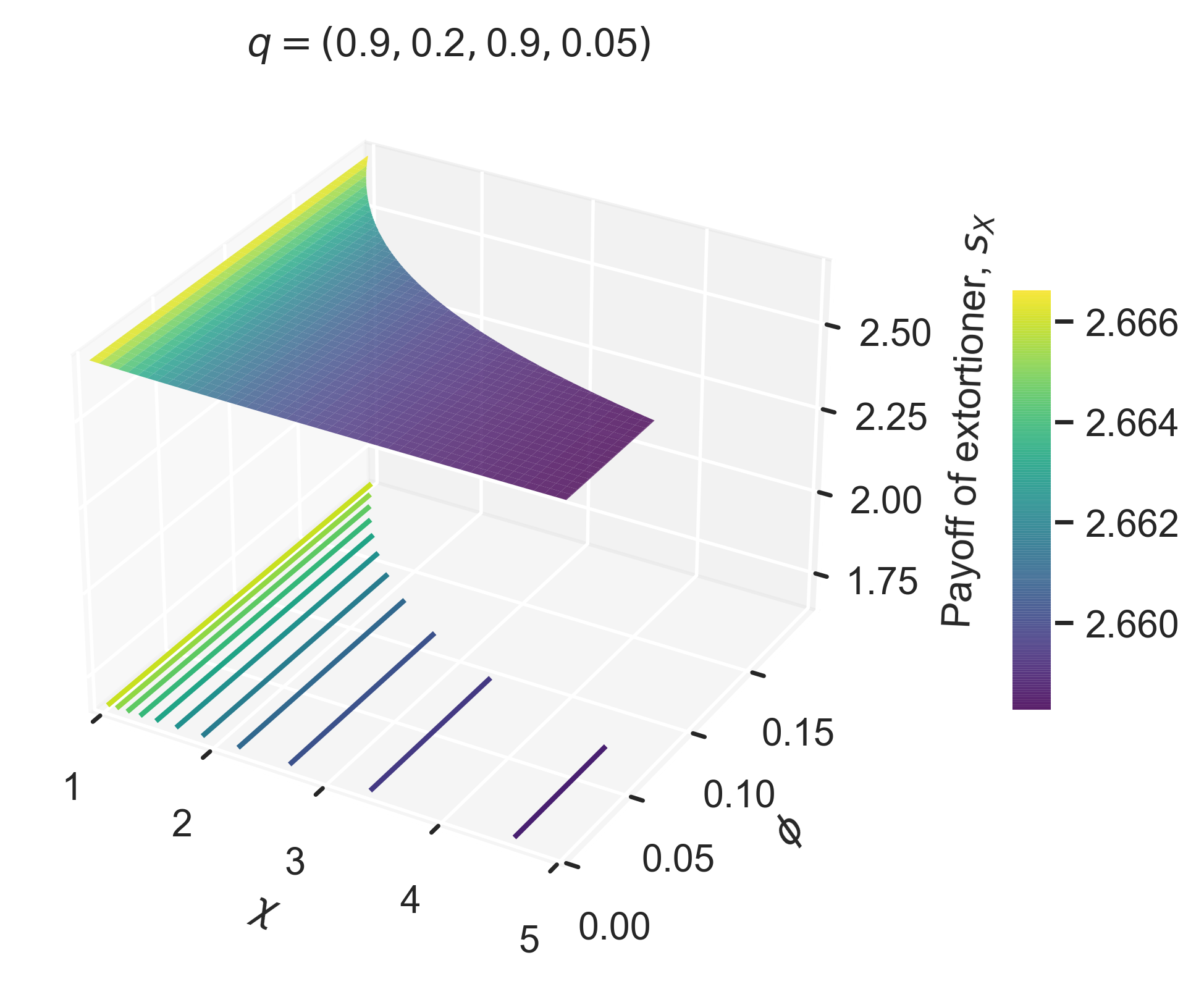}
& \includegraphics[width=2.5cm]{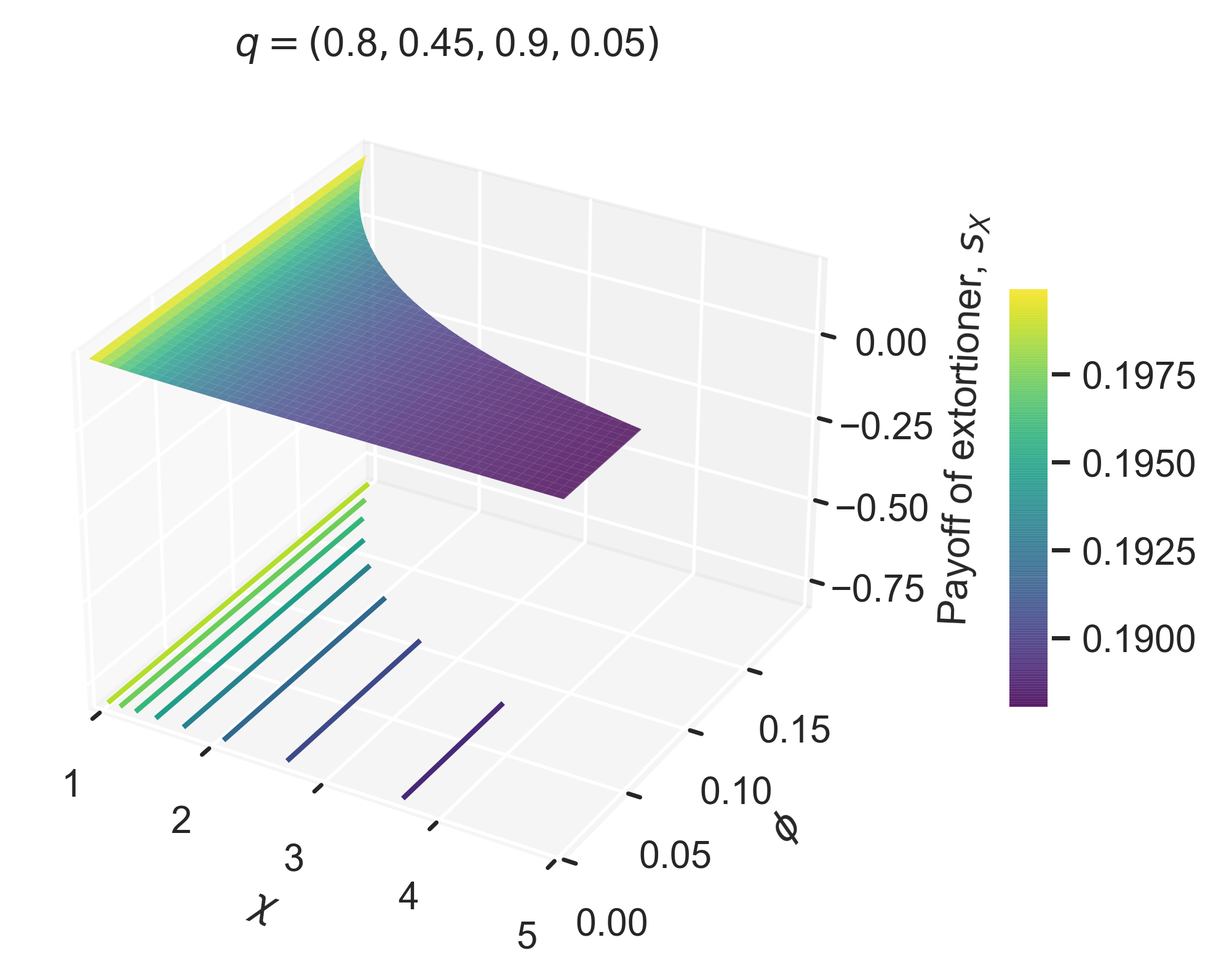} \\
\hline
\end{tabu}
\caption{Surface of $s_X$ when X uses an extortionate ZD strategy and Y uses an unbending strategy. The contour curves on the $xy$-plane are also given and the horizontal plane $z = P$ is added for one particular example on the upper right corner. We use two different colormaps depending on whether it is possible for $s_X$ to be less than $P$. As before, $q_{A} = (2P - T - S)/(P - S)$.}
\label{summary_examples_X}
\end{table}

\begin{table}[H]
\centering
\footnotesize
\tabulinesep=1.5mm
\begin{tabu}{| c | c | c | c |}
\hline
Class & \makecell[cc]{$T + S > 2P$\\ $(R, S, T, P) = (3, 0, 5, 1)$} & \makecell[cc]{$T + S = 2P$\\ $(R, S, T, P) = (3, 0, 5, 2.5)$}  
&  \makecell[cc]{$T + S < 2P$\\ $(R, S, T, P) = (1, -3, 2, 0)$} \\
\hline
\multirow{2}{*}{A} & \multirow{2}{*}{\includegraphics[width=2.5cm]{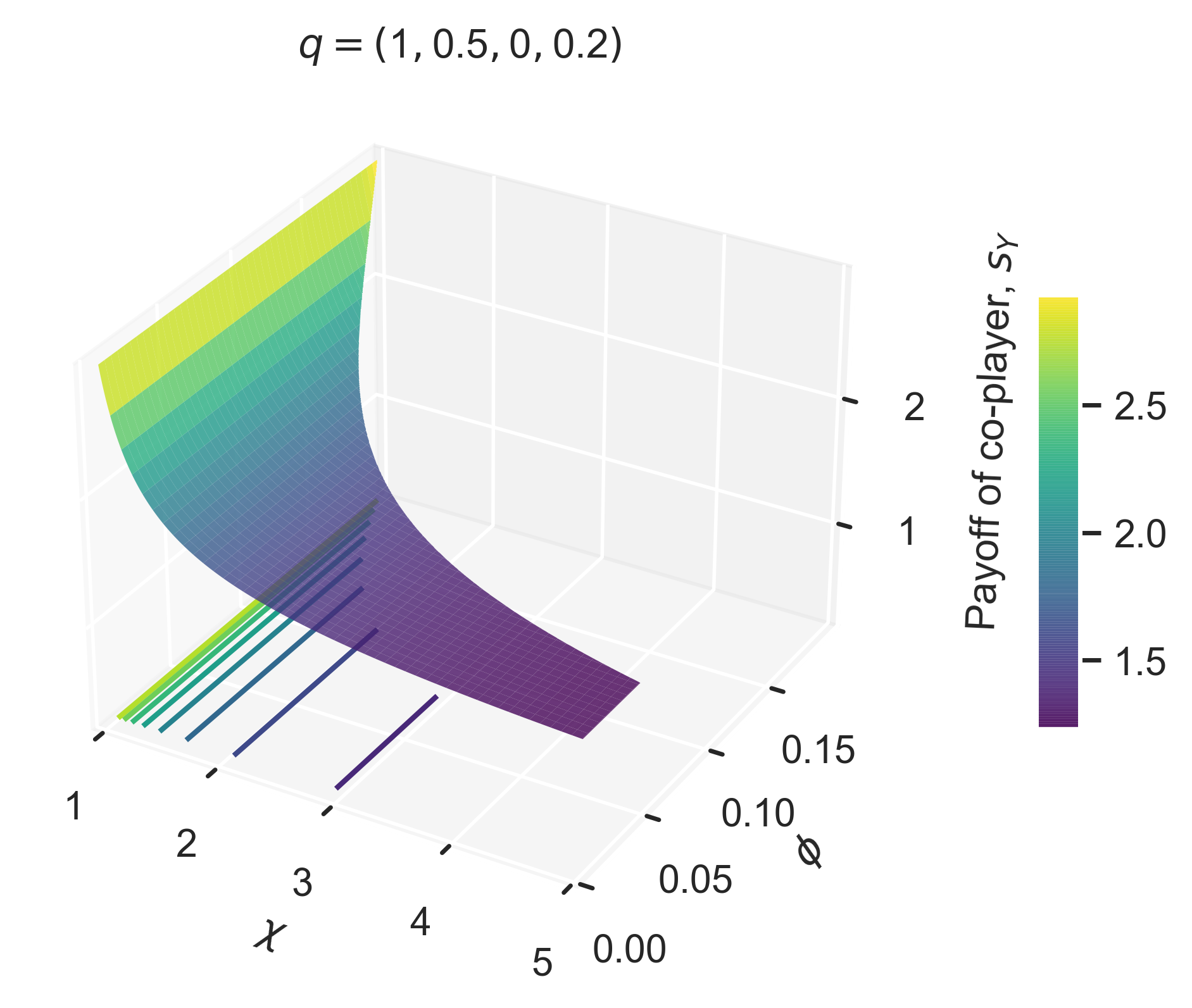}} & \multirow{2}{*}{\includegraphics[width=2.5cm]{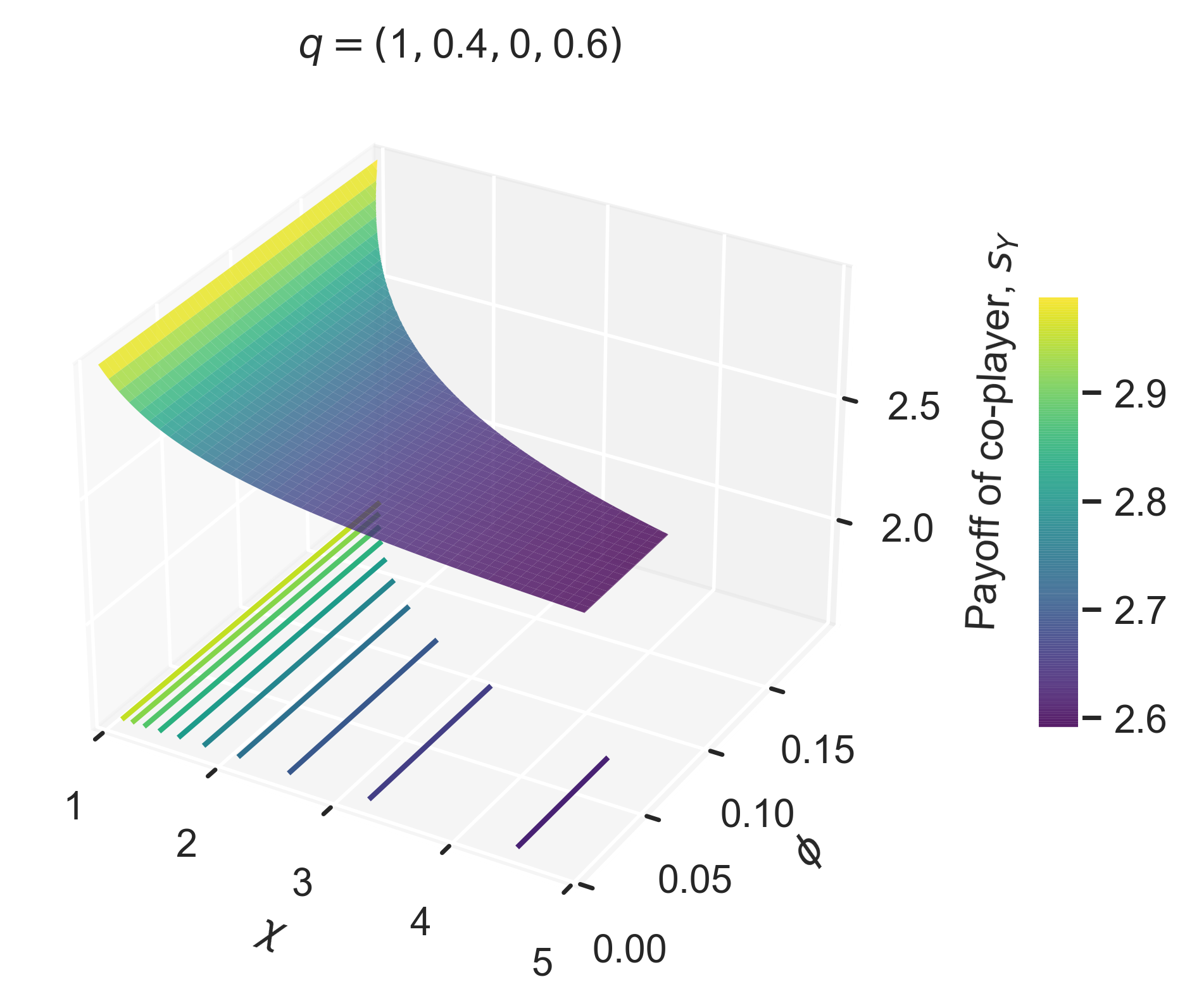}} &
$q_2 < q_A \qquad q_2 > q_A$ \\
& & & \includegraphics[width = 2cm]{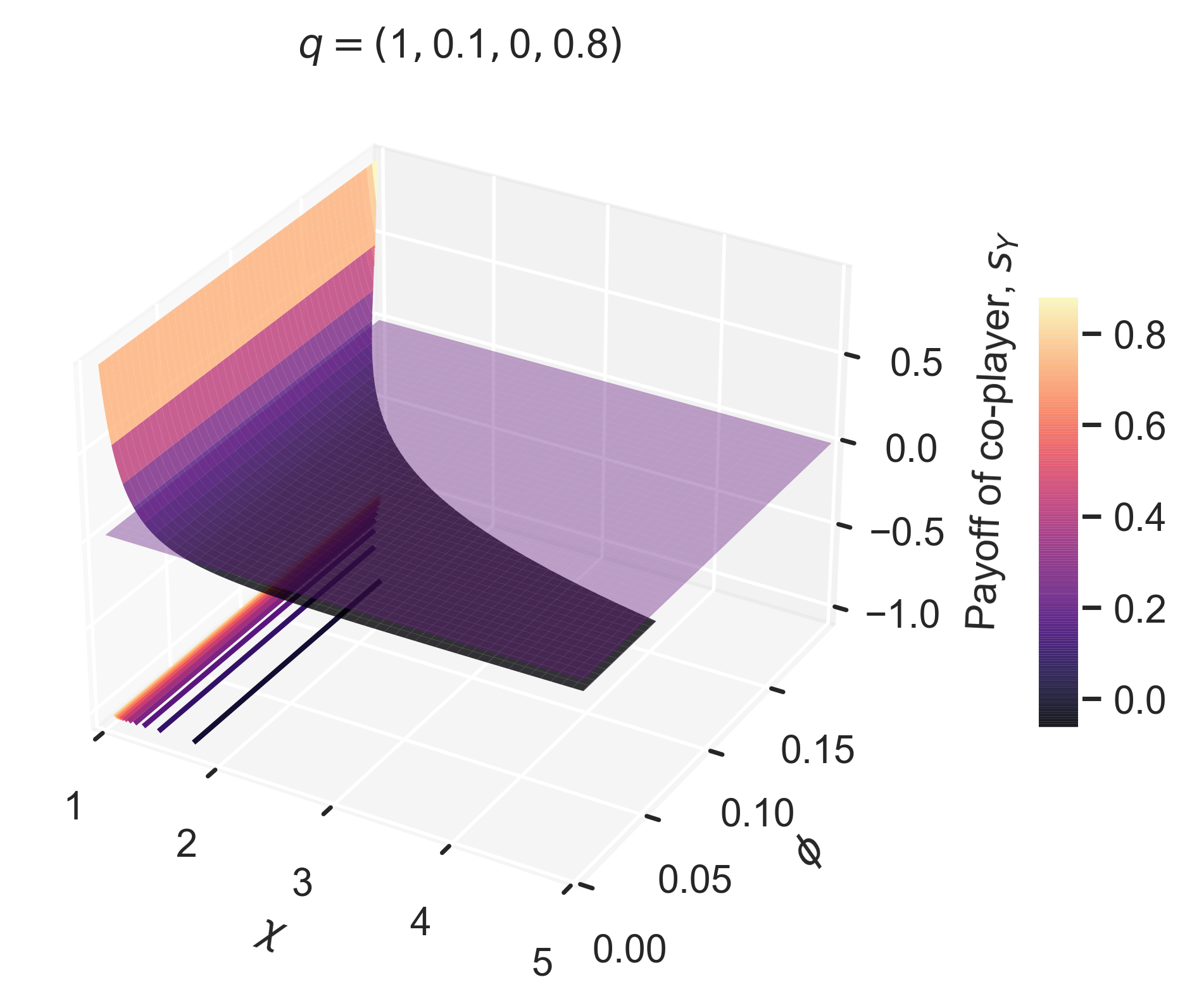}\includegraphics[width = 2cm]{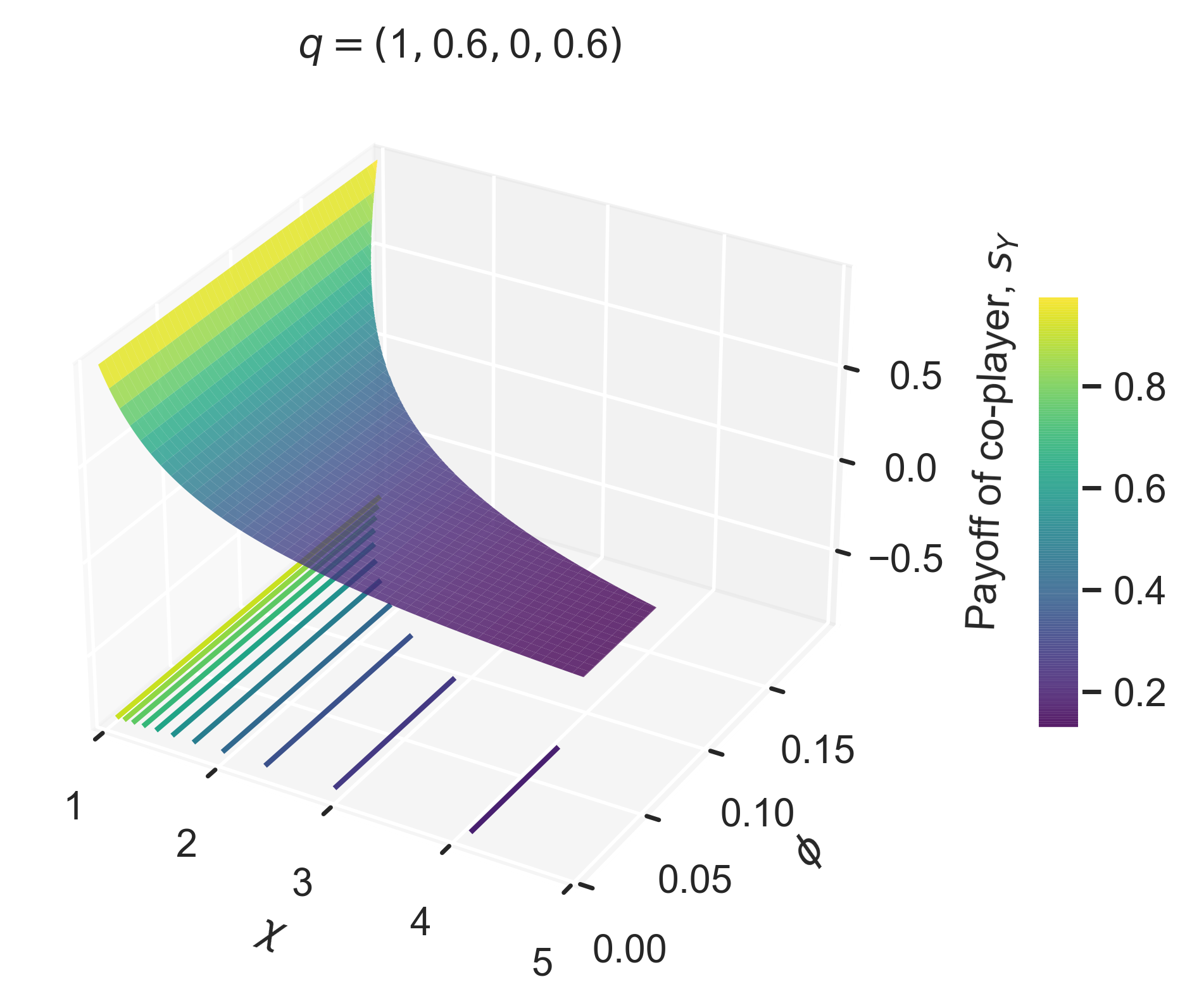} \\
\hline
B & & & \includegraphics[width=2.5cm]{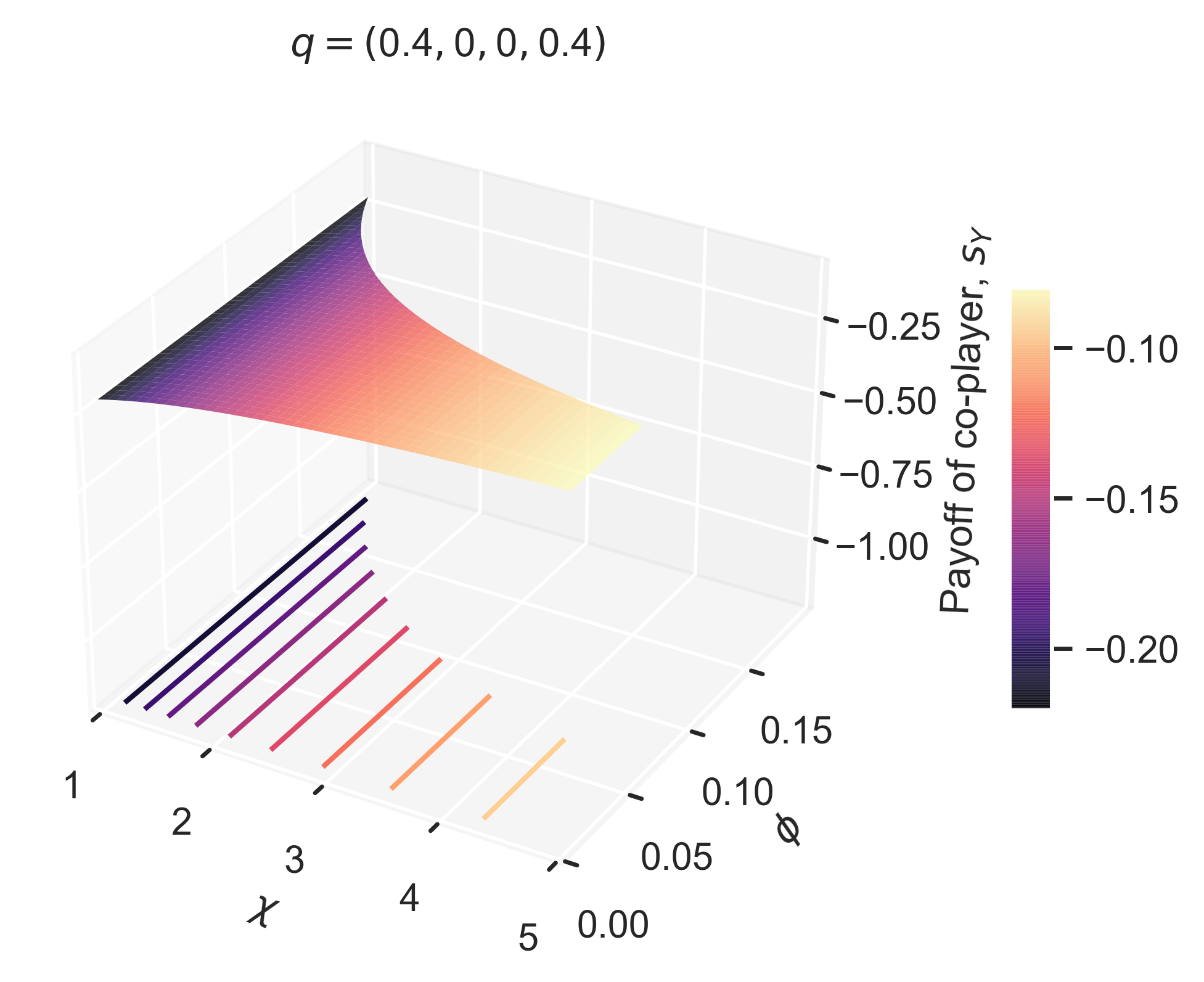} \\
\hline
\multirow{2}{*}{C} & \multirow{2}{*}{\includegraphics[width=2.5cm]{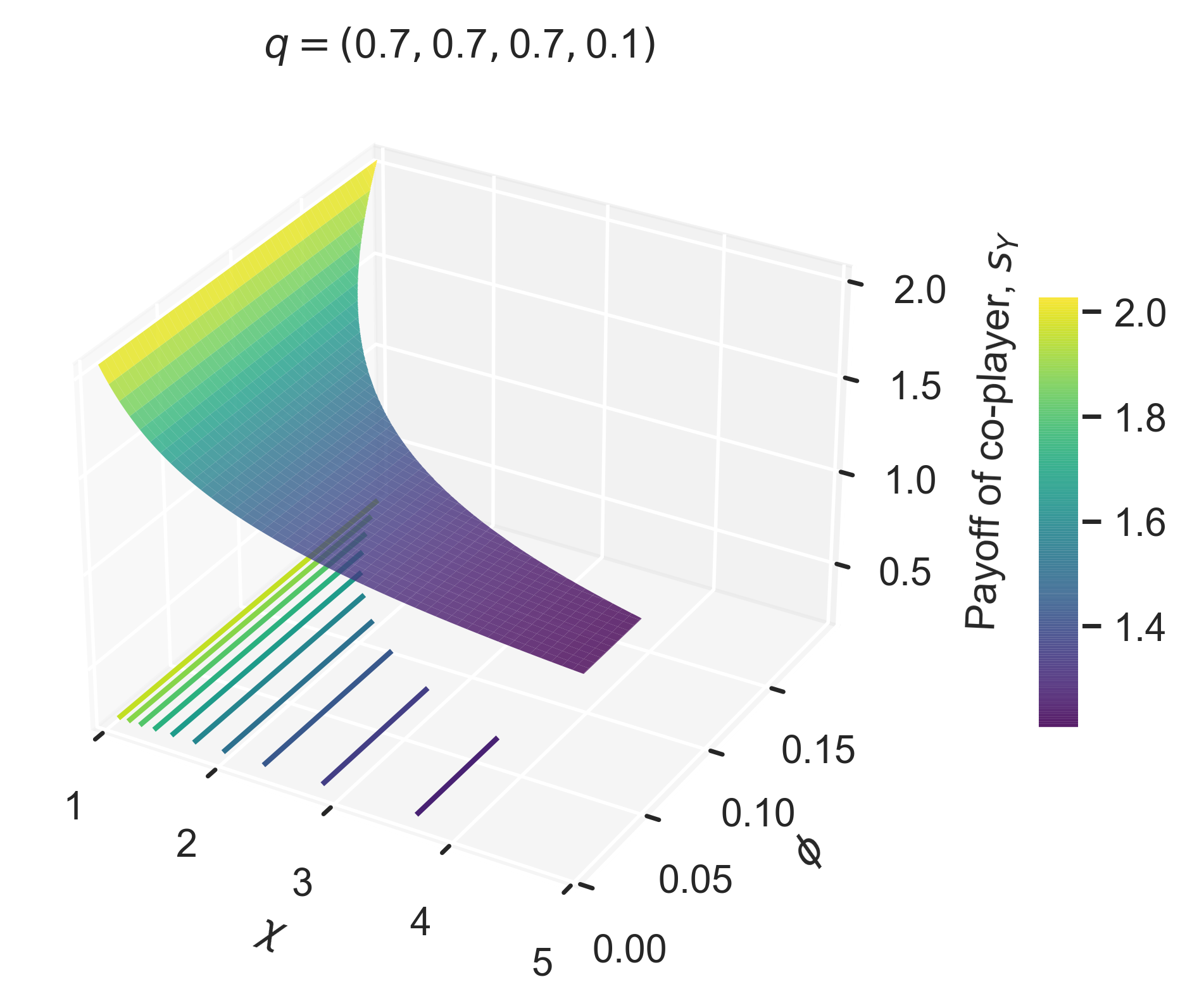}} 
& \multirow{2}{*}{\includegraphics[width=2.5cm]{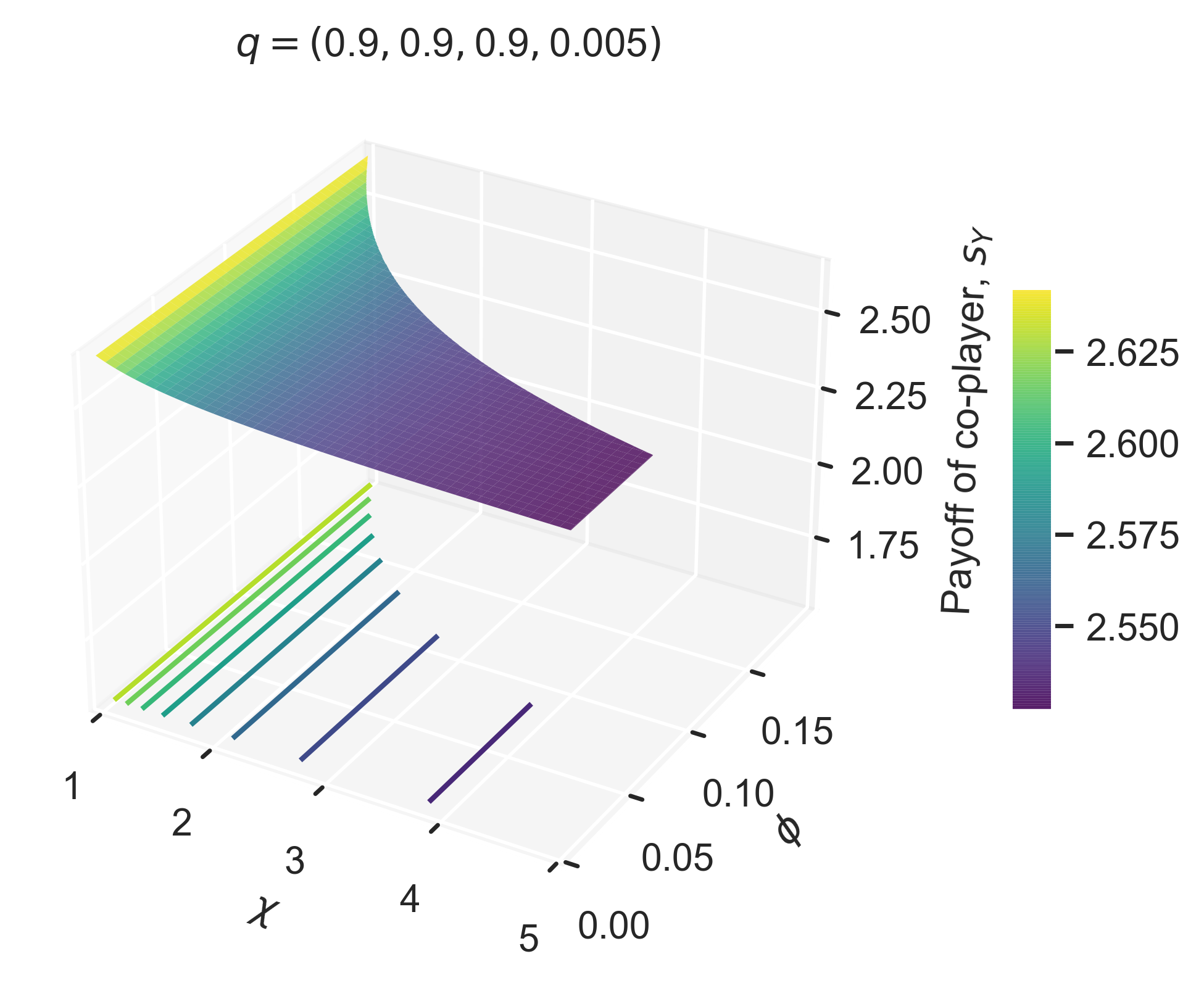}} 
& $q_1 < q_c \qquad q_1 > q_C$ \\
& & & \includegraphics[width=2cm]{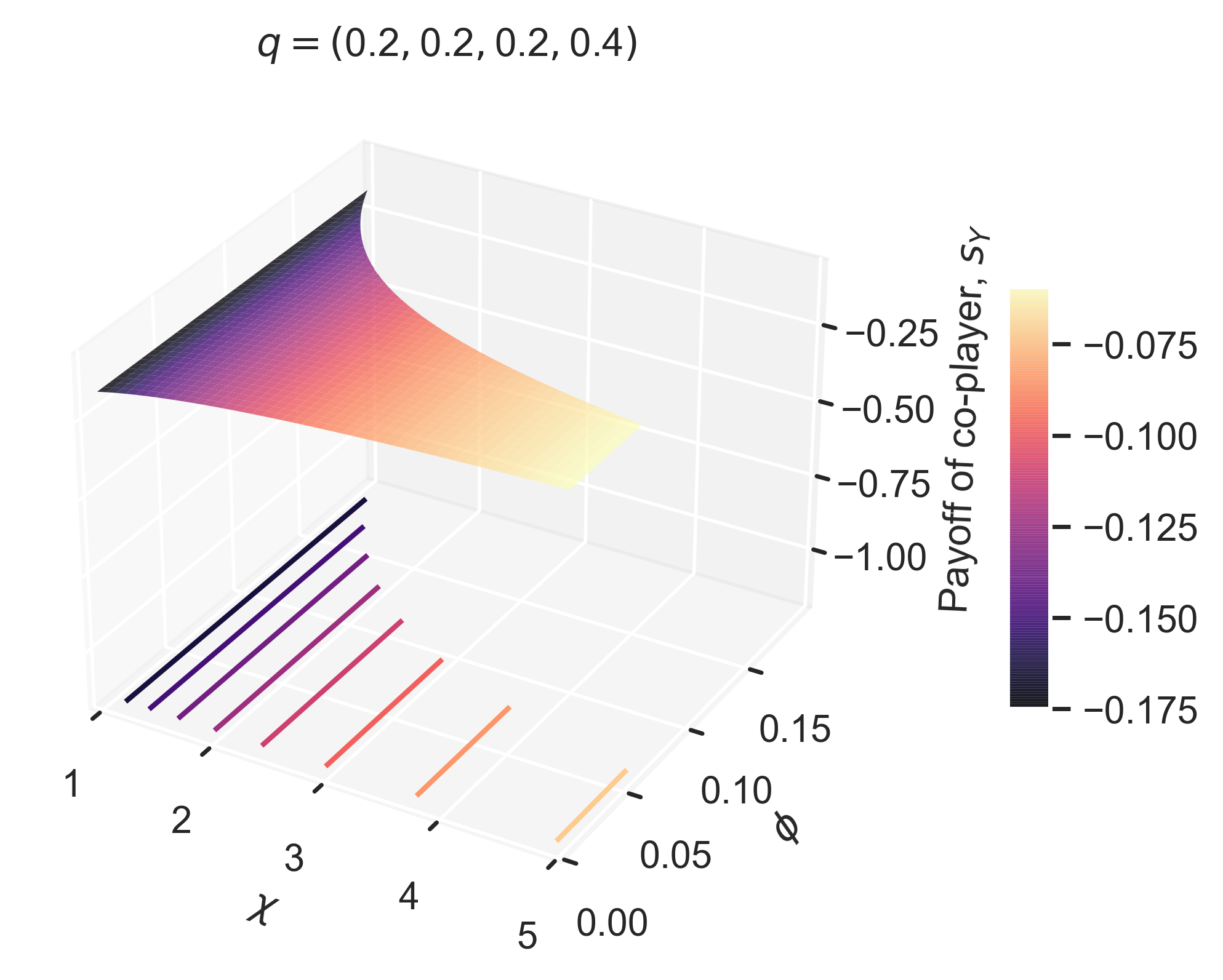}\includegraphics[width=2cm]{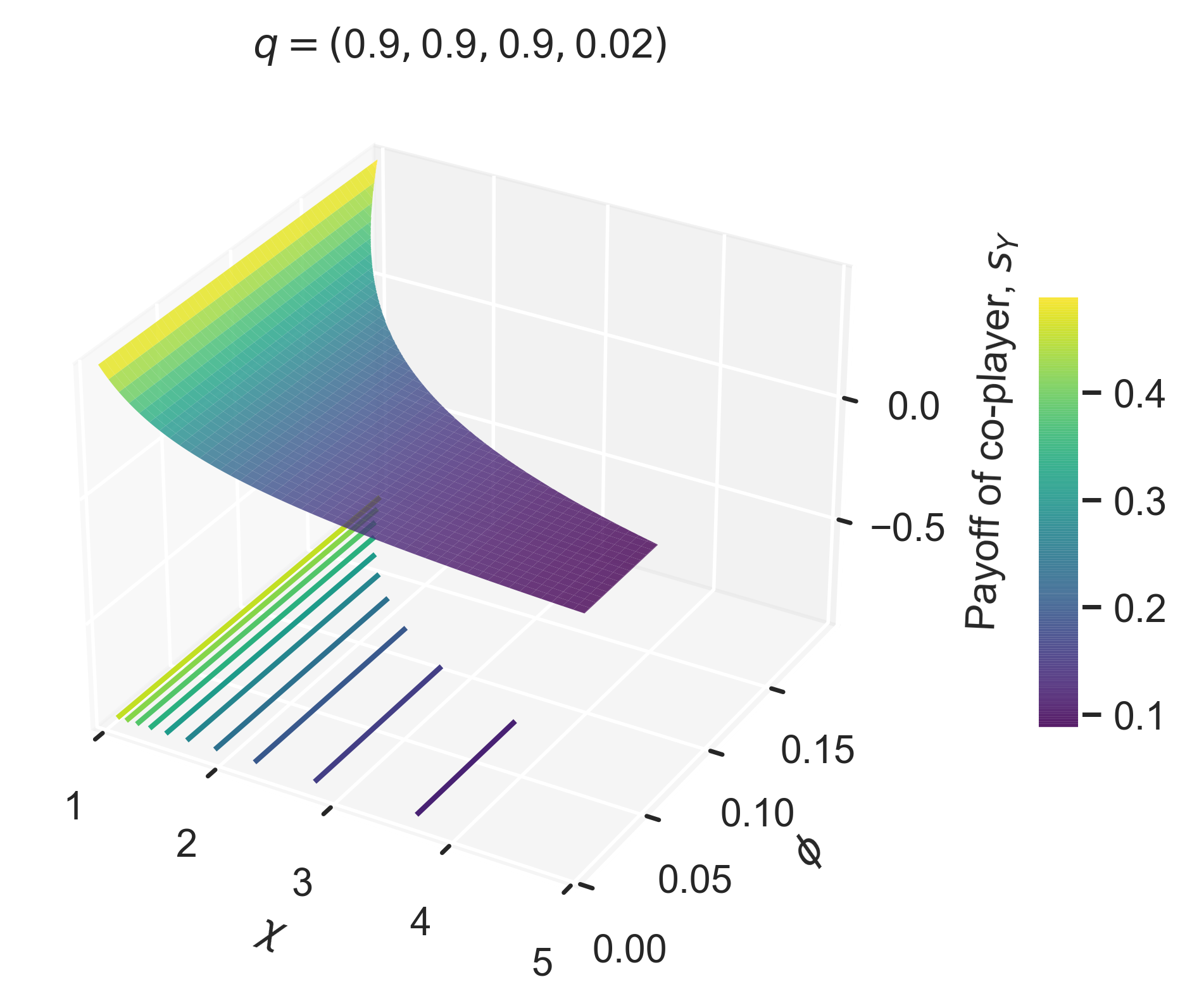} \\
\hline
D & \includegraphics[width=2.5cm]{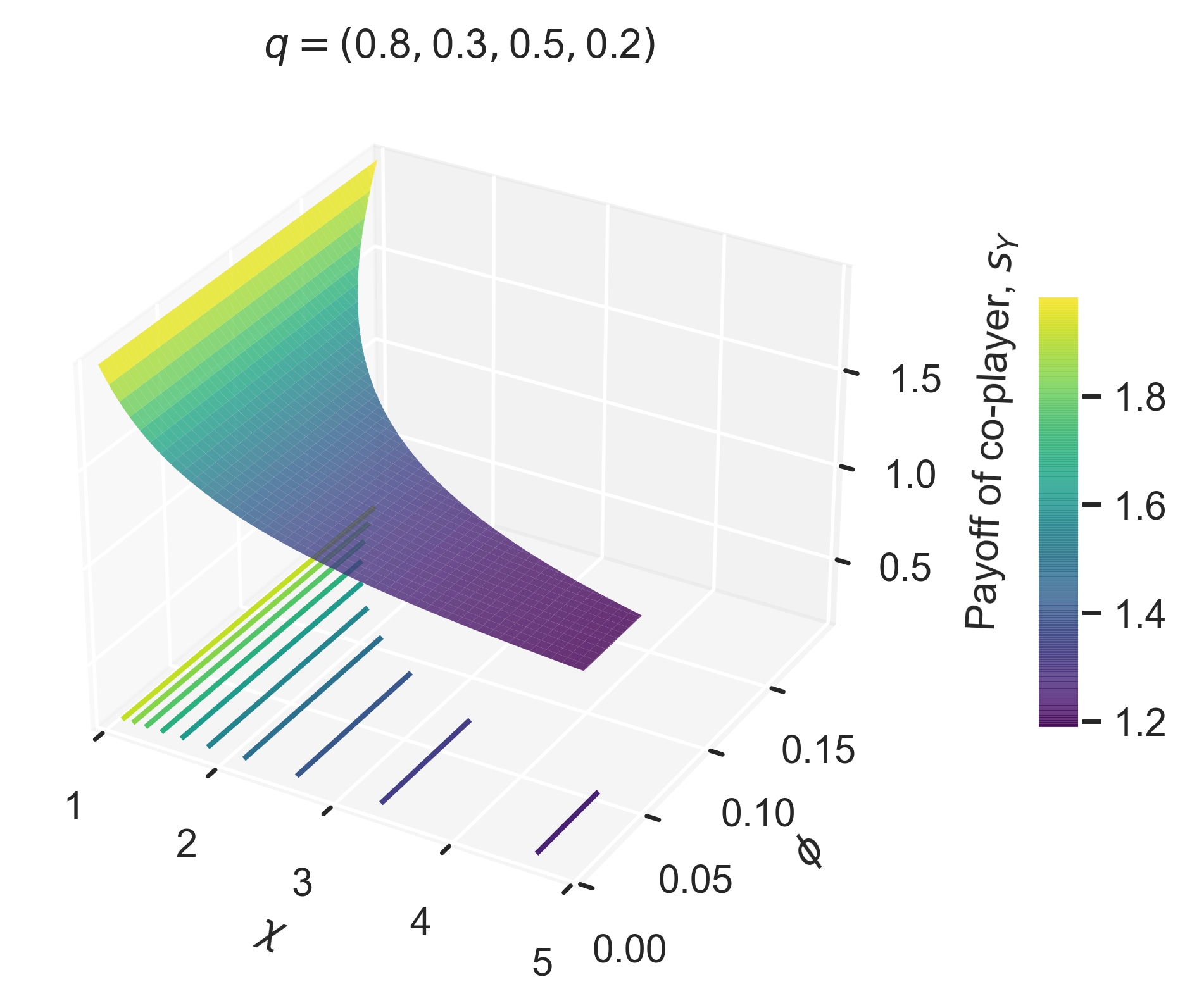} & \includegraphics[width=2.5cm]{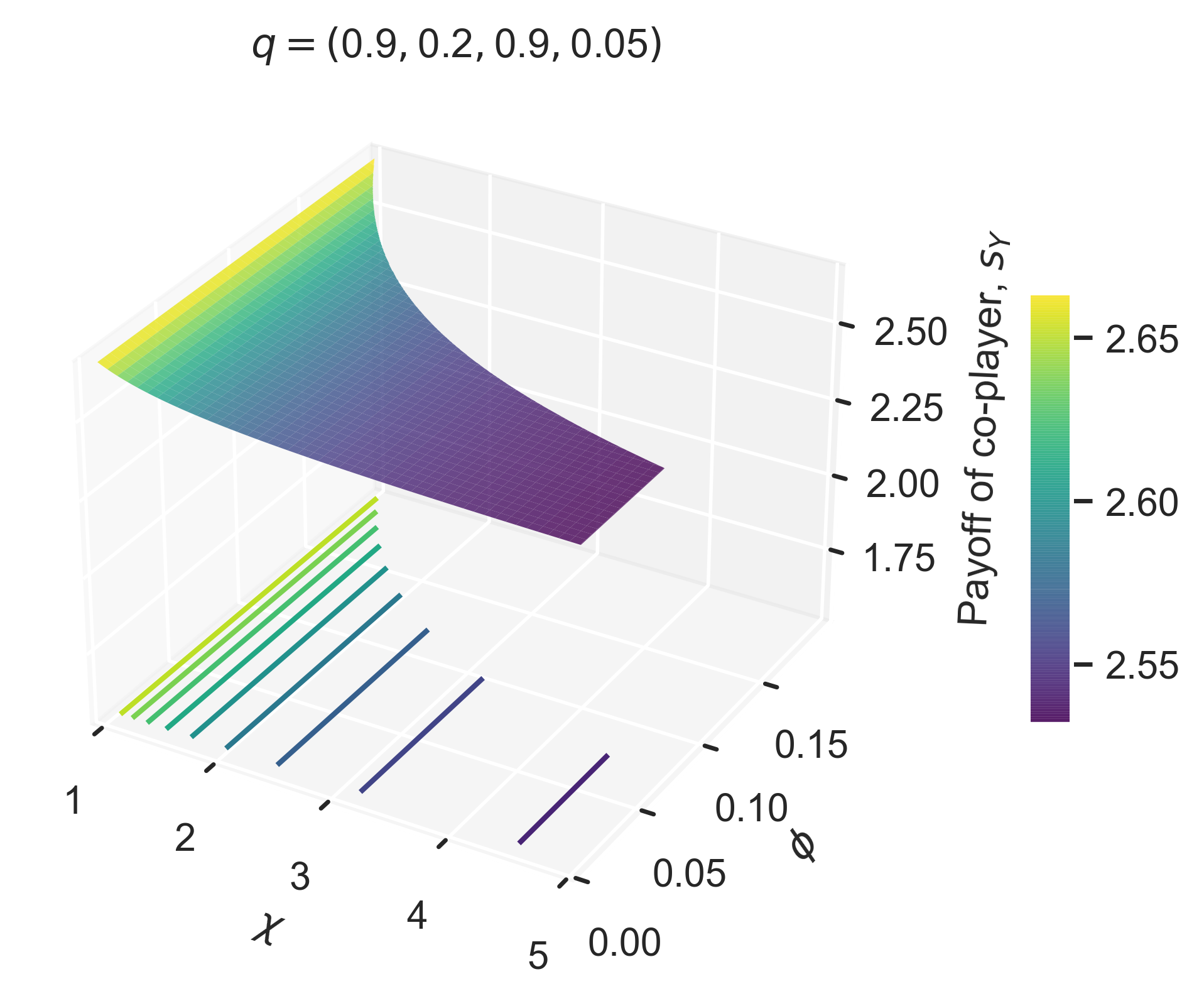}
& \includegraphics[width=2.5cm]{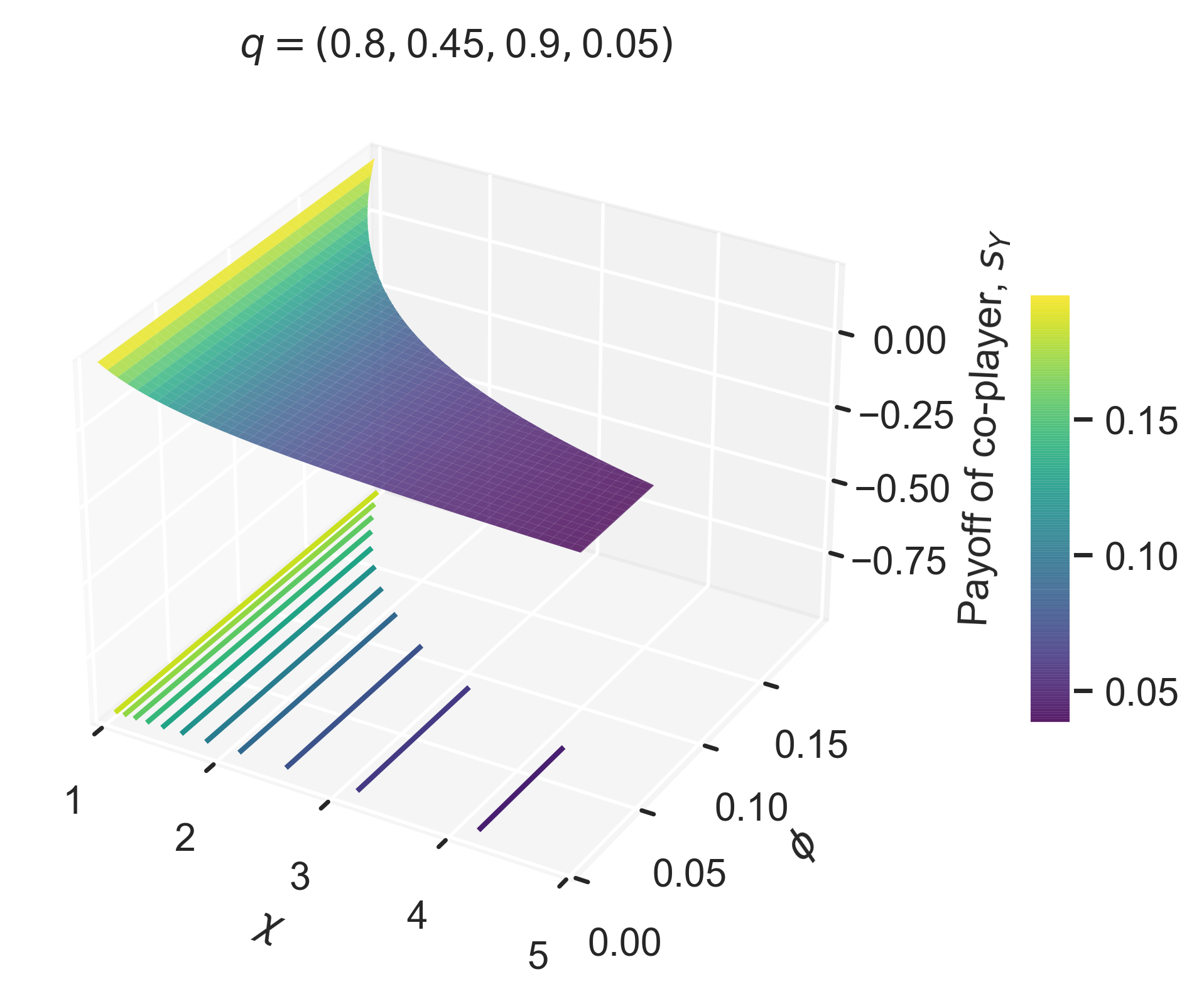} \\
\hline
\end{tabu}
\caption{Surface of $s_Y$ when X uses an extortionate ZD strategy and Y uses an unbending strategy.}
\label{summary_examples_Y}
\end{table}

\begin{table}[H]
\centering
\footnotesize
\tabulinesep=1.5mm
\begin{tabu}{| c | c | c | c |}
\hline
Class & \makecell[cc]{$T + S > 2P$\\ $(R, S, T, P) = (3, 0, 5, 1)$} & \makecell[cc]{$T + S = 2P$\\ $(R, S, T, P) = (3, 0, 5, 2.5)$}  
&  \makecell[cc]{$T + S < 2P$\\ $(R, S, T, P) = (1, -3, 2, 0)$} \\
\hline
\multirow{2}{*}{A} & \multirow{2}{*}{\includegraphics[width=2.5cm]{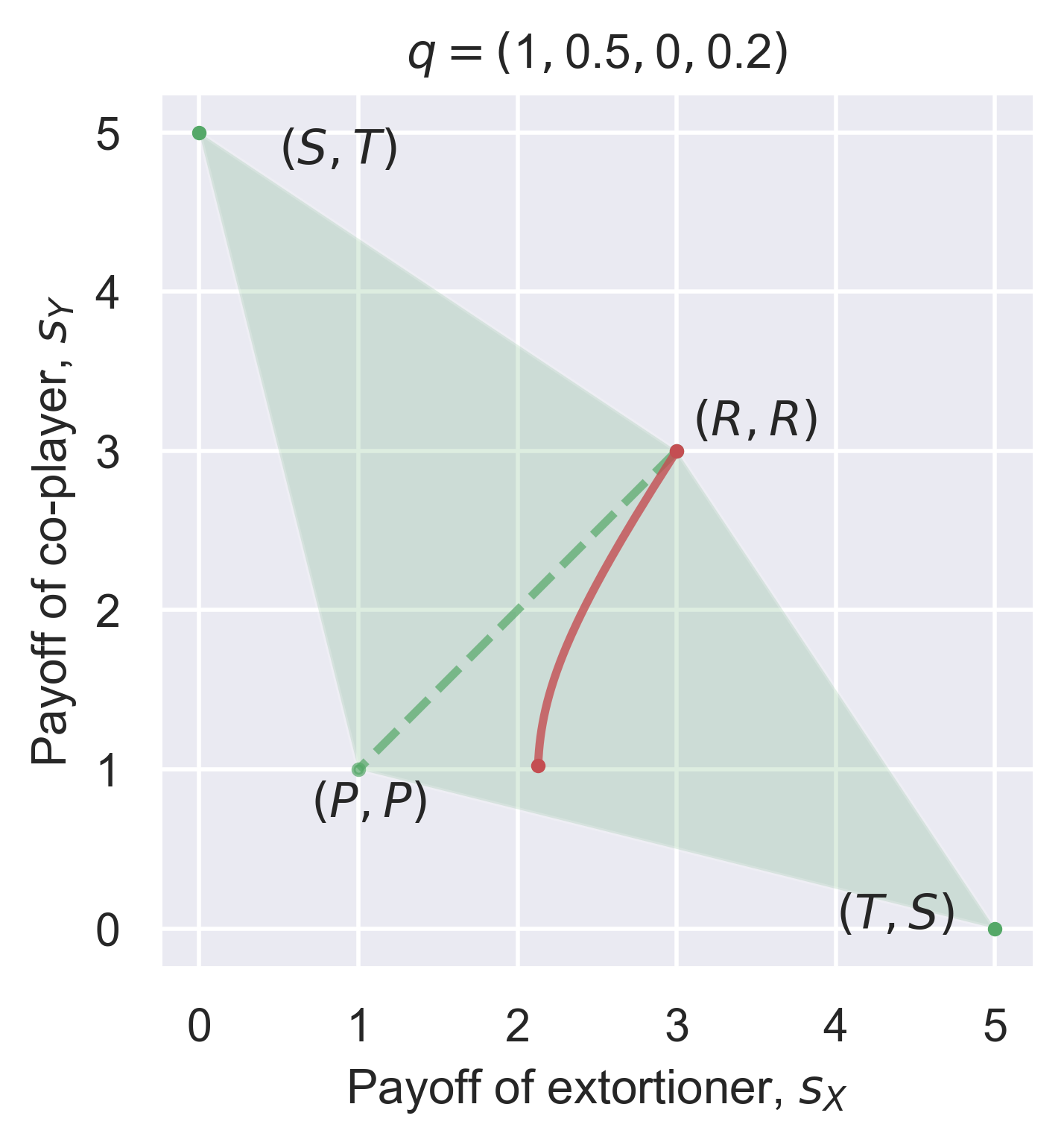}} & \multirow{2}{*}{\includegraphics[width=2.5cm]{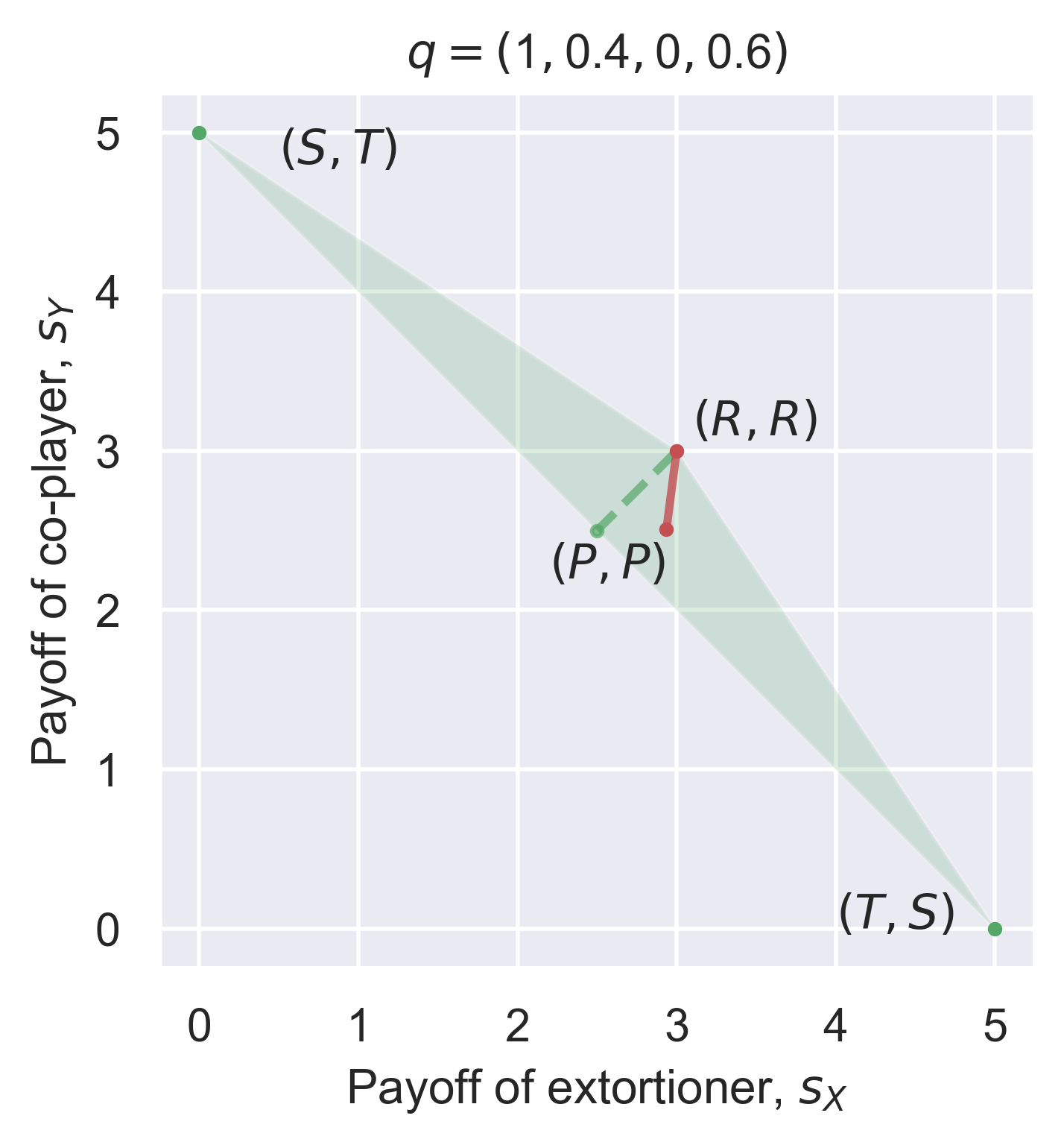}} &
$q_2 < q_A \qquad q_2 > q_A$ \\
& & & \includegraphics[width = 2cm]{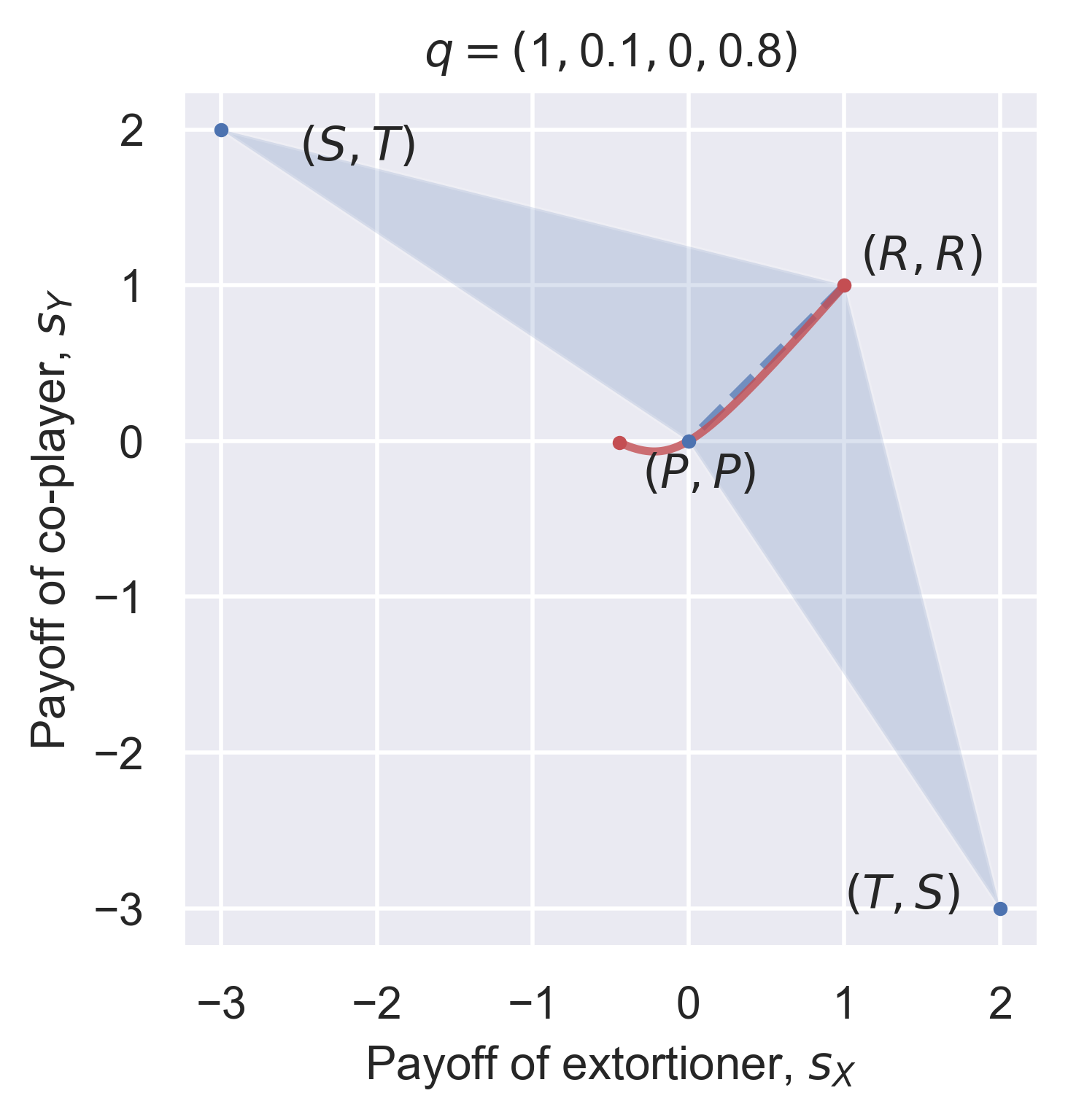}\includegraphics[width = 2cm]{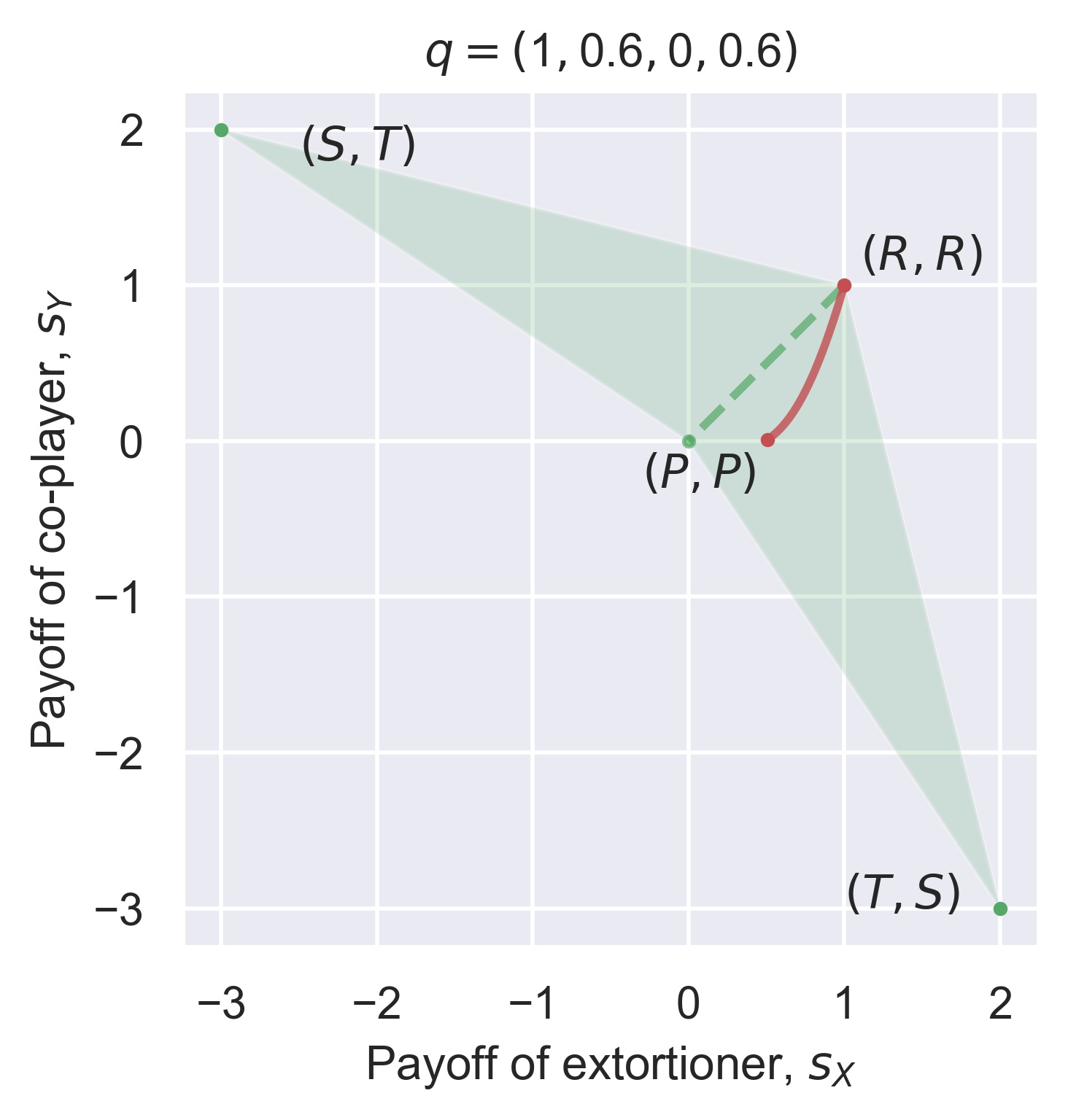} \\
\hline
B & & & \includegraphics[width=2.5cm]{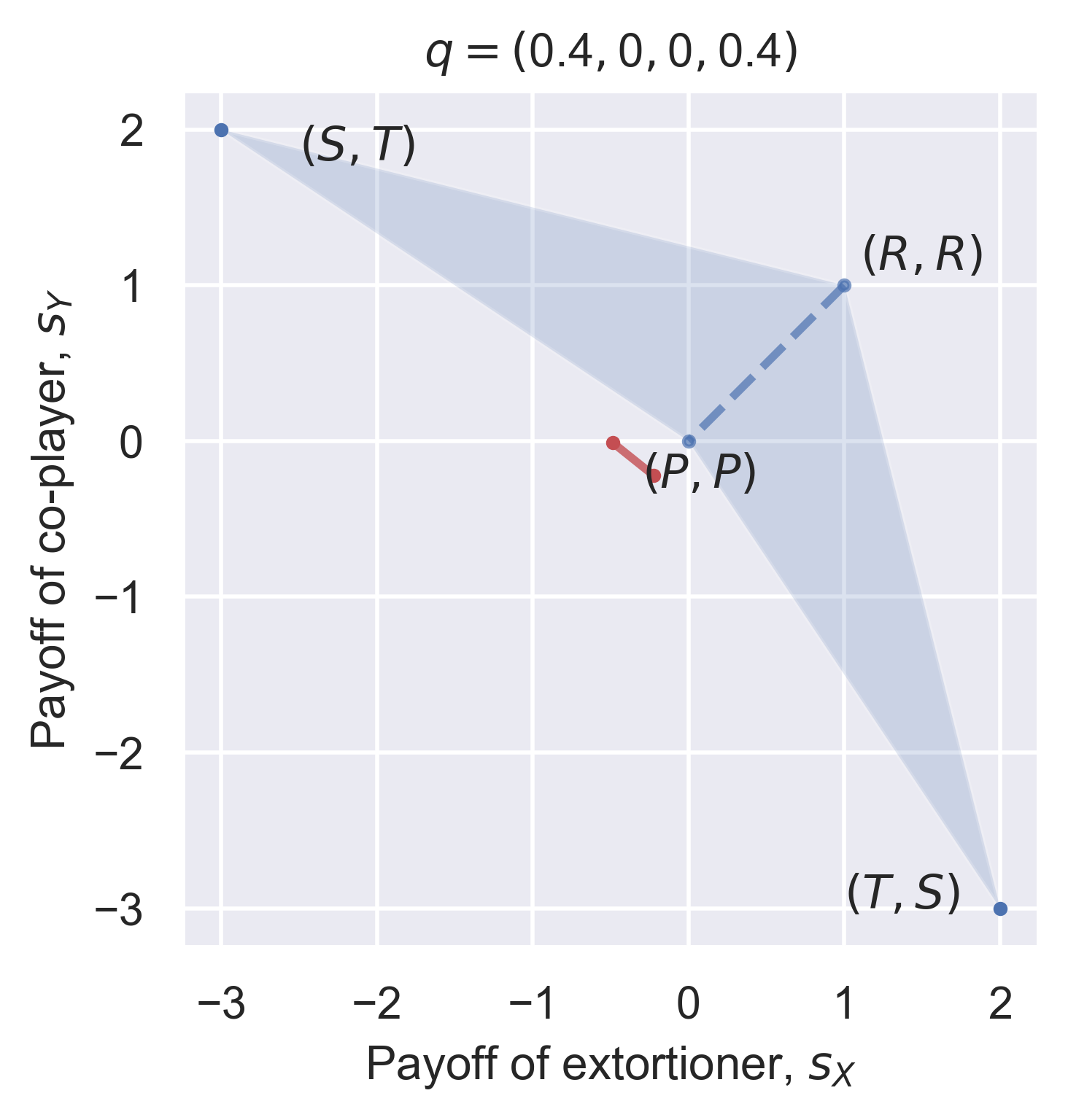} \\
\hline
\multirow{2}{*}{C} & \multirow{2}{*}{\includegraphics[width=2.5cm]{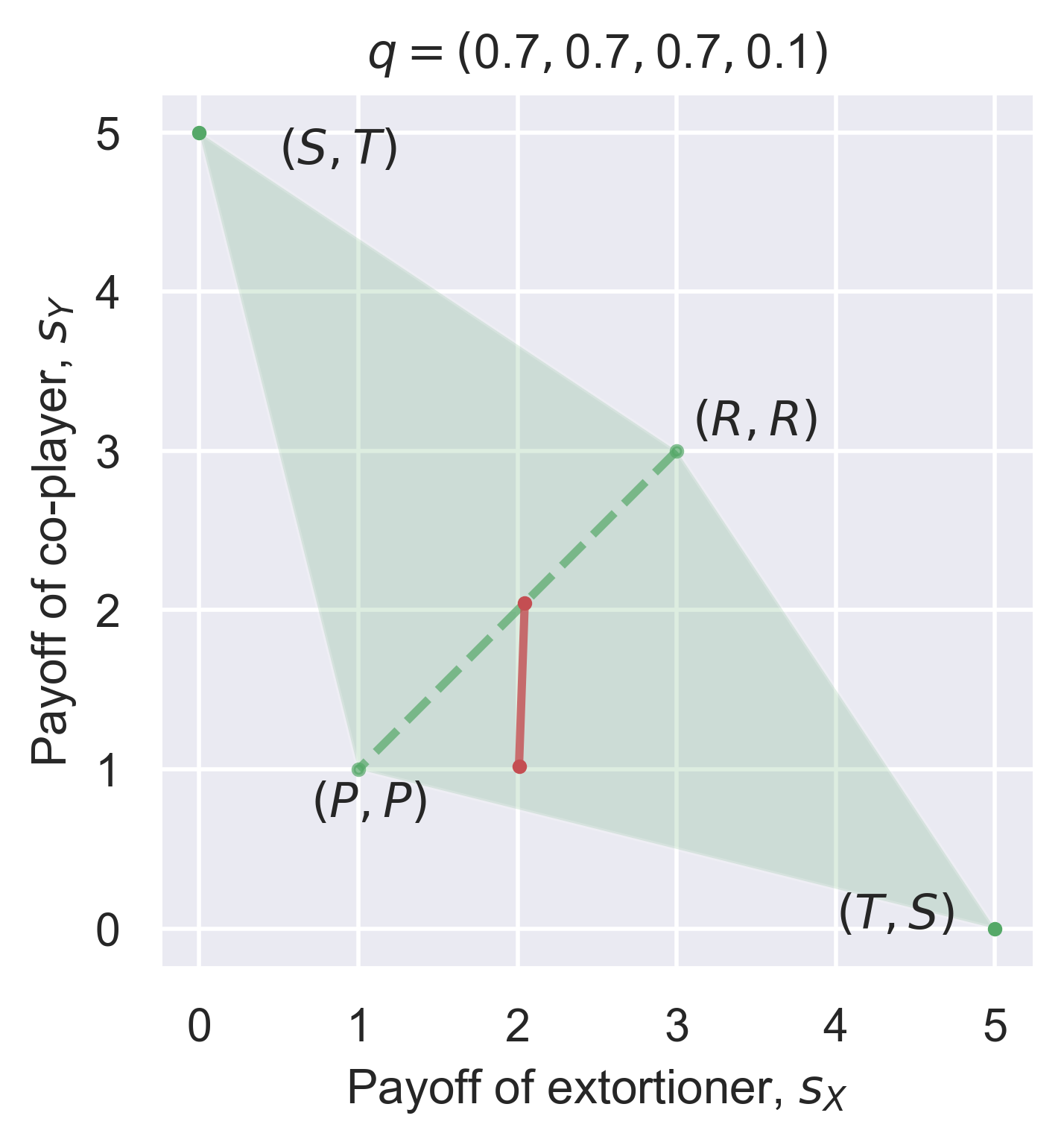}} 
& \multirow{2}{*}{\includegraphics[width=2.5cm]{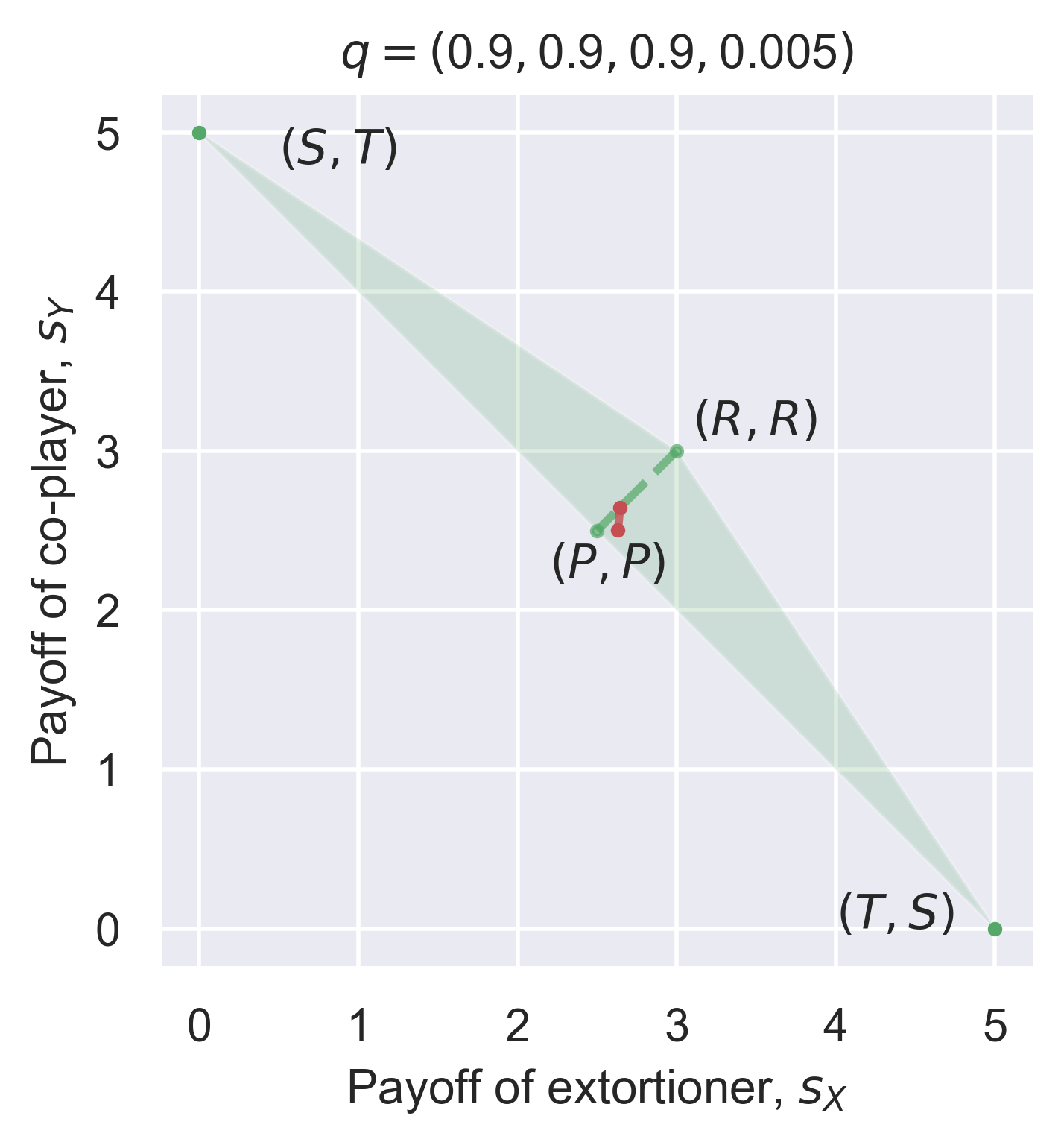}} 
& $q_1 < q_c \qquad q_1 > q_C$ \\
& & & \includegraphics[width=2cm]{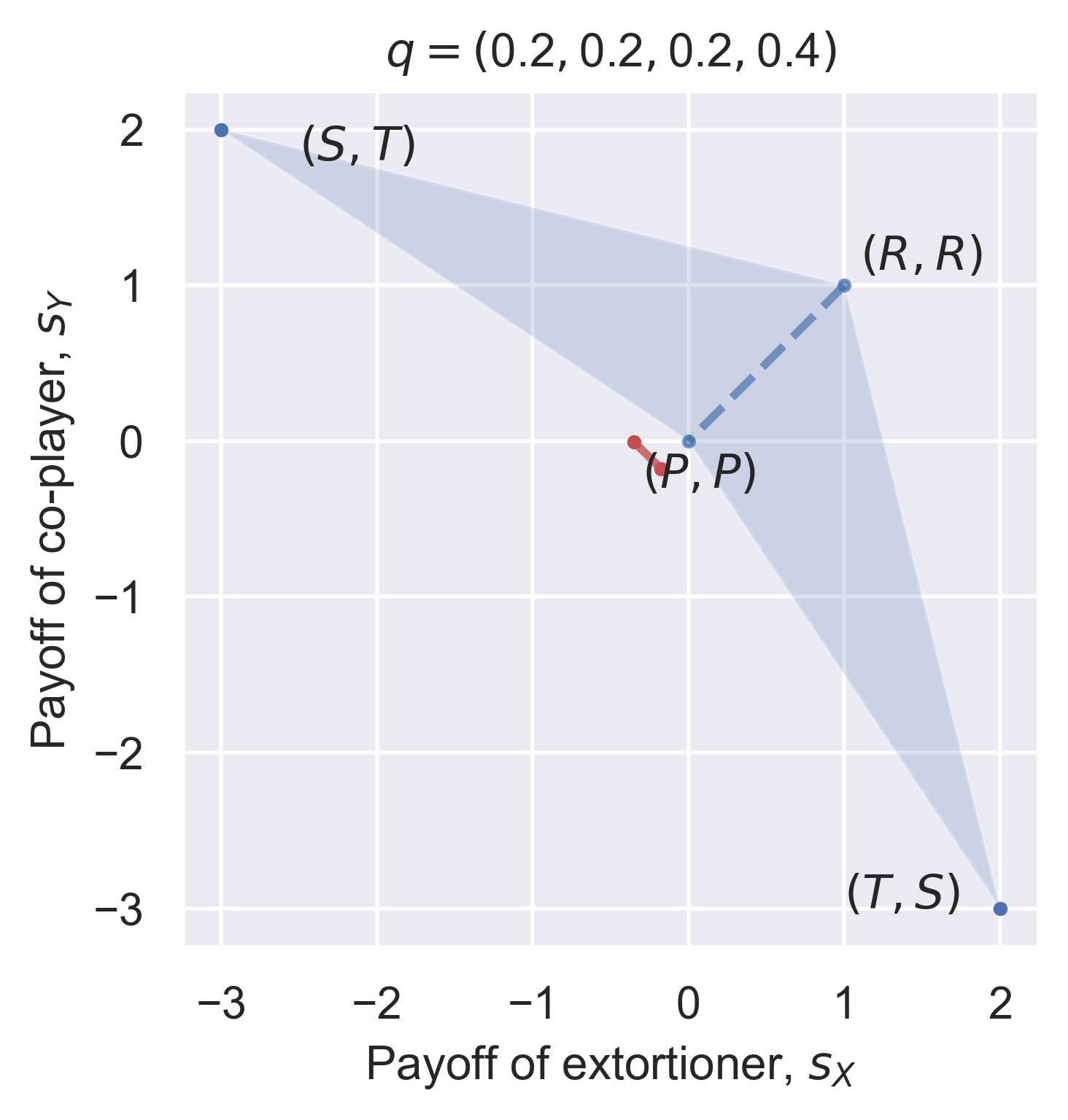}\includegraphics[width=2cm]{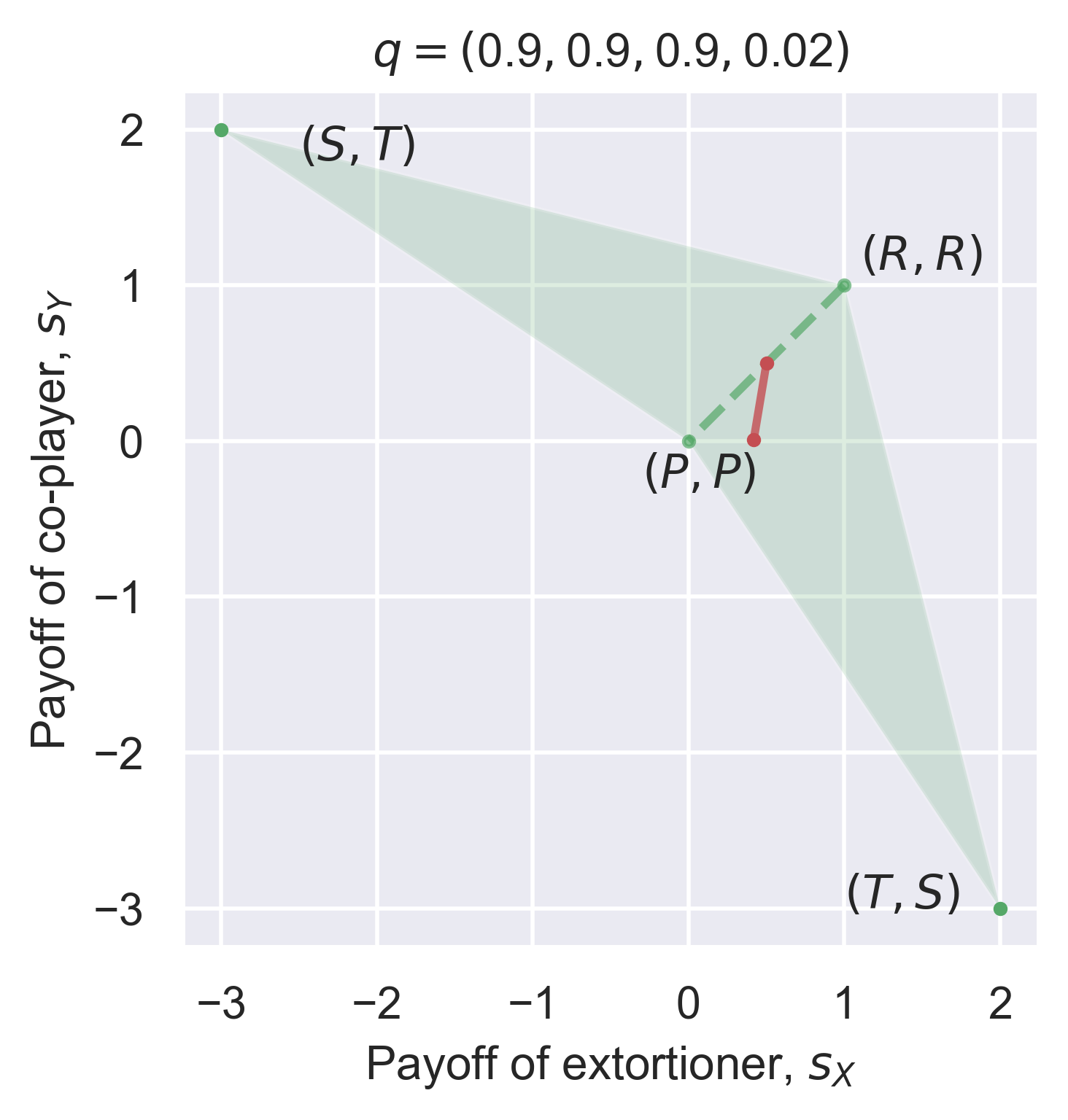} \\
\hline
D & \includegraphics[width=2.5cm]{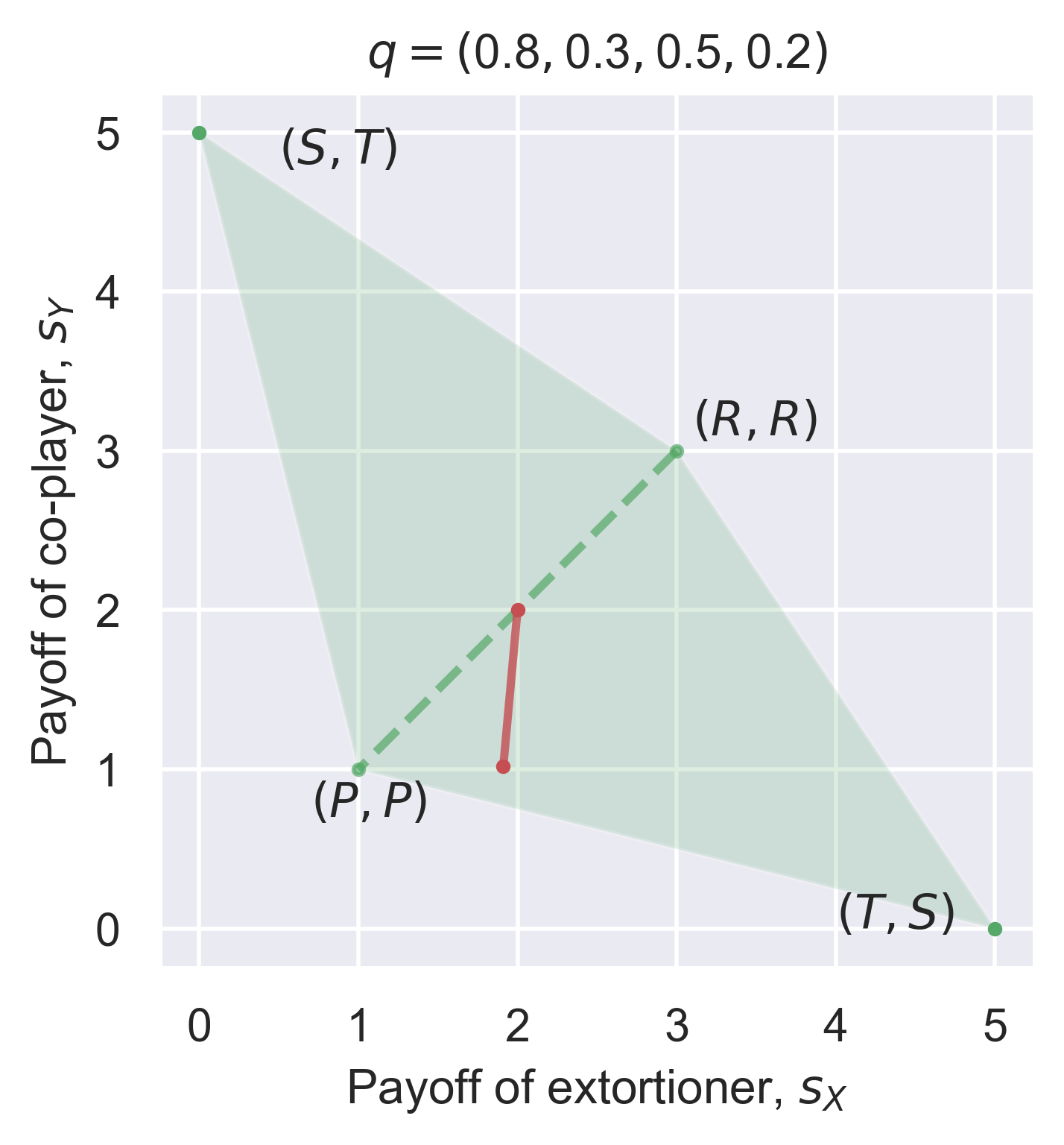} & \includegraphics[width=2.5cm]{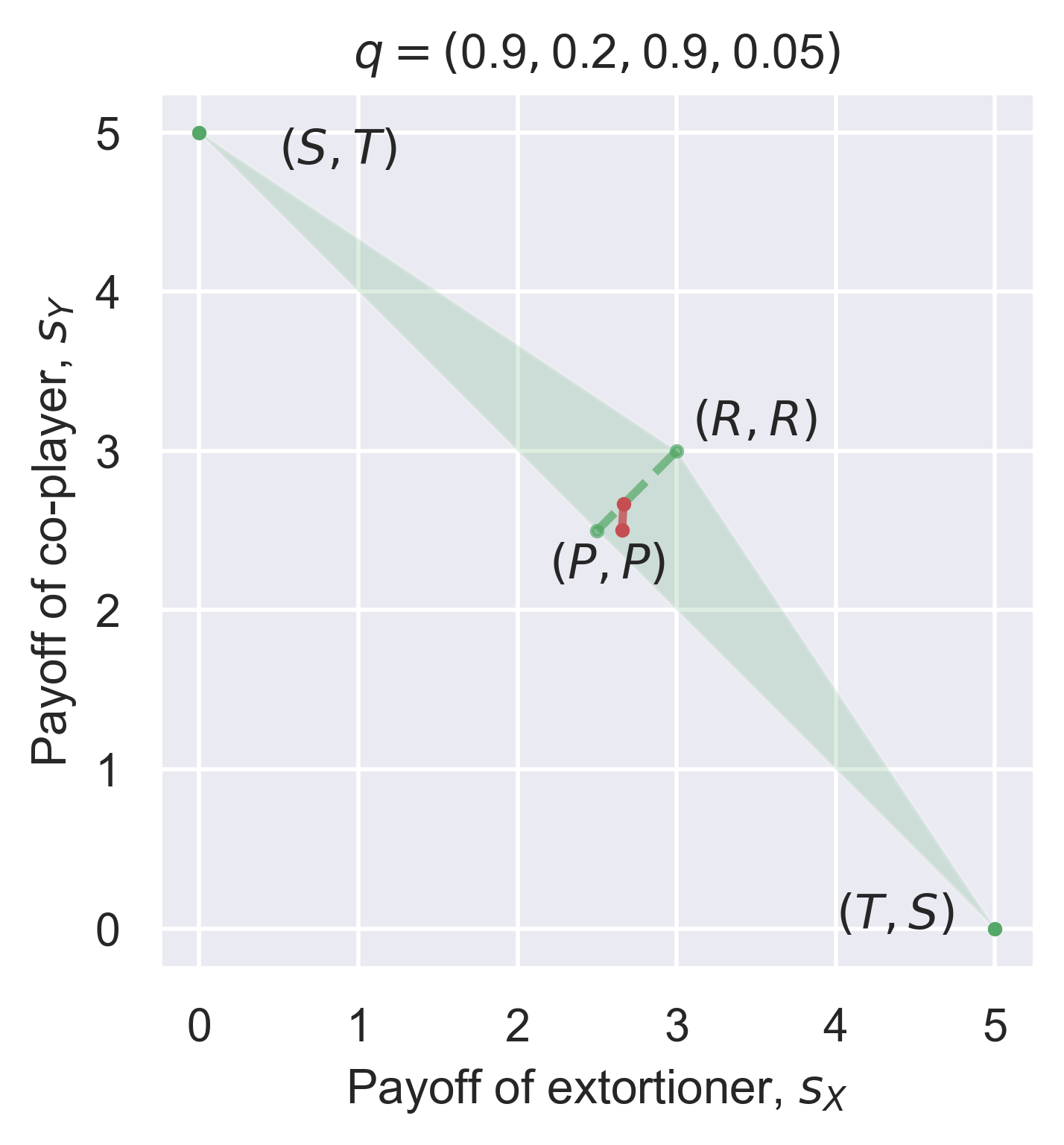}
& \includegraphics[width=2.5cm]{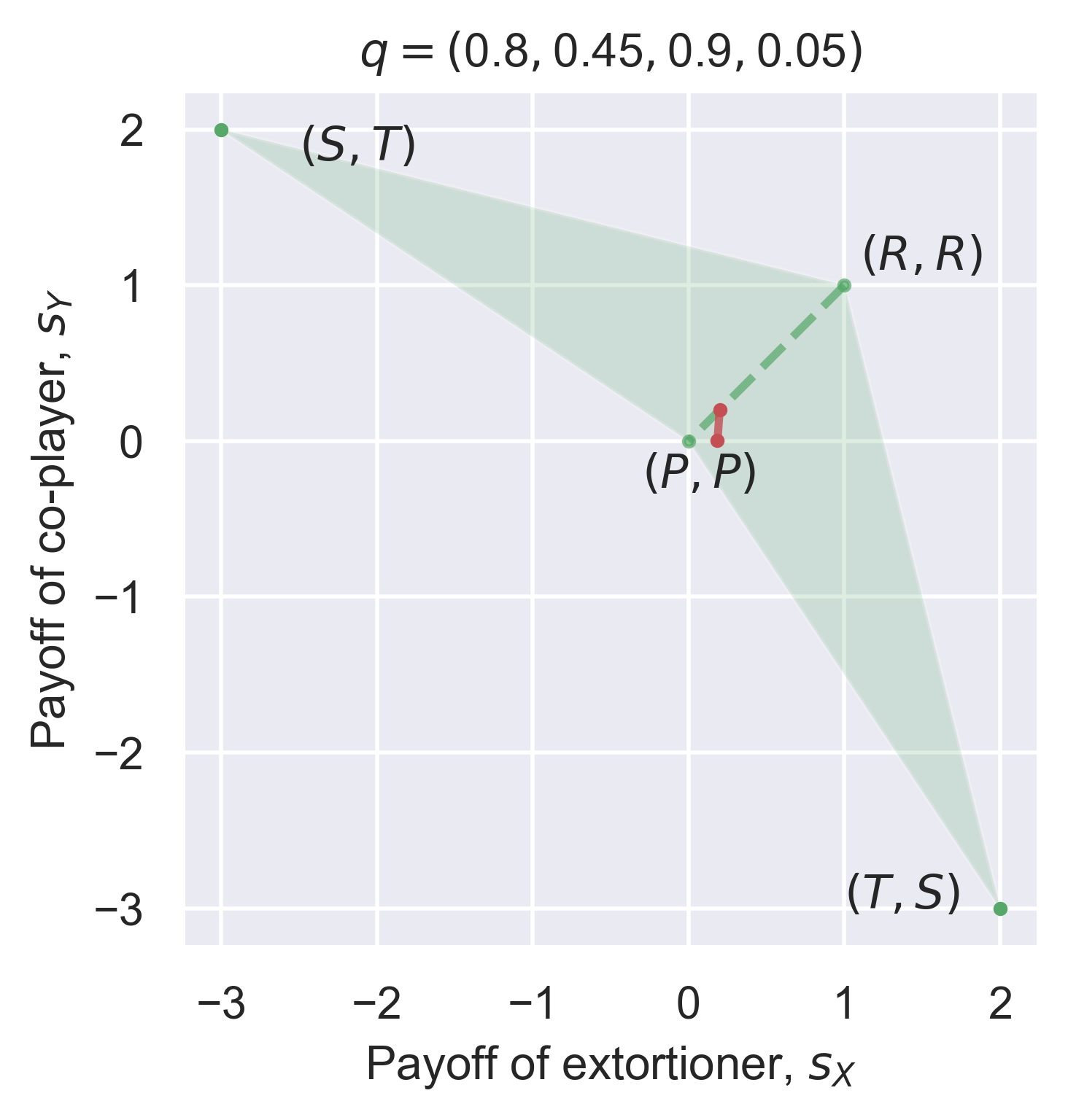} \\
\hline
\end{tabu}
\caption{Parametric curve of $(s_X, s_Y)$ with variable $\chi$ increasing from $1$ to $+\infty$. The coordinates of the vertices are the four possible outcomes of the one-shot PD game. The dashed line connects the two vertices $(P, P)$ and $(R, R)$.}
\label{summary_examples_quad}
\end{table}

\begin{table}[H]
\centering
\footnotesize
\tabulinesep=1.5mm
\begin{tabu}{| c | c | c | c |}
\hline
Class & \makecell[cc]{$T + S > 2P$\\ $(R, S, T, P) = (3, 0, 5, 1)$} & \makecell[cc]{$T + S = 2P$\\ $(R, S, T, P) = (3, 0, 5, 2.5)$}  
&  \makecell[cc]{$T + S < 2P$\\ $(R, S, T, P) = (1, -3, 2, 0)$} \\
\hline
\multirow{2}{*}{A} & \multirow{2}{*}{\includegraphics[width=2.5cm]{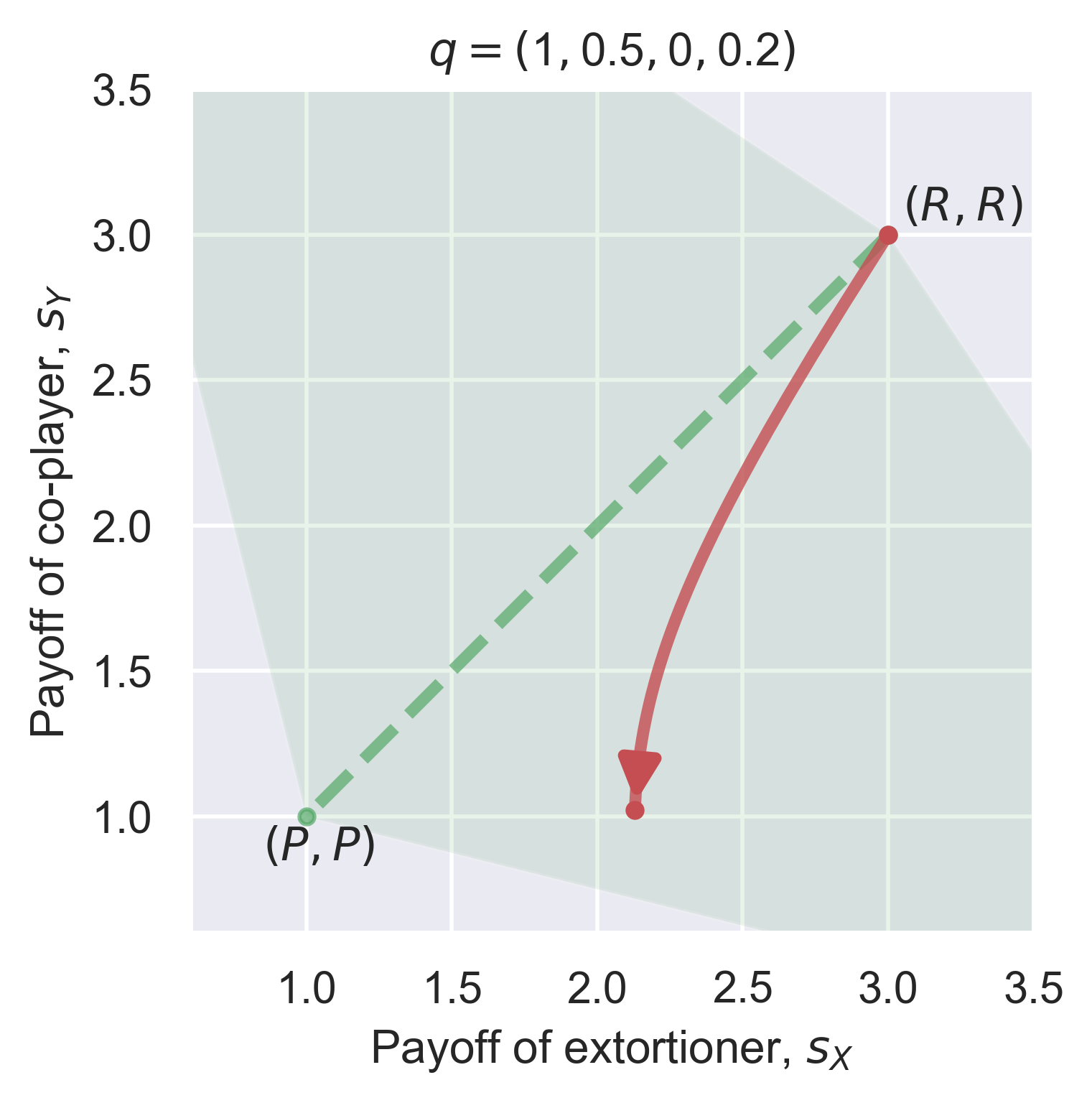}} & \multirow{2}{*}{\includegraphics[width=2.5cm]{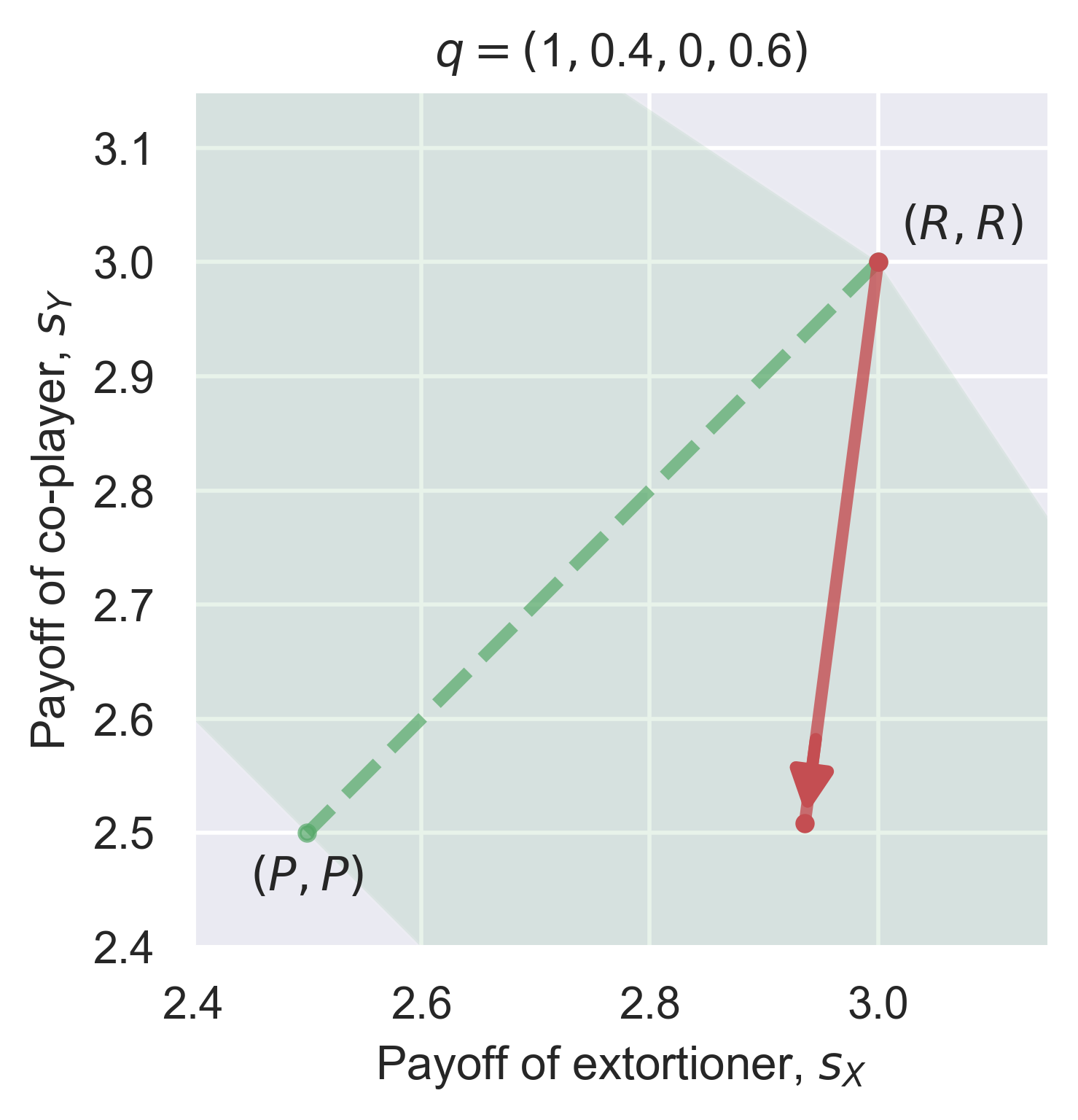}} &
$q_2 < q_A \qquad q_2 > q_A$ \\
& & & \includegraphics[width = 2cm]{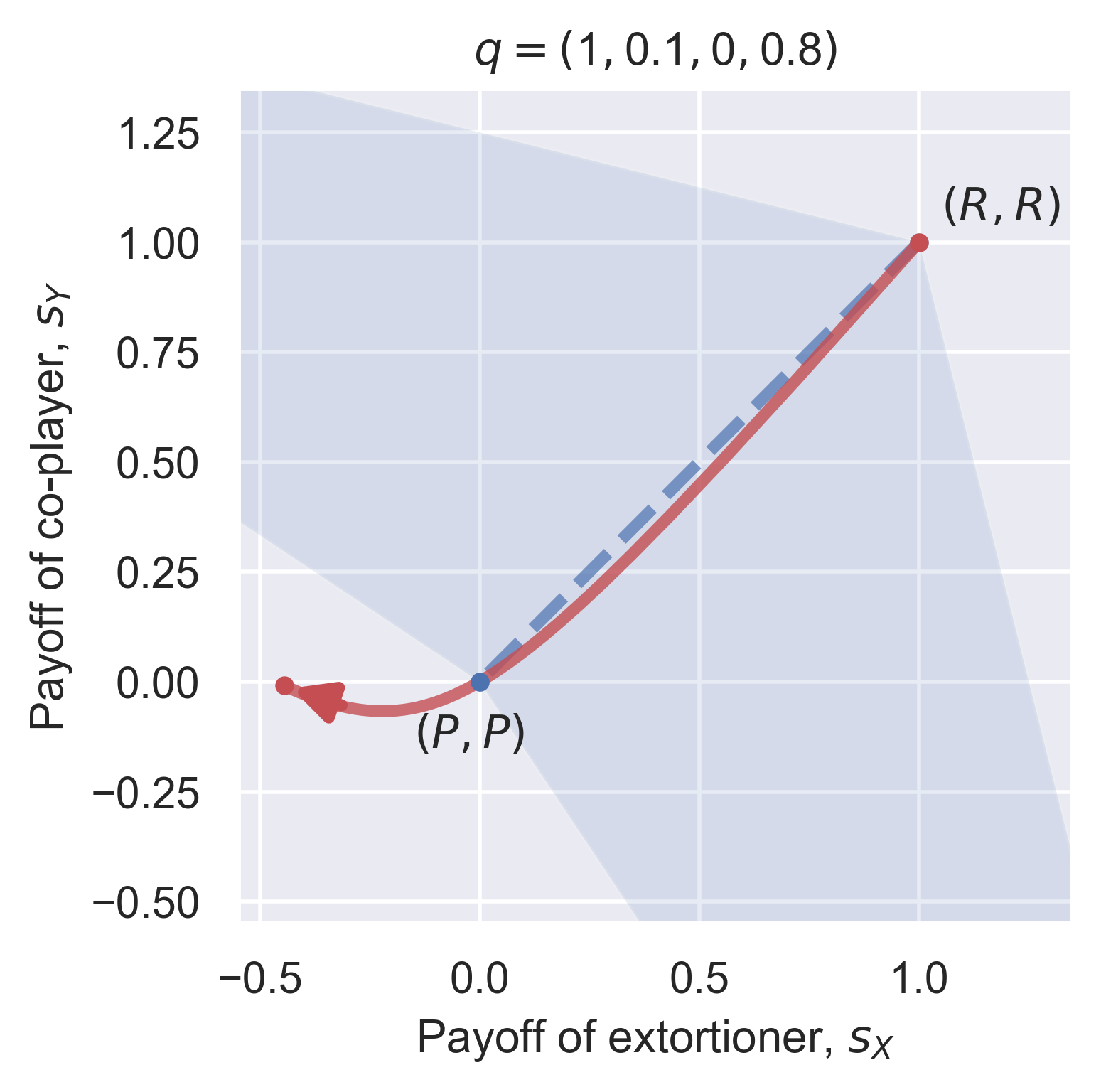}\includegraphics[width = 2cm]{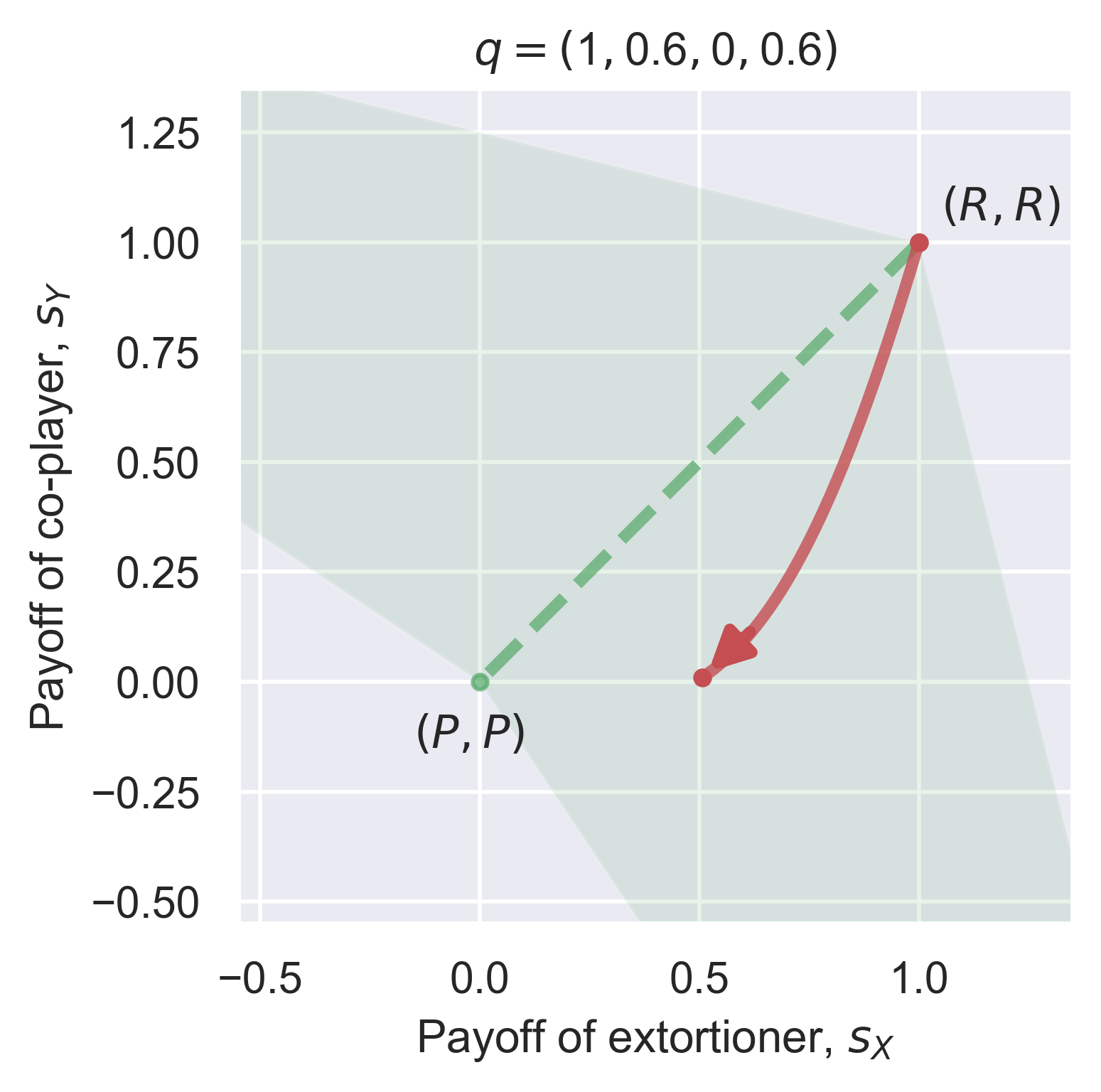} \\
\hline
B & & & \includegraphics[width=2.5cm]{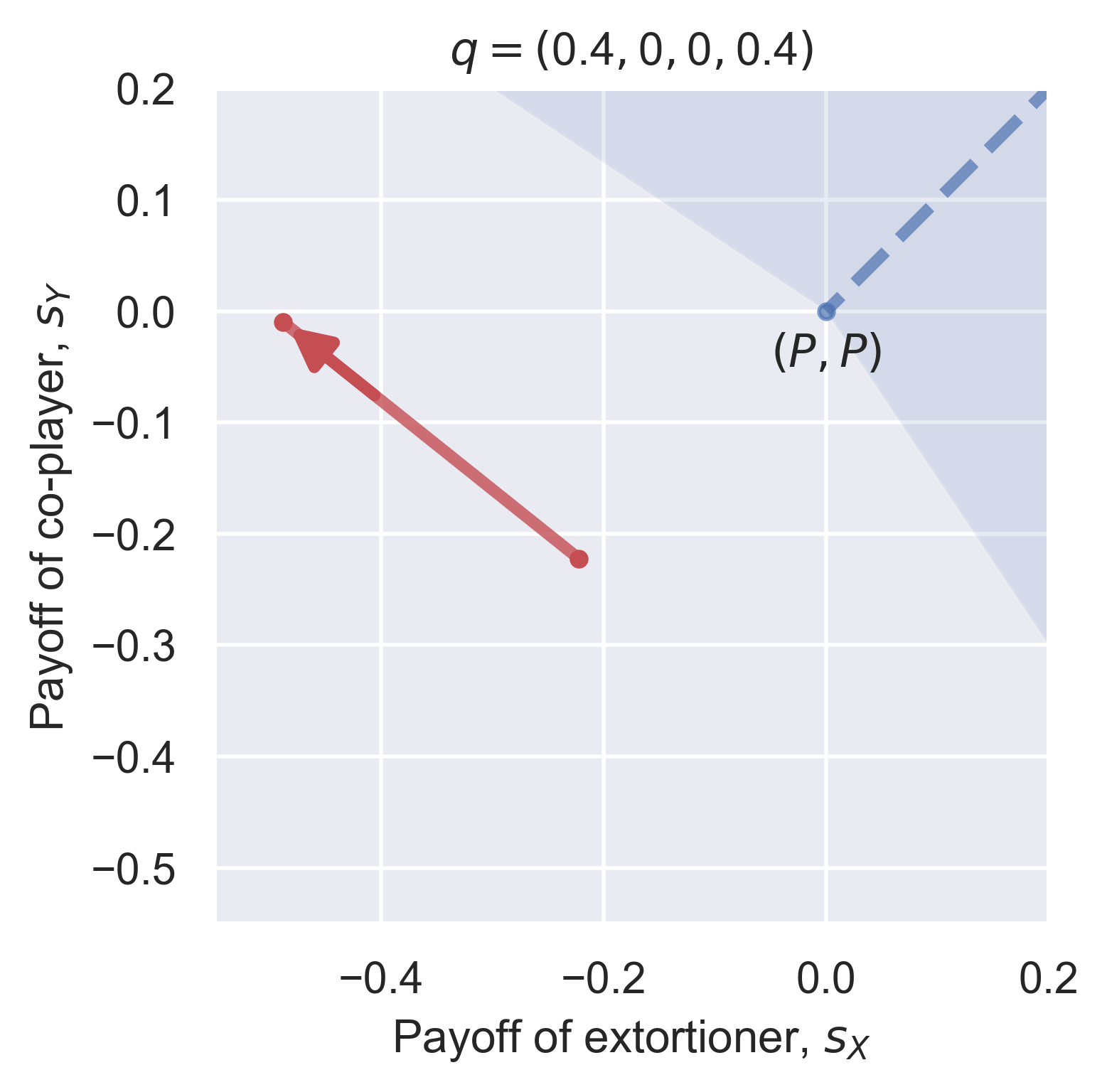} \\
\hline
\multirow{2}{*}{C} & \multirow{2}{*}{\includegraphics[width=2.5cm]{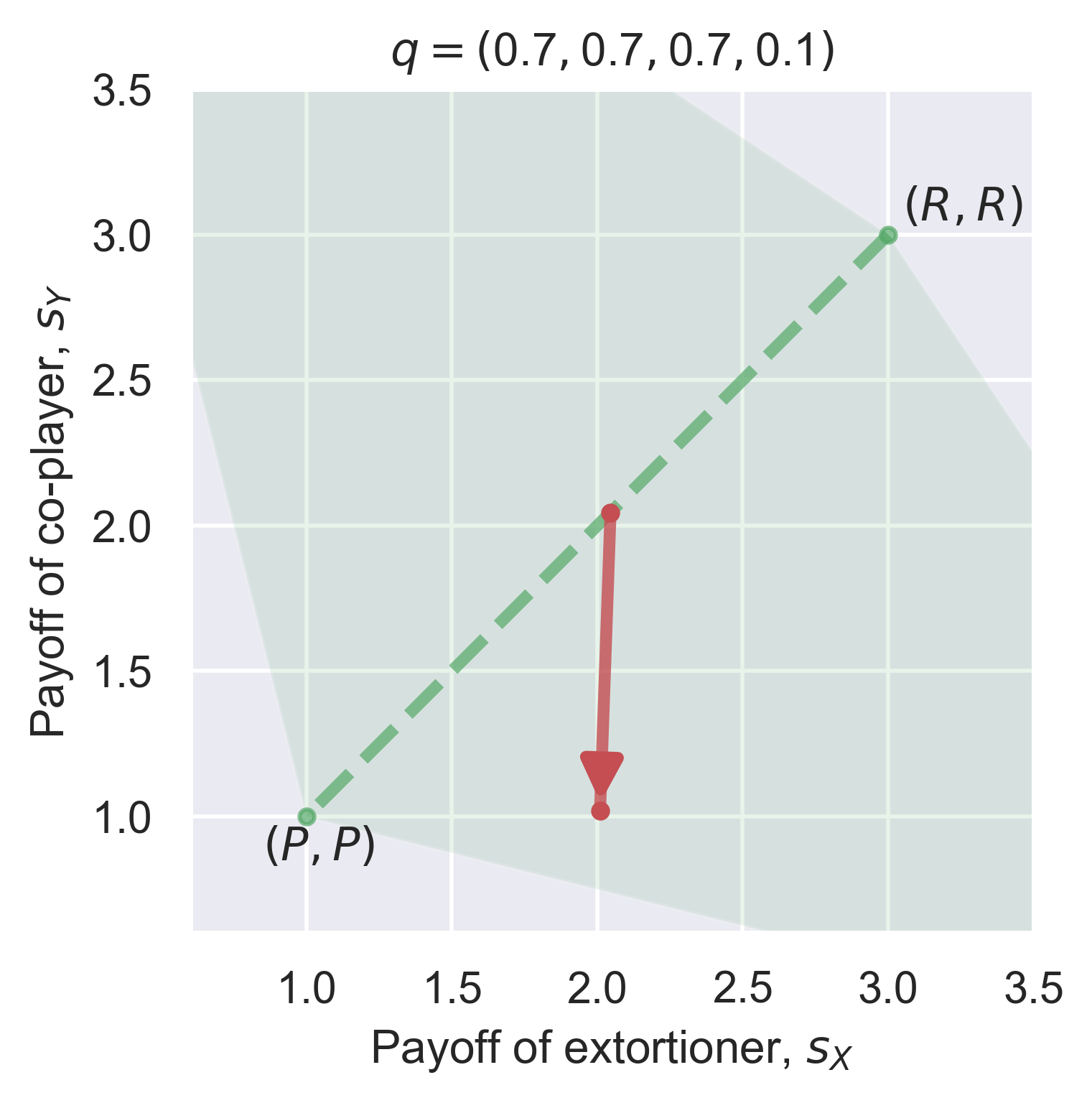}} 
& \multirow{2}{*}{\includegraphics[width=2.5cm]{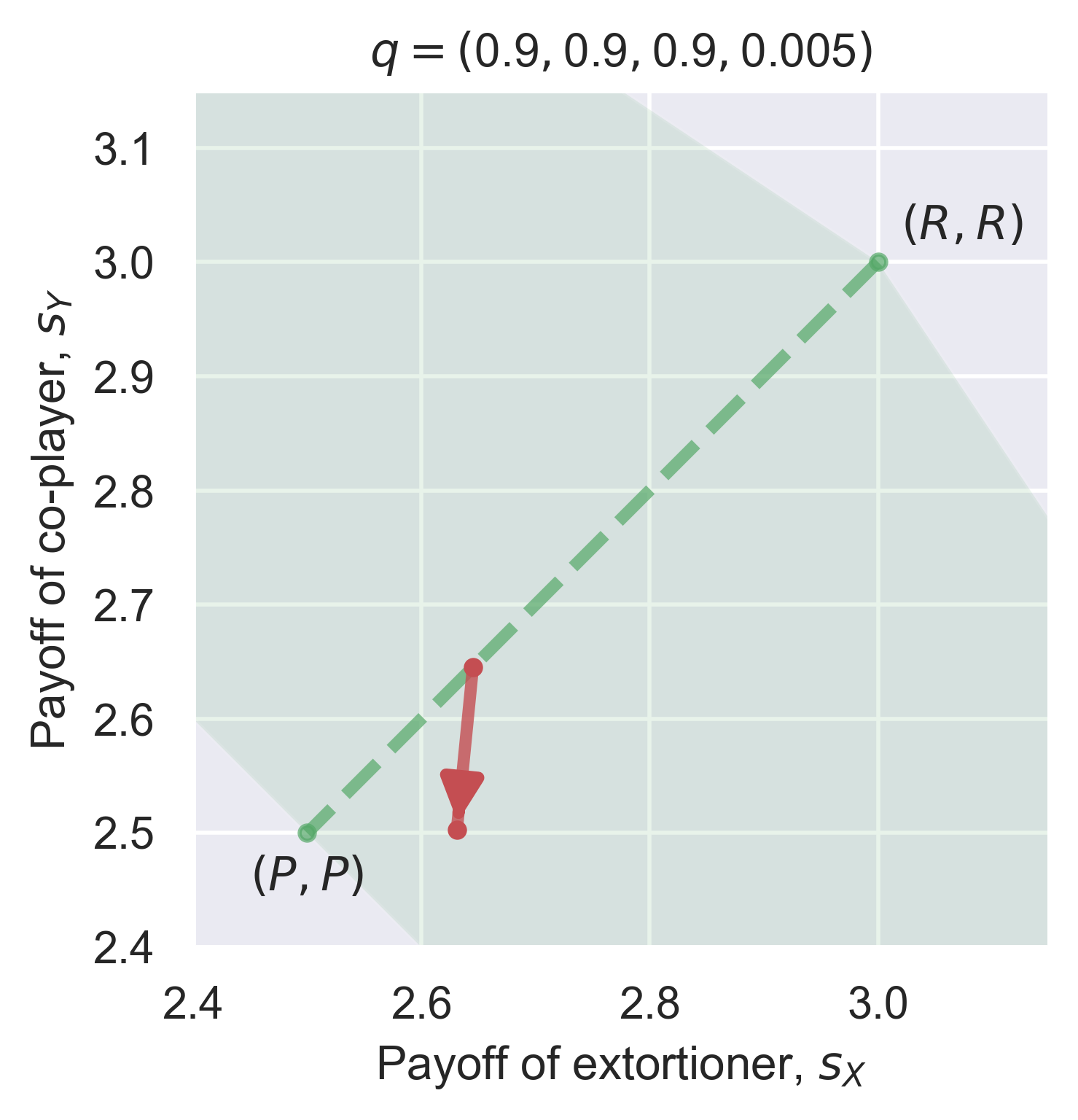}} 
& $q_1 < q_c \qquad q_1 > q_C$ \\
& & & \includegraphics[width=2cm]{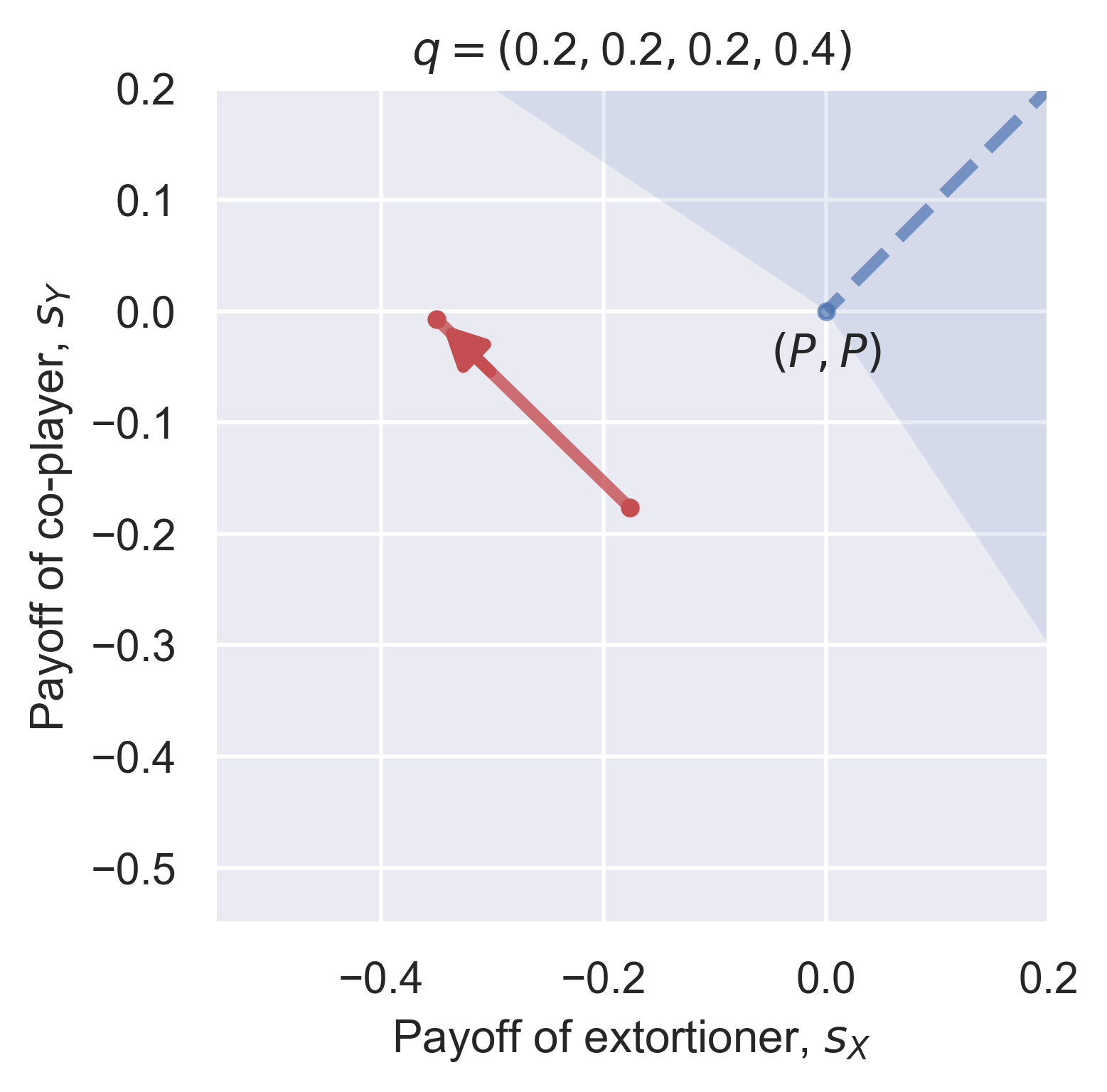}\includegraphics[width=2cm]{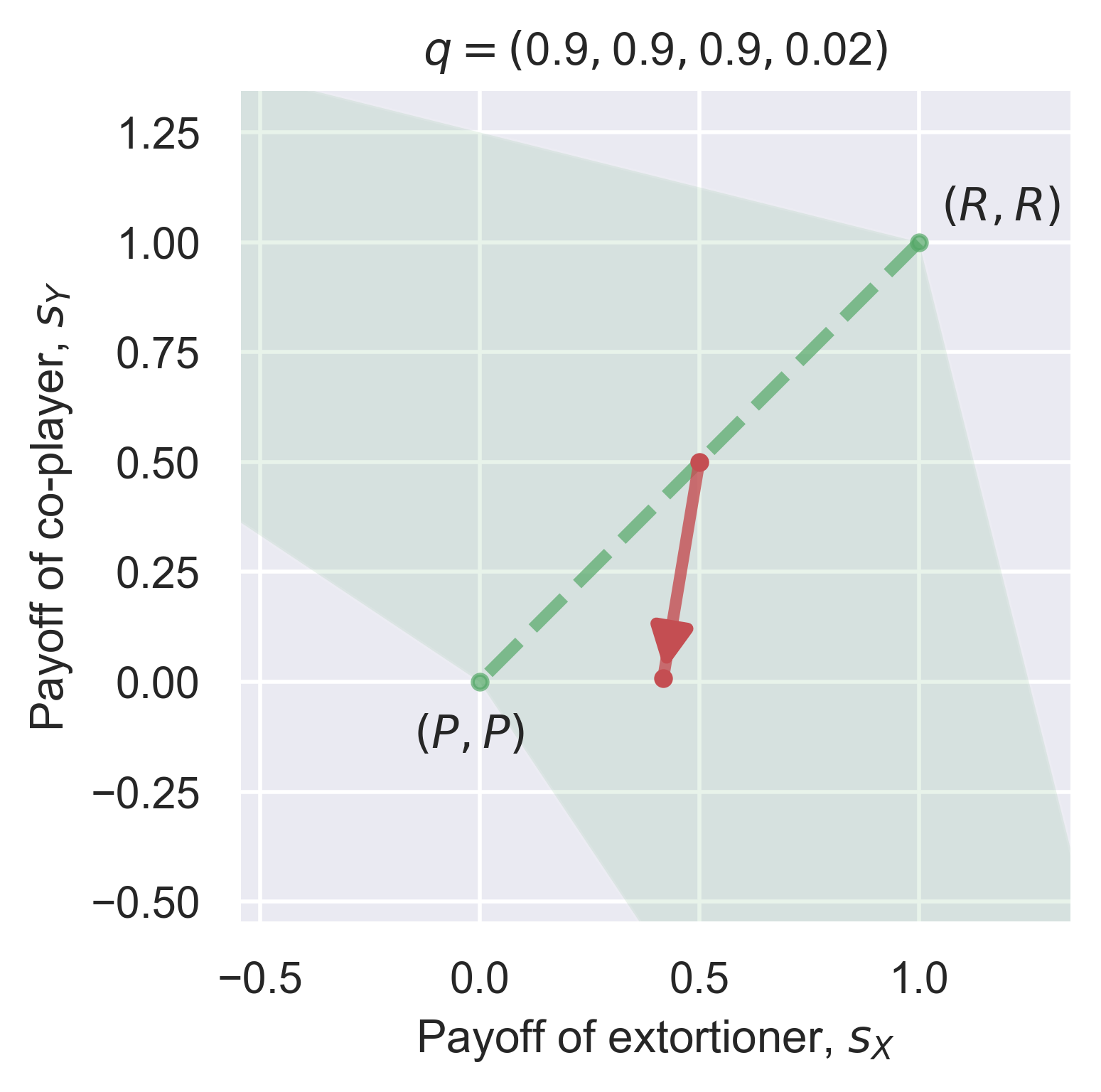} \\
\hline
D & \includegraphics[width=2.5cm]{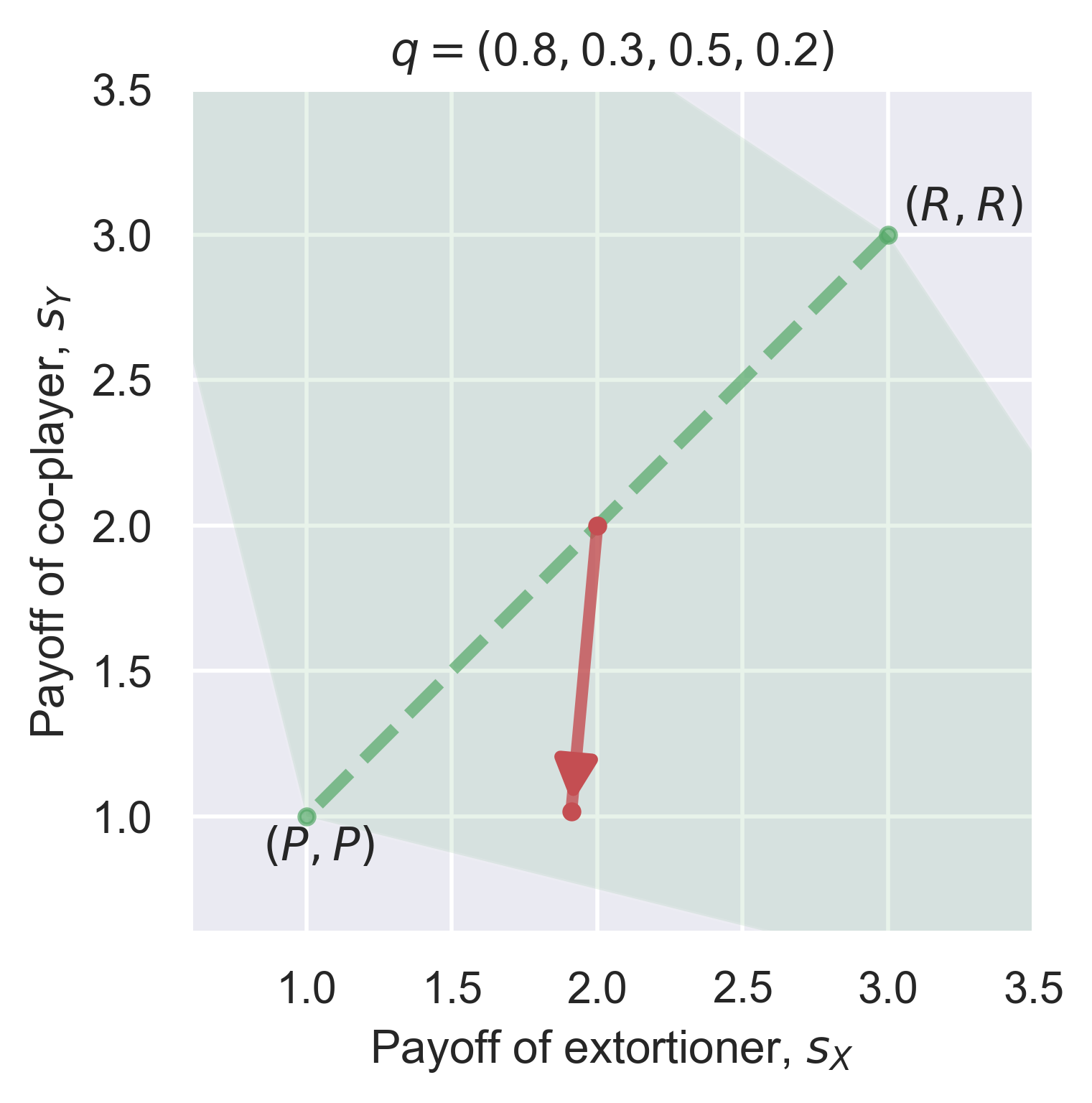} & \includegraphics[width=2.5cm]{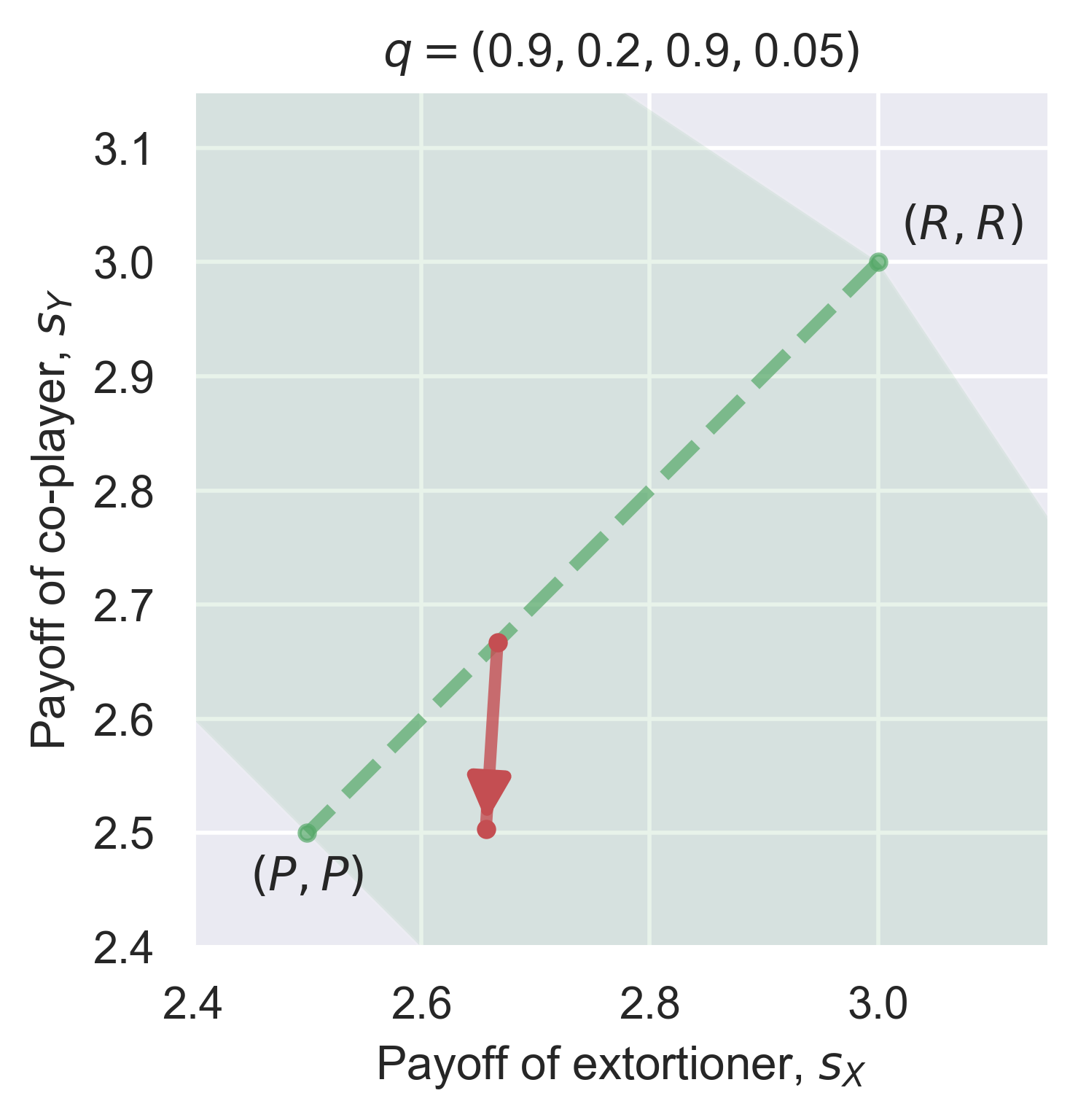}
& \includegraphics[width=2.5cm]{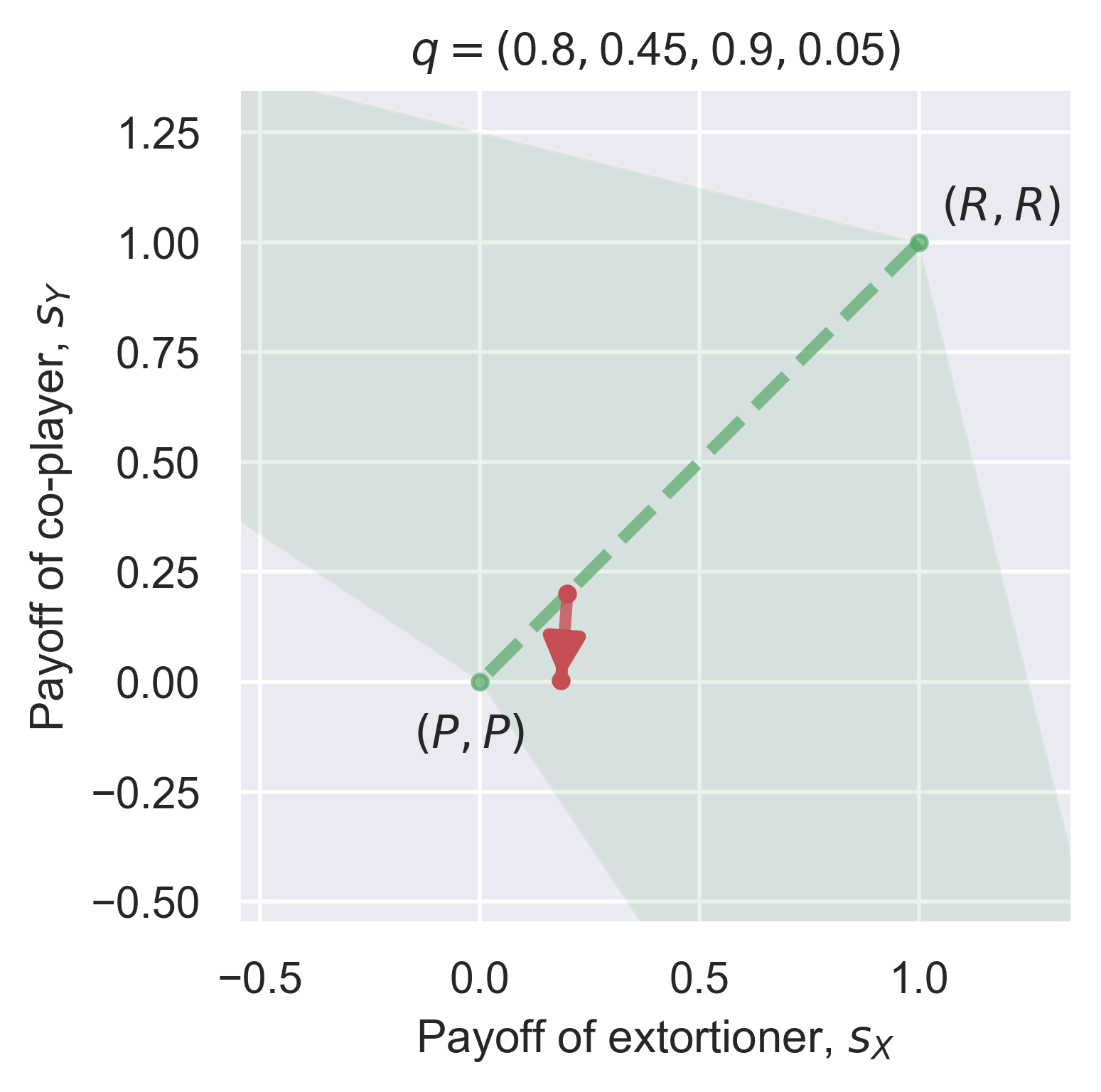} \\
\hline
\end{tabu}
\caption{Zoom-in view of the parametric curve of $(s_X, s_Y)$ with variable $\chi$ increasing from $1$ to $+\infty$. The arrow denotes the direction of motion.}
\label{summary_examples_quad_zoom}
\end{table}

\section{The complete picture of extortionate ZD's superiority: the need for studying $T+S<2P$ }
In what follows, we include a few excerpts from the literature where it is believed that an extortioner (using $O =P$ and $\chi >1$) always gets a higher payoff than the punishment $P$ and that of the co-player and (or) its payoff increases with respect to the extortion factor $\chi$. We also attach a few examples which ignore the role of $\phi$ in shaping the payoffs. As we point out in the main text as well as in the supplementary information here, the superiority of extortionate ZD strategies is typically studied under conventional IPD games where $T+S>2P$, but for the sake of completeness, it is equally important to study the robustness of ZD strategies across Prisoner's Dilemma games even of drastically different nature such as the case of $T+S<2P$, as detailed in our supplementary information. 
\par

\cite{Press_Dyson}
\begin{itemize}
%\item In particular, a player X who is witting of these strategies can \begin{enumerate*} [label=(\roman*\upshape)] \item deterministically set her opponent Y{\textquoteright}s score, independently of his strategy or response, or \item enforce an extortionate linear relation between her and his scores. \end{enumerate*}
\item Under the extortionate strategy, X{\textquoteright}s score depends on Y{\textquoteright}s strategy $\bm{q}$, and both are maximized when Y fully cooperates, with $\bm{q} = (1, 1, 1, 1)$.
\end{itemize}

\cite{Stewart_Plotkin_2012}
\begin{itemize}
\item Extortion strategies, whose existence Press and Dyson report, grant a disproportionate number of high payoffs to X at Y{\textquoteright}s expense.
\end{itemize}

\cite{Stewart_Plotkin_2013}
\begin{itemize}
\item In particular, Press and Dyson highlighted a subset of ZD strategies, called ``extortion strategies" that grants the extorting player a disproportionately high payoff when employed against a naive opponent who blindly adjusts his strategy to maximize his own payoff.
\end{itemize}

\cite{Adami_Hintze}
\begin{itemize}
\item In extortionate games, the strategy being preyed upon can increase their own payoff by modifying their own strategy $\bm{q}$, but this only increases the extortionate strategy{\textquoteright}s payoff. As a consequence, Press and Dyson conclude that a ZD strategy will always dominate any opponent that adapts its own strategy to maximize their payoff, for example, by Darwinian evolution.
\end{itemize}

\cite{Hilbe_Nowak_Sigmund}
\begin{itemize}
%\item A subset of those strategies allows ``extortioners" to ensure that any increase in one player{\textquoteright}s own payoff exceeds that of the other player by a fixed percentage.
\item Another subset consists of the extortion strategies, which guarantee that one player{\textquoteright}s own surplus exceeds the coplayer{\textquoteright}s surplus by a fixed percentage.
\item In this case, player \rom{1} can guarantee that his or her own ``surplus" (over the maximin value P) is the $\chi$-fold of the coplayer{\textquoteright}s surplus.
\end{itemize}

\cite{Hilbe_Nowak_Traulsen}
\begin{itemize}
\item In this way, a player can manipulate and extort his co-player, thereby ensuring that the own payoff never falls below the co-player{\textquoteright}s payoff.
\item Press and Dyson discovered that certain zero-determinant strategies can guarantee that a player always yields at least the opponent{\textquoteright}s payoff.
\item Such extortioner strategies $\bm{p}$ guarantee that the player{\textquoteright}s own surplus (over the maximin value P) exceeds the co-player{\textquoteright}s surplus by a fixed percentage.
\end{itemize}

%\cite{Hilbe_Wu_2014}
%\begin{itemize}
%\item On the other hand, a player who wants to outperform the respective opponent can do so by slightly tweaking the Tit-for-Tat strategy to the own advantage, thereby giving rise to extortionate ZD strategies.
%\end{itemize}

\cite{Hilbe_Traulsen_2015}
\begin{itemize}
\item In particular, ZD strategies can fix the co-player{\textquoteright}s payoff to an arbitrary value between P and R; or ensure that the own {\textquoteleft}surplus{\textquoteright} (over the maximin value P) is twice as large as the co-player{\textquoteright}s surplus; etc.
\end{itemize}

\cite{Hilbe_Wu_2015} % Hilbe_Wu_2015
\begin{itemize}
\item According to the theory, extortioners demand and receive an excessive share of any surplus, which allows them to outperform any adapting co-player.
\item As a result, extortioners are unbeatable: in a pairwise encounter, they cannot be outperformed by any opponent.
\end{itemize}

\cite{Adami_Schossau}
\begin{itemize}
\item Among the selfish strategies there are two main types: the ``Equalizers" that manage to force a fixed payoff onto the opponent, and the ``Extortionists" that fix the relative payoff between player and opponent in such a manner that if the opponent changes its strategy so as to get a better payoff, the extortioner always receives commensurately more.
\end{itemize}

%%%%%%%%%%%%%%%%REBUTTAL%%%%%%%%%%%%%%%%%%%%%%%%

\newpage

\section{The robustness of extortion ability of general ZD strategies}

The introduction of unbending strategies points out the fact that player X may not always get a higher payoff than player Y even if it uses an extortionate ZD strategy with $O = P$ and $\chi > 1$. We can not only extend the above finding to a more general ZD strategy with $P \leq O \leq R$ but also quantify the probability of successful extortion made by a ZD player. \par

\begin{figure}[H]
    \centering
    \subfloat[]{\includegraphics[width=0.33\linewidth]{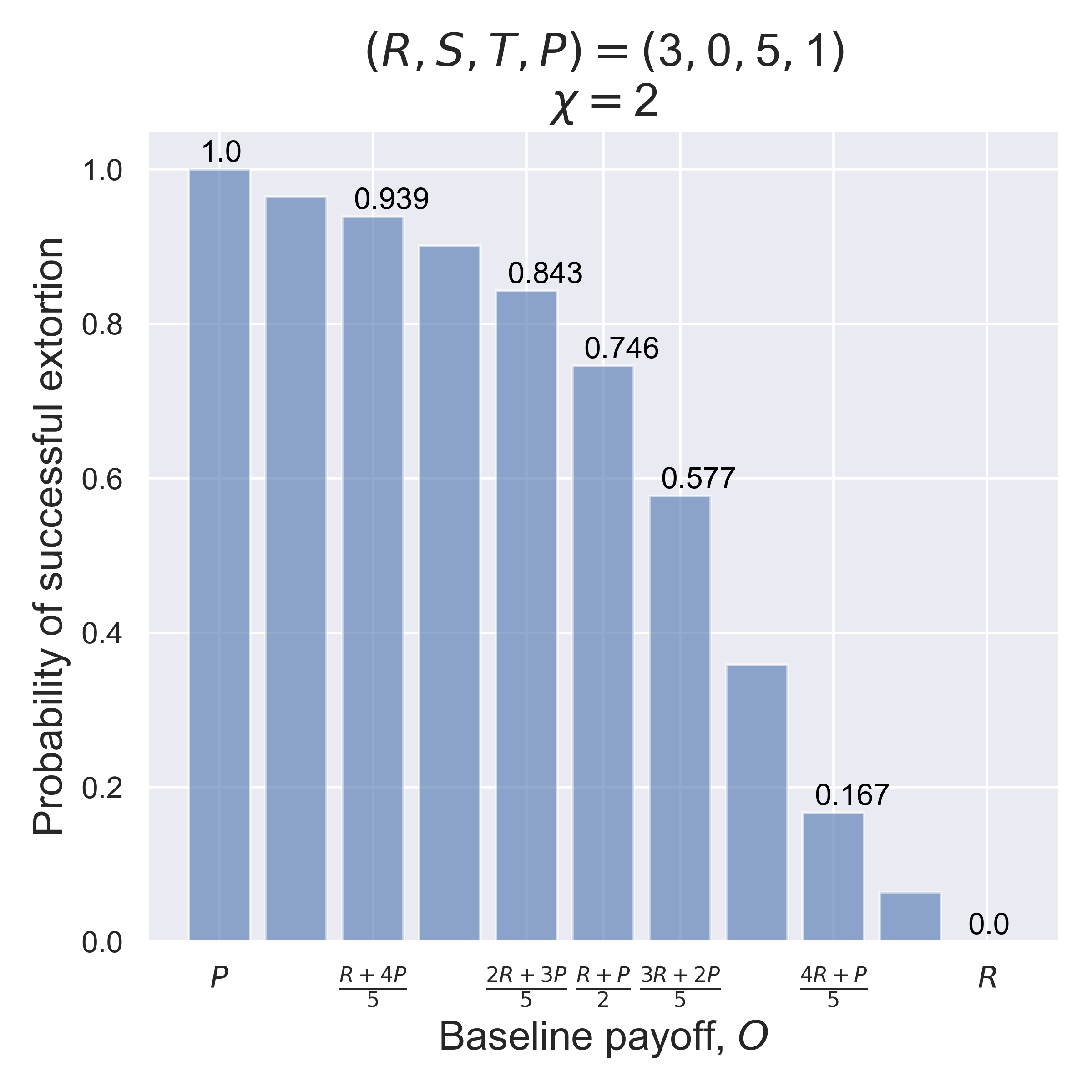}} 
    \subfloat[]{\includegraphics[width=0.33\linewidth]{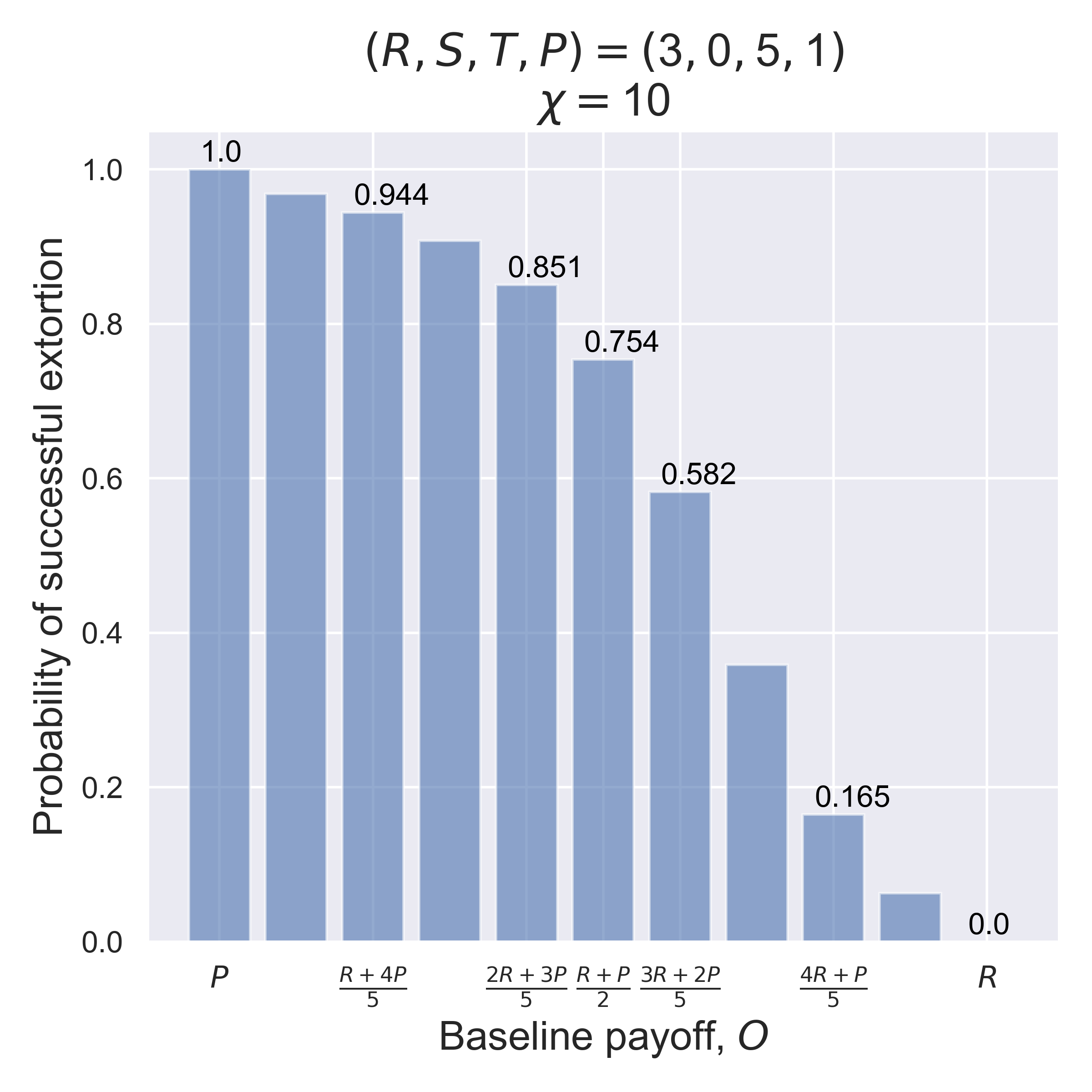}} 
    \subfloat[]{\includegraphics[width=0.33\linewidth]{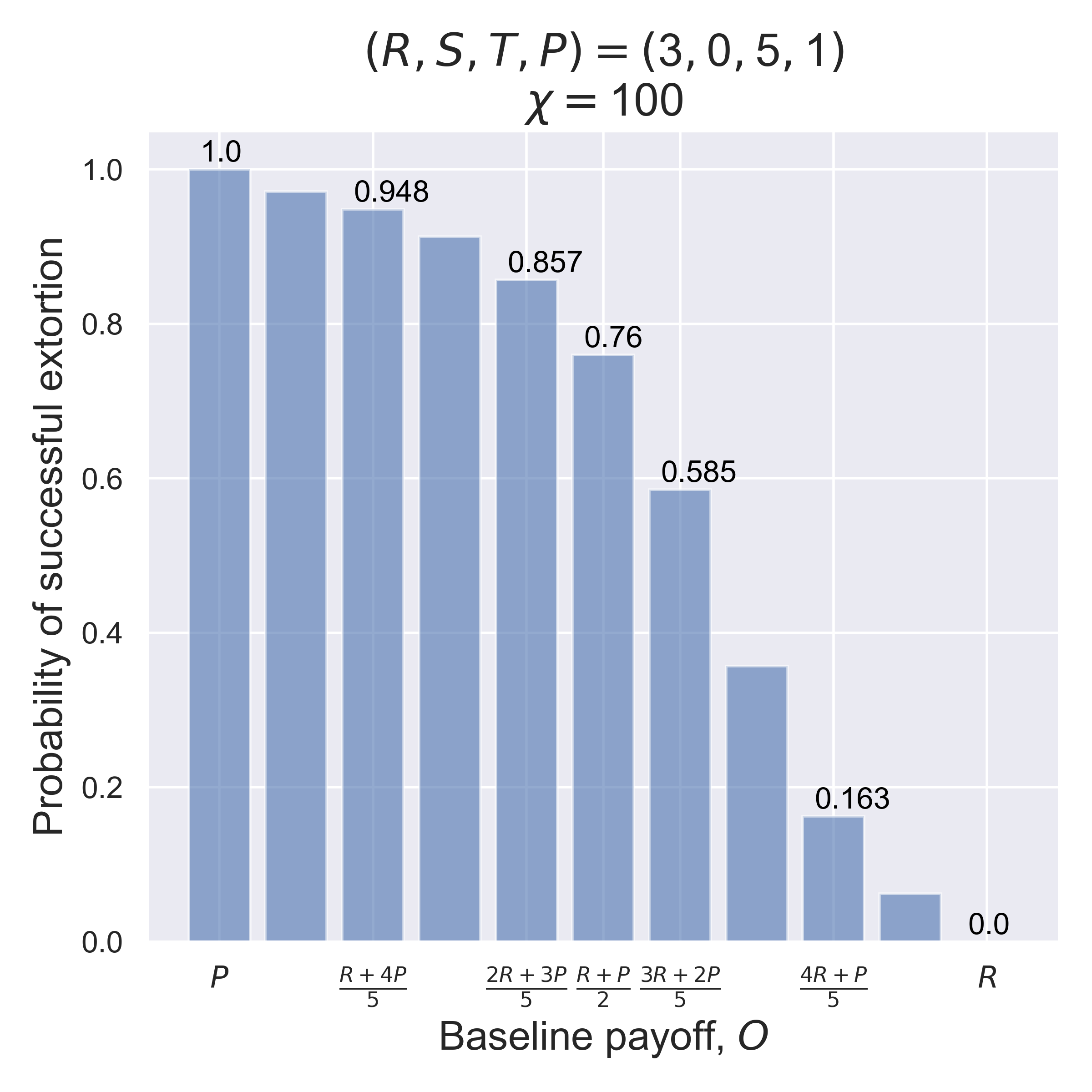}}
    \\
     \subfloat[]{\includegraphics[width=0.33\linewidth]{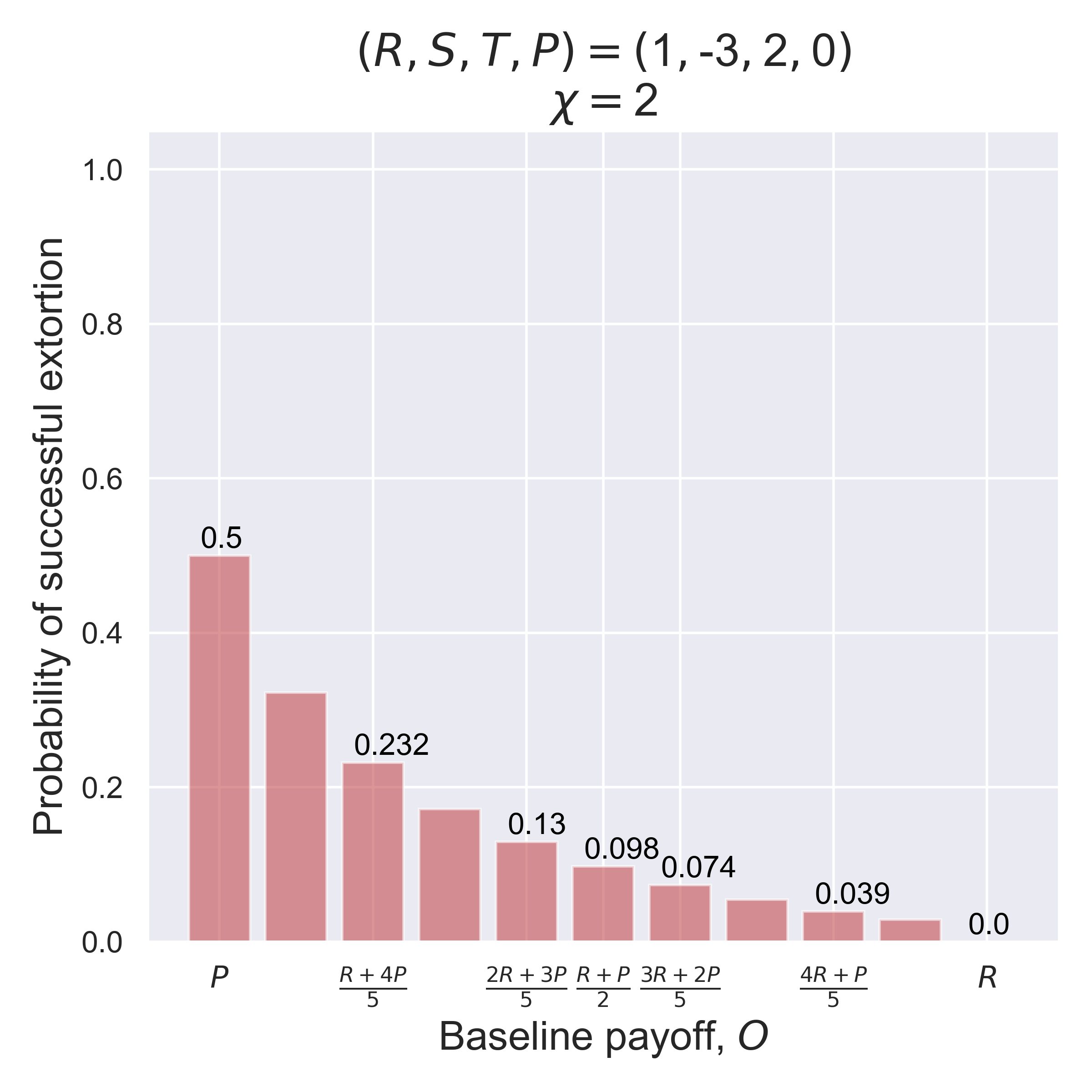}}
     \subfloat[]{\includegraphics[width=0.33\linewidth]{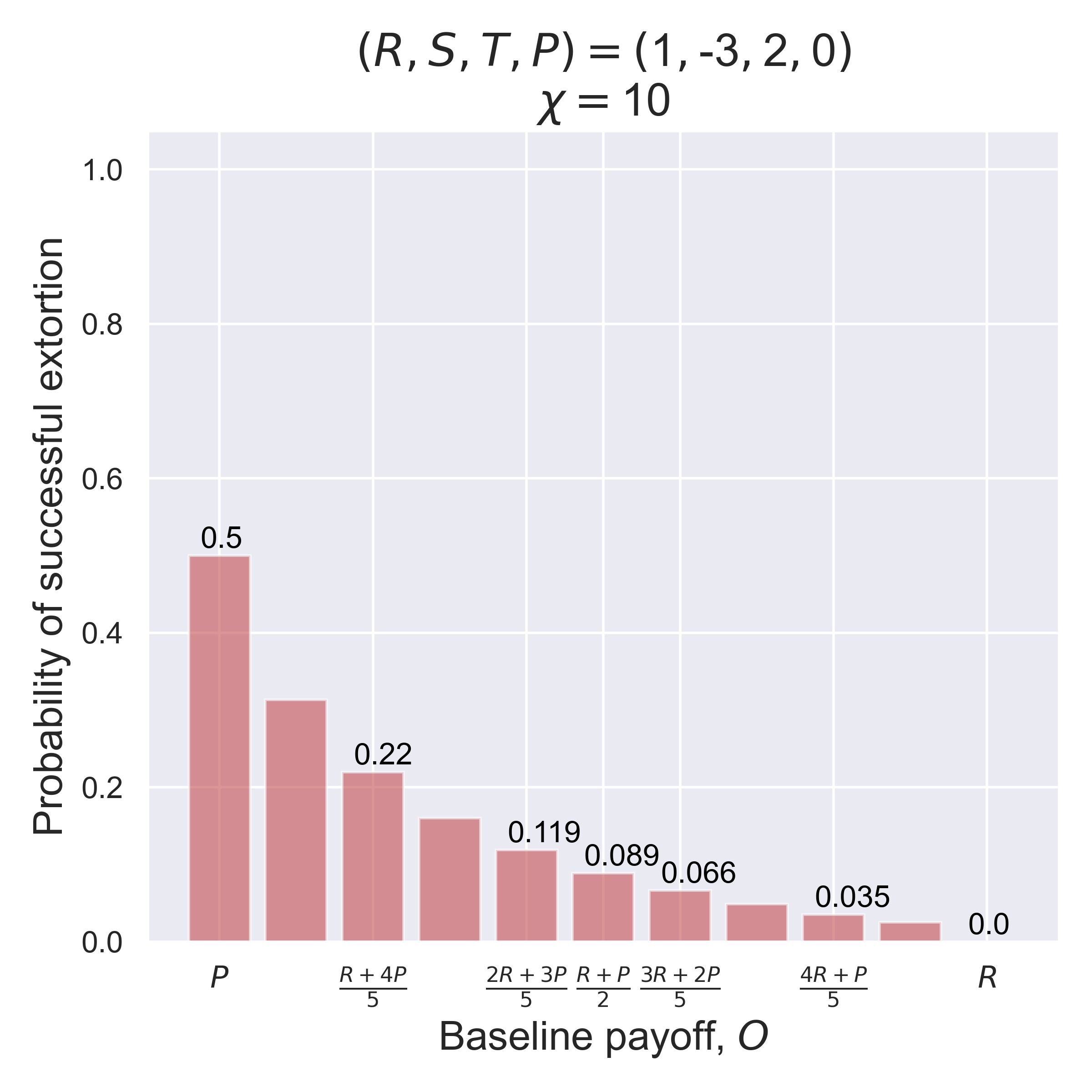}} 
     \subfloat[]{\includegraphics[width=0.33\linewidth]{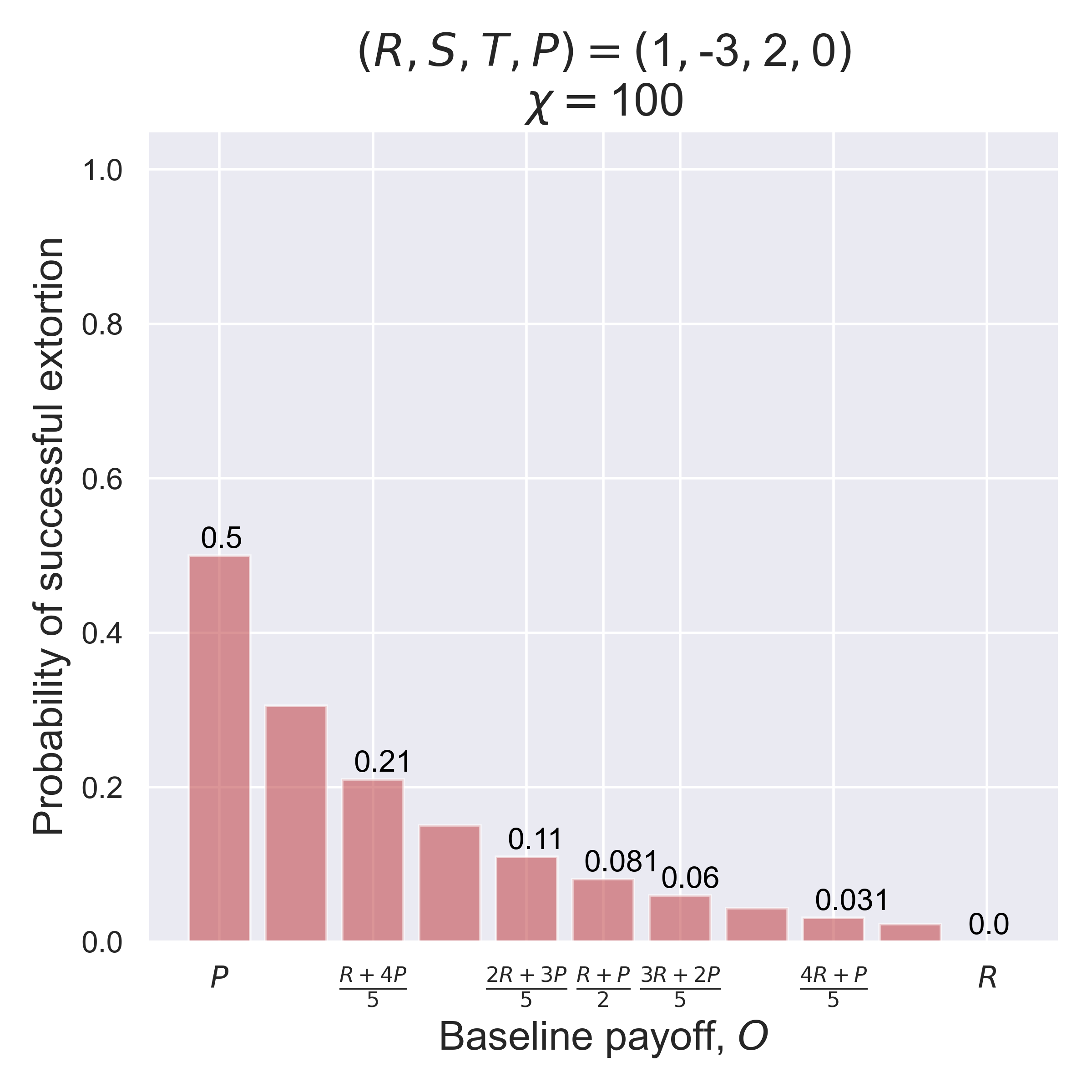}}
     \caption{Probability of ``successful extortion'' that player X actually gets a better payoff than its co-player Y ($s_X > s_Y$). The two players use a general ZD strategy and a random strategy uniformly drawn from memory-one strategies $[0, 1]^4$, respectively. The baseline payoff $O$ varies from $P$ to $R$ and the extortion factor $\chi$ is set as $2$, $10$, and $100$, respectively.}
     \label{robustness}
\end{figure}

Consider different values of $O$ and $\chi$. We present the results in Figure~\ref{robustness}. \par

On the one hand, the baseline payoff $O$ controls the level of generosity of the ZD player and hence impacts its chance to outperform its co-player. Increasing $O$ above $P$ makes an ``extortioner" less likely to be able to ensure the dominance. Noticeably, the payoff structure players an even more pronounced role than does $O$. For $T + S > 2P$, the curvature is concave downwards and the ZD player is able to maintain dominance for most of the time even using intermediate $O$ values. Yet for $T + S < 2P$, the curvature is concave upwards and the ZD player is more likely to lose its dominance for any $P < O \leq R$. \par

On the other hand, the extortion factor $\chi$ determines the level of extortion of the ZD player. It also plays a non-negligible role in making an ``extortioner'' behave like actual one. For $T + S > 2P$, a higher $\chi$ can help the ZD player be dominant if $O$ is small, whereas the same higher $\chi$ can have a counter effect if $O$ is large. As to $T + S < 2P$, the higher $\chi$ is, the more likely an extortion relation would backfire, that is, the more likely ZD fails to dominant its co-player. 

\section{The role played by \texorpdfstring{$\phi$}{e} (continued)}

\begin{figure}[H]
\centering
  \includegraphics[width=0.8\linewidth]{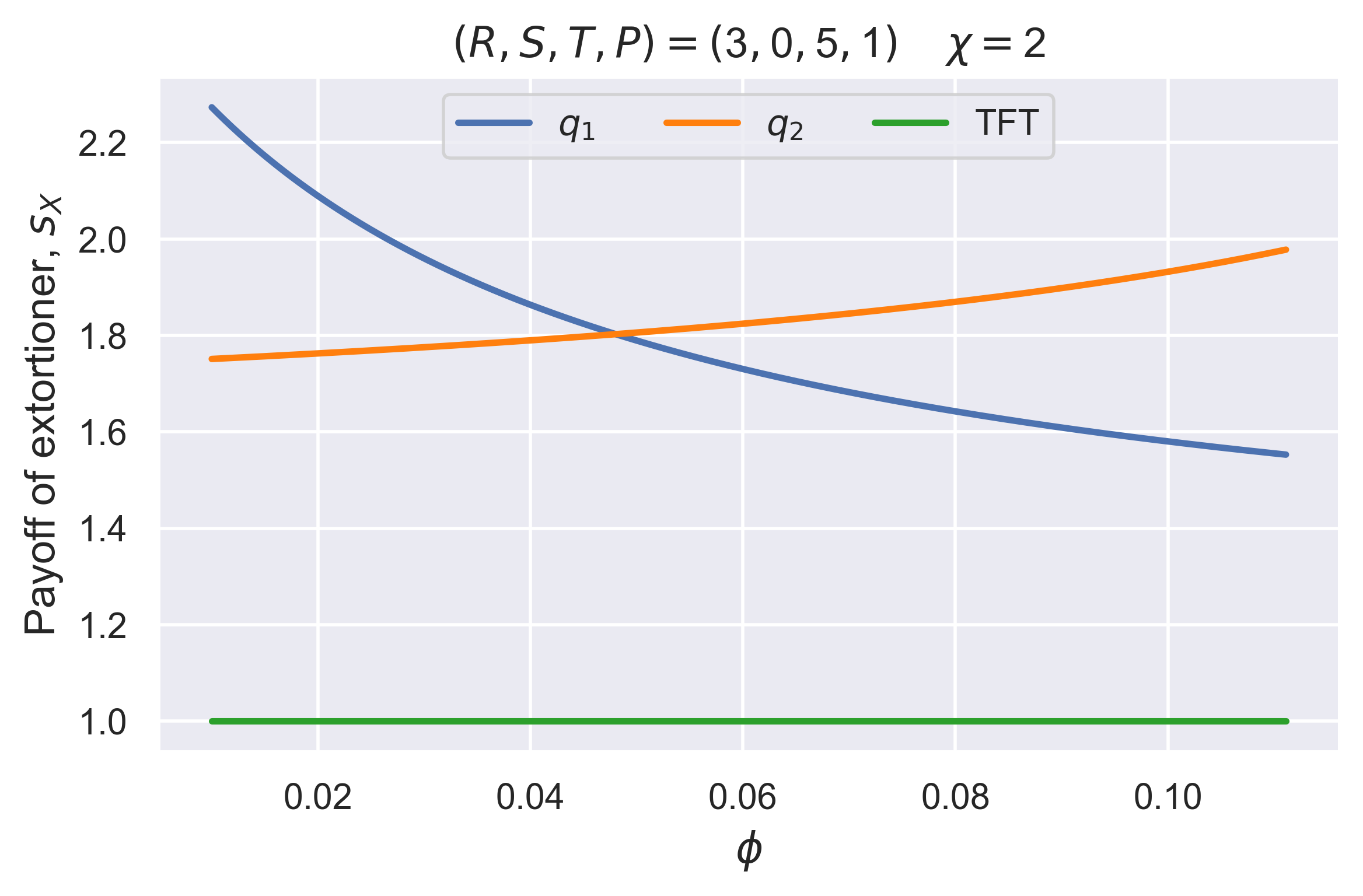}
  \caption{Extortioner's payoff against different strategies. We work on the conventional IPD game and fix $\chi$ as 2. Here, $\bm{q}_1 = (0.05, 0.95, 0.05, 0.1)$ and $\bm{q}_2 = (0.4, 0.1, 0.9, 0.2)$.}
  \label{con_phi}
\end{figure}

\begin{figure}[H]
    \centering
    \subfloat[$s_X(\bm{q}_1, \chi, \phi)$]{\includegraphics[width=0.5\linewidth]{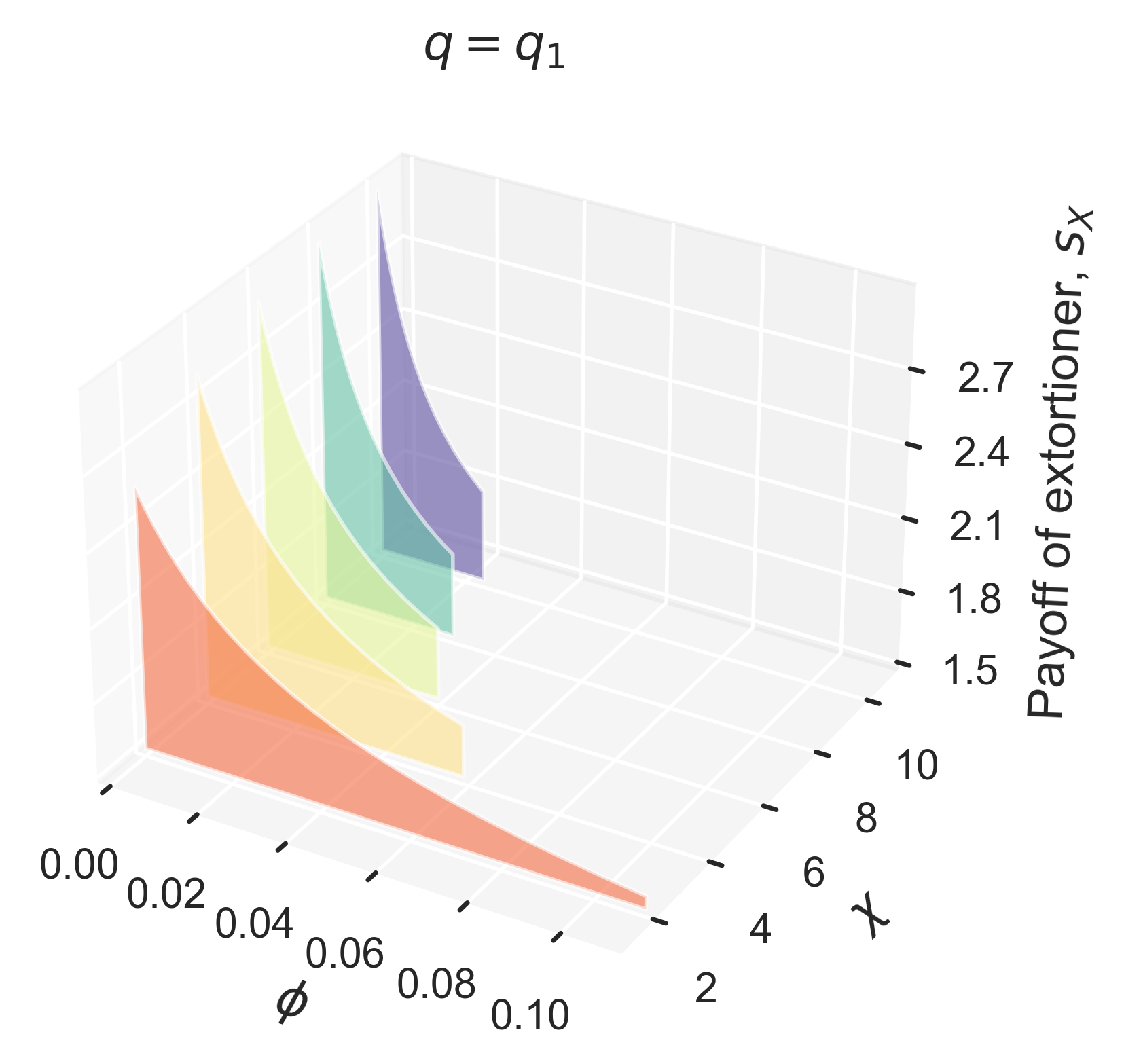}} 
     \subfloat[$s_X(\bm{q}_2, \chi, \phi)$]{\includegraphics[width=0.5\linewidth]{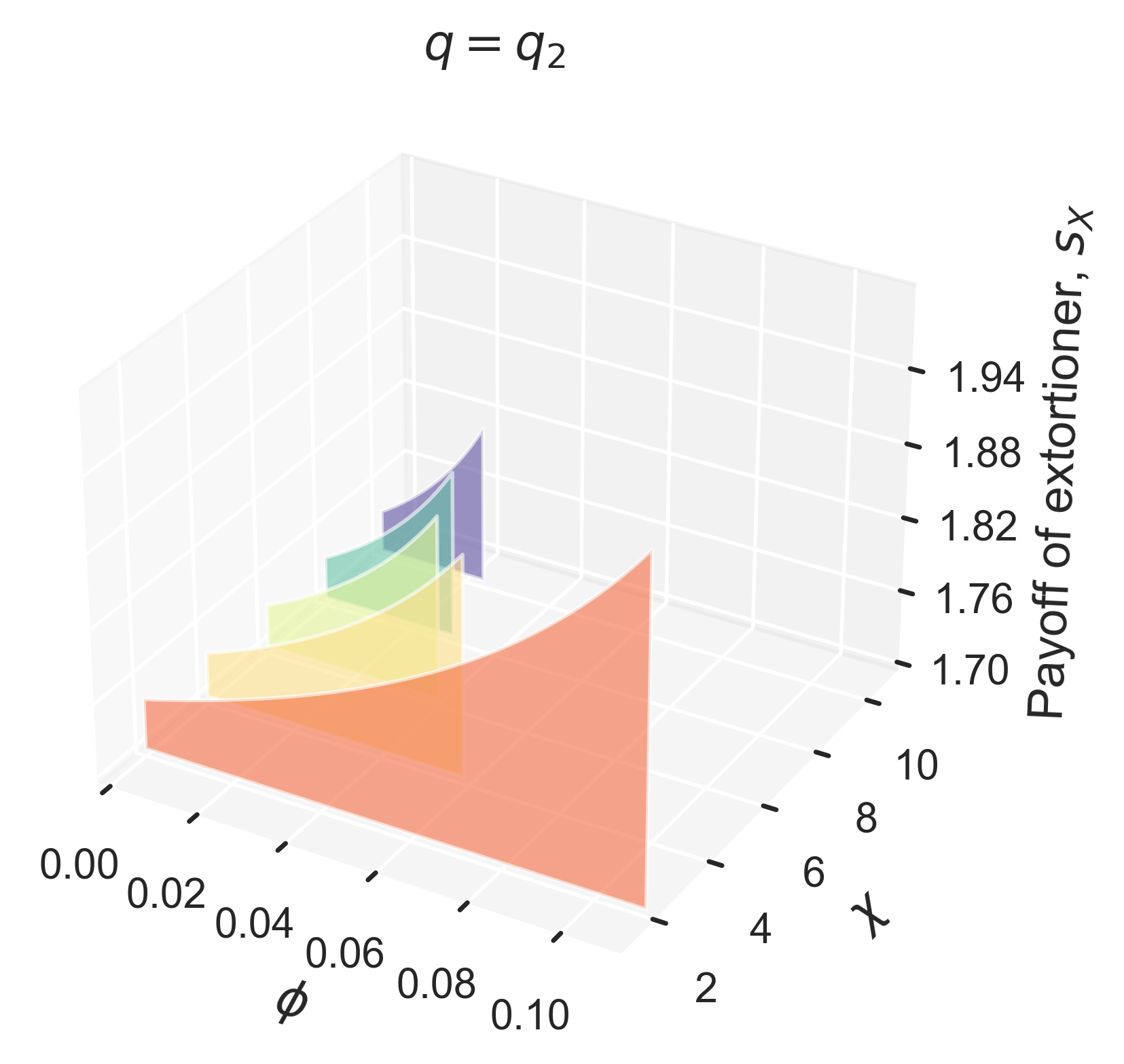}}
     \\
     \subfloat[$s_X(\text{TFT}, \chi, \phi)$]{\includegraphics[width=0.5\linewidth]{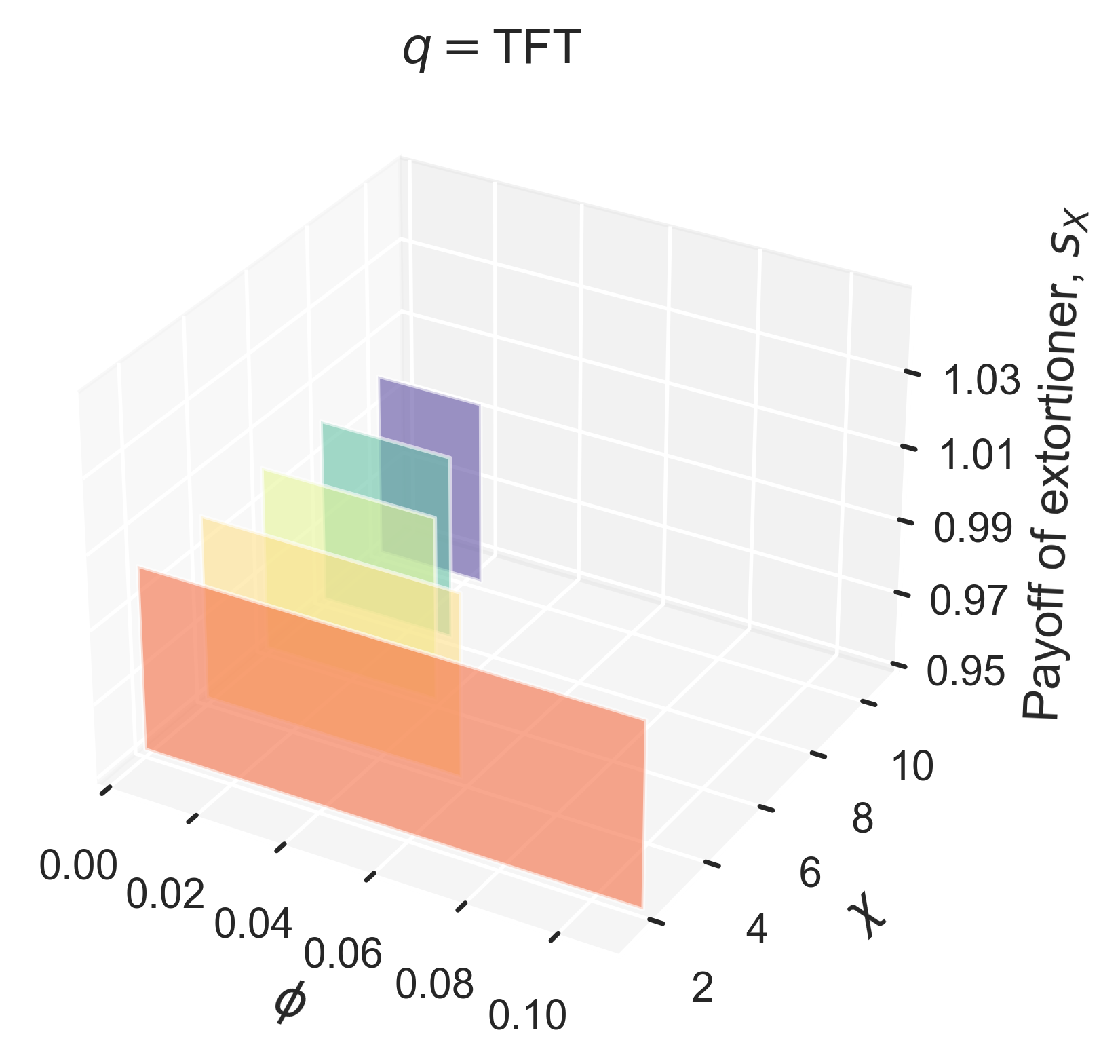}} 
     \caption{Waterfall plots of $s_X$ when playing against different strategies for fixed values of $\chi$. We work on the conventional IPD game. Different colors correspond to different values of $\chi$ (2, 4, 6, 8, and 10). As before, $\bm{q}_1 = (0.05, 0.95, 0.05, 0.1)$ and $\bm{q}_2 = (0.4, 0.1, 0.9, 0.2)$.}
     \label{payoff_poly}
\end{figure}

We have seen previously that tuning the parameter $\phi$ will impact the dependence of $s_X$ on the extortion factor in a non-trivial way. We further show that when $\chi$ is fixed, $s_X$ is a monotonic function of $\phi$. It can be told immediately from the expression of $s_X$, which takes the form of a linear rational function of $\phi$. Recall the three strategies we used in Figure~\ref{phi_dependency}. As a matter of fact, $s_X$ decreases with respect to $\phi$ if $Y$ plays $\bm{q_1}$, increases if $Y$ plays $\bm{q_2}$, and remains a constant if $Y$ plays Tit-for-Tat. The corresponding curves of $s_X$ are given in Figure~\ref{con_phi} for $\chi = 2$. More examples are shown in Figure~\ref{payoff_poly}. 

\section{Steering learning dynamics from extortion to fairness and cooperation: reactive strategies}

\begin{figure}[H]
\centering
   \subfloat[$r = 1.2$]{\includegraphics[width=0.5\linewidth]{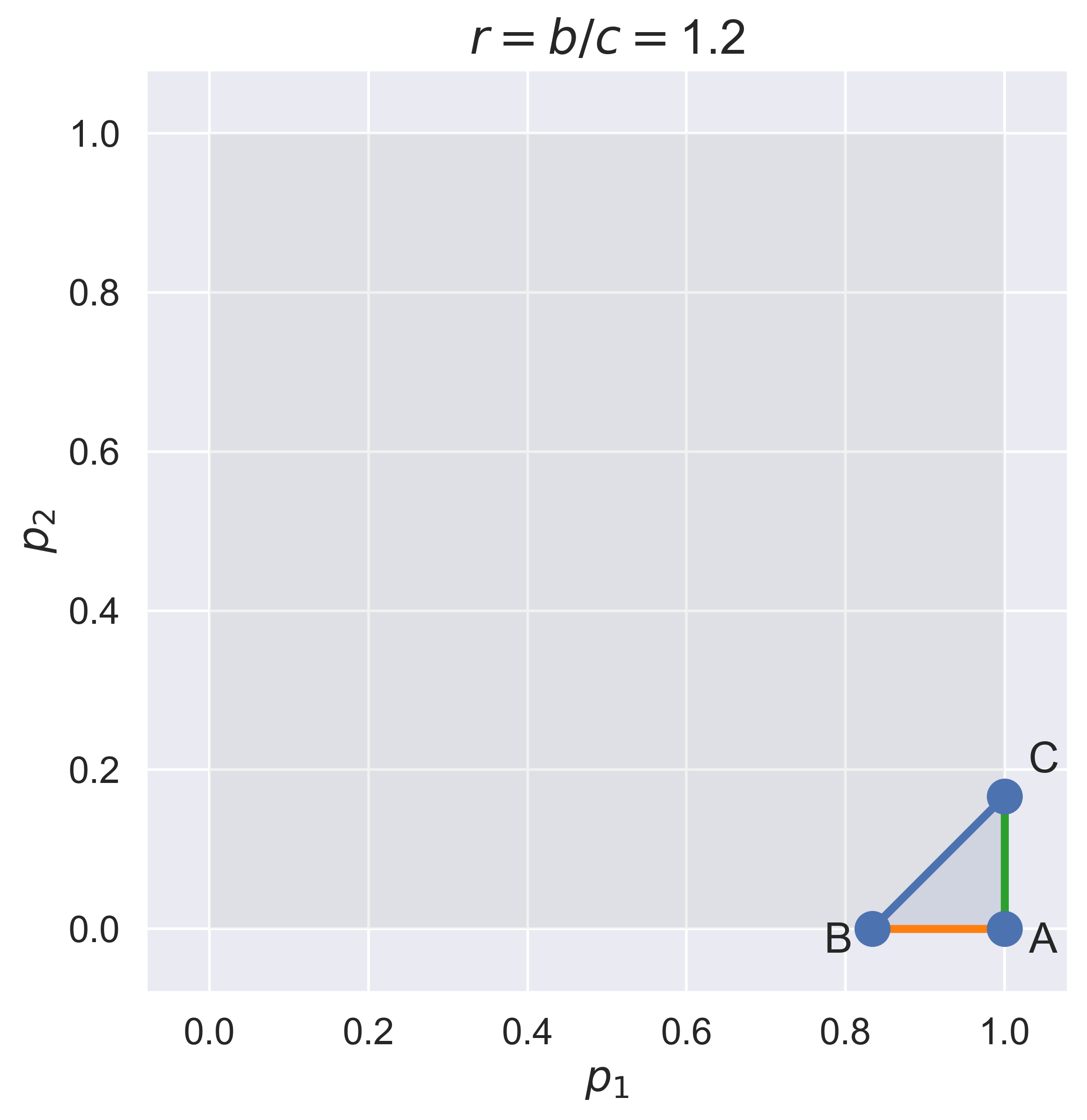}} 
    \subfloat[$r = 4$]{\includegraphics[width=0.5\linewidth]{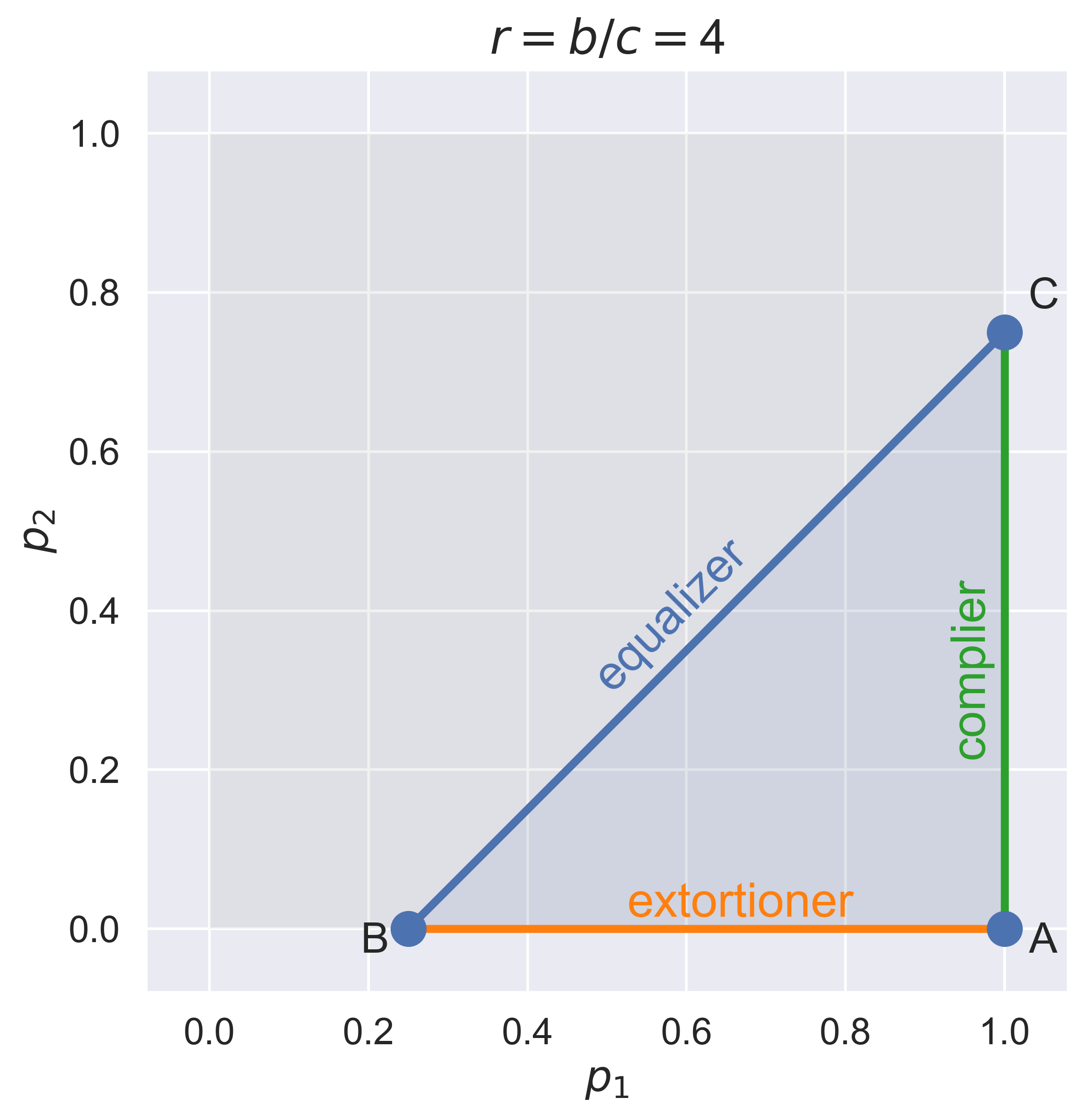}}
\caption{Reactive strategies as a subset of general ZD strategies. Assume that $\phi = 1/(b\chi + 1)$. In each panel, the set of reactive strategies is shaded in gray (the unit square), that of ZD strategies with positive $\chi$ is in blue (the right triangle), and that of ZD strategies with negative $\chi$ is in gray (the pentagon, which ``=" the unit square ``$-$" the triangle). Moreover, the three particular types of ZD strategies: extortioner, complier, and equalizer are highlighted in orange, green, and blue, respectively.}
\label{fig:reactive}
\end{figure}

From now on, we work on the donation game where the four payoffs satisfies $(R, S, T, P) = (b - c, -c, b, 0)$, and $r = b/c$ is the benefit-cost ratio. According to the definition, a strategy $\bm{p} = (p_1, p_2, p_3, p_4)$ is reactive if $p_1 = p_3$ and $p_2 = p_4$. In the donation game, a general ZD strategy (see Equation~\ref{GZD_strategy}) becomes a reactive strategy if $\phi = 1/(b\chi + 1)$. That is, reactive strategies is a subset of general ZD strategies. As a matter of fact, let 
\begin{equation}
\chi \geq 1 \qquad \text{or} \qquad \chi \leq \chi^{\text{upper}} =  \min\{-\frac{b - O}{O + c}, -\frac{O + c}{b - O}\}.
\end{equation}
We can divide the region of reactive strategies into two parts, each of which is a subset of general ZD strategies. The proof is trivial. Two examples are sketched in Figure~\ref{fig:reactive}.

Given the value of $\phi$, we can update the expressions for $p_1$ and $p_2$:
\begin{equation}
\begin{cases}
p_1 = 1 - \frac{(b - c - O)(\chi - 1)}{b\chi + c}, \\
p_2 = \frac{O(\chi - 1)}{b\chi + c}.
\end{cases} \qquad 0 \leq O \leq b - c
\label{eq:reactive}
\end{equation}
The two components satisfy a linear relation 
\begin{equation}
O(1 - p_1) - (b - c - O)p_2 = 0.
\end{equation}
\par

Further, for a few particular types of ZD strategies, their expressions can be obtained by letting the baseline payoff $O$ and (or) the extortion factor $\chi$ be some specific values. The results are summarized in Table~\ref{tab:reactive}. Notice that the generous Tit-for-Tat (GTFT) strategy for the donation game is $(1, \delta, 1, \delta)$, where 
\begin{equation}
\delta = \min\{1 - \frac{T - R}{R - S}, \frac{R - P}{T - P}\} = \min\{\frac{b - c}{b}, \frac{b - c}{b}\} = \frac{b - c}{b}.
\end{equation} 
Therefore, for a generous ZD strategy, if $O = b - c$ and $\chi \to \infty$, or equivalently, $p_1 = 1$ and $p_2 = 1 - c/b$, it would become GTFT. 

\definecolor{gr}{HTML}{EFEFEF}

\begin{table}[H]
\centering
\tabulinesep=1.5mm
\begin{tabu}{c c c c}
\hline
\rowcolor[HTML]{EFEFEF} 
Type & Parameters & Expressions & Remarks \\
\makecell[cc]{extortionate ZD \\ (extortioner)} & \makecell[cc]{$O = P = 0$ \\ $\chi \geq 1$} & 
$\begin{cases}
p_1 = 1 - \frac{(b - c)(\chi - 1)}{b\chi + c}, \\
p_2 = 0.
\end{cases}$
& $\frac{c}{b} \leq p_1 \leq 1$ \\
\cellcolor{gr}{\makecell[cc]{generous ZD \\ (complier)}} & \cellcolor{gr}{\makecell[cc]{$O = R = b - c$ \\ $\chi \geq 1$}} &
\cellcolor{gr}{$\begin{cases}
p_1 = 1, \\
p_2 = \frac{(b - c)(\chi - 1)}{b\chi + c}.
\end{cases}$}
& \cellcolor{gr}{$0 \leq p_2 \leq 1 - \frac{c}{b}$} \\
equalizer & \makecell[cc]{$0 \leq O \leq b - c$ \\ $\chi = +\infty$} &
$\begin{cases}
p_1 = \frac{O + c}{b}, \\
p_2 = \frac{O}{b}.
\end{cases}$
& $p_1 - p_2 = \frac{c}{b}$ \\ 
\cellcolor{gr}{mediocre ZD} & \cellcolor{gr}{\makecell{$O = \frac{T + S}{2} = \frac{b - c}{2}$ \\ $\chi \geq 1$}} & 
\cellcolor{gr}{$\begin{cases}
p_1 = 1 - \frac{(b - c)(\chi - 1)}{2(b\chi + c)}, \\
p_2 = \frac{(b - c)(\chi - 1)}{2(b\chi + c)}.
\end{cases}$}
& \cellcolor{gr}{$p_1 + p_2 - 1 = 0$} \\
\makecell[cc]{fair ZD \\ (Tit-for-Tat)} & $\chi = 1$ & 
$\begin{cases}
p_1 = 1, \\
p_2 = 0.
\end{cases}$ & $s_X = s_Y$ \\
\hline
\end{tabu}
\caption{Different types of ZD strategies in the donation game.}
\label{tab:reactive}
\end{table}

To study extortion and fairness in reciprocal actions, we have considered the scenario where player X uses an extortionate ZD strategy and player Y uses an unbending strategy. To further extend previous results, we now assume that player X uses a reactive strategy in the donation game where X's learning dynamics (payoff optimization through exploring the entire strategy space) is under influence of co-player Y using a fixed unbending strategy from class A or class D.

\section{Under influence of class A of unbending strategies (continued)}

Let $\bm{p} = (p_1, p_2, p_1, p_2)$ (a reactive strategy) and $\bm{q} = (1, q_2, 0, q_4)$ (an unbending strategy from Class A), where $q_2$ and $q_4$ satisfy 
\begin{equation}
\begin{cases}
\frac{c(b - c)}{b^2 + bc - c^2} = q_a < q_2 < 1, \\
0 < q_4 \leq h_A(q_2) = \frac{(b - c)(1 - q_2)[(b^2 + bc - c^2)q_2 - (b - c)c]}{bc^2q_2^2 - (b - c)(b^2 - bc - c^2)q_2 + (b - c)^2(b + c)}.
\end{cases}
\end{equation}
We obtain $s_X(p_1, p_2)$ as a quadratic rational function of $p_1$ and $p_2$. More specifically, we have
\begin{equation}
s_X(p_1, p_2) - (b - c) = \frac{(1 - p_1)\Delta(p_1, p_2)}{f_{A}(p_1, p_2)},
\end{equation} 
where 
\begin{equation}
\Delta(p_1, p_2) = [bq_2q_4 + c(1 - q_2) - (b - c)q_4](1 - p_2) - b(1 - q_2)(q_4p_1 + 1 - q_4),
\end{equation}
and $f_{A}(p_1, p_2) = (1 - q_2)[(1 - p_1) - q_4(p_2 - p_1)^2] + q_4(1 - p_2)$, which is always positive. \par

In particular, we list the values of $s_X$ in Table~\ref{example_A_re} for certain values of $p_1$ and $p_2$ on the boundaries of the unit square $[0, 1]^2$.

\begin{table}[H]
\centering
\tabulinesep=1.5mm
\begin{tabu}{c c c}
\hline
\rowcolor[HTML]{EFEFEF} 
$p_i$ & $\bm{p}$ & $s_X(p_1, p_2)$ \\
$p_1 = p_2 = 0$ & ALLD & $\frac{bq_4}{1 - q_2 + q_4}$ \\
$p_1 = 1$, $0 \leq p_2 < 1$ & $(1, p_2, 1, p_2)$ & $b - c$ \\
$0 \leq p_1 < 1$, $p_2 = 1$ & $(p_1, 1, p_1, 1)$ & $-c$ \\
\hline
\end{tabu}
\caption{Examples of $s_X$. Here, player X uses a reactive strategy $\bm{p}$ while player Y uses an unbending strategy $\bm{q}$ from Class A.}
\label{example_A_re}
\end{table}

\subsection{Maximum value of \texorpdfstring{$s_X$}{e}}

We claim that the maximum value of $s_X$ is either $s_X(1, p_2)$ ($p_2 \neq 1$) or $s_X(0, 0)$. The proof is straightforward as $\Delta$ decreases with respect to both $p_1$ and $p_2$. That is, 
\begin{equation}
\max \Delta(p_1, p_2) = \Delta(0, 0) = cq_4 - (b - c)(1 - q_2).
\end{equation}
Further, if $\Delta(0, 0) \leq 0$, $\max s_X(p_1, p_2) = s_X(1, p_2)$; otherwise, $\max s_X(p_1, p_2) = s_X(0, 0)$. Define 
\begin{equation}
h_{Aa}(q_2) = \frac{(b - c)(1 - q_2)}{c}.
\end{equation} 
We have
\begin{equation}
\max s_X(p_1, p_2) = 
\begin{cases}
s_X(1, p_2), & q_4 \leq h_{Aa} \\
s_X(0, 0). & q_4 > h_{Aa}
\end{cases}
\end{equation}
\par

Compare $h_A(q_2)$ and $h_{Aa}(q_2)$. It is straightforward to obtain the solutions to $h_A(q_2) = h_{Aa}(q_2)$, which are $q_2 = 1$ and $q_2 = b(b - c)/c^2$. Notice that 
\begin{equation}
b(b - c)/c^2 - q_a = \frac{(b + c)^2(b - c)^2}{c^2(b^2 + bc - c^2)} > 0,
\end{equation}
and
\begin{equation}
h_{Aa}(q_a) = \frac{b^2(b - c)}{c(b^2 + bc - c^2)} > 0.
\end{equation}
\par

The relation between $b(b - c)/c^2$ and $1$, that is, the relation between $r$ and $(1 + \sqrt{5})/2$ (the golden ratio), decides whether $s_X$ always takes the maximum value at $p_1 = 1$ or not. More specifically, 
\begin{enumerate}[label=(\roman*)]
\item if $1 < r < \frac{1 + \sqrt{5}}{2}$, 
\begin{equation}
\max s_X(p_1, p_2) = 
\begin{cases}
s_X(1, p_2) = b - c, & 0 < q_4 \leq h_{Aa} \\
s_X(0, 0) = \frac{bq_4}{1 - q_2 + q_4}, & h_{Aa} < q_4 \leq h_A
\end{cases}
\end{equation}
\item if $r \geq \frac{1 + \sqrt{5}}{2}$, $\max s_X(p_1, p_2) = s_X(1, p_2) = b - c$.
\end{enumerate}

\subsection{Monotonicity of \texorpdfstring{$s_X$}{e}}

Consider $s_X(p_1, p_2)$ as a quadratic rational function of $p_1$ and $p_2$. We study the monotonicity of $s_X$ with respect to the two variables. \par

\subsubsection{Partial derivative with respect to \texorpdfstring{$p_1$}{e}}

We first take the partial derivative with respect to $p_1$. In particular, we obtain the expressions of $\partial s_X/\partial p_1$ on the boundaries of the unit square $[0, 1]^2$. The results are summarized in Table~\ref{derivative_A_re_1}.

\begin{table}[H]
\centering
\tabulinesep=1.5mm
\begin{tabu}{c c c}
\hline
\rowcolor[HTML]{EFEFEF} 
$p_i$ & $\bm{p}$ & $\frac{\partial s_X(p_1, p_2)}{\partial p_1}$  \\
& $(p_1, p_2, p_3, p_4)$ & $\frac{q_4(1 - p_2)g_{A1}(p_1, p_2)}{f_{A}^2(p_1, p_2)}$  \\
\rowcolor[HTML]{EFEFEF} 
& & \\
$p_1 = p_2 = 0$ & ALLD & $\frac{\{-[b(1 - q_2) + c]q_4 + (1 - q_2)(bq_2 - c)\}q_4}{(1 - q_2 + q_4)^2}$ \\
\rowcolor[HTML]{EFEFEF} 
& & \\
$p_1 = 0$ and $p_2 = 1$ & $(0, 1, 0, 1)$ & $0$ \\
$p_1 = 0$ and $p_2 = 1 - \varepsilon$ & $(0, 1 - \varepsilon, 0, 1 - \varepsilon)$ & $\frac{\{-[b(1 - q_2) + 2b + c]q_4 + b - c(1 - q_2)\}\varepsilon q_4}{(1 - q_2)(1 - q_4)^2}$ \\
\rowcolor[HTML]{EFEFEF} 
& & \\
$p_2 = 0$ & $(p_1, 0, p_1, 0)$ & $\frac{q_4g_{A1}(p_1, 0)}{f_{A}^2(p_1, 0)}$ \\
\rowcolor[HTML]{EFEFEF} 
& & \\
$p_1 = 1$ & $(1, p_2, 1, p_2)$ & $\frac{[b(1 - q_2) - c](1 - p_2)q_4 + (1 - q_2)[b - c(1 - p_2)]}{[q_2(1 - p_2) + p_2](1 - p_2)q_4}$ \\
\hline 
\end{tabu}
\caption{Partial derivative of $s_X$ with respect to $p_1$. The parameter $\varepsilon$ is an infinitesimal ($\varepsilon \to 0$).}
\label{derivative_A_re_1}
\end{table}

Here, $f_{A}(p_1, p_2)$ is defined as before. Besides, $g_{A1}(p_1, p_2) = e_2p_1^2 + e_1p_1 + e_0$, where
\begin{equation}
\mfootnotesize{
\begin{cases}
e_2 = -(1 - q_2)[b(1 - q_2) + c]q_4 - c(1 - q_2)^2, \\
e_1 = 2(1 - q_2)[b(1 - q_2)p_2 + b + c]q_4 + 2c(1 - q_2)^2,\\
e_0 = [q_2(1 - p_2) + p_2]\{[b(1 - q_2) - c](1 - p_2)q_4 + (1 - q_2)[b - c(1 - p_2)]\} - e_2 - e_1,
\end{cases}}
\end{equation}
Moreover, we have $f_{A}(p_1, 0) = (1 - q_2)[(1 - p_1) - q_4p_1^2] + q_4$ and $g_{A1}(p_1, 0) = e_{20}p_1^2 + e_{10}p_1 + e_{00}$, where
\begin{equation}
\begin{cases}
e_{20} = e_2, \\
e_{10} = 2(b + c)(1 - q_2)q_4 + 2c(1 - q_2)^2,\\
e_{10} = q_2\{[b(1 - q_2) - c]q_4 + (b - c)(1 - q_2)\} - e_2 - e_1.
\end{cases}
\end{equation}
\par

With the expressions in Table~\ref{derivative_A_re_1}, we now consider the sign of the partial derivative. \par

It is straightforward to show that
\begin{equation}
\left.\frac{\partial s_X}{\partial p_1}\right|_{p_1 = 1}
\end{equation}
is always positive, which is equivalent to show that the numerator 
\begin{equation}
(\ast) = [b(1 - q_2) - c](1 - p_2)q_4 + (1 - q_2)[b - c(1 - p_2)]
\end{equation}
is positive. As $(\ast)$ is linear, we only need to consider its values estimated at the boundaries. Letting $p_2 = 0$ and $q_4 = 0$, $p_2 = 0$ and $q_4 = h_A(q_2)$, and $p_2 = 1$, the values are all positive. For example, when $p_2 = 0$ and $q_4 = h_A(q_2)$, $(\ast)$ becomes
\begin{equation}
\frac{b(b - c)^2(1 - q_2)^2[(b + 2c)q_2 + b - c]}{bc^2q_2^2 - (b - c)(b^2 - bc - c^2)q_2 + (b - c)^2(b + c)},
\end{equation}
which is greater than zero. \par

Given that $e_2 < 0$ and $2e_2 + e_1 = 2b(1 - q_2)[q_2(1 - p_2) + p_2]q_4 > 0$, the curve of $g_{A1}$ (as a function of $p_1$) is a parabola opening downwards and the axis of symmetry is on the right of $p_1 = 1$. From the above analysis, we also know that when $p_1 = 1$, $g_{A1} > 0$. Therefore, as $p_1$ increases ($p_2$ being fixed),  
\begin{equation}
\frac{\partial s_X}{\partial p_1}
\end{equation}
is either always nonnegative or first negative and then positive. \par

As to 
\begin{equation}
\left.\frac{\partial s_X(p_1, p_2)}{\partial p_1}\right|_{(0, 0)},
\end{equation}
it is easy to see that the derivative is nonnegative if 
\begin{equation}
0 < q_4 \leq h_{a}(q_2) = \frac{(bq_2 - c)(1 - q_2)}{b(1 - q_2) + c}.
\end{equation}
Otherwise, the derivative is negative. Notice that the difference between $h_{A}(q_2)$ and $h_a(q_2)$ is
\begin{equation}
\frac{bcq_2(1 - q_2)^2[(2b + c)(b - c) + bcq_2]}{[b(1 - q_2) + c][bc^2q_2^2 - (b - c)(b^2 - bc - c^2)q_2 + (b - c)^2(b + c)]},
\end{equation}
which is always positive. For the extreme case where the benefit-cost ratio $r \to +\infty$, the difference approaches zero. Moreover, the difference between $h_{Aa}(q_2)$ and $h_{a}(q_2)$ is
\begin{equation}
\frac{b^2(1 - q_2)^2}{c[b(1 - q_2) + c]},
\end{equation}
being positive as well. 
\par

Similarly, for 
\begin{equation}
\left.\frac{\partial s_X(p_1, p_2)}{\partial p_1}\right|_{(0, 1 - \varepsilon)}, \varepsilon \to 0
\end{equation}
the derivative is nonnegative if
\begin{equation}
0 < q_4 \leq \frac{b - c(1 - q_2)}{b(1 - q_2) + 2b + c}.
\end{equation}
Otherwise, the derivative is negative. Nevertheless, the difference between the right-hand side and $h_A(q_2)$ can be either positive or negative, depending on the values of $r$ and $q_2$. The explicit expressions are tedious and we only list a few examples here:
\begin{enumerate}[label=(\roman*)]
\item $b/c = 1.2$, the difference is always positive,
\item $b/c = 2$, the difference is always nonnegative and is zero when $q_2 = 1/2$,
\item $b/c = 10$, the difference is nonnegative if $9/109 = q_a  < q_2 \leq 3/7$ or $9/10 \leq q_2 < 1$.
\end{enumerate} 
\par

Finally, we consider the sign of
\begin{equation}
\left.\frac{\partial s_X}{\partial p_1}\right|_{p_2 = 0}.
\end{equation}
\par

Since $e_{20} = e_2 < 0$ and $2e_{20} + e_{10} = 2bq_2(1 - q_2)q_4 > 0$, the curve of $g_{A1}(p_1, 0)$ (as a function of $p_1$) is a parabola opening downwards and the axis of symmetry is on the right of $p_1 = 1$. Also, we have $g_{A1}(1, 0) > 0$. Therefore, as $p_1$ increases, the derivative is either always nonnegative or first negative and then positive. \par

Combined with previous results, we claim that 
\begin{enumerate}[label=(\roman*)]
\item $0 < q_4 \leq h_{a}(q_2)$, $\left.\frac{\partial s_X}{\partial p_1}\right|_{p_2 = 0}$ is nonnegative for $0 \leq p_1 \leq 1$,
\item $h_{a}(q_2)  < q_4 < h_{A}(q_2)$, $\left.\frac{\partial s_X}{\partial p_1}\right|_{p_2 = 0}$ is first negative and then positive for $0 \leq p_1 \leq 1$. 
\begin{itemize}
\item e.g. $q_2 = (b - c)/b$ and $q_4 = (2b - 3c)/4b$ ($h_a(q_2) = (b - 2c)/2b$ and $h_A(q_2) = (b - c)/2b$): the partial derivative is negative for $0 \leq p_1 < c/(2b - c)$, zero for $p_1 = c/(2b - c)$, and positive for $c/(2b - c) < p_1 \leq 1$.
\end{itemize}
\end{enumerate}
In particular, we can show that $\left.\frac{\partial s_X}{\partial p_1}\right|_{p_2 = 0}$ is always positive for $p_1 = c/b$ (hence positive for $c/b \leq p_1 \leq 1$). It is straightforward to show as
\begin{equation}
\mfootnotesize{
\left.\frac{\partial s_X}{\partial p_1}\right|_{(c/b, 0)} = \frac{[bc^2q_2^2 - (b - c)(b^2 - bc - c^2)q_2 + (b - c)^2(b + c)]b^2q_4(h_A - q_4)}{[(c^2q_2 + b^2 - c^2)q_4 + b(b - c)(1 - q_2)]^2}.}
\end{equation}
Therefore, if player X uses an extortionate ZD strategy (recall Figure~\ref{fig:reactive}), $s_X$ is free of $p_2$, and its derivative with respect to $p_1$ is always positive. \par

We summarize the above analysis for $\partial s_X/\partial p_1$ in Table~\ref{monotonicity_A_re_1}. 

\begin{table}[H]
\centering
\tabulinesep=1.5mm
\footnotesize
\begin{tabu}{c c | c}
\hline
\rowcolor[HTML]{EFEFEF} 
Partial derivative & Sign &  Monotonicity \\
$\frac{\partial s_X}{\partial p_1}$ & \makecell[cc]{$\bm{+}$ or \\ $\bm{-} \to \bm{+}$} & \makecell[cc]{$\bm{\nearrow}$ or \\ $\bm{\searrow} \to \bm{\nearrow}$} \\
\rowcolor[HTML]{EFEFEF} 
& & \\
$\left.\frac{\partial s_X(p_1, p_2)}{\partial p_1}\right|_{(0, 0)}$ & $\begin{cases} \bm{+}, & 0 < q_4 \leq h_a(q_2) \\ \bm{-}. & h_a(q_2) < q_4 < h_A(q_2) \end{cases}$ & $\begin{cases} \bm{\nearrow}, & 0 < q_4 \leq h_a(q_2) \\ \bm{\searrow}. & h_a(q_2) < q_4 < h_A(q_2) \end{cases}$ \\
\rowcolor[HTML]{EFEFEF} 
& & \\
$\left.\frac{\partial s_X(p_1, p_2)}{\partial p_1}\right|_{(0, 1)}$ & $0$ & $\bm{\rightarrow}$ \\
\rowcolor[HTML]{EFEFEF} 
& & \\
$\left.\frac{\partial s_X}{\partial p_1}\right|_{p_2 = 0}$ & $\begin{cases} \bm{+}, & 0 < q_4 \leq h_a(q_2) \\ \bm{-} \to \bm{+}. & h_a(q_2) < q_4 < h_A(q_2) \end{cases}$ & $\begin{cases} \bm{\nearrow}, & 0 < q_4 \leq h_a(q_2) \\ \bm{\searrow} \to \bm{\nearrow}. & h_a(q_2) < q_4 < h_A(q_2) \end{cases}$ \\
\rowcolor[HTML]{EFEFEF} 
& & \\
$\left.\frac{\partial s_X}{\partial p_1}\right|_{p_1 = 1}$ & $\bm{+}$ & $\bm{\nearrow}$ \\
\hline 
\end{tabu}
\caption{Monotonicity of $s_X$ with respect to $p_1$ as $p_1$ increases from $0$ to $1$. When the sign is $\bm{+}$, it also includes zero.}
\label{monotonicity_A_re_1}
\end{table}

\subsubsection{Partial derivative with respect to \texorpdfstring{$p_2$}{e}}

We then take the partial derivative of $s_X$ with respect to $p_2$. As before, we consider the expressions of $\partial s_X/\partial p_2$ on the boundaries of the unit square $[0, 1]^2$. The results are summarized in Table~\ref{derivative_A_re_2}. 

\begin{table}[H]
\centering
\tabulinesep=1.5mm
\footnotesize
\begin{tabu}{c c c}
\hline
\rowcolor[HTML]{EFEFEF} 
$p_i$ & $\bm{p}$ & $\frac{\partial s_X(p_1, p_2)}{\partial p_2}$  \\
$p_1 = 0$ & $(0, p_2, 0, p_2)$ & $\frac{(1 - q_2)[(1 - q_4p_2)\delta(p_2) - (1 - q_2 + q_4)(1 - q_4)(cp_2 + b)q_4p_2]}{f_{A}^2(0, p_2)}$ \\
\rowcolor[HTML]{EFEFEF} 
& & \\
$p_2 = 0$ & $(p_1, 0, p_1, 0)$ & $\frac{(1 - q_2)(1 - p_1)g_{A2}(p_1, 0)}{f_A^2(p_1, 0)}$ \\
\rowcolor[HTML]{EFEFEF} 
& & \\
$p_1 = 1$ & $(1, p_2, 1, p_2)$ & $0$ \\
\rowcolor[HTML]{EFEFEF} 
& & \\
$p_1 = 1 - \varepsilon$ and & & \\
$p_2 = 0$ & $(1 - \varepsilon, 0, 1 - \varepsilon, 0)$ & $\frac{\epsilon(1 - q_2)[(b - c - bq_2)q_4 - (2b - c)q_2 + (b - c)]}{q_2^2q_4}$ \\
$p_2 = 1 - c/b$ & $(1 - \varepsilon, 1 - c/b, 1 - \varepsilon, 1 - c/b)$ & $\frac{b^2\epsilon(1 - q_2)(\ast)}{c^2(cq_2 + b - c)^2q_4}$ \\
$p_2 = 1$ & $(1 - \varepsilon, 1, 1 - \varepsilon, 1)$ & $-\frac{bq_4}{\epsilon(1 - q_2)}$ \\
\hline 
\end{tabu}
\caption{Partial derivative of $s_X$ with respect to $p_2$. The parameter $\varepsilon$ is an infinitesimal ($\varepsilon \to 0$).}
\label{derivative_A_re_2}
\end{table}

Here, $f_A$ is defined as before. Besides,
\begin{equation}
\delta(p_2) = [(b - c)(1 - q_2) - cq_4]q_4(1 - p_2) - (1 - q_4)[c(1 - q_2) + (b + c)q_4].
\end{equation}
Moreover, we have $g_{A2}(p_1, 0) = e_{20}p_1^2 + e_{10}p_1 + e_{00}$, where
\begin{equation}
\begin{cases}
e_{20} = [b(1 - q_2) + c]q_4^2 + c(1 - q_2)q_4, \\
e_{10} = [(b - c - bq_2)q_4 - (2b - c)q_2 + (b - c)]q_4 - e_{20} - e_{00}, \\
e_{00} = bq_4^2 - (bq_2 + c)q_4 - c(1 - q_2). 
\end{cases}
\end{equation}
In addition,
\begin{equation}
(\ast) = -c(bcq_4 + 2b^2 - c^2)q_2 + (b - c)(c^2q_4 - b^2 + bc + c^2).
\end{equation}
\par

With the expressions in Table~\ref{derivative_A_re_2}, we now consider the sign of the partial derivative. \par

We first show that
\begin{equation}
\left.\frac{\partial s_X(p_1, p_2)}{\partial p_2}\right|_{p_1 = 0}
\end{equation}
is always negative. It can be told that
\begin{equation}
\delta(p_2) < 0 \Rightarrow \left.\frac{\partial s_X(p_1, p_2)}{\partial p_2}\right|_{p_1 = 0} < 0.
\end{equation}
\par

As $\delta(p_2)$ is a linear function of $p_2$, to show that $\delta(p_2) < 0$, we only need to show that $\delta(0) < 0$ and $\delta(1) < 0$. The latter is $- (1 - q_4)[c(1 - q_2) + (b + c)q_4]$ and the former is
\begin{equation}
\delta(0) = bq_4^2 - (bq_2 + c)q_4 - c(1 - q_2).
\end{equation}
Treat it as a quadratic function of $q_4$, corresponding to a parabola opening upwards. Its value at $q_4 = 0$ and $q_4 = q_2 +  c/b$ are both negative. Therefore, to show that $\delta(0) < 0$ always holds for $0 < q_4 \leq h_A(q_2)$, it suffices to show that $h_A(q_2) \leq q_2 + c/b$. The proof is tedious but trivial. In particular, if the benefit-cost ratio $r \to \infty$, we get $c/b \to 0$ and $h_A(q_2) \to q_2$. \par

We then show that 
\begin{equation}
\left.\frac{\partial s_X(p_1, p_2)}{\partial p_2}\right|_{p_2 = 0}
\end{equation}
is either always nonpositive or first negative and then positive as $p_1$ increases. Note that $g_{A2}(p_1, 0)$ is a quadratic function of $p_1$, corresponding to a parabola opening upwards. Also, from the above analysis, we know that $g_{A2}(0, 0) < 0$. Therefore, as $p_1$ increases, if $g_{A2}(1, 0) \leq 0$, the derivative is always nonpositive; otherwise, it is first negative and then positive. \par

We have 
\begin{equation}
g_{A2}(1, 0) = [(b - c - bq_2)q_4 - (2b - c)q_2 + (b - c)]q_4.
\end{equation} 
That is, $g_{A2}(1, 0) \leq 0$ if 
\begin{equation}
\begin{cases}
\frac{b - c}{2b - c} < q_2 < \frac{b - c}{b}, \\
0 < q_4 \leq  \min\{h_A(q_2), \frac{(2b - c)q_2 - (b - c)}{b - c - bq_2}\}, \\
\end{cases}
\qquad \text{or} \qquad
\begin{cases}
\frac{b - c}{b} \leq q_2 < 1, \\
 0 < q_4 \leq h_A(q_2).
\end{cases}
\label{derivative_A_re_2_common}
\end{equation}
 Otherwise, $g_{A2}(1, 0) > 0$.
 \par
 
 Finally, given different values of $p_2$, we consider the sign of 
 \begin{equation}
 \left.\frac{\partial s_X(p_1, p_2)}{\partial p_2}\right|_{p_1 = 1 - \epsilon}, \epsilon \to 0.
 \end{equation}
\par

If $p_2 = 0$, we draw the same conclusion as Equation~\ref{derivative_A_re_2_common}. If $p_2 = 1$, the derivative is negative. Further, if $p_2 = 1 - c/b$ (where $\bm{p}$ approaches GTFT), we show that the derivative is also negative. It is equivalent to show that $(\ast) < 0$. Consider $(\ast)$ as a linear function of $q_2$. We have
\begin{equation}
(\ast)(q_a) = -\frac{(b + c)(b - c)^2(b^2 - c^2q_4)}{b^2 + bc - c^2}, \qquad (\ast)(1) = -b^3 - c^3q_4.
\end{equation}
Therefore, $(\ast) < 0$ always holds for $q_a < q_2 < 1$. Hence the derivative is negative. \par
 
We summarize the above analysis for $\partial s_X/\partial p_2$ in Table~\ref{monotonicity_A_re_2}.

\begin{table}[H]
\centering
\tabulinesep=1.5mm
\begin{tabu}{c c | c}
\hline
\rowcolor[HTML]{EFEFEF} 
Partial derivative & Sign &  Monotonicity \\
$\left.\frac{\partial s_X(p_1, p_2)}{\partial p_2}\right|_{p_1 = 0}$ & $\bm{-}$ & $\bm{\searrow}$ \\
\rowcolor[HTML]{EFEFEF} 
& & \\
$\left.\frac{\partial s_X(p_1, p_2)}{\partial p_2}\right|_{p_2 = 0}$ & \makecell[cc]{$\bm{-}$ or \\ $\bm{-} \to \bm{+}$} & \makecell[cc]{$\bm{\searrow}$ \\ or $\bm{\searrow} \to \bm{\nearrow}$} \\
\rowcolor[HTML]{EFEFEF} 
& & \\
$\left.\frac{\partial s_X(p_1, p_2)}{\partial p_2}\right|_{p_1 = 1}$ & $0$ & $\bm{\rightarrow}$ \\
\rowcolor[HTML]{EFEFEF} 
& & \\
$\left.\frac{\partial s_X(p_1, p_2)}{\partial p_2}\right|_{(1 - \epsilon, 0)}$ & $\bm{-}$ or $\bm{+}$ & $\bm{\searrow}$ or $\bm{\nearrow}$ \\
$\left.\frac{\partial s_X(p_1, p_2)}{\partial p_2}\right|_{(1 - \epsilon, 1 - c/b)}$ & $\bm{-}$ & $\bm{\searrow}$ \\
$\left.\frac{\partial s_X(p_1, p_2)}{\partial p_2}\right|_{(1 - \epsilon, 1)}$ & $\bm{-}$ & $\bm{\searrow}$ \\
\hline 
\end{tabu}
\caption{Monotonicity of $s_X$ with respect to $p_2$ as $p_1$ increases from $0$ to $1$. When the sign is $\bm{-}$, it also includes zero. As before, the parameter $\varepsilon$ is an infinitesimal ($\varepsilon \to 0$).}
\label{monotonicity_A_re_2}
\end{table}

\subsection{Conclusion and example}

\begin{figure}[H]
    \centering
    \subfloat[$r = 1.2$]{\includegraphics[width=0.5\linewidth]{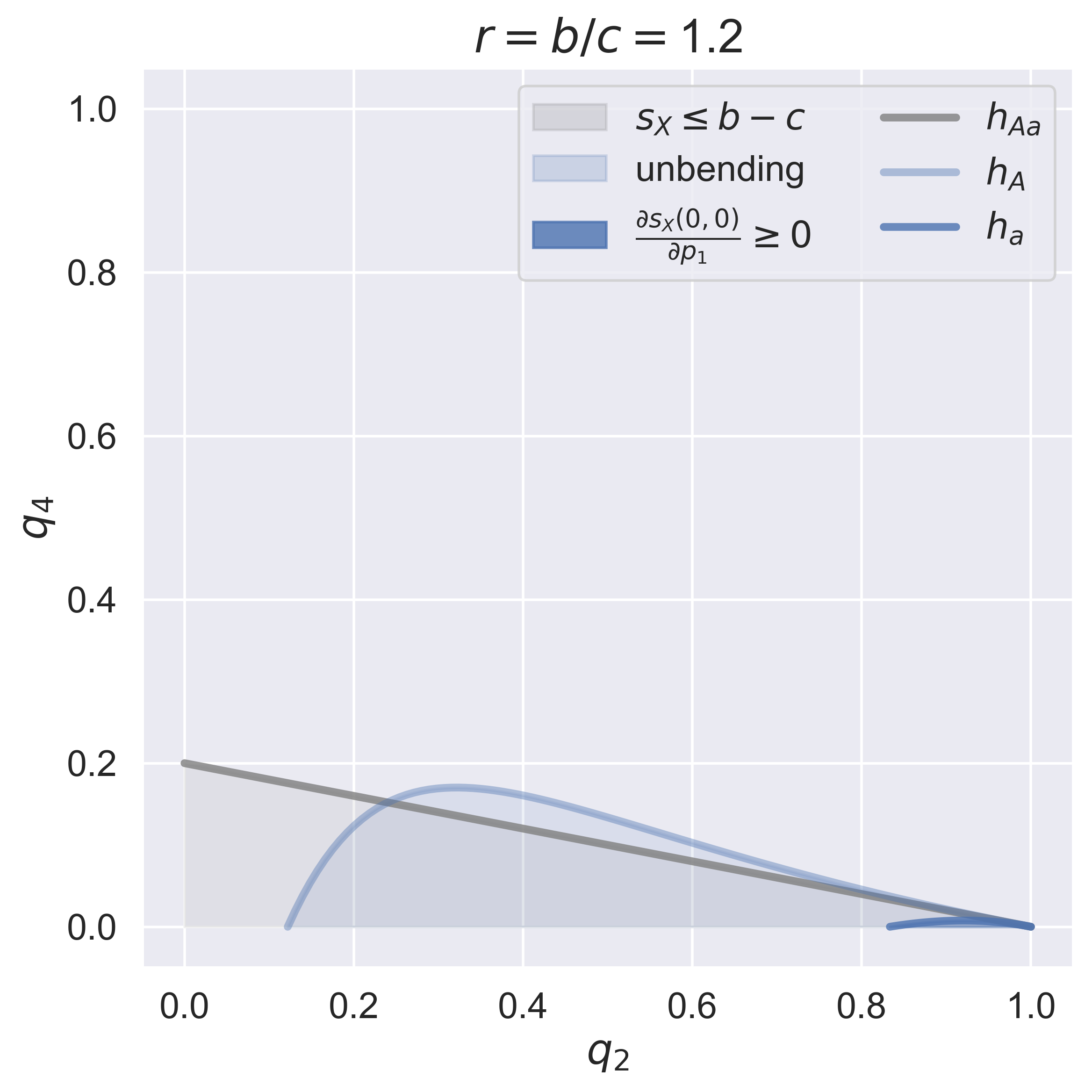}} 
     \subfloat[$r = (1 + \sqrt{5})/2$]{\includegraphics[width=0.5\linewidth]{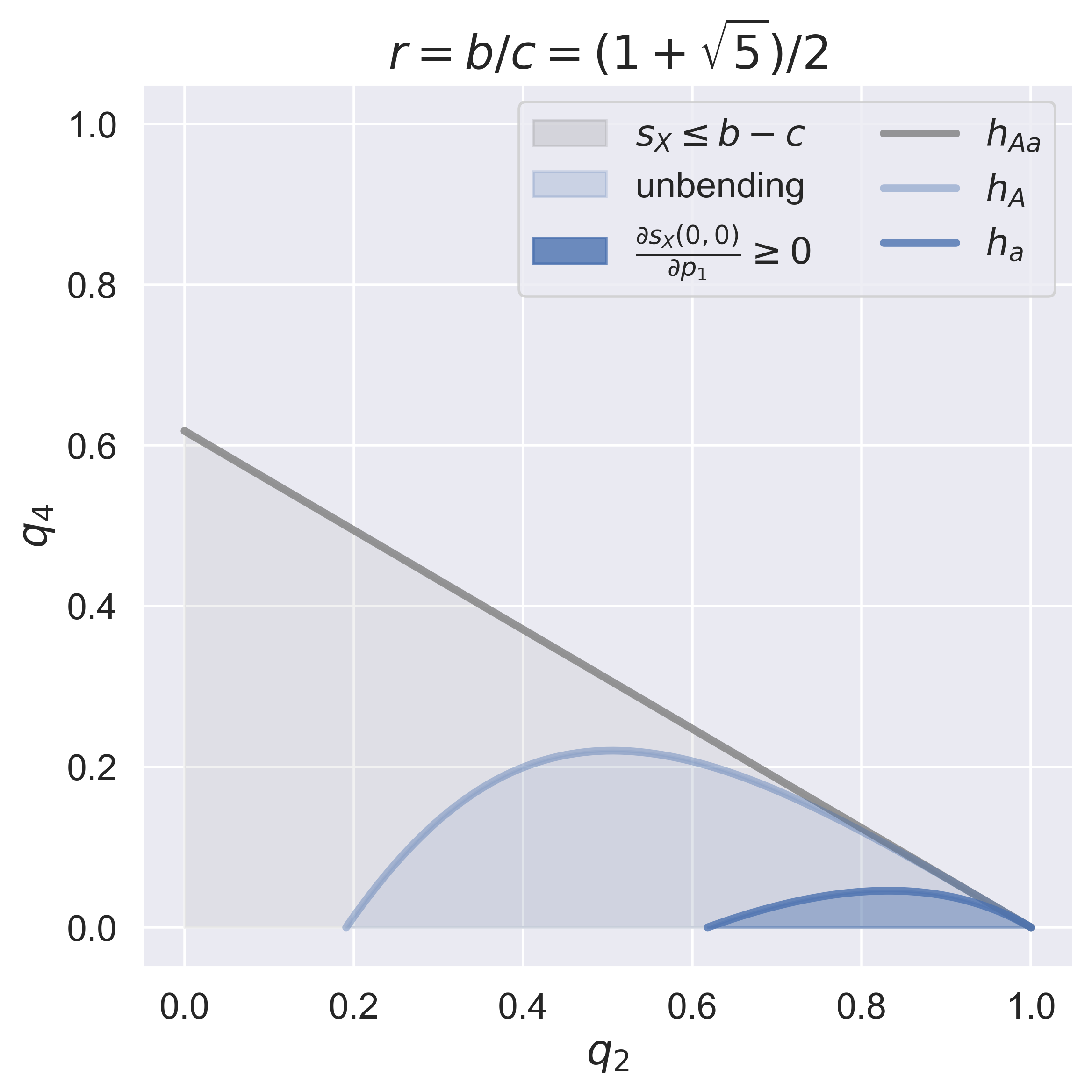}}
     \\
     \subfloat[$r = 4$]{\includegraphics[width=0.5\linewidth]{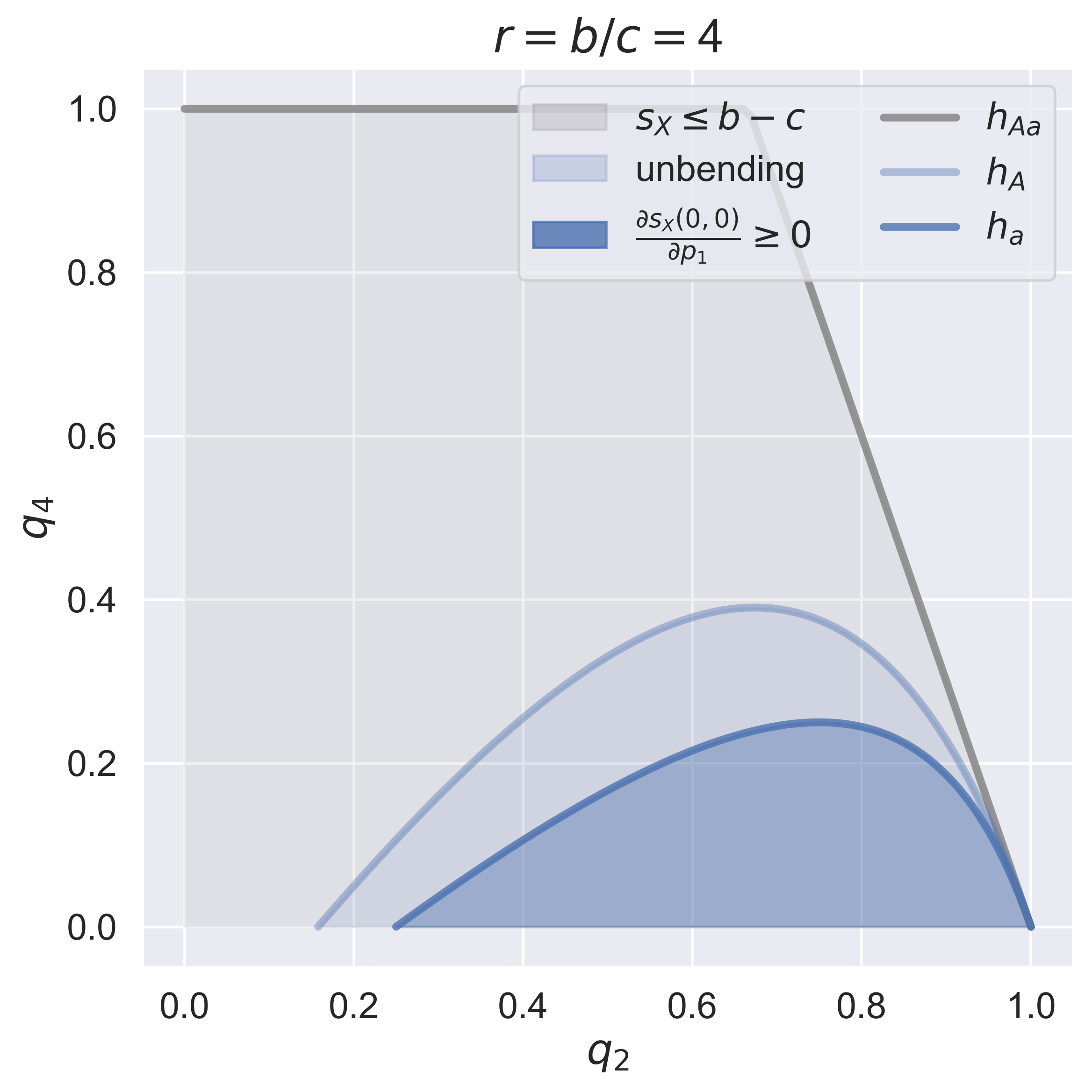}}
     \caption{Regions of $0 < h_4 \leq h_{Aa}$, $0 < h_4 \leq h_A$, and $0 < h_4 \leq h_a$ for the donation game with increasing benefit-cost ratios.}
     \label{target_A_re}
\end{figure}

To conclude, we emphasize how the benefit-cost ratio $r$ would shape the aforementioned regions of
\begin{enumerate}[label=(\roman*)] 
\item $s_X \leq b - c$, that is, $\max s_X(p_1, p_2) = s_X(1, p_2) = b - c$, 
\item unbending strategies, and 
\item $\partial s_X(0, 0)/\partial p_1 \geq 0$, or equivalently, $\partial s_X(p_1, 0)/\partial p_1 \geq 0$.
\end{enumerate} 
Mathematically, the three regions correspond to
\begin{enumerate}[label=(\roman*)] 
\item 
\begin{equation}
q_4 \leq h_{Aa}(q_2) = \frac{(b - c)(1 - q_2)}{c},
\end{equation}
\item 
\begin{equation}
q_4 \leq h_A(q_2) = \frac{(b - c)(1 - q_2)[(b^2 + bc - c^2)q_2 - (b - c)c]}{bc^2q_2^2 - (b - c)(b^2 - bc - c^2)q_2 + (b - c)^2(b + c)},
\end{equation}
and
\item
\begin{equation}
q_4 \leq h_{a}(q_2) = \frac{(bq_2 - c)(1 - q_2)}{b(1 - q_2) + c}.
\end{equation}
\end{enumerate}
It is worth pointing out that $h_{a} \leq h_{Aa}$ and $h_{a} \leq h_{A}$ always hold. Also, 
\begin{equation}
\left.\frac{d h_{A}}{d q_2}\right|_{q_2 = 1} = \left.\frac{d h_{Aa}}{d q_2}\right|_{q_2 = 1} = \left.\frac{d h_{a}}{d q_2}\right|_{q_2 = 1} = - \frac{b - c}{c} = -(r - 1).
\end{equation}
The visualization is given in Figure~\ref{target_A_re}.

Based on the conclusion, we show a few examples of the payoff $s_X$ and its partial derivatives with respect to $p_1$ and $p_2$ in Table~\ref{fig:example_A_re}. \par

\begin{table}[H]
\centering
\tabulinesep=1.5mm
\footnotesize
\begin{tabu}{| c | c | c | c |}
\hline
& & $\partial s_X(0, 0)/\partial p_1 < 0$ & $\partial s_X(0, 0)/\partial p_1 > 0$ \\
\hline
\multirow{2}{*}{$r < r^{\ast}$} & \makecell[cc]{$\max s_X =$ \\ $s_X(0, 0)$} & \raisebox{-.5\height}{\includegraphics[width=4.5cm]{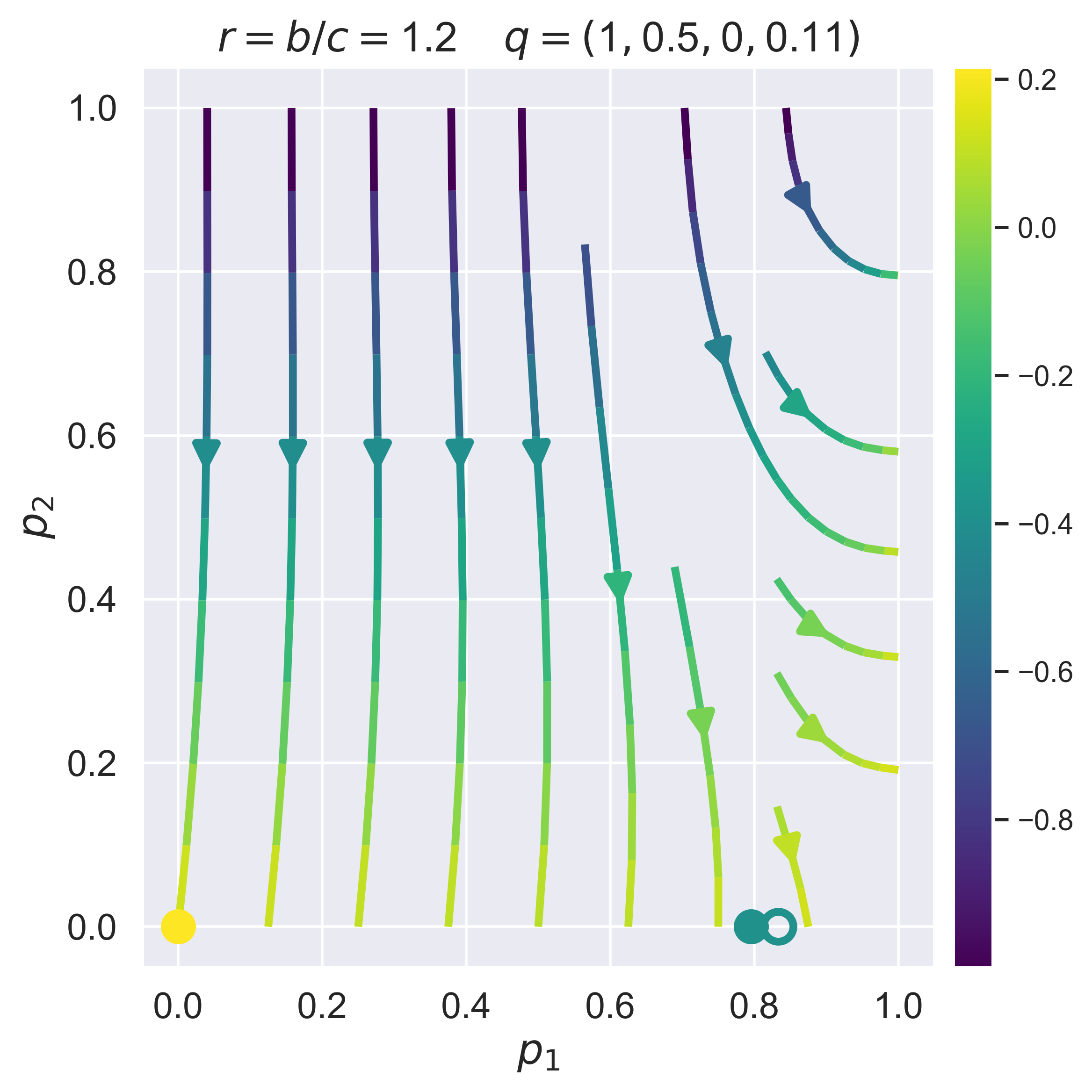}} &  \\
& \makecell[cc]{$\max s_X =$ \\ $s_X(1, p_1)$} & \raisebox{-.5\height}{\includegraphics[width=4.5cm]{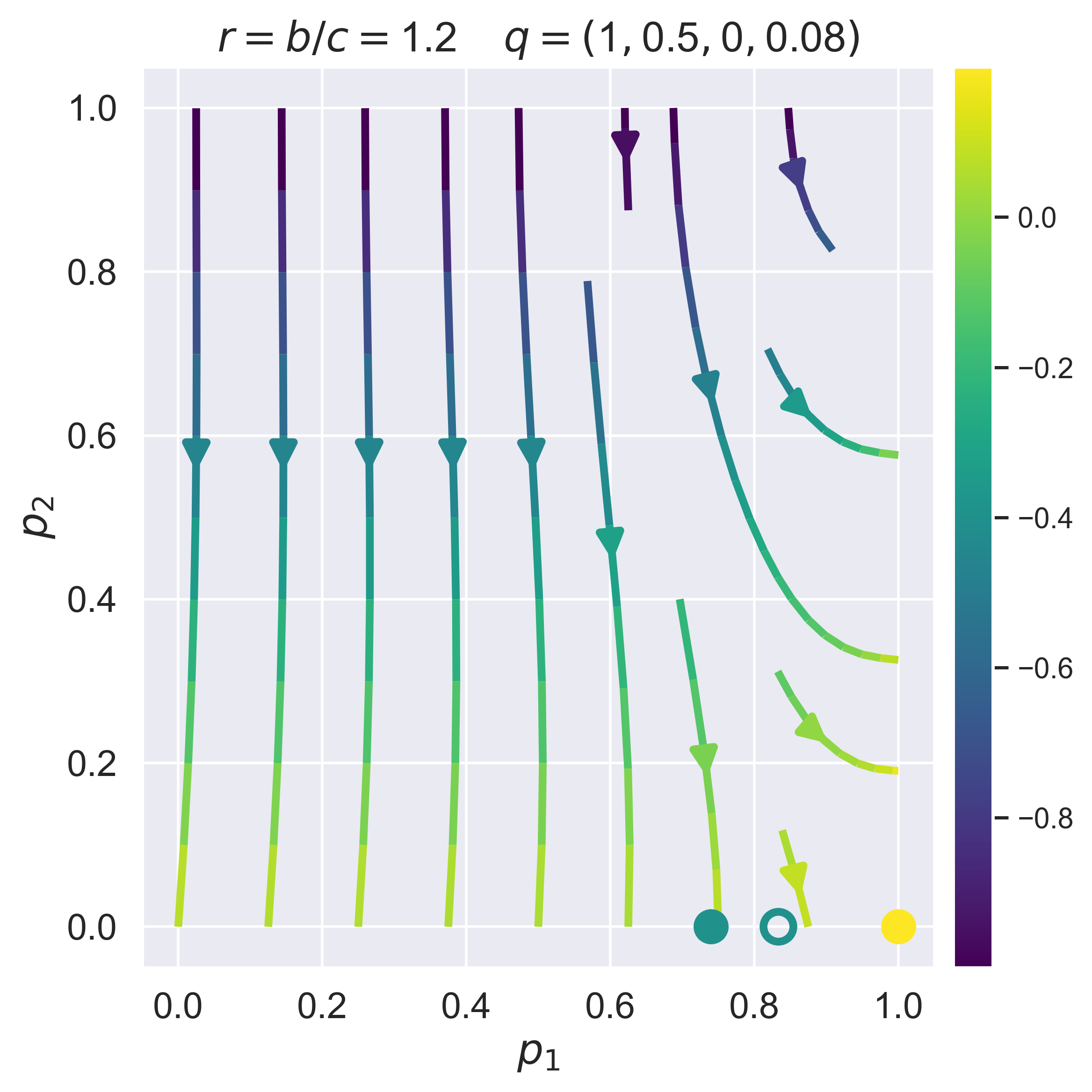}} & 
\raisebox{-.5\height}{\includegraphics[width=4.5cm]{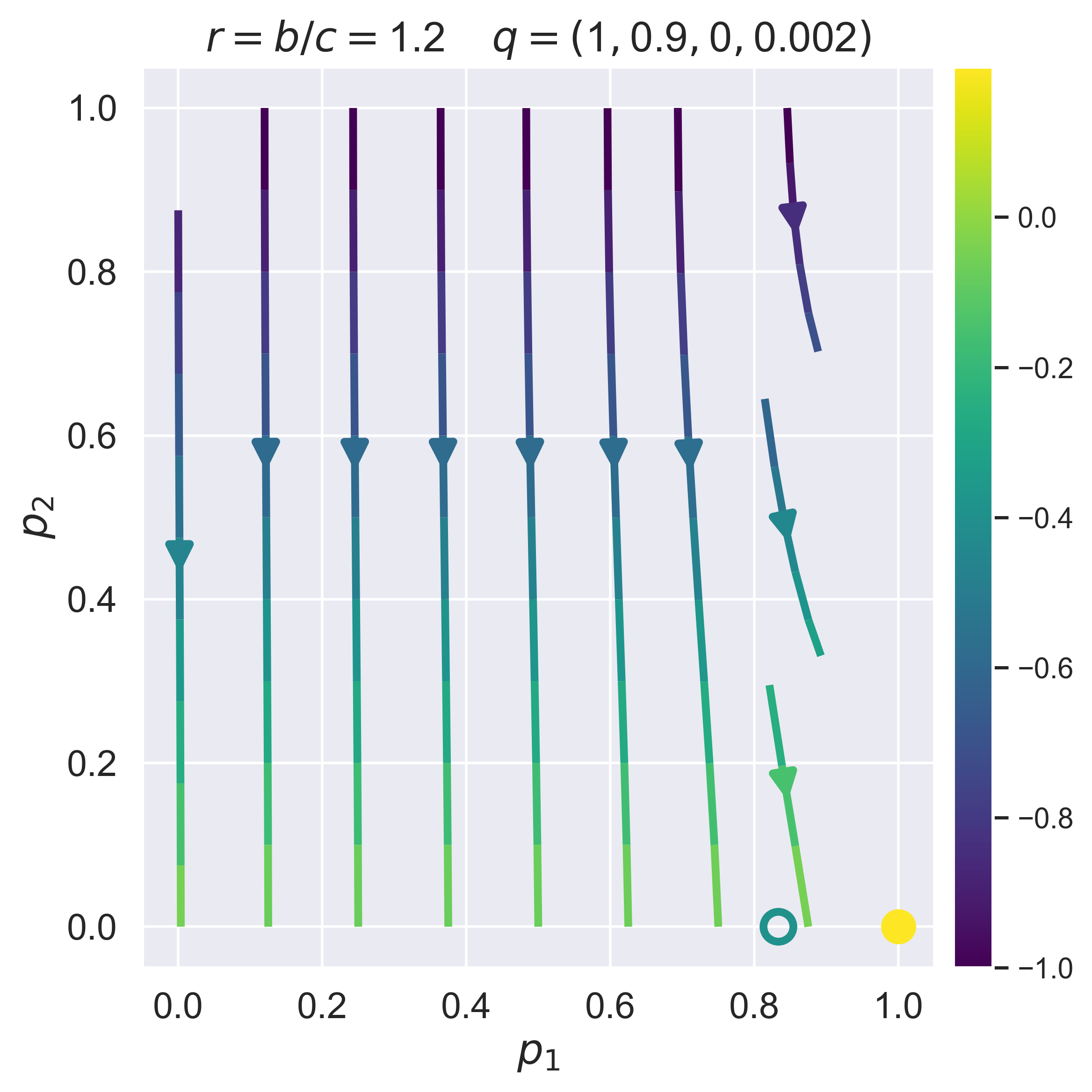}}  \\
\hline
\multicolumn{2}{|c|}{\makecell[cc]{$r > r^{\ast}$ \\ $\max s_X = s_X(1, p_1)$}} & \raisebox{-.5\height}{\includegraphics[width=4.5cm]{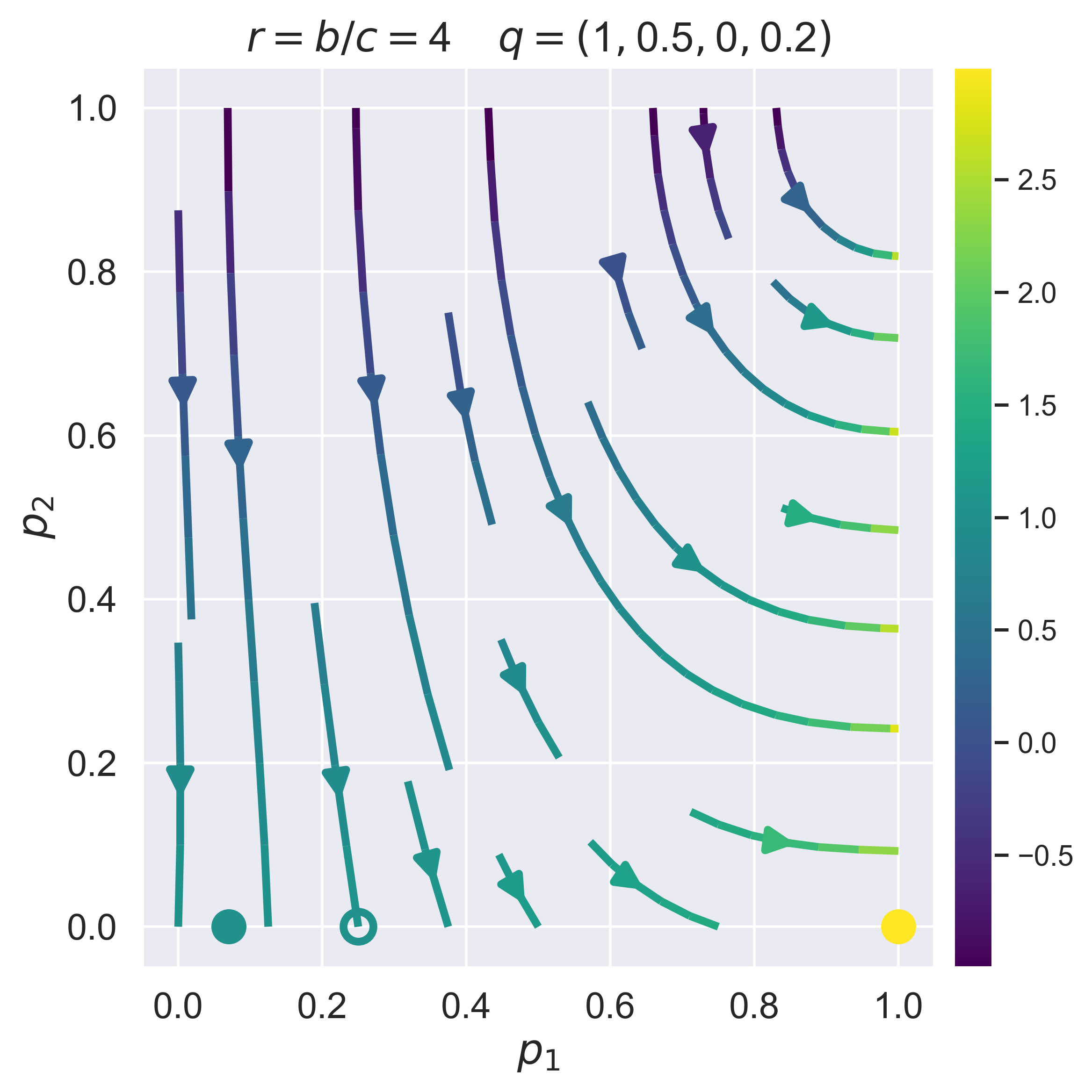}} & 
\raisebox{-.5\height}{\includegraphics[width=4.5cm]{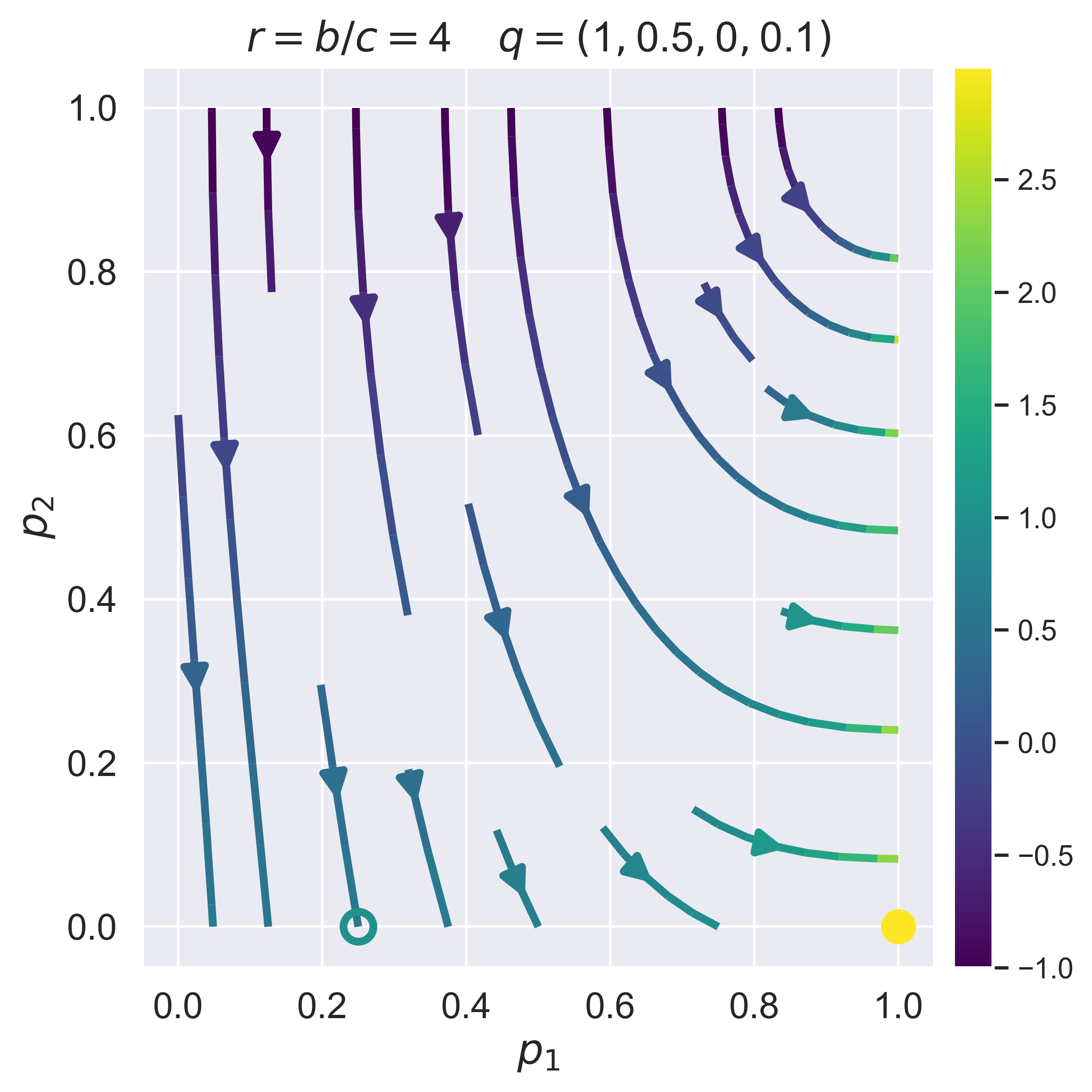}} \\
\hline
\end{tabu}
\caption{Stream plot of $s_X$. Here, $r^{\ast} = (1 + \sqrt{5})/2$ is the golden ratio. The 2-dimensional gradient vector field of $s_X$ is given with respect to $p_1$ and $p_2$. The color at a point $(p_1, p_2)$ reflects the value of $s_X$ at that point (with reference to the color bar). The yellow point denotes where the maximum value of $s_X$ lies (if there is more than one point, only the bottom one with $p_2 = 0$ is shown). The solid green point indicates where $\partial s_X(p_1, 0)/\partial p_1$ becomes zero and the empty green point is $(c/b, 0)$ (the left boundary of extortionate ZD strategies).}
\label{fig:example_A_re}
\end{table}

\section{Under influence of class D of unbending strategies (continued)}

Let $\bm{p} = (p_1, p_2, p_1, p_2)$ (a reactive strategy) and $\bm{q} = (q_1, q_2, q_3,  q_4)$ (an unbending strategy from Class D), where $q_i$'s satisfy
\begin{equation}
q_4 = h_D(q_1, q_2, q_3) = q_2 + q_3 - q_1,
\end{equation}
and 
\begin{equation}
b(q_2 - q_1) + c(q_3 - q_1) + c = d_{D0} < 0 < a_{D0}  = (b - c)(q_2 + q_3 - q_1).
\end{equation}
Notice that $q_3 = 0$ is impossible. \par

We obtain $s_X(p_1, p_2)$ as a linear rational function of $p_1$ and $p_2$.  In particular, when $p_1 = p_2 = 1$, we have
\begin{equation}
s_X(1, 1) = \frac{(b - c)q_3 - c(1 - q_1)}{1 - q_1 + q_3}.
\end{equation}
On the other hand, when $p_1 = q_1 = 1$, $s_X = R = b - c$. Moreover, we get
\begin{equation}
\begin{aligned}
s_X(p_1, p_2) - s_X(1, 1) &= \frac{[b(q_2 - q_1) + c(q_3 - q_1) + c][q_3(1 - p_1) + (1 - q_1)(1 - p_2)]}{(1 - q_1 + q_3)[1 - q_1 + q_3 + (q_1 - q_2)(p_2 - p_1)]} \\
&= \frac{d_{D0}[q_3(1 - p_1) + (1 - q_1)(1 - p_2)]}{(1 - q_1 + q_3)[1 - q_1 + q_3 + (q_1 - q_2)(p_2 - p_1)]}.
\end{aligned}
\end{equation}
\par

\subsection{Maximum value of \texorpdfstring{$s_X$}{e}}

We claim that the maximum value of $s_X$ is $s_X(1, 1)$ if $q_1 \neq 1$ ($\bm{q}$ is not a generous ZD strategy) and $s_X(1, p_2)$ if $q_1 = 1$ ($\bm{q}$ is a generous ZD strategy). \par

Given that $a_{D0} > 0$, $q_1 < q_2 + q_3$, we have
\begin{equation}
1 - q_1 + q_3 + (q_1 - q_2)(p_2 - p_1) > 1 - q_2 + (q_1 - q_2)(p_2 - p_1) \geq 0.
\end{equation}
Therefore, the denominator of $s_X(p_1, p_2) - s_X(1, 1)$ is always positive. Meanwhile, we know that $d_{D0} < 0$. Besides, 
\begin{equation}
q_3(1 - p_1) + (1 - q_1)(1 - p_2) \geq 0,
\end{equation}
where the equality holds if $p_1 = p_2 = 1$ or $p_1 = q_1 = 1$. That is, $s_X(p_1, p_2) - s_X(1, 1) \leq 0$, where the equality holds if $p_1 = p_2 = 1$ or $p_1 = q_1 = 1$. \par

\subsection{Monotonicity of \texorpdfstring{$s_X$}{e}}

Treat $s_X(p_1, p_2)$ as a linear rational function of $p_1$ and $p_2$. We study the monotonicity of $s_X$ with respect to the two variables. 

\subsubsection{Partial derivative with respect to \texorpdfstring{$p_1$}{e}}

We first take the partial derivative with respect to $p_1$:
\begin{equation}
\begin{aligned}
\frac{\partial s_X(p_1, p_2)}{\partial p_1} &= -\frac{[b(q_2 - q_1) + c(q_3 - q_1) + c][(q_1 - q_2)p_2 + q_2 + q_3 - q_1]}{[1 - q_1 + q_3 + (q_1 - q_2)(p_2 - p_1)]^2} \\
&= -\frac{d_{D0}[(q_1 - q_2)p_2 + q_2 + q_3 - q_1]}{[1 - q_1 + q_3 + (q_1 - q_2)(p_2 - p_1)]^2} > 0. \\
\end{aligned}
\end{equation}
Therefore, $s_X$ always increases with respect to $p_1$ for $0 \leq p_1 \leq 1$. 

\subsubsection{Partial derivative with respect to \texorpdfstring{$p_2$}{e}}

We then take the partial derivative with respect to $p_2$:
\begin{equation}
\begin{aligned}
\frac{\partial s_X(p_1, p_2)}{\partial p_2} &= -\frac{[b(q_2 - q_1) + c(q_3 - q_1) + c][(q_2 - q_1)p_1 + 1 - q_2]}{[1 - q_1 + q_3 + (q_1 - q_2)(p_2 - p_1)]^2} \\
&= -\frac{d_{D0}[(q_2 - q_1)p_1 + 1 - q_2]}{[1 - q_1 + q_3 + (q_1 - q_2)(p_2 - p_1)]^2} \geq 0. \\
\end{aligned}
\end{equation}
The equality holds when $p_1 = q_1 = 1$. Therefore, if $p_1 \neq 1$ or $q_1 \neq 1$, $s_X$ always increases with respect to $p_2$; otherwise, $s_X = b - c$, being a constant. 

\subsection{Conclusion and example}

To conclude, we show how the type of general ZD strategies would determine the maximum value and the monotonicity of $s_X$.
\begin{enumerate}[label=(\roman*)]
\item If $q_1 \neq 1$, that is, if $\bm{q}$ is not a generous ZD strategy, 
\begin{equation}
\max s_X(p_1, p_2) = s_X(1, 1) = \frac{(b - c)q_3 - c(1 - q_1)}{1 - q_1 + q_3},
\end{equation}
and 
\begin{equation}
\frac{\partial s_X(p_1, p_2)}{\partial p_1}  > 0, \frac{\partial s_X(p_1, p_2)}{\partial p_2} > 0. 
\end{equation}
\item If $q_1 = 1$, that is, if $\bm{q}$ is a generous ZD strategy, 
\begin{equation}
\max s_X(p_1, p_2) = s_X(1, p_2) = b - c,
\end{equation}
and 
\begin{equation}
\frac{\partial s_X(p_1, p_2)}{\partial p_1}  > 0, \frac{\partial s_X(p_1, p_2)}{\partial p_2}\begin{cases} > 0, & p_1 \neq 1 \\ = 0. & p_1 = 1\end{cases} 
\end{equation}
\end{enumerate}

We also give two examples of the payoff $s_X$ and its partial derivatives with respect to $p_1$ and $p_2$ in Table~\ref{example_D_re}. 

\begin{table}[H]
\centering
\tabulinesep=1.5mm
\footnotesize
\begin{tabu}{| c | c |}
\hline
\makecell[cc]{$\max s_X(p_1, p_2) = s_X(1, 1)$ \\ \\ $\left.\frac{\partial s_X(p_1, p_2)}{\partial p_2}\right |_{p_1 = 1} > 0$} & 
\makecell[cc]{$\max s_X(p_1, p_2) = s_X(1, p_2)$ \\ \\ $\left.\frac{\partial s_X(p_1, p_2)}{\partial p_2}\right |_{p_1 = 1} = 0$} \\
\hline
\includegraphics[width=5.5cm]{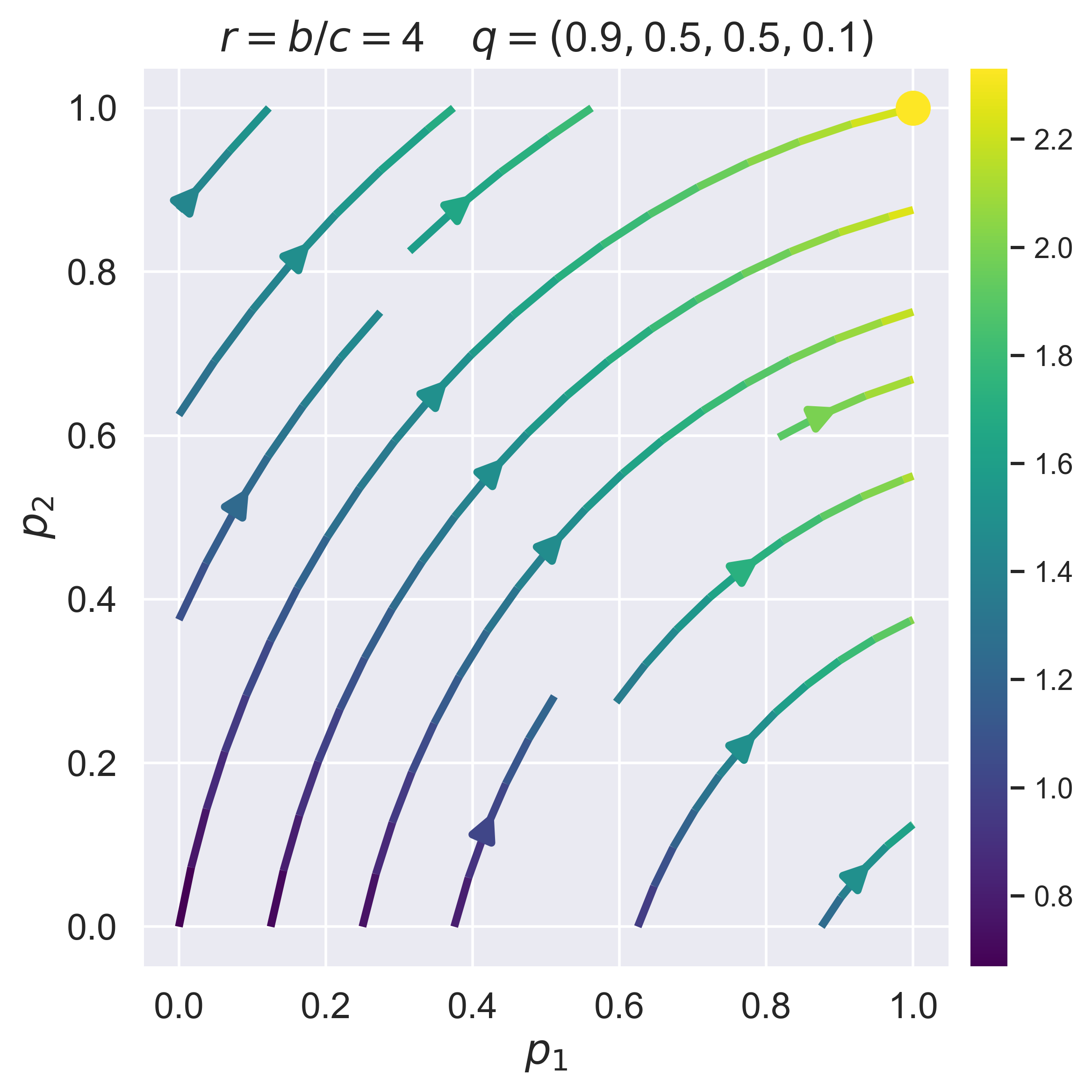} & \includegraphics[width=5.5cm]{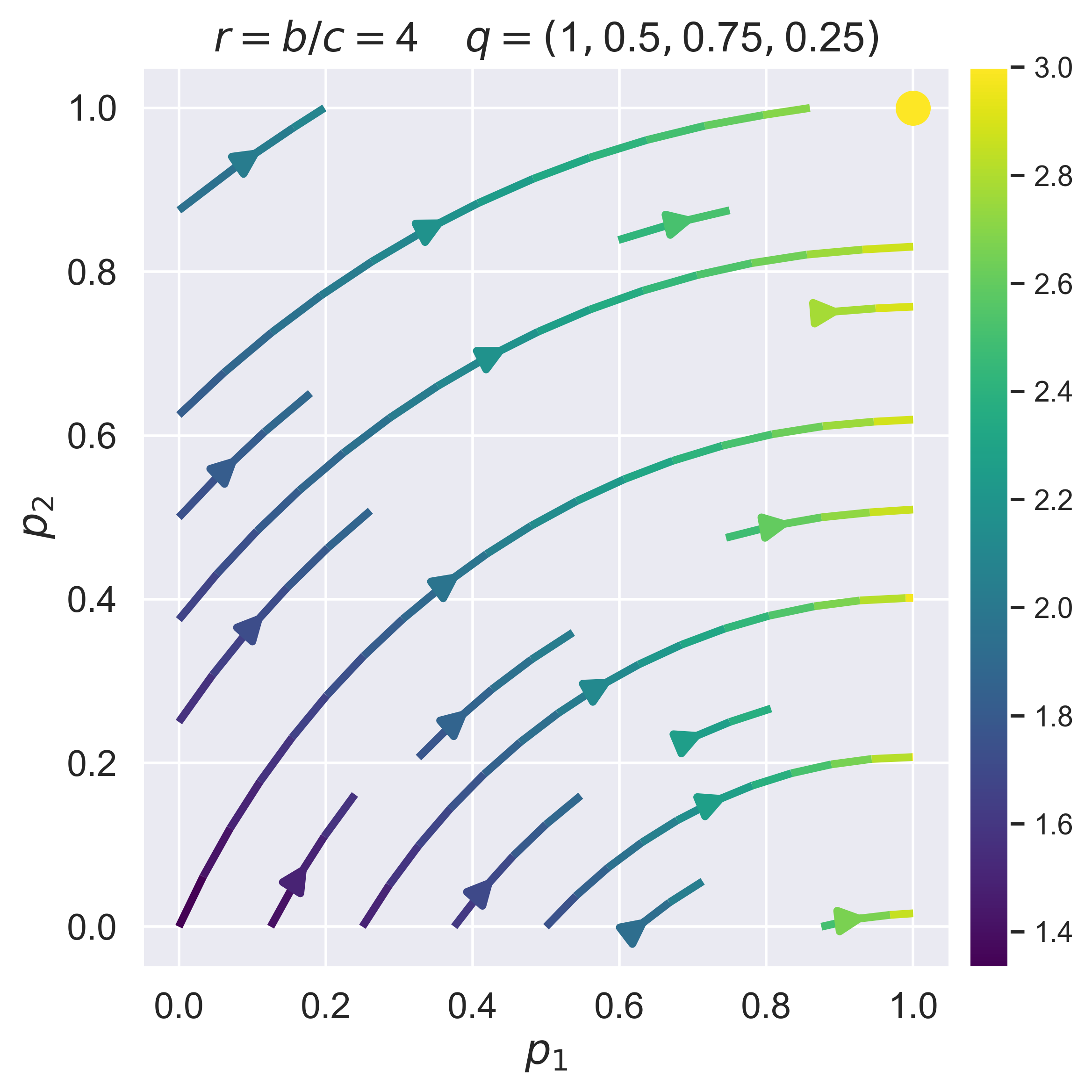} \\
\hline
\end{tabu}
\caption{Stream plot of $s_X$. The 2-dimensional gradient vector field of $s_X$ is given with respect to $p_1$ and $p_2$. The color at a point $(p_1, p_2)$ reflects the value of $s_X$ at that point (with reference to the color bar). The yellow point denotes where the maximum value of $s_X$ lies (if there is more than one point, only the top one with $p_2 = 1$ is shown).}
\label{example_D_re}
\end{table}

\section{Steering learning dynamics from extortion to fairness and cooperation: general ZD strategies}

Recall Figure~\ref{fig:reactive}. Reactive strategies is a subset of general ZD strategies in donation games. In particular, we can consider the set of reactive ZD strategies $\bm{p} = (p_1, p_2, p_1, p_2)$, where $p_1$ and $p_2$ can be written as functions of $O$ and $\chi$ (see Equation~\ref{eq:reactive}). When playing against a fixed unbending co-player, this parametrization in terms of  $O$ and $\chi$ actually includes all possible variations of general ZD strategies that can have a distinct effect on the average payoff $s_X$ since the parameter $\phi$ is already neutralized by unbending strategies in the first place.  \par

We have studied the most general case where player X uses a reactive strategy  (subset of ZD) and player Y uses an unbending strategy in the donation game. We now extend it to general ZD strategies parameterized by $(O, \chi)$ for player X as the following. \par

\subsection{Under influence of class A of unbending strategies (continued)}

Let $\bm{q}$ be an unbending strategy from Class A. We obtain $s_X(O, \chi)$ as a quadratic rational function of $O$ and $\chi$. Its monotonicity with respect to the two variables can be further discussed. Here, we summarize the corresponding analysis on the boundaries. \par

\begin{itemize}
%\item 
%\begin{equation}
%\mfootnotesize{
%\left.\frac{\partial s_X(O, \chi)}{\partial O}\right |_{O = b - c} = 
%\frac{(\chi - 1)(\ast)}{(c\chi + b)\{[b  - c(1 - q_2)]\chi  - [b(1 - q_2) - c]\}q_4}.
%}
%\end{equation}
%Here, $(\ast) = \{(b + c)(b - c)(1 - q_2) + c[b(1 - q_2) - c]q_4\}\chi + b[b(1 - q_2) - c]q_4$.
\item $\left.\frac{\partial s_X(O, \chi)}{\partial O}\right |_{\chi = 1} = 0$.
\item $\left.\frac{\partial s_X(O, \chi)}{\partial \chi}\right |_{O = b - c} = 0$.
\item 
\begin{equation}
\left.\frac{\partial s_X(O, \chi)}{\partial \chi}\right |_{\chi = 1} = \frac{(b - c - O)(\ast)}{(b + c)q_2q_4} < 0.
\end{equation}
Here, $(\ast) = -[(b - c)(1 - q_2 + q_4) - bq_2q_4] < 0$ for $0 < q_4 \leq h_A$. The proof is straightforward. In particular, its value at $q_4 = h_A$ is
\begin{equation}
-\frac{b(b - c)^2[(b + 2c)q_2 + b - c](1 - q_2)^2}{bc^2q_2^2 - (b - c)(b^2 - bc - c^2)q_2 + (b - c)^2(b + c)} > 0.
\end{equation}
\end{itemize}

Two examples of the payoff $s_X$ and its partial derivatives are given in Table~\ref{example_ZD_A_re}.

\begin{table}[H]
\centering
\tabulinesep=1.5mm
\footnotesize
\begin{tabu}{| c | c |}
\hline
\includegraphics[width=5.5cm]{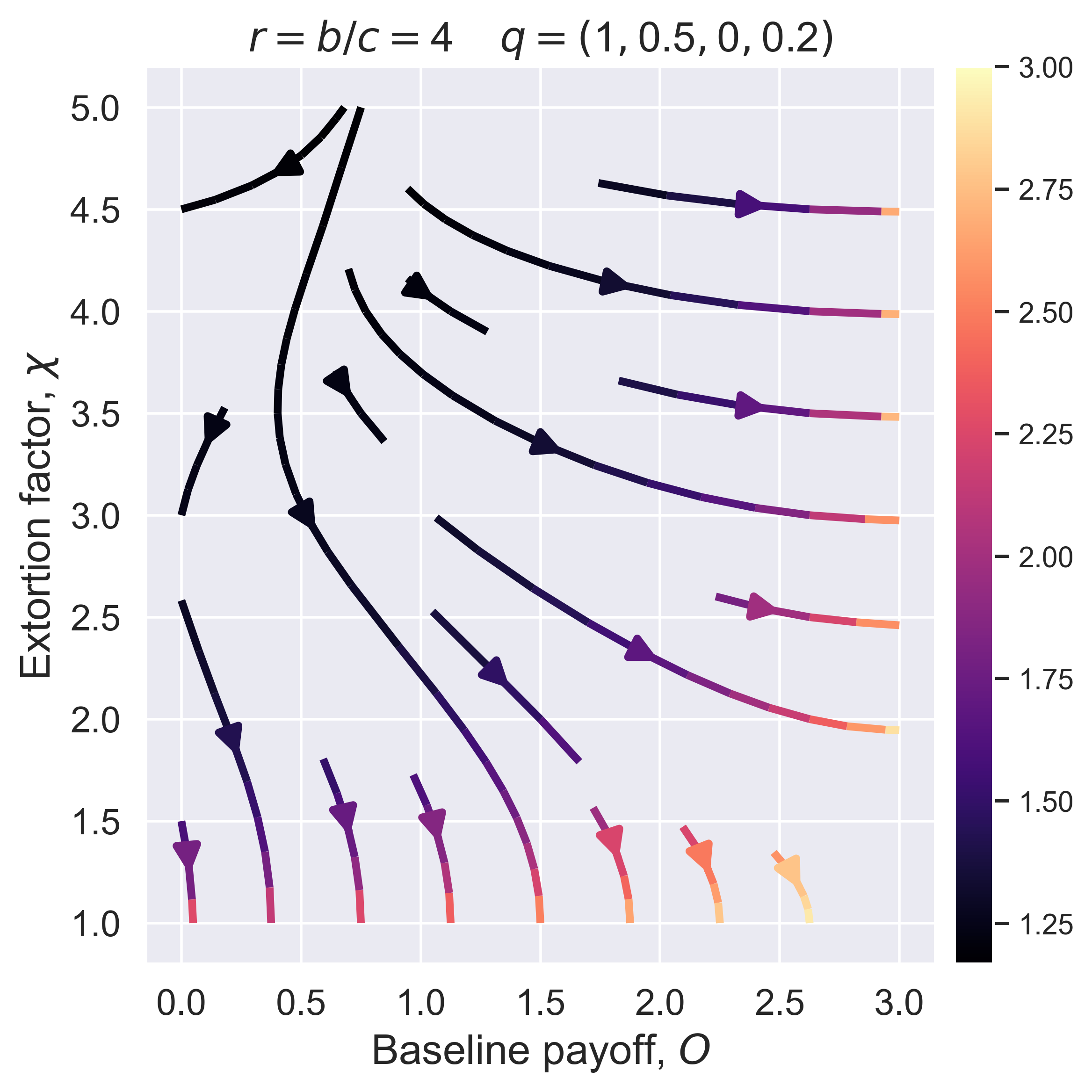} & \includegraphics[width=5.5cm]{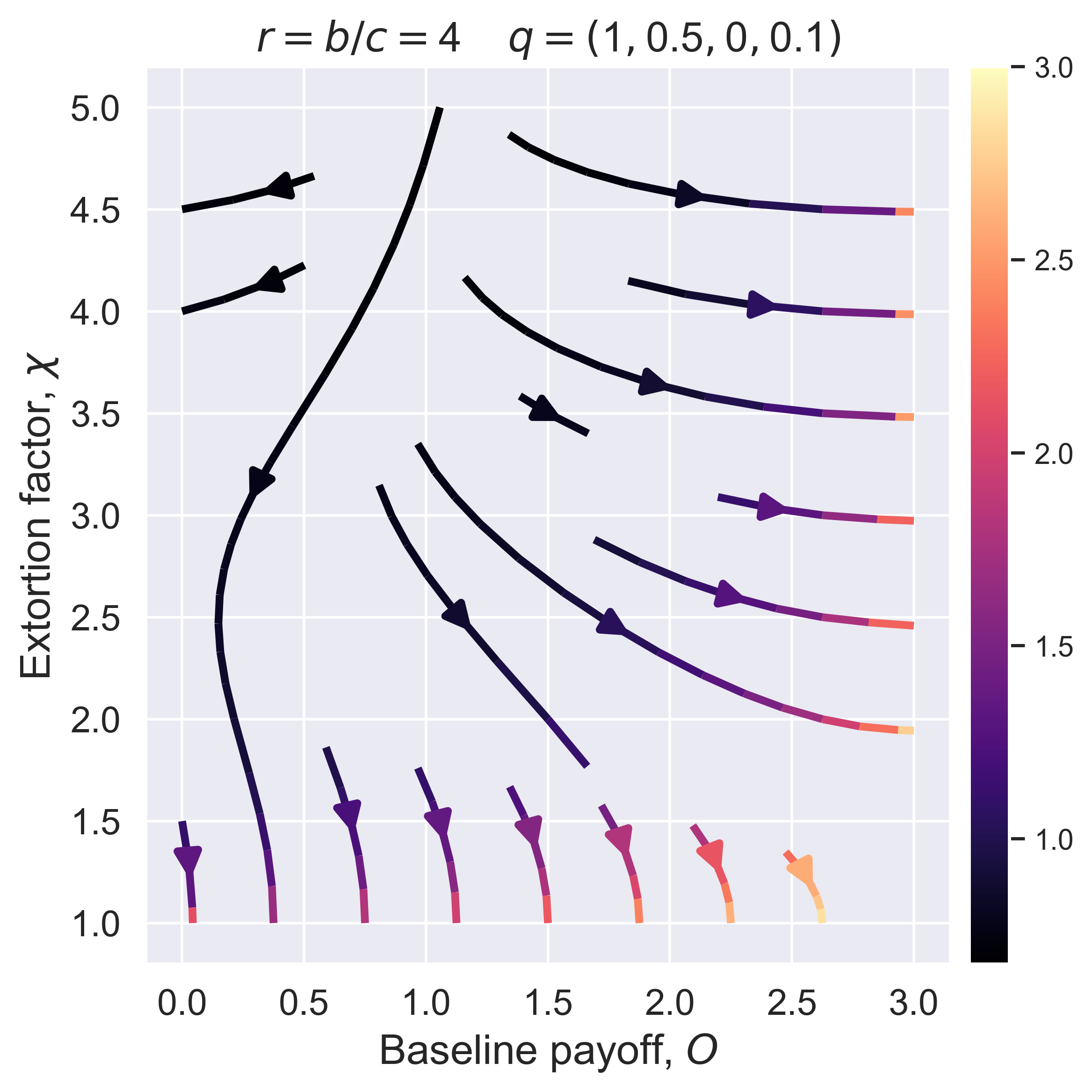} \\
\hline
\end{tabu}
\caption{Stream plot of $s_X$. The 2-dimensional gradient vector field of $s_X$ is given with respect to $O$ and $\chi$. The color at a point $(O, \chi)$ reflects the value of $s_X$ at that point (with reference to the color bar). }
\label{example_ZD_A_re}
\end{table}

\subsection{Under influence of class D of unbending strategies (continued)}

Let $\bm{q}$ be an unbending strategy from Class D. Still, $s_X(O, \chi)$ can be treated a quadratic rational function of $O$ and $\chi$. Its monotonicity with respect to the two variables can be further discussed. \par

\subsubsection{Partial derivative with respect to \texorpdfstring{$O$}{e}}

We first take the partial derivative with respect to the baseline payoff $O$. We get
\begin{equation}
\frac{\partial s_X}{\partial O} = -\frac{(\ast)}{[b(q_2 - q_1) + c(q_3 - q_1) + c](\chi - 1)} = -\frac{(\ast)}{d_{D0}(\chi - 1)}.
\end{equation}
Here, 
\begin{equation}
\begin{aligned}
(\ast) &= [c(q_2 - q_1) + b(q_3 - q_1) + b]\chi + b(q_2 - q_1) + c(q_3 - q_1) + c \\
&= [c(q_2 - q_1) + b(q_3 - q_1) + b](\chi - 1) + (b - c)(q_2 + q_3 - q_1 + 1 - q_1) \\
%&= \{(b + c)(q_2 + q_3 - q_1 + 1 - q_1) - [b(q_2 - q_1) + c(q_3 - q_1) + c]\}(\chi - 1) + (b - c)(q_2 + q_3 - q_1 + 1 - q_1) \\
&= [(b + c)(h_D + 1 - q_1) - d_{D0}](\chi - 1) + (b - c)(h_D + 1 - q_1) \\
& > 0.
\end{aligned} 
\end{equation}
Therefore, $\partial s_X/\partial O \geq 0$ and the equality holds if $\chi = 1$. \par

We then take the partial derivative with respect to the extortion factor $\chi$. We obtain
\begin{equation}
\frac{\partial s_X}{\partial \chi} = -\frac{(b + c)[b(q_2 - q_1) + c(q_3 - q_1) + c](\ast\ast)}{(\ast)^2} = -\frac{(b + c)d_{D0}(\ast\ast)}{(\ast)^2} .
\end{equation}
here, $(\ast)$ is the same as defined above and
\begin{equation}
(\ast\ast) = (q_2 + q_3 - q_1 + 1 - q_1)O - (b - c)(q_2 + q_3 - q_1).
\end{equation}
\par

Notice that Class D is actually the set of general ZD strategies with baseline payoff $O'$ satisfying $0 < O' \leq b - c$ and that
\begin{equation}
O' = \frac{(b - c)(q_2 + q_3 - q_1)}{q_2 + q_3 - q_1 + 1 - q_1}
\end{equation}
is the solution of $(\ast\ast)$. Therefore, we get
\begin{equation}
\frac{\partial s_X}{\partial \chi} 
\begin{cases}
< 0, & O < O' \\
= 0, & O = O' \\
> 0. & O > O'
\end{cases}
\end{equation}
For instance, we always have 
\begin{equation}
\left.\frac{\partial s_X(O, \chi)}{\partial \chi}\right |_{O = 0} < 0 \qquad \text{and} \qquad \left.\frac{\partial s_X(O, \chi)}{\partial \chi}\right |_{O = b - c} \geq 0.
\end{equation}
The equality holds if $O' = b - c$, or equivalently, $q_1 = 1$ (player Y uses a generous ZD strategy).
\par

Two examples of the payoff $s_X$ and its partial derivatives are given in Table~\ref{example_ZD_D_re}.

\begin{table}[H]
\centering
\tabulinesep=1.5mm
\footnotesize
\begin{tabu}{| c | c |}
\hline
\makecell[cc]{$\left.\frac{\partial s_X(O, \chi)}{\partial \chi}\right |_{O = b - c} > 0$} & 
\makecell[cc]{$\left.\frac{\partial s_X(O, \chi)}{\partial \chi}\right |_{O = b - c} = 0$} \\
\hline
\includegraphics[width=5.5cm]{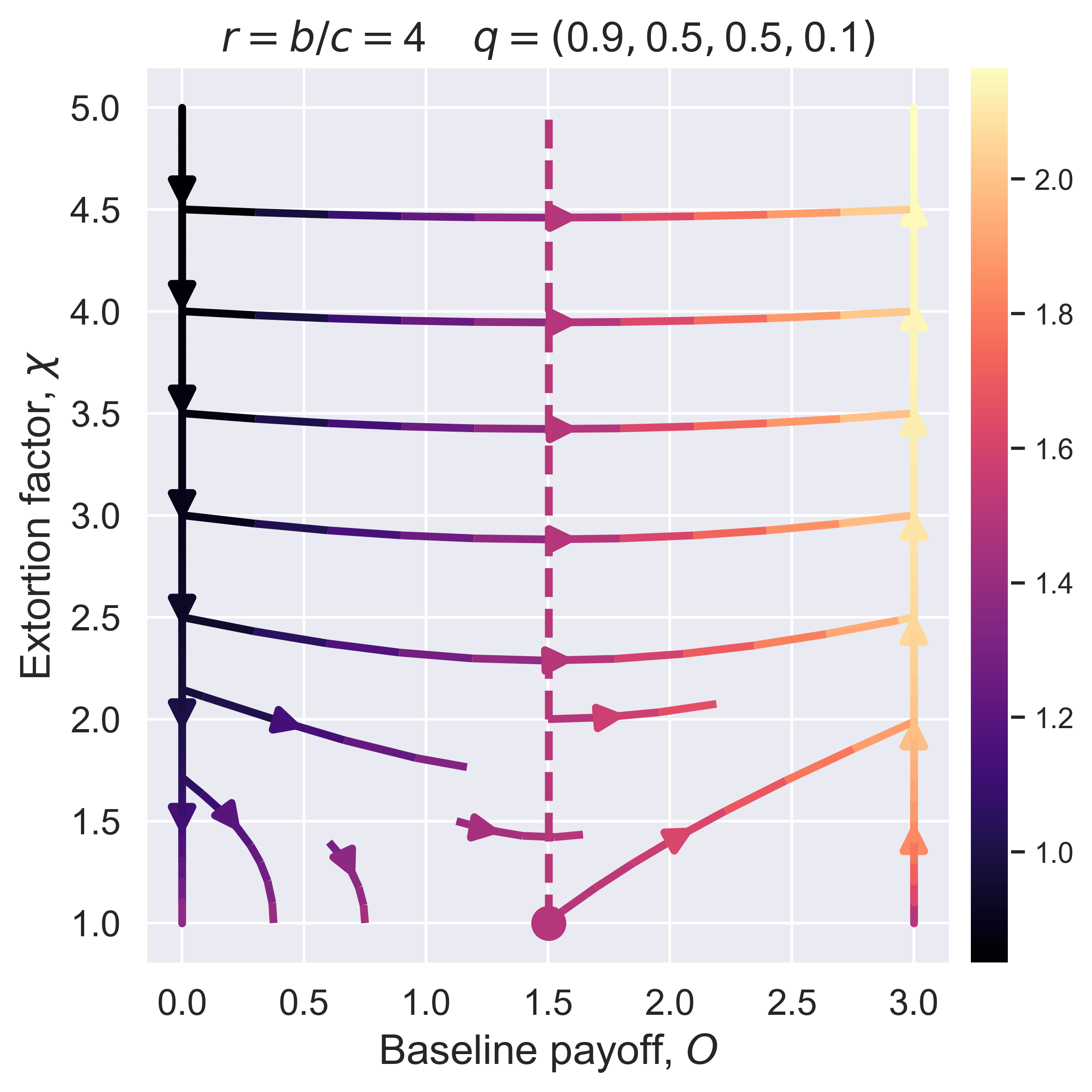} & \includegraphics[width=5.5cm]{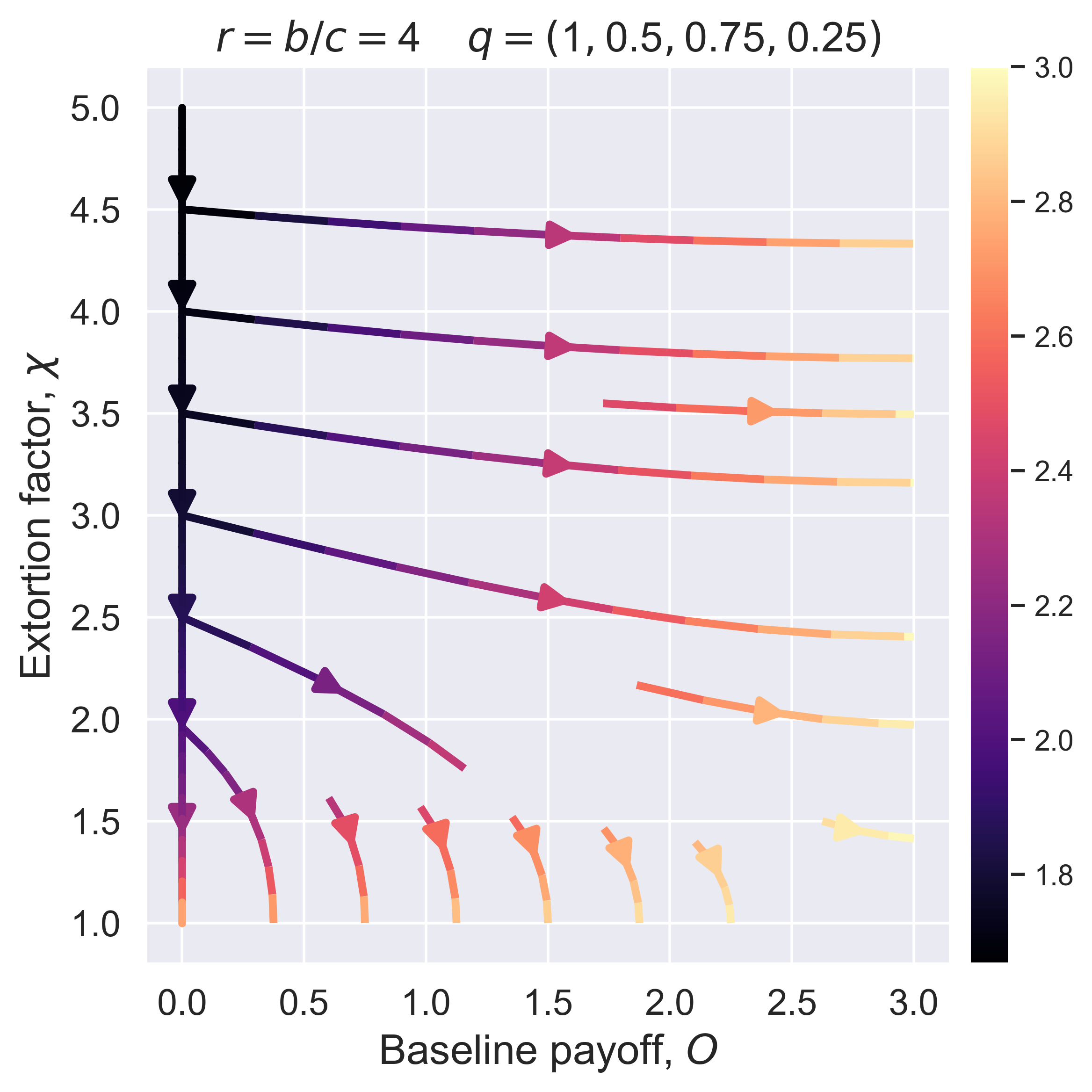} \\
\hline
\end{tabu}
\caption{Stream plot of $s_X$. The 2-dimensional gradient vector field of $s_X$ is given with respect to $O$ and $\chi$. The color at a point $(O, \chi)$ reflects the value of $s_X$ at that point (with reference to the color bar). The dashed line in the first panel indicates where $\partial s_X/\partial \chi$ turns from negative to positive, that is, where the two baseline payoffs $O$ and $O'$ are equal.}
\label{example_ZD_D_re}
\end{table}

\section{Extended search of unbending strategies against general ZD co-players}

Finally, we consider that player X uses a more general ZD strategy with the baseline payoff $O$ satisfies $P < O < R$. For a given $O$, the sets of admissible unbending strategies from the four classes discussed previously need to be updated based on the original definition of unbending properties. \par

As before, if the opponent Y tries a common strategy such as ALLC, the payoff of X will be $(T - S)(R - O)\chi/[(R - S)\chi + T - R] + O$, which does not involve $\phi$. However, when X plays against a more general opponent, its payoff is oftentimes dependent on $\phi$. Moreover, it can be shown that when $\chi$ is fixed, $s_X$ is a monotonic function of $\phi$ (the expression of $s_X$ takes the form of a linear rational function of $\phi$). \par

Similarly, a necessary condition $s_X(\bm{q}, \chi, \phi^{\text{upper}}) = s_X(\bm{q}, \chi, \phi^{\text{upper}}/2)$ yields the solutions of $\bm{q}$ for $s_X$ being independent of $\phi$ (see Table~\ref{phi_independency_gen}, which shares a few solutions with, but is not identical to Table~\ref{phi_independency}).

\begin{table}[H]
\centering
\tabulinesep=1.5mm
\begin{tabu}{c c}
\hline
\rowcolor[HTML]{EFEFEF} 
Solution & $s_X$ \\
$q_1 = q_2 = 1$ & $\frac{(T - S)(R - O)\chi}{(R - S)\chi + (T - R)} + O$ \\
\rowcolor[HTML]{EFEFEF} 
& \\
$q_2 = 1$ and $q_4 = 0$ & $\frac{(T - S)(O - P)(a_{O1}\chi + a_{O0})\chi}{f_O(\chi)} + O$ \\ \hline
$q_3 = q_4 = 0$ & $-\frac{(T - S)(O - P)\chi}{(T - P)\chi + (P - S)} + O$ \\
\rowcolor[HTML]{EFEFEF} 
& \\
$q_1 = 1$ and $q_3 = 0$ & $\frac{(T - S)(R - O)(a_{A1}\chi + a_{A0})\chi}{f_A(\chi)} + O$ \\
\rowcolor[HTML]{EFEFEF} 
& \\
$q_1 = q_2 = q_3 = q_4$ & $\frac{(T - S)[(T + S - R - P)q_1^2  - (T + S - 2P)q_1 + O - P]\chi}{f_C(\chi)} + O$ \\ \hline
\makecell[cc]{$q_1 = \frac{O - P - (T + S - O - P)q_2}{O - P - (T + S - R - P)q_2}$, \\ \\ $q_2 = q_3$, and \\ \\ $q_4 = \frac{(O - P)(1 - q_2)}{(T + S - O - P) - (T + S - R - P)q_2}$} & $O$ \\
\rowcolor[HTML]{EFEFEF} 
& \\
$q_4 = h_D$ & $\frac{(T - S)[-(T + S - 2O)q_1 + (R - O)(q_2 + q_3) + T + S - R - O]\chi}{f_D(\chi)} + O$ \\
\hline
\end{tabu}
\caption{Solutions of $\bm{q}$ for $s_X$ being independent of $\phi$ and the corresponding expressions of $s_X$. Here, $h_D$ is the same multivariate linear function of $q_1$, $q_2$ and $q_3$ as before. Additionally, $f_O$ and $f_A$ are quadratic functions of $\chi$, whereas $f_C$ and $f_D$ are linear functions of $\chi$.}
\label{phi_independency_gen}
\end{table}

It is straightforward to see that if $q_1 = q_2 = 1$, the derivative $ds_X/d\chi = (T - R)(T - S)(R - O)/[(R - S)\chi + (T - R)]^2$ will always be positive. On the other hand, if $q_3 = q_4 = 0$, the derivative $ds_X/d\chi = -(T - S)(P - S)(O - P)/[(T - P)\chi + (P - S)]^2$ will always be negative. \par

The two solutions $q_2 = 1$ and $q_4 = 0$ and $q_1 = q_2 = q_3 = q_4$ also allow unbending properties. \par

Assume that $q_2 = 1$ and $q_4 = 0$. If we further let $q_3 = 0$ (a subset of the strategies satisfying $q_3 = q_4 = 0$), $s_X$ will be a decreasing function of $\chi$. If we let $q_3 = 1$ instead,  whether $s_X$ can be a decreasing function of $\chi$ depends on the relation between $T + S$ and $2O$. To be more specific, if $T + S < 2O$ (that is, $\max\{T + S, 2P\} < 2O$), $ds_X/d\chi$ will be negative for $0 \leq q_1 < (2O - T - S)/(R + O - T - S)$. \par

On the other hand, assume that $q_1 = q_2 = q_3 = q_4$. If we further let $q_1 = 0$ ($\bm{q} = (0, 0, 0, 0)$), $s_X$ will be a decreasing function of $\chi$. More generally, $ds_X/d\chi$ will be negative for $0 \leq q_1 < (\ast)$, where $(\ast)$ is the root between $0$ and $1$ of the equation
\begin{equation}
(T + S - R - P)q_1^2 - (T + S - 2P)q_1 + O - P = 0.
\end{equation}
\par

\begin{figure}[H]
    \centering
     \subfloat[$T + S > 2P$]{\includegraphics[width=0.33\linewidth]{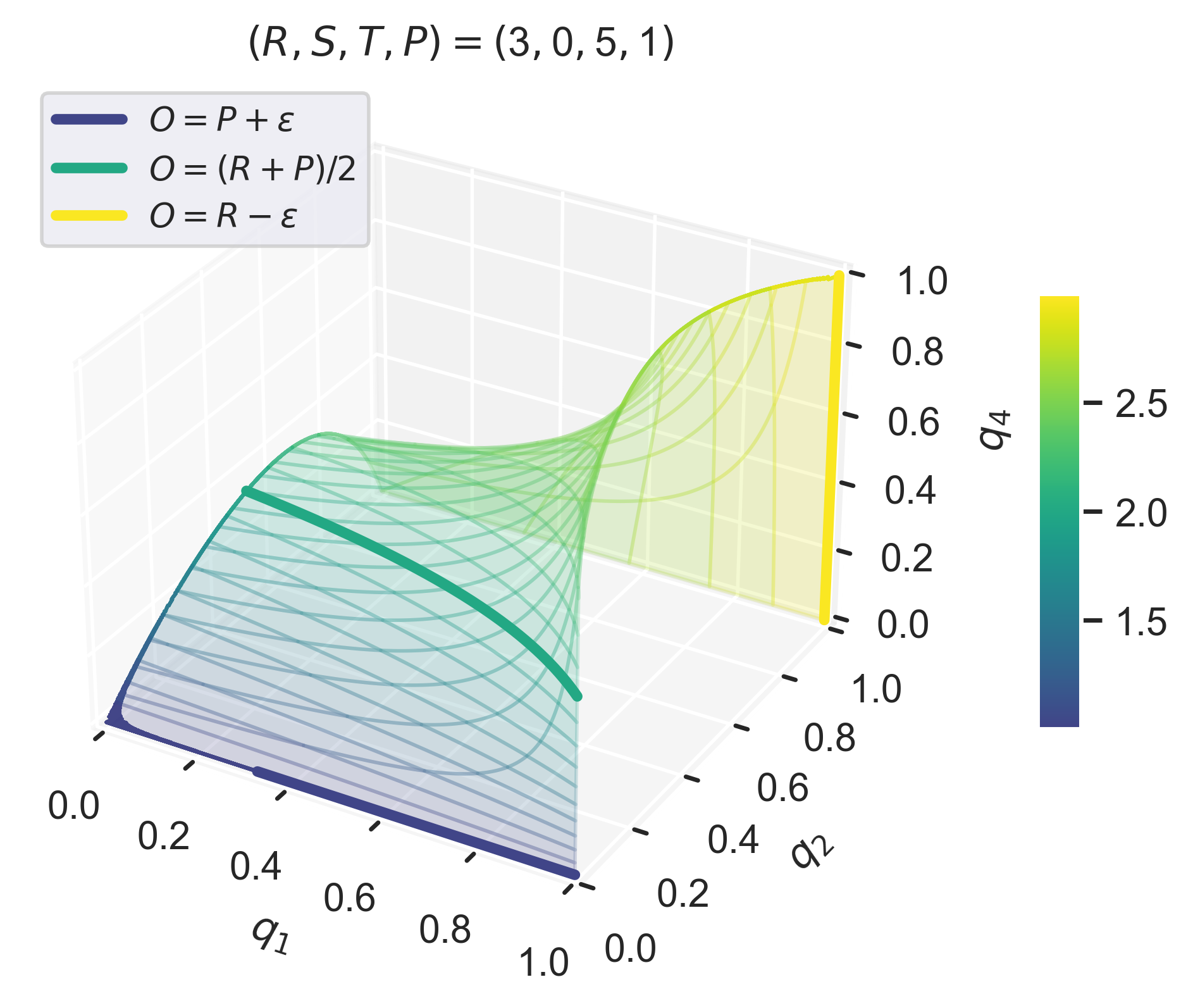}}
     \subfloat[$T + S = 2P$]{\includegraphics[width=0.33\linewidth]{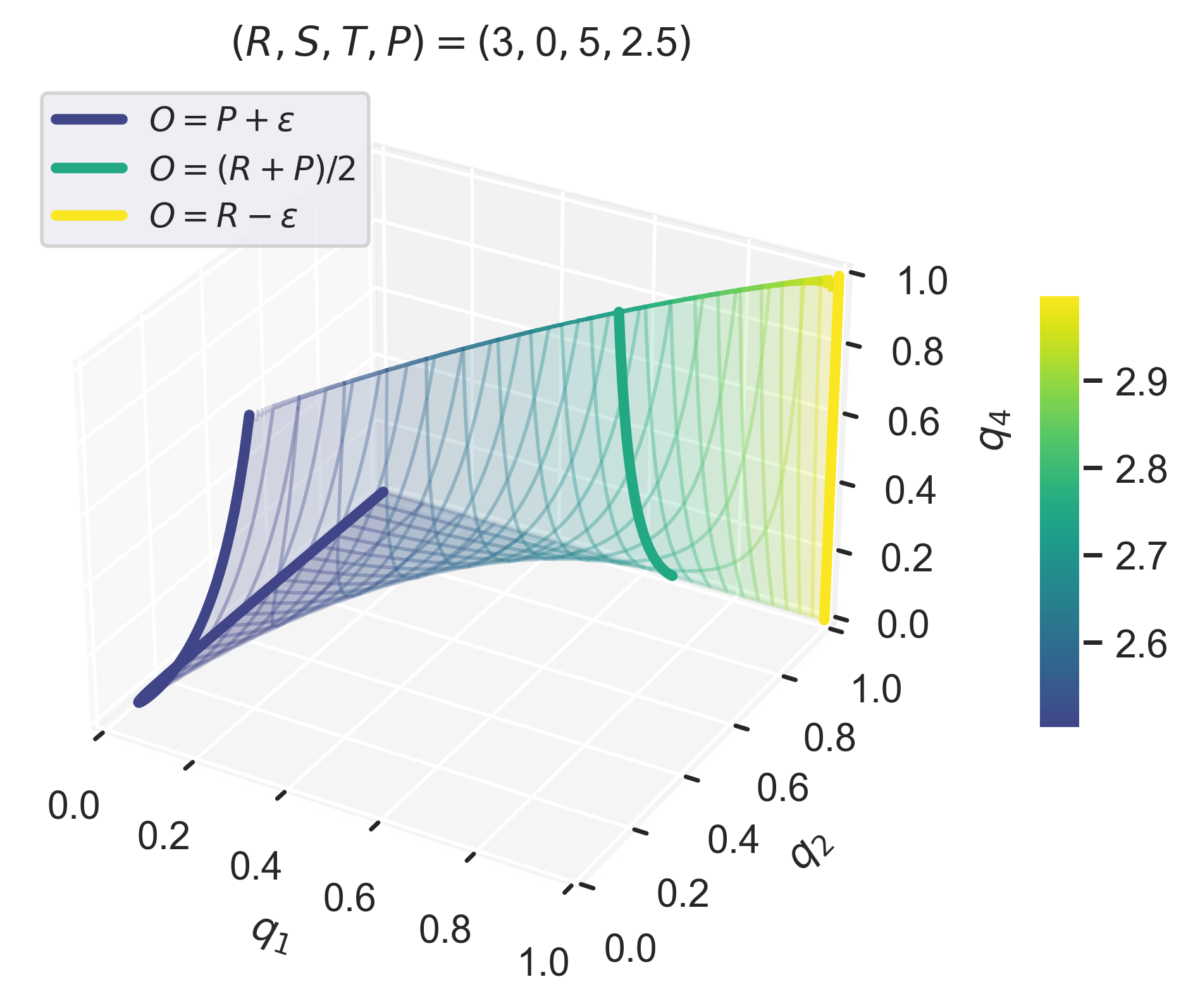}} 
     \subfloat[$T + S < 2P$]{\includegraphics[width=0.33\linewidth]{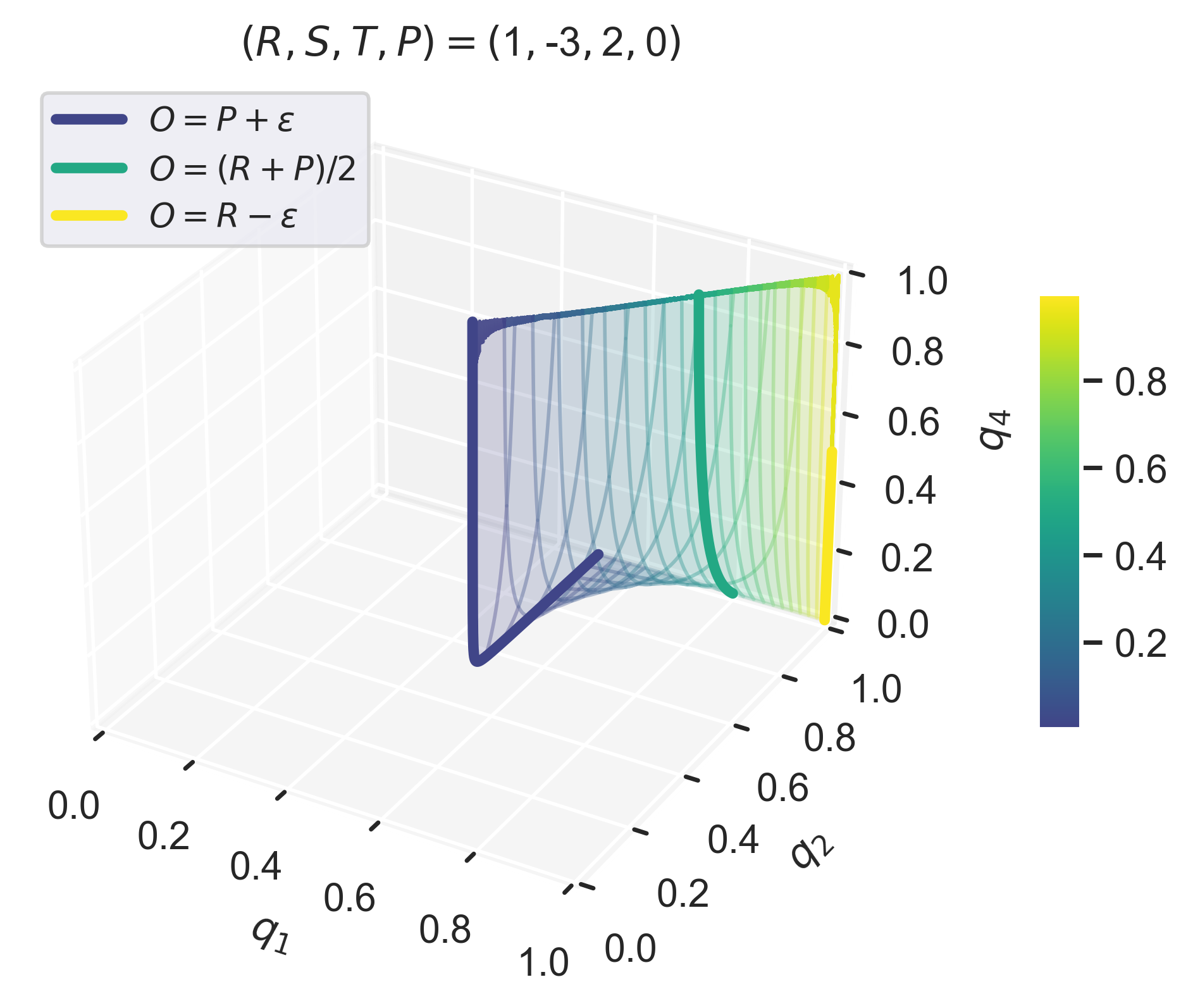}} 
     \caption{Set of points satisfying $q_1 = q_{1s}$ and $q_4 = q_{4s}$. The color map indicates the value of $O$. For a given $O$, $(q_1, q_2, q_4)$ forms a curve in the 3-dimensional space. The three curves corresponding to $O = P + \varepsilon$, $O = (R + P)/2$ and $O = R - \varepsilon$ are highlighted. Here, $\varepsilon = 0.001$. }
     \label{target_S}
\end{figure}

More interestingly, if $\bm{q} = (q_1, q_2, q_3, q_4)$ satisfies 
\begin{equation}
q_1 = q_{1s} = \frac{O - P - (T + S - O - P)q_2}{O - P - (T + S - R - P)q_2}, 
\end{equation}
\begin{equation}
q_2 = q_3,
\end{equation}
and
\begin{equation}
q_4 = q_{4s} = \frac{(O - P)(1 - q_2)}{(T + S - O - P) - (T + S - R - P)q_2},
\end{equation}
we have $s_X = s_Y = O$. That is, as long as player Y is aware of the baseline payoff $O$ used by player X, it can apply such a strategy $\bm{q}$ that the payoffs of both players will be fixed to $O$ no matter how large the extortion factor $\chi$ is. The set of $\bm{q}$ in different cases (depending on the relation between $T + S$ and $2P$) is given in Figure~\ref{target_S}. 

In particular, if $O = P$, the expressions of $\bm{q}$ in different cases can be further simplified.
\begin{enumerate}[label = (\roman*)]
\item $T + S > 2P$: there is no admissible $\bm{q}$.
\item $T + S = 2P$: $\bm{q} = (0, q_2, q_2, 0)$, where $0 \leq q_2 \leq 1$.
\item $T + S < 2P$: $\bm{q} = (\frac{2P - T - S}{R + P - T - S}, q_2, q_2, 0)$, where $0 \leq q_2 \leq 1$.
\end{enumerate}

And if $O = P + \varepsilon$ ($\varepsilon \to 0$), we get the following sets of $\bm{q}$ in different cases.
\begin{enumerate}[label = (\roman*)]
\item $T + S > 2P$: 
\begin{equation}
\mfootnotesize{
\bm{q} = (\frac{\varepsilon - (T + S - 2P - \varepsilon)q_2}{\varepsilon - (T + S - R - P)q_2}, q_2, q_3, \frac{\varepsilon(1 - q_2)}{(T + S - 2P - \varepsilon) - (T + S - R - P)q_2}),}
\end{equation} 
where 
\begin{equation}
\mfootnotesize{
0 \leq q_1 \leq 1, 0 \leq q_2 \leq \frac{\varepsilon}{T + S - 2P - \varepsilon},}
\end{equation}
and
\begin{equation}
\mfootnotesize{
\frac{(T + S - 2P - 2\varepsilon)\varepsilon}{(T + S - 2P - \varepsilon)^2 - (T + S - R - P)\varepsilon} \leq q_4 \leq \frac{\varepsilon}{T + S - 2P - \varepsilon}.}
\end{equation}
\\
For example, for the conventional IPD game where $(R, S, T, P) = (3, 0, 5, 1)$, we have
\begin{equation}
\mfootnotesize{
\bm{q} = (\frac{\varepsilon - (3 - \varepsilon)q_2}{\varepsilon - q_2}, q_2, q_2, \frac{\varepsilon(1 - q_2)}{(3 - \varepsilon) - q_2}),}
\end{equation}
where
\begin{equation}
\mfootnotesize{
0 \leq q_1 \leq 1, 0 \leq q_2 \leq \frac{\varepsilon}{3 - \varepsilon}, \,\text{and} \, \frac{(3 - 2\varepsilon)\varepsilon}{(3 - \varepsilon)^2 - \varepsilon} \leq q_4 \leq \frac{\varepsilon}{3 - \varepsilon}.}
\end{equation}

\item $T + S = 2P$:
\begin{equation}
\mfootnotesize{
\bm{q} = (\frac{\varepsilon(1 + q_2)}{\varepsilon + (R - P)q_2}, q_2, q_3, \frac{\varepsilon(1 - q_2)}{(R - P)q_2 - \varepsilon}),}
\end{equation}
where
\begin{equation}
\mfootnotesize{
\frac{2\varepsilon}{R - P + \varepsilon} \leq q_1 \leq \frac{R - P + 3\varepsilon}{3(R - P) + \varepsilon}, \frac{2\varepsilon}{R - P + \varepsilon} \leq q_2 \leq 1, \, \text{and} \, 0 \leq q_4 \leq 1.}
\end{equation}

\item $T + S < 2P$:
\begin{equation}
\mfootnotesize{
\bm{q} = (\frac{\varepsilon + (2P - T - S + \varepsilon)q_2}{\varepsilon + (R + P - T - S)q_2}, q_2, q_3, \frac{\varepsilon(1 - q_2)}{(R + P - T - S)q_2 - (2P - T - S + \varepsilon)}),}
\end{equation}
where
\begin{equation}
\mfootnotesize{
\frac{2P - T - S + 2\varepsilon}{R + P - T - S + \varepsilon} \leq q_1 \leq \frac{(2P - T - S + 3\varepsilon)(2P - T - S) + (R + P - T - S)\varepsilon + 3\varepsilon^2}{(R + P - T - S)(2P - T - S) + 3(R + P - T - S)\varepsilon + \varepsilon^2},}
\end{equation}
\begin{equation}
\mfootnotesize{
\frac{2P - T - S + 2\varepsilon}{R + P - T - S + \varepsilon} \leq q_2 \leq 1, \, \text{and} \, 0 \leq q_4 \leq 1.}
\end{equation}
\end{enumerate}

Likewise, we can try to figure out the set of unbending strategies based on Table~\ref{phi_independency_gen}. However, for a general $O$ between $P$ and $R$, it is not always possible to get the explicit expression of $\bm{q}$. For the rest of the analysis, we will focus on Class A, where $q_1 = 1$ and $q_3 = 0$ and Class D, where $q_4 = h_D$. \par

\section{Class A of unbending strategies (continued)}

As before, Class A is under the restriction of $q_1 = 1$ and $q_3 = 0$. The ZD player's payoff $s_X$ is a quadratic rational function of $\chi$ (recall Table~\ref{phi_independency_gen}). We have
\begin{equation}
s_X = \frac{(T - S)(R - O)(a_{A1}\chi + a_{A0})\chi}{f_A(\chi)} + O,
\end{equation}
where
\begin{equation}
\mfootnotesize{
\begin{cases}
a_{A1} = [(P - S)q_2 + (T + S - O - P)]q_4 - (O - P)(1 - q_2), \\
a_{A0} = [(T - P)q_2 - (T + S - O - P)]q_4  + (O - P)(1 - q_2).
\end{cases}}
\end{equation}
Moreover, the denominator $f_A(\chi) = d_{A2}\chi^2 + d_{A1}\chi + d_{A0}$ is a quadratic function of $\chi$ with 
\begin{equation}
\mfootnotesize{
\begin{cases}
d_{A2} = [(T - R)(P - S)q_2  + (T - S)(R - P) - (R - S)(O - P)]q_4 + (T - P)(R - O)(1 - q_2), \\
d_{A1} = \{[T(T - P) - S(P - S) - R(T + S - 2P)]q_2 + (2R - T - S)(O - P)\}q_4 - (T + S - 2P)(R - O)(1 - q_2), \\
d_{A0} = [(T - P)(R - S)q_2 + (O - P)(T - R) - (T - S)(R - P)]q_4 - (R - O)(P - S)(1 - q_2).
\end{cases}}
\end{equation}

If $q_2 = 0$, we have
\begin{equation}
\mfootnotesize{
\begin{aligned}
s_X &= \frac{a_{A0}\chi}{d_{A1}\chi + d_{A0}} + O,  \\
\frac{d s_X}{d\chi} &= \frac{a_{A0}d_{A0}}{(d_{A1}\chi + d_{A0})^2},
\end{aligned}}
\end{equation}
where
\begin{equation}
\mfootnotesize{
\begin{cases}
a_{A0} = (T - S)(R - O)[(T + S - O - P)q_4 - (O - P)], \\
d_{A1} = [(T - S)(R - O) + (T - R)(O - P)]q_4 + (T - P)(R - O), \\
d_{A0} = [(T - S)(R - O) + (R - S)(O - P)]q_4 + (R - O)(P - S).
\end{cases}}
\end{equation}
\par

The set of unbending strategies depends on the relation between $T + S$ and $2P$. We present closed-form analysis in two different cases: $T + S = 2P$ and $T + S < 2P$, whereas for $T+S >2P$ (see Fig.~3a in the main text) we resort to numerical solutions to find unbending strategies. \par

\subsection{Case \rom{2}: \texorpdfstring{$T + S = 2P$}{e}}

If $T + S = 2P$, we have

\begin{equation}
\mfootnotesize{
s_X = \frac{(T - S)(R - O)(a_{A1}\chi + a_{A0})\chi}{f_A(\chi)} + O,}
\end{equation}
where
\begin{equation}
\mfootnotesize{
\begin{cases}
a_{A1} = [(T - S)q_2 - (2O - T - S)]q_4 - (2O - T - S)(1 - q_2), \\
a_{A0} = [(T - S)q_2 + (2O - T - S)]q_4 + (2O - T - S)(1 - q_2),\\
\end{cases}}
\end{equation}
and $f_A(\chi) = d_{A2}\chi^2 + d_{A1}\chi + d_{A0}$ with
\begin{equation}
\mfootnotesize{
\begin{cases}
d_{A2} = [(T - R)(T - S)q_2  + (T - S)(2R - T - S) - (R - S)(2O - T - S)]q_4 + (T - S)(R - O)(1 - q_2),\\
d_{A1} = [(T - S)^2q_2 + (2R - T - S)(2O - T - S)]q_4,\\
d_{A0} = [(T - S)(R - S)q_2 + (2O - T - S)(T - R) - (T - S)(2R - T - S)]q_4 - (T - S)(R - O)(1 - q_2).
\end{cases}}
\end{equation}
\par

Further, we obtain
\begin{equation}
\mfootnotesize{
\frac{d s_X}{d\chi} = \frac{(T - S)(R - O)g_A(\chi)}{f^2_A(\chi)},}
\end{equation}
where $g_A(\chi) = e_2\chi^2 + e_1\chi + e_0$ with $e_i$'s being functions of $q_2$ and $q_4$. We show that $g_A(1) = e_2 + e_1 + e_0 \leq 0$ is the necessary and sufficient condition for $s_X$ to be monotonically decreasing with respect to $\chi$. \par

Routine calculation gives
\begin{equation}
\mfootnotesize{
g_A(1) = e_2 + e_1 + e_0 = 2(T - S)^2q_2q_4\Gamma(q_4),}
\end{equation}
of which the last factor on the right-hand side is
\begin{equation}
\footnotesize
\Gamma(q_4) = 2[(R - S)q_2 - (2R - T - S)]q_4 - (2R - T - S)(1 - q_2).
\end{equation}
\par

If $q_2 = 0$ or $q_4 = 0$ (that is, $q_3 = q_4 = 0$), we immediately have  $ds_X/d\chi < 0$. Otherwise, the sign of $ds_X/d\chi$ is decided by $g_A(\chi)$ and $g_A(1) \leq 0$ is a necessary condition for $s_X$ to be a decreasing function of $\chi$. It suffices to show that $g_A(1) \leq 0$, or equivalently, $\Gamma(q_4) \leq 0$ is also a sufficient condition. \par

Similarly, we let
\begin{equation}
\mfootnotesize{
2e_2 + e_1 = 4(T - S)q_2q_4\gamma(q_4),}
\end{equation}
of which the last factor on the right-hand side is
\begin{equation}
\mfootnotesize{
\gamma(q_4) = [(T - S)(R - S)q_2 - (T - S)(2R - T - S) - (T - R)(2O - T - S)]q_4 - (T - S)(R + O - T - S)(1 - q_2).}
\end{equation}
Notice that
\begin{equation}
\mfootnotesize{
(T - O)\Gamma - \gamma = (R - S)(2O - T - S)(1 - q_2)(1 + q_4) > 0.}
\end{equation}
We get $\gamma < 0$. \par

On the other hand, since
\begin{equation}
\mfootnotesize{
\frac{2(T - O)^2q_2q_4\Gamma - e_2}{(2O - T - S)(1 - q_2)(1 + q_4)} = (\ast)q_4 + (\ast\ast) > 0,}
\end{equation}
where
\begin{equation}
\mfootnotesize{
\begin{cases}
(\ast) = (R - S)[2(T - O) + T - S]q_2 + (T - S)(2R - T - S) - (T - R)(2O - T - S), \\
(\ast\ast) = (T - S)(R - O)(1 - q_2), \\
\end{cases}}
\end{equation}
we have $e_2 < 0$. \par

To conclude, $g_A(1) \leq 0$ always implies $e_2 < 0$ and $2e_2 + e_1 < 0$. As a result, we have $g_A(\chi) < 0$ for $\chi > 1$. The set of $q_2$ and $q_4$ is therefore:
\begin{equation} 
0 \leq q_4 \leq
\begin{cases}
1, & 0 \leq q_2 < \frac{3(2R - T - S)}{4R - T - 3S} \\
\frac{(1 - q_2)(2R - T - S)}{2[(R - S)q_2 - (2R - T - S)]}. & \frac{3(2R - T - S)}{4R - T - 3S} \leq q_2 < 1
\end{cases}
\end{equation}
It is worth pointing out that for a general $O$ between $P$ and $R$, this set of unbending strategies is essentially the same as that for $O = P$ (that is, when player X uses an extortionate ZD strategy). \par

\subsection{Case \rom{3}: \texorpdfstring{$T + S < 2P$}{e}}

For a general IPD game where $T + S < 2P$, we obtain (once again)
\begin{equation}
\mfootnotesize{
ds_X/d\chi = \frac{(T - S)(R - O)g_A(\chi)}{f^2_A(\chi)},}
\end{equation}
where $g_A(\chi) = e_2\chi^2 + e_1\chi + e_0$ with $e_i$'s being functions of $q_2$ and $q_4$. We show that (once again) $g_A(1) = e_2 + e_1 + e_0 \leq 0$ is the necessary and sufficient condition for $s_X$ to be monotonically decreasing with respect to $\chi$. \par

As before, let
\begin{equation}
\mfootnotesize{
g_A(1) = (T - S)^2q_2q_4\Gamma(q_4),}
\end{equation}
of which the last factor on the right-hand side is
\begin{equation}
\mfootnotesize{
\Gamma(q_4) = [(R - S)q_2 - (2R - T - S)]q_4 - (R - P)(1 - q_2).}
\end{equation}
\par

If $q_2 = 0$ or $q_4 = 0$ (that is, $q_3 = q_4 = 0$), we immediately have $ds_X/d\chi < 0$. Otherwise, the sign of $ds_X/d\chi$ is decided by $g_A(\chi)$ and $g_A(1) \leq 0$ is a necessary condition for $s_X$ to decrease. It suffices to show that $g_A(1) \leq 0$, or equivalently, $\Gamma(q_4)$ is also a sufficient condition. \par

Similarly, we let
\begin{equation}
\mfootnotesize{
2e_2 + e_1 = 2(T - S)q_2q_4\gamma(q_4),}
\end{equation}
of which the last factor on the right-hand side is
\begin{equation}
\mfootnotesize{
\gamma(q_4) = [(R - S)(P - S)q_2 - (T - S)(R + P - T - S) - (T - R)(O - P)]q_4 - [(T - S)(R - P) - (R - O)(P - S)](1 - q_2).}
\end{equation}
Notice that
\begin{equation}
\mfootnotesize{
(T - O)\Gamma - \gamma = (R - S)(1 - q_2)[(O + P - T - S)q_4 + O - P] > 0.}
\end{equation}
We get $\gamma < 0$. \par

On the other hand, since
\begin{equation}
\mfootnotesize{
\frac{(T - O)^2q_2q_4\Gamma - e_2}{(1 - q_2)[(O + P - T - S)q_4 + O - P]} = (\ast)q_4 + (\ast\ast) > 0,}
\end{equation}
where
\begin{equation}
\mfootnotesize{
\begin{cases}
(\ast) = (R - S)(T + P - O - S)q_2 + (T - S)(R - P) - (T - R)(O - P), \\
(\ast\ast) = (R - O)(P - S)(1 - q_2),
\end{cases}}
\end{equation}
we have $e_2 < 0$. \par

To conclude, $g_A(1) \leq 0$ always implies $e_2 < 0$ and $2e_2 + e_1 < 0$. As a result, we have $g_A(\chi) < 0$ for $\chi > 1$. The set of $q_2$ and $q_4$ is therefore:
\begin{equation}
\begin{cases}
0 \leq q_2 < 1,\\
0 \leq q_4 \leq h_A(q_2),\\
\end{cases}
\end{equation}
where
\begin{equation}
h_A(q_2) = 
\begin{cases}
1, & 0 \leq q_2 < \frac{(R - P) + (2R - T - S)}{2R - P - S}\\
\frac{(R - P)(1 - q_2)}{(R - S)q_2 - (2R - T - S)}. & \frac{(R - P) + (2R - T - S)}{2R - P - S} \leq q_2 < 1
\end{cases}
\end{equation}
Once again, for a general $O$ between $P$ and $R$, this set of unbending strategies is essentially the same as that for $O = P$ (that is, when player X uses an extortionate ZD strategy). \par

\section{Class D of unbending strategies (continued)}

The last class requires 
\begin{equation}
\mfootnotesize{
q_4 = h_D(q_1, q_2, q_3) = \frac{a_{D00}}{2R - T - S},}
\end{equation}
under which we obtain
\begin{equation}
\begin{dcases}
s_X = \frac{(T - S)a_{D0}\chi}{f_D(\chi)} + P,\\
\frac{\partial s_X}{\partial O} = \frac{-d_{D0}(\chi - 1)}{f_D(\chi)}, \\
\frac{\partial s_X}{\partial \chi} = \frac{(T - S)a_{D0}d_{D0}}{f_D^2(\chi)}.
\end{dcases}
\end{equation}
Here, $f_D(\chi) = d_{D1}\chi + d_{D0}$ and 
\begin{equation}
\begin{cases}
a_{D00} = -(T + S - 2P)q_1 + (R - P)(q_2 + q_3) + T + S - R - P, \\
a_{D0} = -(T + S - 2O)q_1 + (R - O)(q_2 + q_3) + T + S - R - O, \\
d_{D1} = -(T - S)q_1 + (T - R)q_2 + (R - S)q_3 + R - S, \\
d_{D0} = -(T - S)q_1 + (R - S)q_2 + (T - R)q_3 + T - R.
\end{cases}
\end{equation}
\par

To determine the monotonicity of $s_X$, we need to figure out the sign of $a_{D0}d_{D0}$. For the set of unbending strategies, we need $a_{D0}d_{D0} < 0$. That is, $a_{D0} < 0 < d_{D0}$ or $d_{D0} < 0 < a_{D0}$. Combined with the fact that $0 \leq q_4 \leq 1$, or equivalently, $0 \leq a_{D00} \leq 2R - T - S$, the set of unbending strategies is 
\begin{equation}
a_{D0} < 0 < d_{D0} \qquad \text{or} \qquad d_{D0} < 0 < a_{D0} \qquad \text{and} \qquad 0 \leq a_{D00} \leq 2R - T - S.
\end{equation}
We are more interested in the subset (which we consider as Class D)
\begin{equation}
d_{D0} < 0 < a_{D0} \qquad \text{and} \qquad 0 \leq a_{D00} \leq 2R - T - S,
\end{equation}
which can be simplified as $d_{D0} < 0 < a_{D0}$.  The proof is given below. \par

On the one hand, it can be shown that $a_{D00} \geq a_{D0}$. Since $a_{D00} - a_{D0} = (O - P)(q_2 + q_3 + 1 - 2q_1)$, if there exist such a point $(p_1, p_2, p_3)$ that $a_{D0} > a_{D00}$, then $q_1 > (q_2 + q_3 + 1)/2$. On the other hand, $a_0$ evaluated at the point is $(2R - T - S)(q_2 + q_3 - 1)/2$. As $a_0 > 0$, we get $q_2 + q_3 > 1$ and hence $q_1 > 1$, a contradiction. Therefore, $a_{D00} \geq a_{D0}$ always holds and the plane $a_{D00} = 0$ is never above the plane $a_{D0} = 0$ inside the unit cube. \par

On the other hand, we have shown previously that if $d_{D0} < 0 < a_{D00}$, we always have $a_{D00} \leq 2R - T - S$. Therefore, for Class D of unbending strategies, the implicit condition for $s_X$ to be a decreasing function with respect to $\chi$ is $d_{D0} < 0 < a_{D0}$. \par

Again, we point out that Class D is actually the set of general ZD strategies with baseline payoff $O'$ satisfying $P < O' \leq R$.  For a general IPD game, assume that player X and player Y use two ZD strategies with baseline payoffs $O$ and $O'$, respectively.  A conclusion can be drawn intuitively.
\begin{enumerate}[label = (\roman*)]
\item If $O' > O$, it is better off for player X to increase O and decrease $\chi$.
\item If $O' < O$, it is better off for player X to increase O and increase $\chi$.
\end{enumerate}

%\nocite{*}
%\bibliographystyle{naturemag}
%\bibliography{ref}